\documentclass[a4paper,fleqn,usenatbib]{mnras}

\usepackage{multicol}
\usepackage[T1]{fontenc}
\usepackage{ae,aecompl}

\usepackage{graphicx}	
\usepackage{amsmath}	
\usepackage{amssymb}	

\title[The strength of flickering in CVs]
{A comparative study of the strength of flickering in cataclysmic 
variables}

\author[A. Bruch]{Albert Bruch
\\
Laborat\'orio Nacional de Astrof\'{\i}sica, Rua Estados Unidos, 154,
CEP 37500-364, Itajub\'a, MG, Brazil
}

\date{Accepted XXX. Received YYY; in original form ZZZ}

\pubyear{2020}

\newcommand{\hochpunkt}[1]{\mbox{$^{\raisebox{.3ex}{\scriptsize #1}}_{\raisebox{.6ex}{\hspace{.17em}.}}$}}
\begin{document}
\label{firstpage}
\pagerange{\pageref{firstpage}--\pageref{lastpage}}
\maketitle

\begin{abstract}
Flickering is a universal phenomenon in accreting astronomical systems which
still defies detailed physical understanding. It is particularly evident
in cataclysmic variables (CVs). Attempting to define boundary conditions 
for models, the strength of the flickering is measured in several thousand 
light curves of more than 100 CVs. The flickering amplitude is parameterized
by the FWHM of a Gaussian fit to the magnitude distribution of data points 
in a light curve. This quantity requires several corrections before a 
comparison between
different sources can be made. While no correlations of the flickering
strength with simple parameters such as component masses, orbital inclination 
or period were detected, a dependence  on the 
absolute magnitude of the primary component and on the CV subtype is found.
In particular, flickering in VY~Scl tpye novalike variables is systematically
stronger than in UX~UMa type novalikes. The broadband spectrum of the 
flickering light source can be fit by simple models  
but shows excess in the $U$ band. When the data permitted to investigate the
flickering strength as a function of orbital phase in eclipsing CVs, such a 
dependence was found, but it is different for different systems. Surprisingly,
there are also indications for variations of the flickering strength with
the superhump phase in novalike variables with permanent 
superhumps. In dwarf novae, the flickering amplitude is high 
during quiescence, drops quickly at an intermediate magnitude when the
system enters into (or returns from) an outburst and, on average, remains 
constant above a given brightness threshold.
\end{abstract}

\begin{keywords}
stars: activity -- {\it (stars:)} binaries: close --
{\it (stars:)} novae, cataclysmic variables 

\end{keywords}

\paragraph*{Remark on the text}
{\it The main body of this paper is identical to MNRAS 503, 953 (2021). However,
I present here extended versions of Appendices A and B. Appendix D is not
part of the MNRAS article.}

\section{Introduction}
\label{Introduction}

Flickering is a phenomenon normally associated to the accretion of mass
onto a central object. It occurs to a more or less obvious degree in
astronomical systems as diverse as Active Galactic Nuclei 
\citep[][and references therein]{Garcia99}, certain stages of star formation
such as T~Tau stars \citep[][and references therein]{Herbst99, Kenyon00}, 
X-ray binaries \citep{vanderKlis04} or some (but not all) symbiotic stars
\citep{Dobrzycka96, Sokoloski01}.
In the optical range flickering is by far most conspicuous in cataclysmic 
variables (CVs) which thus provide the most suitable test bed for studies of the
properties and the origin of this phenomenon.

CVs are close binary stars where a late type companion -- the secondary --
which is in most cases on or close to the main sequence 
fills its Roche lobe and 
transfers matter to a white dwarf primary. In general, conservation of 
angular momentum forces this material to form an accretion disk around 
the white dwarf where viscosity causes it to slowly move inwards and
to finally settle on the surface of the compact star. In optical light
this disk is almost always the most luminous part of the system. If, however,
the white dwarf possesses a strong magnetic field, the disk may be
truncated in its inner part or its formation may be inhibited altogether, 
and the matter is guided along the field lines to regions close to the 
magnetic poles of the white dwarf. For a comprehensive description of most 
aspects of CVs and their classification into subgroups, see, e.g., 
\citet{Warner95}. 

\begin{figure}
	\includegraphics[width=\columnwidth]{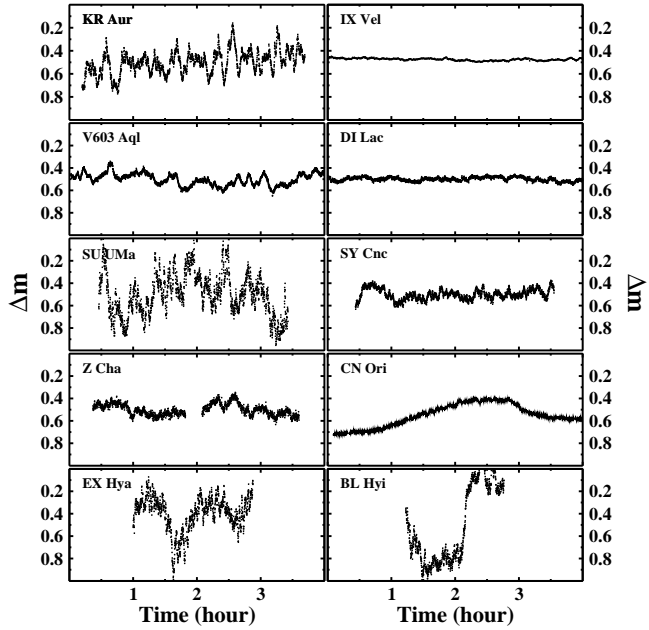}
      \caption[]{Examples of flickering light curves, all drawn on the same
                 time and magnitude scale in order to illustrate their
                 diversity. Each row contains two examples of (from top to
                 bottom) novalike variables, old novae, dwarf novae in
                 quiescence and outburst, and magnetic CVs.}
\label{example-lc}
\end{figure}

Flickering manifests itself most obviously as a continuous series of apparently
stochastically distributed overlapping flares which, in CVs, occur on time
scales of the order of minutes (often superposed upon variations on longer
time scales and of different origin). Their amplitudes depend heavily on 
the individual system and its momentary photometric state, and can
range from a few millimagnitudes to more than an entire magnitude. A few
examples, giving an impression of its diversity, are shown in 
Fig.~\ref{example-lc}, drawn all on the same time and magnitude scale. From
top to bottom each row shows two light curves of novalike variables, old
novae, dwarf novae in quiescence and in outburst, and magnetic CVs [an 
intermediate polar (left) and a AM~Her star (right)]. Many more examples
are shown in \ref{atlas} which contains an atlas of flickering light curves
\input epsf

Following the pioneering study of \citet{Bruch92} an increasing number of
papers dealing with phenomenological aspects of flickering have been
published.
An overview of the different topics covered by these efforts is given in the 
introduction of \citet{Bruch15} to which the reader is referred for more
details and further references.

All the divers objects which exhibit flickering have in common some kind of
mass accretion onto a central object. Most often, albeit not always, 
it occurs by means of an 
accretion disk around a central star. Models 
aimed to explain the origin of flickering in disk dominated CVs have 
been put forward repeatedly \citep{Kley97, Lyubarskii97, Yonehara97, 
Pavlidou01, Pearson05, Dobrotka12, Scaringi14}. Many of them 
are conceptually similar, interpreting the flickering frequency spectrum in 
terms of the propagation of disturbances within the accretion disk. 
However, the origin and the nature of the supposed 
disturbances are 
not identified, and neither do we have a good insight in
the emission mechanism which leads to the observed flickering. 

Although the efforts of the past decades have widely enlarged 
our knowledge about flickering, a deep understanding has still not been 
achieved. Many of the publications of the past deal 
with the temporal (frequency) behaviour of the 
flickering. While this is an important approach, another dimension
characterizing flickering has by far not as thoroughly been explored, 
i.e., its strength as given by its amplitude, which can 
be drastically different from one system to the next as is evident from 
Fig.~\ref{example-lc}.
An exception is the wavelet study of \citet{Fritz98} 
which constitutes an attempt to indirectly quantify simultaneously
the temporal behaviour of the flickering and its strength by analyzing the
scalegram \citep{Scargle93} properties of light curves. Here, a new approach
is introduced to measure objectively the flickering amplitude in a variety of 
light curves with divers characteristics and to search for systematic 
properties in a large number of CVs.

The present study is based on several
thousand light curves of more than 100 systems of most kinds of cataclysmic
variables and deals exclusively with optical data. Since it is not
trivial to measure the strength of the flickering in an objective manner I
first address this question in 
Sect.~\ref{How to measure the flickering amplitude?} before briefly describing
the vast observational material used in this investigation in 
Sect.~\ref{The observational data}. Results are then presented and analysed in
Sect.~\ref{Results and analysis}, and an assessment is attempted
in Sect.~\ref{Assessment of the results}.
Finally, the main findings of this study are summarized in
Sect.~\ref{Summary}. 

\section{How to measure the flickering amplitude?}
\label{How to measure the flickering amplitude?}

There may be many ways to define what exactly is to be understood under the 
term ``flickering amplitude''. The simplest approach, taking it to be the 
difference between the maximum and the minimum in a flickering light 
curve is not appropriate even in the absence of some disturbing effects 
which will be enumerated below, because it depends strongly on individual
data points which mark the extremes. A more suitable way is an analysis of 
the distribution of data points in a flickering light curve which considers
all data. 

\begin{figure}
	\includegraphics[width=\columnwidth]{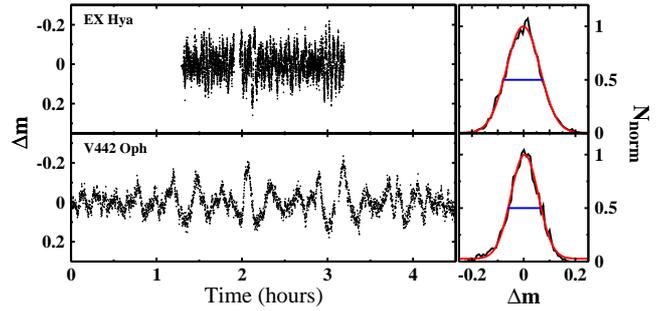}
      \caption[]{{\it Left:} Representative light curves of the intermediate 
                 polar EX~Hya (after subtraction of variations on time scales 
                 of the white dwarf rotation period) and the novalike variable 
                 V442~Oph (after subtraction of variations on orbital time 
                 scales).{\it Right:} Normalized distribution of the magnitude
                 values in the light curves (black) and a Gaussian fit
                 (red). The FWHM of the Gaussian is indicated as a blue
                 horizontal line.}
\label{def-amp}
\end{figure}

As an example, Fig.~\ref{def-amp} shows in the
left frames the observed light curves of the intermediate polar EX~Hya and 
the novalike variable V442~Oph. Variations on 
orbital (V442~Oph) and white dwarf rotation time scales (EX~Hya) have been 
subtracted. The flickering characteristics in the two
systems are markedly different: while EX~Hya exhibits a rapid succession of
flares of similar amplitude and time scales, the flare amplitudes in V442~Oph
encompass a much wider range and their time scales increase with growing
amplitude\footnote{This underlines the importance of the temporal 
dimension of the flickering behaviour -- not to be investigated in this 
study -- in addition to the flickering strength. It should be noted, however, 
that EX~Hya is a rather extreme case, the difference of the temporal 
characteristics are much less expressed in the light curves of the large
majority of CVs. Unlike the bulk of objects regarded here, it is a magenetic
systems in which flickering is likely caused by reprocessing of X-rays 
generated at the footpoint of accretion columns \citep{Semena14}. 
EX~Hya was chosen here in order to illustrate the important point
that adopting as flickering amplitude the simple difference between the 
highest and lowest data points can be misleading.}. 
The right frames of the figure contain the normalized distribution 
function of the magnitude points in the light curves (black) and the best fit 
Gauss function (red). The latter excellently follows the observed distribution,
indicating that the width of the Gaussian is a good and robust measure of
the flickering strength. To be definite, I will regard the FWHM (full width 
at half maximum) of a Gaussian fit to the distribution of magnitude
points in a flickering light curve, i.e., the blue bars in the examples
shown in the figure, as the parameter describing the flickering
amplitude. I will hitherto denote this quantity as $A$, and it is expressed
in magnitudes. It must be stressed that $A$ is, of course, not the true 
difference between the brightest and the faintest magnitude in the light 
curve but rather a proxy. Indeed, $A$ is significantly smaller for V442~Oph
(0.12) than for EX~Hya (0.15), 
although the difference between the highest and lowest data
points in the light curve are about the same in both cases. While $A$ is 
significantly smaller than the true difference, it is a much more robust 
value. 

Simple as the measurement of $A$ might appear at first glance, there are some
pitfalls which introduce errors or uncertainties into the results. Therefore,
some precautions have to be taken which will now be addressed in turn.

\subsection{The time base of the flickering light curve}
\label{The time base of the flickering light curve}

Due to the finite length of an observed light curve one can never be sure 
that the measured amplitude is not only a lower limit to the real amplitude 
since the star may exhibit stronger flares which just were not seen in
the sampled time interval. Similarly, $A$ may be overestimated if by chance
some unusually strong flares happen to fall within the observing window.
This effect can be avoided if only light curves are selected which have a
time base sufficiently long for the flickering within this interval to be
representative.

Extensive tests with artificial light curves simulating flickering along the
lines detailed in section~2 of \citet{Bruch15} show that using a time base
of the order of 1 hour or longer is sufficient to avoid a systematic error
of $A$ due to the currently discussed effect. This is the reason why only 
light curves longer than this limit were chosen for this study 
(see Sect.~\ref{The observational data}).

\subsection{Separating variations unrelated to flickering}
\label{Separating variations unrelated to flickering}

Apart from variations due to flickering, a CV light curve often contains
modulations caused by other effects. This may be the bright spot (or
accretion columns in magnetic systems) rotating into view, superhumps in
SU~UMa stars in superoutburst or in permanent superhumpers, ellipsoidal
variations in systems where the secondary star contributes significantly to
the total light, variations due to transition phases at the onset or the
decline from outbursts or low states, etc. Such effects must be separated from
flickering before its amplitude can be measured.

As long as the time scale of such unrelated modulations are longer than
the flickering time scale, they can be removed by subtracting a low pass
filtered version of a light curve from the original one. A suitable filter
should be able to follow strong and sometimes steep variations such as 
superhumps in SU~UMa type dwarf novae, or variations caused by the rotation 
into and out of view of particular system components, but it should leave 
the flickering flares untouched.

After tests with different filters I found that for the present purpose
a Savitzky-Golay filter \citep{Savitzky64} is most appropriate. It
has basically two free parameters, which is the cutoff time scale $\Delta t$
and the degree of the smoothing polynomial used by the algorithm. The higher 
the degree of the latter, the more details are preserved in the smoothed light
curve. I found that a 4$^{\rm th}$ degree polynomial yields satisfactory 
results. It is thus used throughout. 

\begin{figure}
	\includegraphics[width=\columnwidth]{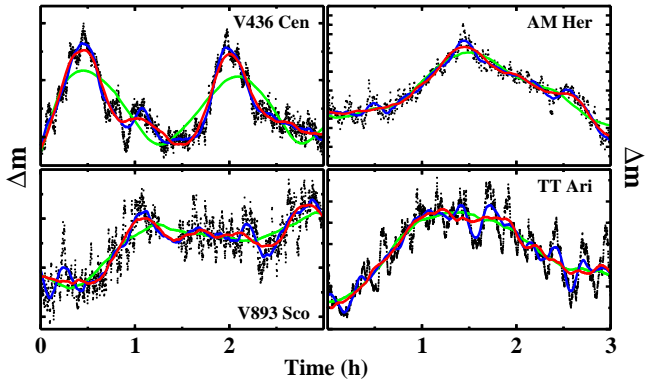}
      \caption[]{Three hour sections of four CVs of
                 different types and photometric states together with
                 their filtered versions, applying a Savitzky-Golay filter
                 with a cut-off time scale of 30~min (blue), 60~min (red) and
                 120~min (green).}
\label{savitzky-golay}
\end{figure}

The performance of the Savitzky-Golay filter is demonstrated on four
examples in Fig.~\ref{savitzky-golay}. The frames show a 3~h
section of the light curves of four CVs of different types and photometric 
states, together with their filtered versions, using $\Delta t = 30$~min (blue),
60~min (red) and 120~min (green). 
The light curve of the SU~UMa type dwarf nova V436~Cen was taken during 
superoutburst and contains two superhumps. The polar AM~Her has a pronounced
hump caused by the varying aspect of the accretion pole on the white dwarf.
The short period eclipsing dwarf nova V893~Sco (eclipses have been removed
from the light curve) shows erratic variations on longer time scales 
superposed on rapid flickering. Finally, the novalike variable TT~Ari
exhibits strong flickering flares on a comparatively long time scale, apart
from a clear modulation due to the permanent superhump observed in this
system. The 30~min cut-off filter (blue) follows the variations occurring on 
longer time scales satisfactorily in most light curves, but in TT~Ari it also 
follows some stronger flickering flares. The 120~min cut-off 
filter (green), while appearing to be satisfactory in the cases of
AM~Her and TT~Ari demonstrates some deficiencies in the case of V893~Sco,
and completely fails to follow the superhumps in V436~Cen. On the other hand
the 60~min cut-off filter (red) appears to be a good compromise. It is able to
satisfactorily follow the variations apparently not caused by flickering,
but leaves the flickering largely untouched. 

The same behaviour is observed in the light curves of other CVs included in
this study. Thus, I will adopt $\Delta t = 60$~min as the standard,
and will henceforth denominate the flickering amplitude as $A_{60}$ in
order to indicate that it only refers to variations on time scales shorter
than 60~min. It must therefore be kept in mind that only
flickering on time scales below this limit will be seen. 
 
\subsection{Data noise correction}
\label{Data noise correction}

Data noise represents a serious problem. It will increase the measured 
amplitude because on the
statistical mean it leads to an apparent increase of the brightness of at 
least some of the data points in the vicinity of a flare maximum beyond its
true value. Similarly, noise will decrease the magnitude of some of the
data points close to the minima between flickering flares below the true
minima. Thus, the apparent total amplitude is overestimated.

Assuming data noise to be distributed normally with a standard deviation
$\sigma_{\rm n}$, a correction is easy. Since flickering can also be 
approximated by a Gaussian distribution, the standard deviation 
$\sigma_{\rm fl}$ of the data points due to flickering and $\sigma_{\rm n}$ add 
in quadrature to the observed standard deviation $\sigma_{\rm o}$. Thus,
the noise-corrected value of $A$ is:
\begin{equation}
\label{Equation noise correction}
A = \kappa \sigma_{\rm fl} = 
    \kappa \sqrt{\sigma_{\rm o}^2 - \sigma_{\rm n}^2}
\end{equation}
where $\kappa = 2 \sqrt{2 \ln 2}$ is the ratio between the FWHM and
the standard deviation of a Gaussian.

The problem is, of course, the determination of the noise level which, in
general, is composed of different components such as Poisson noise, detector
readout noise, atmospheric perturbations, etc. In particular
at low flickering levels its amplitude may be comparable to random
noise and a corresponding correction to $A$ is important. 

For the large majority of light curves used here no direct information about
the noise level is available. However, the well known red noise properties
of flickering offer a way to solve this problem. Since the pioneering studies
of \citet{Elsworth82} it is common knowledge that at high frequencies the power 
spectra of cataclysmic variable light curves are in most cases characterized 
by red noise caused by flickering, meaning that $P \propto f^{-\zeta}$, where 
$P$ is the power, $f$ the frequency and $\zeta$ the spectral index. At very 
high frequencies white noise due to (approximately) Gaussian measurement 
errors takes over. On the double logarithmic scale, red noise causes a 
linear drop of the power with increasing frequency with a slope of $-\zeta$, 
and white noise results in a constant power level. Thus, the power spectrum
of a simulated pure noise light curve, sampled in the same way as the real 
one, and adjusting the noise amplitude such that the power matches the white 
noise part of the power spectrum of the real data, provides the noise 
amplitude $\sigma_{\rm n}$ of the latter.

\begin{figure}
	\includegraphics[width=\columnwidth]{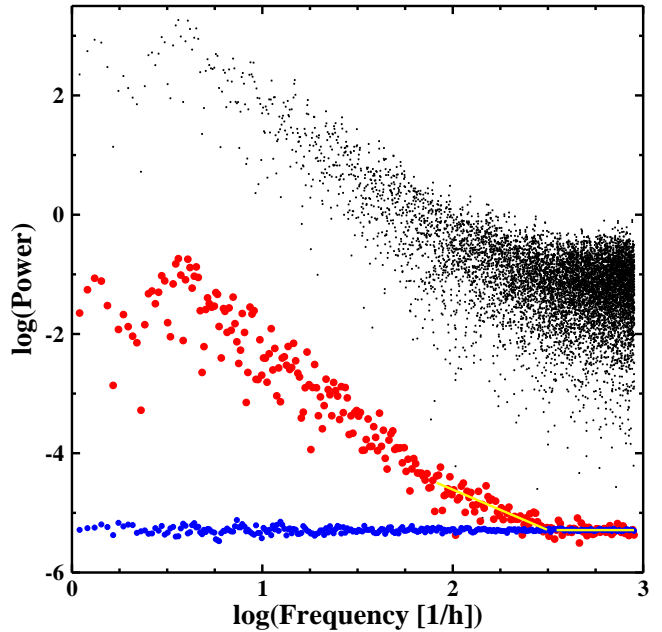}
      \caption[]{Power spectrum (small back dots) on a double
                logarithmic scale of a light curve of KR~Aur
                together with a version of the same data
                binned in constant intervals of $\log(f)$
                (where $f$ is the frequency; red dots, vertically offset
                for clarity). The yellow solid line is a least
                squares fit of a straight line and a constant to the data in
                the high frequency part of the spectrum, where the
                division between the straight line and the constant was
                chosen such that the overall $\chi^2$ is minimized. The
                blue dots represent the average of the power spectra of
                50 light curves of pure Gaussian noise, sampled in the same 
                way as the real light curve, with the standard deviation
                chosen such that the level of the noise power spectrum matches
                that of the constant high
                frequency part of the power spectrum of KR~Aur.} 
\label{noise-method-schematic}
\end{figure}

As an example, Fig.~\ref{noise-method-schematic} shows the power spectrum 
of the KR~Aur light curve (upper left frame of Fig.~\ref{example-lc}) on a
double logarithmic scale as black dots, calculated using a constant (linear)
step width in frequency. It shows the expected red noise -- white noise
properties mentioned above, but is seen to scatter over several orders of 
magnitude in power. Therefore, binning the
data in constant intervals of $\log(f)$ is appropriate. The results are shown 
as red dots (vertically offset from the original power spectrum for clarity). 
While at low frequencies the scatter remains quite large, the higher 
frequency red noise part now becomes more obvious, as does the white noise 
part at very high frequencies. The blue dots represent the average of several
power spectra of light curves containing pure Gaussian noise, sampled in the
same way as the original data. The standard deviation $\sigma_{\rm n}$ of the 
noise has been chosen such that the power spectrum matches the white noise 
part of the KR~Aur power spectrum. $\sigma_{\rm n}$ can thus be identified 
with the noise level in the real data.

For this to work, the power spectra must not be normalized.
Therefore, the often used Lomb-Scargle algorithm \citep{Lomb76, Scargle82}
with the normalization of \citet{Horne86} is not appropriate. Instead, the 
power spectra are calculated following \citet{Deeming75}.

For light curves with a comparatively low time resolution and a low noise
level the Nyquist frequency may be lower than the frequency where white noise
starts to dominate in the power spectra. Then, the level of the high 
frequency end of the power spectra provides only an upper limit for the noise.
Using this limit to correct $A$ (Eq.~\ref{Equation noise correction}) leads
to a systematic underestimation of the flickering amplitude. 
It is found that on average $A_{60}$ derived from different light curves of 
the same object in the same photometric state is reduced by 5\% when an upper
limit instead of the true noise level is adopted. This is much less than 
the natural scatter of $A_{60}$ between light curves. In order to take this
systematic effect into account at least on the statistical mean, if not
individually, a corresponding correction is applied to the final values of 
the flickering amplitude in these cases.

\subsection{Time resolution correction}
\label{Time resolution correction}

The time resolution of a light curve may also influence the measured
flickering amplitude. If the integration time for the data points in a light
curve is not short compared to the time scale of rapid flickering
variations, the finite length of the integration 
will act as a low pass filter, smoothing the light curve, reducing the
apparent magnitude of flickering peaks and diminishing the depth of
the valleys between them. Thus, the apparent amplitude of flickering gets 
smaller. Therefore, in order to 
compare the amplitude measured in light curves taken with different
time resolutions, a correction must be applied in order to reduce the
amplitude to a standard time resolution. I arbitrarily choose this to be
5~s because many of the highest quality light curves used
in this study were sampled with this cadence. 

\begin{figure}
	\includegraphics[width=\columnwidth]{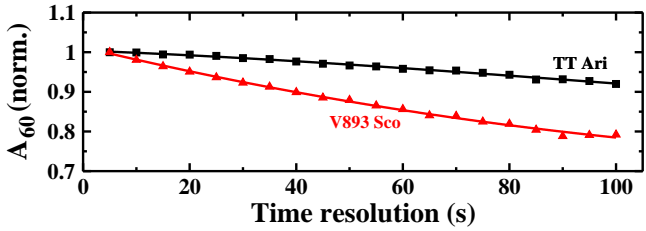}
      \caption[]{$A_{60}$ as a function of time resolution, normalized
                 to $A_{60}$ measured at a time resolution of 5~s, of two 
                 representative light curves of TT~Ari (black squares) and 
                 V893~Sco (red triangles). The
                 solid lines represent cubic polynomial fits to the data.}
\label{time-res-dep}
\end{figure}

The dependence of the measured flickering amplitude on the time resolution
$\tau$ of the investigated light curves is not simply a function of $\tau$ but
depends also on the temporal behaviour of the flickering which can be 
different for different objects. This is demonstrated in
Fig.~\ref{time-res-dep}. Here, the residual light curves (i.e., after
subtraction of their filtered version) of TT~Ari and
V893~Sco (Fig.~\ref{savitzky-golay}) were binned in time intervals $\tau$,
thus degrading their time resolution. $A_{60}$, normalized to its value at 
$\tau = 5$~s, is plotted as black squares (TT~Ari) and red triangles 
(V893~Sco) as a function of $\tau$ in the figure. The solid lines are 
cubic polynomials fitted to the data points. The dependence of $A_{60}$
on $\tau$ is much stronger for V893~Sco because the dominant flickering
variations occur on
shorter time scales than in TT~Ari (see Fig.~\ref{savitzky-golay}).

The reduction of $A_{60}$ to the standard time resolution is performed as
follows: For each object and photometric state a representative
light curve (ideally: long, low noise, high time resolution)
is selected. Binned versions of this light curve are then used to calculate 
$A_{60}$ as a function of $\tau$. This relationship then provides the reduction 
factor $1/A_{\rm 60,norm} (\tau)$ for a light curve with time resolution 
$\tau$ which must be multiplied with the measured flickering amplitude. 

\subsection{Secondary star correction}
\label{Secondary star correction}

Flickering arises in the primary component of a CV, considered here to 
consist of anything within the Roche lobe of the white dwarf. Thus, any
contribution of the secondary star to the total light of the system
dilutes the flickering and the measured amplitude is smaller
that it would be in the absence of the secondary. Therefore, in order to
be able to compare the flickering amplitude in different systems it makes
sense to remove the diluting effect of the secondary if its contribution to
the total light is significant.

In order to reduce $A_{60}$ to the value that it would have in the absence of 
the secondary, let us transform the light of the system from magnitudes into
flux units (on an arbitrary scale). Let $F_{\rm m}$ be the average flux 
observed in a light curve, and $\Delta F_{\rm f}$ the flux difference of the 
flickering light source at two instances in time. Let $\Delta m_{\rm f}$ be
the magnitude difference of the system at these instances. Without loss of
generality we may choose $\Delta m_{\rm f} = A_{60}$. Then,

\begin{equation}
\label{Equation delta m_f}
A_{60} \equiv \Delta m_{\rm f} = 
-2.5 \log \left( \frac{F_{\rm m} - 0.5 \Delta F_{\rm f}}
                      {F_{\rm m} + 0.5 \Delta F_{\rm f}} \right) 
\end{equation}
After some arithmetics this yields
\begin{equation}
\label{Equation F_f}
\Delta F_{\rm f} = 2 \frac{F_{\rm m} \left( 1-X\right)}{1+X}
\end{equation}
where
\begin{equation}
\label{Equation X}
X \equiv 10^{-0.4 \Delta m_{\rm f}} = \frac{F_{\rm m} - 0.5 \Delta F_{\rm f}}
                                     {F_{\rm m} + 0.5 \Delta F_{\rm f}}
\end{equation}

The flickering amplitude in the absence of the secondary star,
$\Delta m_{\rm f,c}$, is given by Eq.~\ref{Equation delta m_f} after 
substitution of $F_{\rm m}$ by $F_{\rm p}$, i.e., the average flux of the 
primary component. Let $F_{\rm s}$ be the flux of the secondary star and $Y$ 
its fractional contribution to the total light 
(i.e., $Y = F_{\rm s}/F_{\rm m} = F_{\rm s}/(F_{\rm p} + F_{\rm s})$). \
Then $F_{\rm p} = (1-Y) F_{\rm m}$. Since the flux unit is arbitrary, 
let $F_{\rm m} \equiv 1$. The corrected flickering amplitude is then

\begin{equation}
\label{Equation delta m_f,c}
A_{\rm 60,c} \equiv \Delta m_{f,c} =  
-2.5 \log \left( \frac{\left( 1-Y \right) - 0.5 \Delta F_{\rm f}}
                      {\left( 1-Y \right) + 0.5 \Delta F_{\rm f}} \right) 
\end{equation}
where $\Delta F_{\rm f}$ can be calculated from Eq.~\ref{Equation X}. 

\begin{figure}
	\includegraphics[width=\columnwidth]{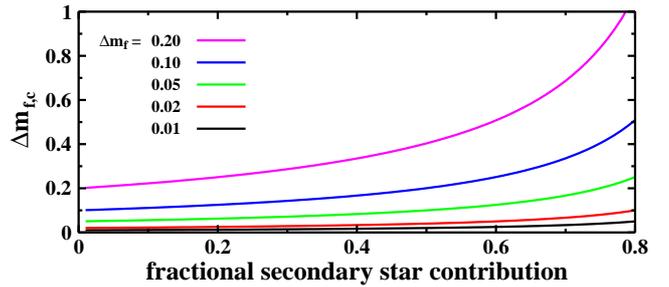}
      \caption[]{Flickering amplitude $\Delta m_{f,c}$ after subtraction of the
                 flux of the secondary star as a function of the fractional
                 contribution of the secondary to the total system light for
                 various values of $\Delta m_f$, i.e., the flickering 
                 amplitude observed in the presence of the secondary star
                 light.}
\label{secon-amp-corr}
\end{figure}

To illustrate the effect, Fig.~\ref{secon-amp-corr} shows $\Delta m_{f,c}$
as a function of the fractional contribution of the secondary star to the
total light of the system for various values of $\Delta m_{\rm f}$.

In most CVs this contribution is negligibly. However, if the orbital period 
exceeds a few hours this may not be true any more. In order to
perform the correction in these cases the literature was searched for
quantitative statements about the contribution of the secondary star. If
available, these refer to a wavelength and a system magnitude which are
normally different from those of the light curves employed in this study. 
Therefore a reduction to the wavelength and the magnitude level of the data 
used here is required.

A corresponding correction for different magnitude levels is straight
forward if the magnitude of the secondary star is assumed to be 
constant. Brightness modulations of the secondary due to 
ellipsoidal variations or illumination effects are neglected. Correcting
for wavelength differences is more complicated because it involves
knowledge or assumptions about the spectra of both, secondary and
primary components. An assessment of the involved uncertainties is
made in Appendix~A.

In principle, just as for the secondary star contribution, a correction for
the light of the white dwarf should also be applied. However, in almost all
cases the white dwarf does not contribute much to the optical light of a CV.
This may be different in very short orbital period systems, but even then
quantitative estimates are rarely available. Therefore, this correction is
not considered here.

\section{The observational data}
\label{The observational data}

The data used in the present study consist of more than 3600 light curves
referring to 107 different CVs. They were collected 
from a variety of sources. As a general rule only light curves extending 
over a time interval of at least 60 minutes and having a time resolution
of better then 100 seconds were selected. Eclipses occurring in the light 
curves of eclipsing systems were masked. Almost all CV subtypes are 
represented. However, the investigated systems do not
include intermediate polars with white dwarf rotation periods of the order
of 10~min because of difficulties to separate flickering from rotation 
induced variations on such time scales. But systems with short rotation
periods (and with low amplitudes of the related variability) such as DQ~Her 
and long periods (where the flickering can be well separated) such as EX~Hya 
are considered. Although many light curves of AE~Aqr are available this system
is disregarded because it exhibits a very peculiar type of flickering
\citep{Bruch91}.

A large quantity of high quality light curves were provided by 
B.\ Warner, collected by himself and his collaborators at SAAO. With few
exceptions they were observed in white light. Their time resolution ranges
from 1~s to 10~s, with the bulk of them observed at 5~s resolution.  Likewise,
many light curves of superb quality were put at the author's disposal by 
R.E.\ Nather. They were
mostly observed at McDonald Observatory and at Wise Observatory and were also
obtained in white light. Their time resolution is similar to that of the
SAAO data. E.\ Robinson provided many light curves observed in the Stiening 
$UBVR*$ photometric system \citep{Horne85}\footnote{The asterisk ($*$) is
used here to distinguish this (uncalibrated) system from the better defined
Johnson $UBVRI$ or Cousins $UBVR_{\rm C}I_{\rm C}$ system.} at McDonald 
Observatory. They consist of simultaneously observed data in four passbands 
and have a very high time resolution between 0.2~s and 2~s.

Numerous other light curves were observed by the author and his collaborators
at the Wise Observatory in the $UBV$ system, using a time resolution of
15.5~s, at the Observatory do Pico dos Dias (OPD) in the $UBVR_{\rm C}I_{\rm C}$ 
system and later in white light, employing time resolutions in general of 
the order of 5~s, and at La Silla in the Walraven $VBLUW$ system 
\citep{Walraven60}. Light curves in the latter system were also provided 
by A.\ Hollander. All observations taken in this system have a time 
resolution of $\approx$21~s. Since the data in the $L$ and $W$ passbands
proofed to be too noisy for the present purpose, they were ignored.

Many more light curves were provided by researchers worldwide or were retrieved
from various data archives, in particular from the OPD data 
bank\footnote{https://www.gov.br/mcti/pt-br/rede-mcti/lna/composicao/coast/obs/opd} 
and from the American Association of 
Variable Star Observers (AAVSO) International 
Database\footnote{https://www.aavso.org}. They refer to
various photometric systems or passbands and encompass a variety of time
resolutions (in most cases not as short as the data described above). A 
substantial part of the data has been published previously, however, almost
always emphasizing aspects not related to flickering. 

In order to compare the white light observations with others obtained in a
specific photometric band, it will be assumed that they are roughly 
equivalent to observations in the $B$ band if they were taken with a 
photoelectric multiplier (i.e., essentially the observations taken by 
B.\ Warner and his group and by R.E.\ Nather in the 1970ies and 1980ies). 
In contrast, white light observations taken with CCD cameras appear to be 
closer to $V$ band observations \citep{Bruch18}. 

In order to assess differences in flickering behaviour depending on the
brightness of a system (i.e., around the outburst cycle of dwarf novae) it
is important to know the average magnitude at the epoch of the light curve. 
This is no particular problem in the case 
of observations in a calibrated photometric system \{for light 
curves in the Walraven system a transformation into Johnson $V$ magnitudes
was performed using 
$V = 6.885 - 2.5[V_{\rm W}+0.030(V-B)_{\rm W}]$
(J.\ Lub, private communication), where the index W indicates values in
Walraven system\}.
In the other cases a rough estimate was obtained from the visual AAVSO long 
term light curves whenever possible, or -- as a final refuge -- from visual
observations of the observer as recorded in the observation logs. Even so,
for a part of the light curves it was not possible to get an estimate of the
brightness of the system at the time of the observations. 

\section{Results and analysis}
\label{Results and analysis}

The flickering amplitude $A_{60}$ was measured in all light curves of the
107 different CVs. The results are summarized in Table~1.
The photometric type and the subtype, adopting the standard nomenclature, are
indicated in the second and third column (SU = SU~UMa type includes the
WZ~Sge stars) of the table. In the case of dwarf novae two photometric states 
are distinguished: O = outburst, including superoutbursts of SU~UMa stars and
Q = quiescence (forth column). If both were observed, the magnitude 
separating the two states is also given, recognizing that the separation is
somewhat arbitrary and that both, outburst and quiescent states, encompass a
range of magnitudes. 
Regarding VY~Scl stars and some magnetic
variables which also exhibit low states, $A_{60}$ always refers to the high
state. The subsequent columns of the table contain the average values of the
flickering amplitude, expressed in magnitudes, observed in the individual 
light curves of the various stars and photometric states for different 
photometric bands. The small differences of 
the effective wavelengths of similar bands in the different photometric 
systems are ignored. 
Numbers in brackets are the standard deviation of $A_{60}$ and the number of 
light curves upon which the average is based.

\begin{table*}
\label{Table: A60}
	\centering
	\caption{Average values of the flickering amplitude $A_{60}$ measured 
                 in the $UBVRI$ bands for 107 CVs. In the case of dwarf novae
                 the photometric states [outburst (O) and quiescence (Q)] are 
                 distinguished and, when both
                 states were observed, the magnitude dividing them is 
                 indicated. The numbers in brackets are the standard
                 deviation of $A_{60}$ derived from different light curves
                 and the number of light curves upon which the average values
                 are based. For more details, see text.}

\begin{tabular}{lllllllll}
\hline
Name     & \multicolumn{2}{l}{Type} & ph.\ state & 
$A_{60} (U)$ & $A_{60} (B)$ & $A_{60} (V)$ & $A_{60} (R)$ & $A_{60} (I)$ \\
\hline
RX And    & DN & Z  & (m<13.0)            & 
0.043 \phantom{(0.000)} (\phantom{0}1)    & 
0.029 (0.003)           (\phantom{0}3)    & 
0.018 (0.013)           (\phantom{0}9)    & 
0.017 (0.001)           (\phantom{0}3)    & 
      \phantom{(0.000)}                   \\
RX And    & DN & Z  & (m>13.0)            & 
      \phantom{(0.000)}  \phantom{0}      & 
      \phantom{(0.000)}  \phantom{0}      & 
0.093 (0.044)           (19)              & 
      \phantom{(0.000)}  \phantom{0}      & 
      \phantom{(0.000)}                   \\
AR And    & DN & UG & (m<14.4)            & 
      \phantom{(0.000)}  \phantom{0}      & 
0.011 \phantom{(0.000)} (\phantom{0}1)    & 
0.009 \phantom{(0.000)} (\phantom{0}1)    & 
      \phantom{(0.000)}  \phantom{0}      & 
      \phantom{(0.000)}                   \\
AR And    & DN & UG & (m>14.4)            & 
      \phantom{(0.000)}  \phantom{0}      & 
      \phantom{(0.000)}  \phantom{0}      & 
0.139 (0.088)           (\phantom{0}3)    & 
      \phantom{(0.000)}  \phantom{0}      & 
      \phantom{(0.000)}                   \\
DX And    & DN & UG &                     & 
      \phantom{(0.000)}  \phantom{0}      & 
      \phantom{(0.000)}  \phantom{0}      & 
0.027 (0.008)           (\phantom{0}2)    & 
      \phantom{(0.000)}  \phantom{0}      & 
      \phantom{(0.000)}                   \\
V455 And  & DN & SU & (m<14.0)            & 
      \phantom{(0.000)}  \phantom{0}      & 
      \phantom{(0.000)}  \phantom{0}      & 
      \phantom{(0.000)}  \phantom{0}      & 
      \phantom{(0.000)}  \phantom{0}      & 
0.066 (0.042)           (3)               \\
V455 And  & DN & SU & (m>14.0)            & 
      \phantom{(0.000)}  \phantom{0}      & 
      \phantom{(0.000)}  \phantom{0}      & 
0.088 (0.027)           (61)              & 
0.073 (0.039)           (\phantom{0}7)    & 
      \phantom{(0.000)}                   \\
V704 And  & NL & VY &                     & 
      \phantom{(0.000)}  \phantom{0}      & 
      \phantom{(0.000)}  \phantom{0}      & 
0.103 (0.064)           (51)              & 
      \phantom{(0.000)}  \phantom{0}      & 
      \phantom{(0.000)}                   \\
UU Aqr    & NL & UX &                     & 
      \phantom{(0.000)}  \phantom{0}      & 
0.151 (0.023)           (16)              & 
0.136 (0.027)           (14)              & 
      \phantom{(0.000)}  \phantom{0}      & 
      \phantom{(0.000)}                   \\
VY Aqr    & DN & UG & (m<14.0)            & 
      \phantom{(0.000)}  \phantom{0}      & 
      \phantom{(0.000)}  \phantom{0}      & 
0.023 (0.012)           (\phantom{0}4)    & 
      \phantom{(0.000)}  \phantom{0}      & 
      \phantom{(0.000)}                   \\
VY Aqr    & DN & UG & (m>14.0)            & 
      \phantom{(0.000)}  \phantom{0}      & 
0.115 \phantom{(0.000)} (\phantom{0}1)    & 
0.067 \phantom{(0.000)} (\phantom{0}1)    & 
      \phantom{(0.000)}  \phantom{0}      & 
      \phantom{(0.000)}                   \\
HL Aqr    & NL & UX &                     & 
0.063 (0.047)           (\phantom{0}5)    & 
0.050 (0.026)           (\phantom{0}4)    & 
0.025 (0.026)           (\phantom{0}2)    & 
      \phantom{(0.000)}  \phantom{0}      & 
      \phantom{(0.000)}                   \\
CZ Aql    & DN &    &                     & 
      \phantom{(0.000)}  \phantom{0}      & 
      \phantom{(0.000)}  \phantom{0}      & 
0.110 (0.024)           (13)              & 
      \phantom{(0.000)}  \phantom{0}      & 
      \phantom{(0.000)}                   \\
DH Aql    & DN & SU &                     & 
      \phantom{(0.000)}  \phantom{0}      & 
      \phantom{(0.000)}  \phantom{0}      & 
0.026 \phantom{(0.000)} (\phantom{0}1)    & 
      \phantom{(0.000)}  \phantom{0}      & 
      \phantom{(0.000)}                   \\
V603 Aql  & N  &    &                     & 
0.080 (0.031)           (18)              & 
0.062 (0.026)           (22)              & 
0.058 (0.025)           (58)              & 
0.063 (0.020)           (\phantom{0}9)    & 
0.085 (0.045)           (4)               \\
V725 Aql  & DN & SU &                     & 
      \phantom{(0.000)}  \phantom{0}      & 
      \phantom{(0.000)}  \phantom{0}      & 
0.031 (0.007)           (\phantom{0}4)    & 
      \phantom{(0.000)}  \phantom{0}      & 
      \phantom{(0.000)}                   \\
V794 Aql  & NL & VY &                     & 
      \phantom{(0.000)}  \phantom{0}      & 
      \phantom{(0.000)}  \phantom{0}      & 
0.076 (0.036)           (\phantom{0}3)    & 
      \phantom{(0.000)}  \phantom{0}      & 
      \phantom{(0.000)}                   \\
V1315 Aql & NL & UX &                     & 
      \phantom{(0.000)}  \phantom{0}      & 
0.075 (0.013)           (\phantom{0}2)    & 
      \phantom{(0.000)}  \phantom{0}      & 
      \phantom{(0.000)}  \phantom{0}      & 
      \phantom{(0.000)}                   \\
AT Ara    & DN & UG &                     & 
      \phantom{(0.000)}  \phantom{0}      & 
      \phantom{(0.000)}  \phantom{0}      & 
0.171 (0.042)           (\phantom{0}3)    & 
0.228 (0.018)           (\phantom{0}3)    & 
      \phantom{(0.000)}                   \\
TT Ari    & NL & VY &                     & 
0.124 (0.044)           (25)              & 
0.090 (0.034)           (36)              & 
0.086 (0.039)           (245)             & 
0.090 (0.039)           (62)              & 
0.086 (0.052)           (5)               \\
T Aur     & N  &    &                     & 
      \phantom{(0.000)}  \phantom{0}      & 
0.054 (0.003)           (\phantom{0}2)    & 
0.042 (0.021)           (\phantom{0}6)    & 
0.071 \phantom{(0.000)} (\phantom{0}1)    & 
      \phantom{(0.000)}                   \\
KR Aur    & NL & VY &                     & 
      \phantom{(0.000)}  \phantom{0}      & 
0.175 (0.027)           (\phantom{0}2)    & 
      \phantom{(0.000)}  \phantom{0}      & 
      \phantom{(0.000)}  \phantom{0}      & 
      \phantom{(0.000)}                   \\
V363 Aur  & NL & UX &                     & 
      \phantom{(0.000)}  \phantom{0}      & 
0.036 (0.003)           (\phantom{0}2)    & 
      \phantom{(0.000)}  \phantom{0}      & 
0.034 \phantom{(0.000)} (\phantom{0}1)    & 
      \phantom{(0.000)}                   \\
SY Cnc    & DN & Z  & (m<12.2)            & 
      \phantom{(0.000)}  \phantom{0}      & 
0.012 (0.006)           (\phantom{0}6)    & 
0.000 (0.000)           (\phantom{0}0)    & 
      \phantom{(0.000)}  \phantom{0}      & 
      \phantom{(0.000)}                   \\
SY Cnc    & DN & Z  & (m>12.2)            & 
      \phantom{(0.000)}  \phantom{0}      & 
0.062 (0.029)           (\phantom{0}4)    & 
0.127 \phantom{(0.000)} (\phantom{0}1)    & 
      \phantom{(0.000)}  \phantom{0}      & 
      \phantom{(0.000)}                   \\
YZ Cnc    & DN & SU & (m<13.3)            & 
      \phantom{(0.000)}  \phantom{0}      & 
0.028 (0.009)           (10)              & 
0.019 (0.012)           (19)              & 
      \phantom{(0.000)}  \phantom{0}      & 
      \phantom{(0.000)}                   \\
YZ Cnc    & DN & SU & (m>13.3)            & 
      \phantom{(0.000)}  \phantom{0}      & 
0.108 (0.058)           (\phantom{0}2)    & 
0.106 (0.038)           (37)              & 
      \phantom{(0.000)}  \phantom{0}      & 
      \phantom{(0.000)}                   \\
AC Cnc    & NL & UX &                     & 
      \phantom{(0.000)}  \phantom{0}      & 
      \phantom{(0.000)}  \phantom{0}      & 
0.098 (0.011)           (\phantom{0}4)    & 
      \phantom{(0.000)}  \phantom{0}      & 
      \phantom{(0.000)}                   \\
OY Car    & DN & SU & (m<14.8)            & 
0.072 (0.312)           (\phantom{0}2)    & 
0.055 (0.012)           (\phantom{0}7)    & 
0.044 (0.011)           (\phantom{0}5)    & 
      \phantom{(0.000)}  \phantom{0}      & 
      \phantom{(0.000)}                   \\
OY Car    & DN & SU & (m>14.8)            & 
      \phantom{(0.000)}  \phantom{0}      & 
0.146 (0.048)           (\phantom{0}5)    & 
0.084 (0.036)           (\phantom{0}7)    & 
      \phantom{(0.000)}  \phantom{0}      & 
      \phantom{(0.000)}                   \\
QU Car    & NL &    &                     & 
0.053 (0.054)           (\phantom{0}2)    & 
0.037 (0.022)           (\phantom{0}2)    & 
0.033 (0.014)           (33)              & 
      \phantom{(0.000)}  \phantom{0}      & 
      \phantom{(0.000)}                   \\
HT Cas    & DN & SU & (m<15.4)            & 
0.092 (0.012)           (\phantom{0}2)    & 
0.051 (0.019)           (\phantom{0}2)    & 
0.035 (0.017)           (27)              & 
      \phantom{(0.000)}  \phantom{0}      & 
      \phantom{(0.000)}                   \\
HT Cas    & DN & SU & (m>15.4)            & 
      \phantom{(0.000)}  \phantom{0}      & 
0.164 (0.040)           (13)              & 
0.126 (0.045)           (45)              & 
0.142 \phantom{(0.000)} (\phantom{0}1)    & 
      \phantom{(0.000)}                   \\
WX Cen    & NL &    &                     & 
      \phantom{(0.000)}  \phantom{0}      & 
      \phantom{(0.000)}  \phantom{0}      & 
0.031 (0.006)           (\phantom{0}9)    & 
      \phantom{(0.000)}  \phantom{0}      & 
      \phantom{(0.000)}                   \\
BV Cen    & DN & UG & (m<12.0)            & 
      \phantom{(0.000)}  \phantom{0}      & 
0.018 (0.003)           (\phantom{0}2)    & 
0.012 (0.004)           (\phantom{0}4)    & 
      \phantom{(0.000)}  \phantom{0}      & 
      \phantom{(0.000)}                   \\
BV Cen    & DN & UG & (m>12.0)            & 
0.236 (0.046)           (\phantom{0}5)    & 
0.165 (0.012)           (\phantom{0}4)    & 
0.124 (0.052)           (\phantom{0}6)    & 
      \phantom{(0.000)}  \phantom{0}      & 
      \phantom{(0.000)}                   \\
MU Cen    & DN & UG &                     & 
      \phantom{(0.000)}  \phantom{0}      & 
      \phantom{(0.000)}  \phantom{0}      & 
0.181 (0.030)           (\phantom{0}6)    & 
      \phantom{(0.000)}  \phantom{0}      & 
      \phantom{(0.000)}                   \\
V436 Cen  & DN & SU & (m<14.0)            & 
      \phantom{(0.000)}  \phantom{0}      & 
0.036 (0.010)           (11)              & 
      \phantom{(0.000)}  \phantom{0}      & 
      \phantom{(0.000)}  \phantom{0}      & 
      \phantom{(0.000)}                   \\
V436 Cen  & DN & SU & (m>14.0)            & 
      \phantom{(0.000)}  \phantom{0}      & 
0.168 (0.068)           (\phantom{0}6)    & 
      \phantom{(0.000)}  \phantom{0}      & 
      \phantom{(0.000)}  \phantom{0}      & 
      \phantom{(0.000)}                   \\
V442 Cen  & DN & UG & (m<13.5)            & 
      \phantom{(0.000)}  \phantom{0}      & 
0.010 (0.004)           (\phantom{0}9)    & 
      \phantom{(0.000)}  \phantom{0}      & 
      \phantom{(0.000)}  \phantom{0}      & 
      \phantom{(0.000)}                   \\
V442 Cen  & DN & UG & (m>13.5)            & 
      \phantom{(0.000)}  \phantom{0}      & 
0.175 (0.046)           (\phantom{0}5)    & 
      \phantom{(0.000)}  \phantom{0}      & 
      \phantom{(0.000)}  \phantom{0}      & 
      \phantom{(0.000)}                   \\
V504 Cen  & NL & VY &                     & 
      \phantom{(0.000)}  \phantom{0}      & 
      \phantom{(0.000)}  \phantom{0}      & 
0.198 (0.036)           (12)              & 
      \phantom{(0.000)}  \phantom{0}      & 
      \phantom{(0.000)}                   \\
V834 Cen  & NL & AM &                     & 
      \phantom{(0.000)}  \phantom{0}      & 
      \phantom{(0.000)}  \phantom{0}      & 
0.160 (0.055)           (10)              & 
      \phantom{(0.000)}  \phantom{0}      & 
      \phantom{(0.000)}                   \\
V1033 Cen & NL & AM &                     & 
      \phantom{(0.000)}  \phantom{0}      & 
      \phantom{(0.000)}  \phantom{0}      & 
0.083 (0.008)           (\phantom{0}3)    & 
      \phantom{(0.000)}  \phantom{0}      & 
      \phantom{(0.000)}                   \\
WW Cet    & DN & Z  & (m<14.5)            & 
0.135 (0.042)           (\phantom{0}3)    & 
0.101 (0.070)           (\phantom{0}5)    & 
0.095 (0.045)           (\phantom{0}4)    & 
0.147 \phantom{(0.000)} (\phantom{0}1)    & 
0.144 \phantom{(0.000)} (1)               \\
WW Cet    & DN & Z  & (m>14.5)            & 
0.135 (0.080)           (\phantom{0}3)    & 
0.289 (0.165)           (\phantom{0}3)    & 
0.207 (0.047)           (\phantom{0}3)    & 
      \phantom{(0.000)}  \phantom{0}      & 
      \phantom{(0.000)}                   \\
WX Cet    & DN & SU &                     & 
      \phantom{(0.000)}  \phantom{0}      & 
0.025 (0.005)           (\phantom{0}5)    & 
      \phantom{(0.000)}  \phantom{0}      & 
      \phantom{(0.000)}  \phantom{0}      & 
      \phantom{(0.000)}                   \\
BO Cet    & NL & UX &                     & 
      \phantom{(0.000)}  \phantom{0}      & 
      \phantom{(0.000)}  \phantom{0}      & 
0.077 (0.018)           (13)              & 
      \phantom{(0.000)}  \phantom{0}      & 
      \phantom{(0.000)}                   \\
Z Cha     & DN & SU & (m<15.0)            & 
      \phantom{(0.000)}  \phantom{0}      & 
0.051 (0.018)           (26)              & 
0.036 (0.020)           (26)              & 
      \phantom{(0.000)}  \phantom{0}      & 
      \phantom{(0.000)}                   \\
Z Cha     & DN & SU & (m>15.0)            & 
      \phantom{(0.000)}  \phantom{0}      & 
0.082 (0.040)           (18)              & 
0.052 (0.020)           (24)              & 
0.140 \phantom{(0.000)} (\phantom{0}1)    & 
      \phantom{(0.000)}                   \\
ST Cha    & DN & SU &                     & 
      \phantom{(0.000)}  \phantom{0}      & 
      \phantom{(0.000)}  \phantom{0}      & 
0.090 \phantom{(0.000)} (\phantom{0}1)    & 
      \phantom{(0.000)}  \phantom{0}      & 
      \phantom{(0.000)}                   \\
TV Col    & NL & IP &                     & 
      \phantom{(0.000)}  \phantom{0}      & 
0.092 (0.034)           (35)              & 
      \phantom{(0.000)}  \phantom{0}      & 
0.064 (0.025)           (\phantom{0}7)    & 
      \phantom{(0.000)}                   \\
T CrB     & RN &    &                     & 
0.126 (0.057)           (12)              & 
      \phantom{(0.000)}  \phantom{0}      & 
      \phantom{(0.000)}  \phantom{0}      & 
      \phantom{(0.000)}  \phantom{0}      & 
      \phantom{(0.000)}                   \\
SS Cyg    & DN & UG & (m<10.4)            & 
0.022 (0.007)           (\phantom{0}3)    & 
0.019 (0.005)           (\phantom{0}3)    & 
0.013 (0.008)           (14)              & 
      \phantom{(0.000)}  \phantom{0}      & 
      \phantom{(0.000)}                   \\
SS Cyg    & DN & UG & (m>10.4)            & 
0.140 (0.046)           (\phantom{0}5)    & 
0.137 (0.073)           (\phantom{0}5)    & 
0.086 (0.042)           (67)              & 
      \phantom{(0.000)}  \phantom{0}      & 
      \phantom{(0.000)}                   \\
EM Cyg    & DN & Z  & (m<13.0)            & 
      \phantom{(0.000)}  \phantom{0}      & 
0.038 (0.018)           (\phantom{0}2)    & 
0.050 (0.000)           (\phantom{0}2)    & 
      \phantom{(0.000)}  \phantom{0}      & 
      \phantom{(0.000)}                   \\
EM Cyg    & DN & Z  & (m>13.0)            & 
0.032 \phantom{(0.000)} (\phantom{0}1)    & 
0.029 \phantom{(0.000)} (\phantom{0}1)    & 
0.105 (0.012)           (\phantom{0}2)    & 
      \phantom{(0.000)}  \phantom{0}      & 
      \phantom{(0.000)}                   \\
V751 Cyg  & NL & VY &                     & 
      \phantom{(0.000)}  \phantom{0}      & 
0.099 (0.004)           (\phantom{0}2)    & 
0.059 (0.031)           (\phantom{0}5)    & 
      \phantom{(0.000)}  \phantom{0}      & 
      \phantom{(0.000)}                   \\
\hline
\end{tabular}

\end{table*} 	

\begin{table*}
	\centering
{\bf Table 1} (continued)
\vspace{1em}

\begin{tabular}{lllllllll}
\hline
Name     & \multicolumn{2}{l}{Type} & ph.\ state & 
$A_{60} (U)$ & $A_{60} (B)$ & $A_{60} (V)$ & $A_{60} (R)$ & $A_{60} (I)$ \\
\hline
HR Del    & N  &    &                     & 
0.031 (0.016)           (16)              & 
0.023 (0.009)           (20)              & 
0.019 (0.018)           (18)              & 
0.020 (0.007)           (\phantom{0}4)    & 
0.020 (0.009)           (2)               \\
DO Dra    & DN & SU & (m<13.0)            & 
      \phantom{(0.000)}  \phantom{0}      & 
      \phantom{(0.000)}  \phantom{0}      & 
0.066 (0.012)           (\phantom{0}4)    & 
0.090 (0.015)           (\phantom{0}2)    & 
      \phantom{(0.000)}                   \\
DO Dra    & DN & SU & (m>13.0)            & 
0.075 (0.042)           (\phantom{0}2)    & 
0.135 (0.053)           (\phantom{0}6)    & 
0.251 (0.243)           (58)              & 
0.147 (0.052)           (\phantom{0}9)    & 
      \phantom{(0.000)}                   \\
AQ Eri    & DN & SU &                     & 
      \phantom{(0.000)}  \phantom{0}      & 
      \phantom{(0.000)}  \phantom{0}      & 
0.031 (0.014)           (\phantom{0}7)    & 
      \phantom{(0.000)}  \phantom{0}      & 
      \phantom{(0.000)}                   \\
KT Eri    & N  &    &                     & 
      \phantom{(0.000)}  \phantom{0}      & 
      \phantom{(0.000)}  \phantom{0}      & 
0.026 (0.001)           (\phantom{0}2)    & 
      \phantom{(0.000)}  \phantom{0}      & 
      \phantom{(0.000)}                   \\
U Gem     & DN & UG & (m<12.5)            & 
      \phantom{(0.000)}  \phantom{0}      & 
0.020 (0.013)           (\phantom{0}6)    & 
0.024 (0.012)           (18)              & 
      \phantom{(0.000)}  \phantom{0}      & 
      \phantom{(0.000)}                   \\
U Gem     & DN & UG & (m>12.5)            & 
      \phantom{(0.000)}  \phantom{0}      & 
0.022 (0.007)           (\phantom{0}6)    & 
0.117 (0.047)           (58)              & 
      \phantom{(0.000)}  \phantom{0}      & 
      \phantom{(0.000)}                   \\
DM Gem    & N  &    &                     & 
      \phantom{(0.000)}  \phantom{0}      & 
0.047 \phantom{(0.000)} (\phantom{0}1)    & 
      \phantom{(0.000)}  \phantom{0}      & 
      \phantom{(0.000)}  \phantom{0}      & 
      \phantom{(0.000)}                   \\
IR Gem    & DN & SU & (m<14.0)            & 
      \phantom{(0.000)}  \phantom{0}      & 
      \phantom{(0.000)}  \phantom{0}      & 
0.018 (0.013)           (\phantom{0}4)    & 
      \phantom{(0.000)}  \phantom{0}      & 
      \phantom{(0.000)}                   \\
IR Gem    & DN & SU & (m>14.0)            & 
      \phantom{(0.000)}  \phantom{0}      & 
0.265 \phantom{(0.000)} (\phantom{0}1)    & 
0.224 (0.130)           (\phantom{0}2)    & 
      \phantom{(0.000)}  \phantom{0}      & 
      \phantom{(0.000)}                   \\
RZ Gru    & NL & UX &                     & 
      \phantom{(0.000)}  \phantom{0}      & 
      \phantom{(0.000)}  \phantom{0}      & 
0.067 (0.011)           (\phantom{0}4)    & 
      \phantom{(0.000)}  \phantom{0}      & 
      \phantom{(0.000)}                   \\
AH Her    & DN & Z  & (m<12.5)            & 
      \phantom{(0.000)}  \phantom{0}      & 
0.015 (0.001)           (\phantom{0}3)    & 
0.022 (0.004)           (\phantom{0}3)    & 
      \phantom{(0.000)}  \phantom{0}      & 
      \phantom{(0.000)}                   \\
AH Her    & DN & Z  & (m>12.5)            & 
      \phantom{(0.000)}  \phantom{0}      & 
0.029 \phantom{(0.000)} (\phantom{0}1)    & 
0.069 (0.028)           (\phantom{0}3)    & 
0.063 \phantom{(0.000)} (\phantom{0}1)    & 
      \phantom{(0.000)}                   \\
AM Her    & NL & AM &                     & 
      \phantom{(0.000)}  \phantom{0}      & 
      \phantom{(0.000)}  \phantom{0}      & 
0.147 (0.050)           (54)              & 
0.146 (0.051)           (\phantom{0}2)    & 
      \phantom{(0.000)}                   \\
DQ Her    & N  & IP &                     & 
      \phantom{(0.000)}  \phantom{0}      & 
      \phantom{(0.000)}  \phantom{0}      & 
0.040 (0.018)           (165)             & 
0.042 \phantom{(0.000)} (\phantom{0}1)    & 
      \phantom{(0.000)}                   \\
V533 Her  & N  & IP &                     & 
0.085 \phantom{(0.000)} (\phantom{0}1)    & 
0.068 \phantom{(0.000)} (\phantom{0}1)    & 
0.083 (0.038)           (13)              & 
0.046 \phantom{(0.000)} (\phantom{0}1)    & 
      \phantom{(0.000)}                   \\
V795 Her  & NL &    &                     & 
0.136 \phantom{(0.000)} (\phantom{0}1)    & 
0.068 (0.015)           (\phantom{0}9)    & 
0.063 (0.027)           (53)              & 
      \phantom{(0.000)}  \phantom{0}      & 
      \phantom{(0.000)}                   \\
EX Hya    & DN & IP &                     & 
0.108 \phantom{(0.000)} (\phantom{0}1)    & 
0.205 (0.026)           (47)              & 
0.198 (0.052)           (59)              & 
      \phantom{(0.000)}  \phantom{0}      & 
      \phantom{(0.000)}                   \\
VW Hyi    & DN & SU & (m<10.9)            & 
      \phantom{(0.000)}  \phantom{0}      & 
0.082 (0.033)           (31)              & 
0.023 (0.002)           (\phantom{0}2)    & 
      \phantom{(0.000)}  \phantom{0}      & 
      \phantom{(0.000)}                   \\
VW Hyi    & DN & SU & (m>10.9)            & 
      \phantom{(0.000)}  \phantom{0}      & 
0.013 (0.008)           (16)              & 
0.053 (0.026)           (23)              & 
      \phantom{(0.000)}  \phantom{0}      & 
      \phantom{(0.000)}                   \\
WX Hyi    & DN & SU & (m<13.6)            & 
0.094 (0.057)           (\phantom{0}5)    & 
0.057 (0.019)           (14)              & 
0.036 (0.018)           (\phantom{0}5)    & 
0.053 \phantom{(0.000)} (\phantom{0}1)    & 
0.073 \phantom{(0.000)} (1)               \\
WX Hyi    & DN & SU & (m>13.6)            & 
0.184 (0.076)           (10)              & 
0.248 (0.117)           (14)              & 
0.191 (0.055)           (\phantom{0}9)    & 
      \phantom{(0.000)}  \phantom{0}      & 
      \phantom{(0.000)}                   \\
BL Hyi    & NL & AM &                     & 
      \phantom{(0.000)}  \phantom{0}      & 
      \phantom{(0.000)}  \phantom{0}      & 
0.201 (0.069)           (10)              & 
      \phantom{(0.000)}  \phantom{0}      & 
      \phantom{(0.000)}                   \\
DI Lac    & N  &    &                     & 
      \phantom{(0.000)}  \phantom{0}      & 
0.022 \phantom{(0.000)} (\phantom{0}1)    & 
      \phantom{(0.000)}  \phantom{0}      & 
      \phantom{(0.000)}  \phantom{0}      & 
      \phantom{(0.000)}                   \\
X Leo     & DN & UG & (m<14.0)            & 
      \phantom{(0.000)}  \phantom{0}      & 
      \phantom{(0.000)}  \phantom{0}      & 
0.025 \phantom{(0.000)} (\phantom{0}1)    & 
      \phantom{(0.000)}  \phantom{0}      & 
      \phantom{(0.000)}                   \\
X Leo     & DN & UG & (m>14.0)            & 
      \phantom{(0.000)}  \phantom{0}      & 
0.075 \phantom{(0.000)} (\phantom{0}1)    & 
      \phantom{(0.000)}  \phantom{0}      & 
      \phantom{(0.000)}  \phantom{0}      & 
      \phantom{(0.000)}                   \\
GW Lib    & DN & SU & (m<12.0)            & 
      \phantom{(0.000)}  \phantom{0}      & 
0.021 \phantom{(0.000)} (\phantom{0}1)    & 
0.017 (0.010)           (15)              & 
      \phantom{(0.000)}  \phantom{0}      & 
0.012 (0.003)           (3)               \\
GW Lib    & DN & SU & (m>12.0)            & 
      \phantom{(0.000)}  \phantom{0}      & 
      \phantom{(0.000)}  \phantom{0}      & 
0.037 (0.025)           (\phantom{0}3)    & 
      \phantom{(0.000)}  \phantom{0}      & 
      \phantom{(0.000)}                   \\
BR Lup    & DN & SU &                     & 
      \phantom{(0.000)}  \phantom{0}      & 
0.046 (0.040)           (\phantom{0}6)    & 
0.024 \phantom{(0.000)} (\phantom{0}1)    & 
      \phantom{(0.000)}  \phantom{0}      & 
      \phantom{(0.000)}                   \\
AY Lyr    & DN & SU & (m<15.0)            & 
      \phantom{(0.000)}  \phantom{0}      & 
0.016 (0.007)           (\phantom{0}3)    & 
0.015 \phantom{(0.000)} (\phantom{0}1)    & 
      \phantom{(0.000)}  \phantom{0}      & 
      \phantom{(0.000)}                   \\
AY Lyr    & DN & SU & (m>15.0)            & 
      \phantom{(0.000)}  \phantom{0}      & 
0.111 \phantom{(0.000)} (\phantom{0}1)    & 
      \phantom{(0.000)}  \phantom{0}      & 
      \phantom{(0.000)}  \phantom{0}      & 
      \phantom{(0.000)}                   \\
MV Lyr    & NL & VY &                     & 
0.216 (0.198)           (\phantom{0}2)    & 
0.086 (0.003)           (\phantom{0}2)    & 
0.089 (0.054)           (56)              & 
      \phantom{(0.000)}  \phantom{0}      & 
      \phantom{(0.000)}                   \\
AQ Men    & NL & UX &                     & 
      \phantom{(0.000)}  \phantom{0}      & 
      \phantom{(0.000)}  \phantom{0}      & 
0.052 (0.020)           (\phantom{0}7)    & 
      \phantom{(0.000)}  \phantom{0}      & 
      \phantom{(0.000)}                   \\
BT Mon    & N  &    &                     & 
      \phantom{(0.000)}  \phantom{0}      & 
0.075 (0.030)           (\phantom{0}4)    & 
0.044 (0.014)           (10)              & 
      \phantom{(0.000)}  \phantom{0}      & 
      \phantom{(0.000)}                   \\
KQ Mon    & NL & UX &                     & 
      \phantom{(0.000)}  \phantom{0}      & 
0.053 (0.008)           (\phantom{0}3)    & 
      \phantom{(0.000)}  \phantom{0}      & 
      \phantom{(0.000)}  \phantom{0}      & 
      \phantom{(0.000)}                   \\
RS Oph    & RN &    &                     & 
0.114 (0.046)           (\phantom{0}8)    & 
0.109 (0.049)           (12)              & 
0.139 (0.055)           (16)              & 
0.269 (0.225)           (\phantom{0}2)    & 
      \phantom{(0.000)}                   \\
V380 Oph  & NL & VY &                     & 
      \phantom{(0.000)}  \phantom{0}      & 
      \phantom{(0.000)}  \phantom{0}      & 
0.141 (0.050)           (\phantom{0}6)    & 
      \phantom{(0.000)}  \phantom{0}      & 
      \phantom{(0.000)}                   \\
V426 Oph  & DN & Z  & (m<11.9)            & 
0.080 (0.025)           (\phantom{0}2)    & 
0.045 (0.001)           (\phantom{0}2)    & 
0.044 (0.013)           (\phantom{0}4)    & 
      \phantom{(0.000)}  \phantom{0}      & 
      \phantom{(0.000)}                   \\
V426 Oph  & DN & Z  & (m>11.9)            & 
0.219 \phantom{(0.000)} (\phantom{0}1)    & 
0.194 \phantom{(0.000)} (\phantom{0}1)    & 
0.140 (0.050)           (24)              & 
      \phantom{(0.000)}  \phantom{0}      & 
      \phantom{(0.000)}                   \\
V442 Oph  & NL & VY &                     & 
      \phantom{(0.000)}  \phantom{0}      & 
0.128 (0.055)           (\phantom{0}8)    & 
0.080 (0.027)           (\phantom{0}6)    & 
      \phantom{(0.000)}  \phantom{0}      & 
      \phantom{(0.000)}                   \\
V841 Oph  & N  &    &                     & 
      \phantom{(0.000)}  \phantom{0}      & 
0.018 (0.008)           (\phantom{0}9)    & 
      \phantom{(0.000)}  \phantom{0}      & 
      \phantom{(0.000)}  \phantom{0}      & 
      \phantom{(0.000)}                   \\
V2051 Oph & DN & SU & (m<14.5)            & 
      \phantom{(0.000)}  \phantom{0}      & 
0.089 (0.034)           (\phantom{0}7)    & 
0.093 (0.031)           (\phantom{0}7)    & 
      \phantom{(0.000)}  \phantom{0}      & 
      \phantom{(0.000)}                   \\
V2051 Oph & DN & SU & (m>14.5)            & 
0.342 (0.047)           (\phantom{0}2)    & 
0.184 (0.087)           (39)              & 
0.234 (0.129)           (\phantom{0}5)    & 
      \phantom{(0.000)}  \phantom{0}      & 
      \phantom{(0.000)}                   \\
CN Ori    & DN & UG & (m<13.8)            & 
      \phantom{(0.000)}  \phantom{0}      & 
0.012 (0.006)           (15)              & 
0.013 (0.006)           (22)              & 
      \phantom{(0.000)}  \phantom{0}      & 
      \phantom{(0.000)}                   \\
CN Ori    & DN & UG & (m>13.8)            & 
      \phantom{(0.000)}  \phantom{0}      & 
0.027 \phantom{(0.000)} (\phantom{0}1)    & 
0.040 (0.018)           (19)              & 
      \phantom{(0.000)}  \phantom{0}      & 
      \phantom{(0.000)}                   \\
CZ Ori    & DN & UG &                     & 
      \phantom{(0.000)}  \phantom{0}      & 
0.037 \phantom{(0.000)} (\phantom{0}1)    & 
      \phantom{(0.000)}  \phantom{0}      & 
      \phantom{(0.000)}  \phantom{0}      & 
      \phantom{(0.000)}                   \\
V1193 Ori & NL & UX &                     & 
      \phantom{(0.000)}  \phantom{0}      & 
      \phantom{(0.000)}  \phantom{0}      & 
0.105 (0.034)           (21)              & 
      \phantom{(0.000)}  \phantom{0}      & 
      \phantom{(0.000)}                   \\
GS Pav    & NL & UX &                     & 
      \phantom{(0.000)}  \phantom{0}      & 
      \phantom{(0.000)}  \phantom{0}      & 
0.032 (0.025)           (\phantom{0}5)    & 
      \phantom{(0.000)}  \phantom{0}      & 
      \phantom{(0.000)}                   \\
V345 Pav  & NL & UX &                     & 
      \phantom{(0.000)}  \phantom{0}      & 
      \phantom{(0.000)}  \phantom{0}      & 
0.049 (0.010)           (\phantom{0}9)    & 
      \phantom{(0.000)}  \phantom{0}      & 
      \phantom{(0.000)}                   \\
IP Peg    & DN & UG & (m<14.3)            & 
      \phantom{(0.000)}  \phantom{0}      & 
0.056 (0.020)           (13)              & 
0.084 (0.054)           (\phantom{0}7)    & 
      \phantom{(0.000)}  \phantom{0}      & 
      \phantom{(0.000)}                   \\
IP Peg    & DN & UG & (m>14.3)            & 
0.323 \phantom{(0.000)} (\phantom{0}1)    & 
0.126 (0.026)           (\phantom{0}4)    & 
0.154 (0.059)           (24)              & 
0.362 (0.219)           (\phantom{0}3)    & 
      \phantom{(0.000)}                   \\
LQ Peg    & NL & VY &                     & 
      \phantom{(0.000)}  \phantom{0}      & 
      \phantom{(0.000)}  \phantom{0}      & 
0.028 (0.009)           (31)              & 
0.022 (0.017)           (\phantom{0}2)    & 
      \phantom{(0.000)}                   \\
TZ Per    & DN & Z  &                     & 
      \phantom{(0.000)}  \phantom{0}      & 
      \phantom{(0.000)}  \phantom{0}      & 
0.013 (0.005)           (\phantom{0}2)    & 
      \phantom{(0.000)}  \phantom{0}      & 
      \phantom{(0.000)}                   \\
GK Per    & N  & UG &                     & 
0.161 (0.055)           (14)              & 
0.095 (0.031)           (17)              & 
0.079 (0.028)           (53)              & 
0.059 (0.022)           (32)              & 
0.095 \phantom{(0.000)} (1)               \\
KT Per    & DN & SU &                     & 
      \phantom{(0.000)}  \phantom{0}      & 
0.018 (0.016)           (\phantom{0}9)    & 
0.028 (0.011)           (\phantom{0}4)    & 
      \phantom{(0.000)}  \phantom{0}      & 
      \phantom{(0.000)}                   \\
RR Pic    & N  &    &                     & 
      \phantom{(0.000)}  \phantom{0}      & 
0.043 (0.010)           (21)              & 
0.030 (0.011)           (15)              & 
      \phantom{(0.000)}  \phantom{0}      & 
      \phantom{(0.000)}                   \\
TY PsA    & DN & SU & (m<14.3)            & 
      \phantom{(0.000)}  \phantom{0}      & 
0.036 (0.010)           (\phantom{0}7)    & 
      \phantom{(0.000)}  \phantom{0}      & 
      \phantom{(0.000)}  \phantom{0}      & 
      \phantom{(0.000)}                   \\
TY PsA    & DN & SU & (m>14.3)            & 
      \phantom{(0.000)}  \phantom{0}      & 
0.151 (0.034)           (\phantom{0}9)    & 
      \phantom{(0.000)}  \phantom{0}      & 
      \phantom{(0.000)}  \phantom{0}      & 
      \phantom{(0.000)}                   \\
VV Pup    & NL & AM &                     & 
      \phantom{(0.000)}  \phantom{0}      & 
0.112 \phantom{(0.000)} (\phantom{0}1)    & 
0.193 (0.107)           (\phantom{0}2)    & 
      \phantom{(0.000)}  \phantom{0}      & 
      \phantom{(0.000)}                   \\
CP Pup    & N  &    &                     & 
      \phantom{(0.000)}  \phantom{0}      & 
0.048 (0.012)           (21)              & 
0.052 (0.011)           (41)              & 
      \phantom{(0.000)}  \phantom{0}      & 
      \phantom{(0.000)}                   \\
V348 Pup  & NL & UX &                     & 
      \phantom{(0.000)}  \phantom{0}      & 
0.088 (0.013)           (\phantom{0}4)    & 
      \phantom{(0.000)}  \phantom{0}      & 
      \phantom{(0.000)}  \phantom{0}      & 
      \phantom{(0.000)}                   \\
\hline
\end{tabular}

\end{table*}

\begin{table*}
	\centering
{\bf Table 1} (continued)
\vspace{1em}

\begin{tabular}{lllllllll}
\hline
Name     & \multicolumn{2}{l}{Type} & ph.\ state & 
$A_{60} (U)$ & $A_{60} (B)$ & $A_{60} (V)$ & $A_{60} (R)$ & $A_{60} (I)$ \\
\hline
T Pyx     & RN &    &                     & 
      \phantom{(0.000)}  \phantom{0}      & 
0.052 (0.008)           (\phantom{0}3)    & 
0.033 (0.016)           (67)              & 
      \phantom{(0.000)}  \phantom{0}      & 
      \phantom{(0.000)}                   \\
WZ Sge    & DN & SU &                     & 
      \phantom{(0.000)}  \phantom{0}      & 
0.056 (0.024)           (33)              & 
      \phantom{(0.000)}  \phantom{0}      & 
      \phantom{(0.000)}  \phantom{0}      & 
      \phantom{(0.000)}                   \\
V3885 Sgr & NL & UX &                     & 
0.019 (0.011)           (\phantom{0}8)    & 
0.011 (0.006)           (26)              & 
0.019 (0.010)           (\phantom{0}7)    & 
0.031 \phantom{(0.000)} (\phantom{0}1)    & 
0.043 \phantom{(0.000)} (1)               \\
V4140 Sgr & DN & SU &                     & 
      \phantom{(0.000)}  \phantom{0}      & 
0.123 (0.064)           (33)              & 
      \phantom{(0.000)}  \phantom{0}      & 
      \phantom{(0.000)}  \phantom{0}      & 
      \phantom{(0.000)}                   \\
V893 Sco  & DN & SU & (m<13.4)            & 
      \phantom{(0.000)}  \phantom{0}      & 
      \phantom{(0.000)}  \phantom{0}      & 
0.068 (0.021)           (\phantom{0}7)    & 
      \phantom{(0.000)}  \phantom{0}      & 
      \phantom{(0.000)}                   \\
V893 Sco  & DN & SU & (m>13.4)            & 
      \phantom{(0.000)}  \phantom{0}      & 
      \phantom{(0.000)}  \phantom{0}      & 
0.157 (0.044)           (58)              & 
0.190 \phantom{(0.000)} (\phantom{0}1)    & 
      \phantom{(0.000)}                   \\
VY Scl    & NL & VY &                     & 
0.118 (0.024)           (\phantom{0}4)    & 
0.080 (0.011)           (\phantom{0}4)    & 
0.075 (0.019)           (\phantom{0}4)    & 
      \phantom{(0.000)}  \phantom{0}      & 
      \phantom{(0.000)}                   \\
VZ Scl    & NL & VY &                     & 
      \phantom{(0.000)}  \phantom{0}      & 
0.107 (0.052)           (12)              & 
0.106 (0.025)           (24)              & 
      \phantom{(0.000)}  \phantom{0}      & 
      \phantom{(0.000)}                   \\
LX Ser    & NL & VY &                     & 
      \phantom{(0.000)}  \phantom{0}      & 
      \phantom{(0.000)}  \phantom{0}      & 
0.082 (0.030)           (50)              & 
      \phantom{(0.000)}  \phantom{0}      & 
      \phantom{(0.000)}                   \\
RW Sex    & NL & UX &                     & 
0.057 (0.024)           (\phantom{0}4)    & 
0.034 (0.017)           (\phantom{0}8)    & 
0.042 (0.021)           (\phantom{0}4)    & 
0.028 \phantom{(0.000)} (\phantom{0}1)    & 
0.062 \phantom{(0.000)} (1)               \\
KK Tel    & DN & SU &                     & 
      \phantom{(0.000)}  \phantom{0}      & 
      \phantom{(0.000)}  \phantom{0}      & 
0.019 \phantom{(0.000)} (\phantom{0}1)    & 
      \phantom{(0.000)}  \phantom{0}      & 
      \phantom{(0.000)}                   \\
RW Tri    & NL & UX &                     & 
0.101 (0.027)           (\phantom{0}3)    & 
0.069 (0.036)           (\phantom{0}8)    & 
0.034 (0.027)           (58)              & 
0.049 (0.008)           (\phantom{0}3)    & 
      \phantom{(0.000)}                   \\
EF Tuc    & DN & UG & (m<13.2)            & 
      \phantom{(0.000)}  \phantom{0}      & 
      \phantom{(0.000)}  \phantom{0}      & 
0.058 (0.017)           (\phantom{0}3)    & 
      \phantom{(0.000)}  \phantom{0}      & 
      \phantom{(0.000)}                   \\
EF Tuc    & DN & UG & (m>13.2)            & 
      \phantom{(0.000)}  \phantom{0}      & 
      \phantom{(0.000)}  \phantom{0}      & 
0.112 (0.024)           (22)              & 
      \phantom{(0.000)}  \phantom{0}      & 
      \phantom{(0.000)}                   \\
SU UMa    & DN & SU & (m<12.7)            & 
      \phantom{(0.000)}  \phantom{0}      & 
0.059 \phantom{(0.000)} (\phantom{0}1)    & 
0.020 (0.012)           (\phantom{0}3)    & 
      \phantom{(0.000)}  \phantom{0}      & 
      \phantom{(0.000)}                   \\
SU UMa    & DN & SU & (m>12.7)            & 
      \phantom{(0.000)}  \phantom{0}      & 
0.186 (0.061)           (\phantom{0}3)    & 
0.105 \phantom{(0.000)} (\phantom{0}1)    & 
      \phantom{(0.000)}  \phantom{0}      & 
0.125 (0.002)           (2)               \\
SW UMa    & DN & SU &                     & 
      \phantom{(0.000)}  \phantom{0}      & 
0.022 (0.006)           (\phantom{0}4)    & 
0.028 (0.015)           (\phantom{0}9)    & 
      \phantom{(0.000)}  \phantom{0}      & 
      \phantom{(0.000)}                   \\
UX UMa    & NL & UX &                     & 
0.046 \phantom{(0.000)} (\phantom{0}1)    & 
0.035 (0.025)           (10)              & 
0.052 (0.023)           (188)             & 
0.036 (0.006)           (\phantom{0}2)    & 
      \phantom{(0.000)}                   \\
IX Vel    & NL & UX &                     & 
0.027 (0.007)           (\phantom{0}5)    & 
0.014 (0.006)           (38)              & 
0.013 (0.003)           (\phantom{0}5)    & 
0.021 \phantom{(0.000)} (\phantom{0}1)    & 
      \phantom{(0.000)}                   \\
HV Vir    & DN & SU &                     & 
      \phantom{(0.000)}  \phantom{0}      & 
0.020 (0.003)           (\phantom{0}2)    & 
0.027 (0.019)           (\phantom{0}6)    & 
0.026 (0.001)           (\phantom{0}2)    & 
      \phantom{(0.000)}                   \\
CTCV 2056-3014 & NL & IP &                     & 
      \phantom{(0.000)}  \phantom{0}      & 
      \phantom{(0.000)}  \phantom{0}      & 
0.255 (0.051)           (\phantom{0}5)    & 
      \phantom{(0.000)}  \phantom{0}      & 
      \phantom{(0.000)}                   \\
EC 21178-5417 & NL & UX &                     & 
      \phantom{(0.000)}  \phantom{0}      & 
      \phantom{(0.000)}  \phantom{0}      & 
0.038 (0.004)           (\phantom{0}5)    & 
      \phantom{(0.000)}  \phantom{0}      & 
      \phantom{(0.000)}                   \\
LS IV -08$^{\rm o}$ 3 & NL & UX &                     & 
      \phantom{(0.000)}  \phantom{0}      & 
      \phantom{(0.000)}  \phantom{0}      & 
0.020 (0.009)           (\phantom{0}9)    & 
      \phantom{(0.000)}  \phantom{0}      & 
      \phantom{(0.000)}                   \\
\hline
\end{tabular}

\end{table*} 	

\subsection{Correlations with system parameters}
\label{Correlations with system parameters}

Starting the analysis of the results summarized in Table~1,
I investigate possible correlations of $A_{60}$ with dynamical, 
geometrical and photometrical system parameters in order to look for
a systematic behaviour of the flickering strength depending on these
quantities. In particular, I study
the dependence of the flickering on the absolute magnitude,
the orbital period, component masses and the orbital inclination. 
Appendix~B contains a table with relevant numerical values.

In brief, the absolute magnitude $M_V$ is calculated from the apparent 
magnitude as
measured at the epoch of observations and the distance of the star as quoted 
by \citet{Bailer-Jones18}, based on its Gaia parallax, considering the 
interstellar extinction quoted in numerous references (Table B1). 
Concerns about systematic errors of the distances 
\citep[see, e.g.,][]{Tappert20} are not an issue because they only apply
if the fractional parallax error is much larger than encountered in the
present sample of stars. Component
masses and system inclinations are also taken from the literature, many
times requiring a possibly subjective choice between contradicting numerical
values cited by different authors. The latter problem does not apply to the
orbital period which in most cases is known to a much higher precision than
necessary in the present context. In order to take into account the 
dependence of the magnitude on the orbital inclination, in disk dominated 
systems $M_V$ has been reduced to a standard inclination of 
56\hochpunkt{o}7 using the correction formula given by \citet{Paczynski80}. 
No such correction has been applied if the inclination is not known. 
This is equivalent to assuming the average inclination of a statistical
sample of binaries.
Moreover, for systems with a significant contribution of the secondary star 
to the total light, a correction yielding the magnitude $M_{V,{\rm pr}}$ of 
only the primary component has been applied.

In no case -- even delimiting the analysis to specific CV subtypes, photometric
states, or combinations thereof -- a trustworthy correlation of $A_{60}$ with
the orbital period, the inclination, the primary star mass or the mass ratio
could be detected. There is, however, a correlation with the absolute
magnitude of the primary component.

\begin{figure}
	\includegraphics[width=\columnwidth]{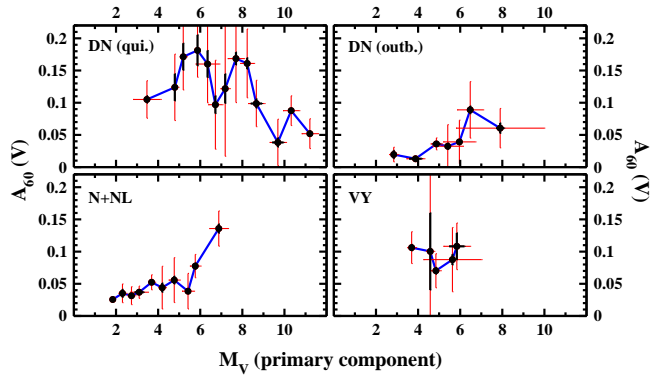}
      \caption[]{The flickering amplitude $A_{60}$ as a function 
                 of the absolute magnitude $M_V$ of the primary component, 
                 first averaged over all light curves of a given system
                 and then binned in intervals of $\Delta M_V = 0.5$ for four
                 different CV populations: dwarf novae in quiescence (upper
                 left) and outburst (upper right), novae plus UX~UMa type
                 novalike variables (lower left) and VY~Scl type stars
                 (lower right). The red error bars are standard deviations
                 calculated based on all individual light curves. The black
                 bars are the standard errors of the means, representing a
                 more meaningful measure of the precision of the ensemble
                 average than the standard deviation.}
\label{amp-mag-dep}
\end{figure}

This is shown in Fig.~\ref{amp-mag-dep}, where $A_{60}$ is shown as a 
function of the absolute magnitude $M_{V,{\rm pr}}$. In order not to give
exaggerated weight to systems with a large number of light curves the
average values of $A_{60}$ were first calculated for each individual system,
and the results were then binned in intervals of $\Delta M_{V,{\rm pr}} = 0.5$.
The red error bars in the figure represent the standard deviation, while the
black bars (often hardly larger than the plot symbols) are the standard errors
of the mean. The latter is a more meaningful measure for the precision of the 
ensemble average compared to the true mean of the population. However, a small
number of systems or light curve 
contributing to the average may lead to a subestimation of the true error of
some points in the graphs.

Four different populations are regarded, all drawn on the same scale in 
Fig.~\ref{amp-mag-dep} in order to facilitate a comparison. Results for
dwarf novae in quiescence are shown in the upper left frame. At the 
faint end, DO~Dra (average values: $M_{V,{\rm pr.}} = 11.6$; 
$A_{60} = 0.67$) has been ignored since the flickering amplitude is 
extremely high and unreliable, possibly because of an overestimation of the
contribution of the secondary star. These data will therefore not be used
henceforth. The upper right frame contains the results for dwarf novae
in outburst. In the lower left frame novae (including recurrent novae) 
and UX~UMa type novalike variables are regarded. Two long period systems 
are excluded as untypical, namely the symbiotic recurrent nova RS~Oph 
\citep[$P = 453.6$~d;][]{Brandi09} 
and the old nova GK~Per
\citep[$P = 1.9968$~d;][]{Morales-Rueda02}. 
The latter also exhibits
dwarf nova outburst, making its classification ambiguous. Both systems
have also the highest absolute magnitude of all stars regarded here.
Novae and UX~UMa type novalike variables
are lumped together because no systemantic differences in their flickering
behaviour could be detected. I include in this group also novalike variables 
which, for lack of sufficient observations, are not firmly established as 
being of the UX~UMa subtype. However, a Kolmogorov-Smirnow tests shows with a 
probability of $\ge$0.994 that these, as well as the confirmed UX~UMa stars 
and the quiescent novae all belong to the same populations when it comes to the 
distribution of $A_{60}$. Finally, the lower right frame of 
Fig.~\ref{amp-mag-dep} contains the results for the VY~Scl type novalike
variables, i.e., those novalikes which in contrast to the UX~UMa stars
are observed to sometimes go into a low state, significantly fainter than
normal.

There is a tendency for a decline of the flickering amplitude of dwarf novae 
in quiescence at very faint magnitudes (but note that statistics are not good 
in this range). Dwarf novae in outburst have a significant trend
to increase
their flickering amplitude with diminishing brightness of the primary 
component. The strong difference of the average level of $A_{60}$ between
quiescence and outburst reflects the long known fact that flickering in
dwarf novae diminishes strongly during outburst. This will be investigated
in more detail in 
Sect.~\ref{Flickering evolution around the dwarf nova outburst cycle}. 
The novae and UX~UMa stars exhibit, at a slightly higher level, 
a similar trend of $A_{60}$ with $M_{V,{\rm pr}}$ as the outbursting
dwarf novae. In contrast, the VY~Scl type novalike variables 
do not show such a tendency
over the limited range in $M_{V,{\rm pr}}$ where they are observed. But it is
striking that their flickering is systematically stronger than in systems
which are not observed to have low states (see also 
Sect.~\ref{Distribution functions}). A systematic difference of the
flickering properties of VY~Scl and UX~UMa type novalike variables was
also noted by \citet{Fritz98} in parameters based on a wavelet analysis.

\subsection{Distribution functions}
\label{Distribution functions}

\begin{figure}
	\includegraphics[width=\columnwidth]{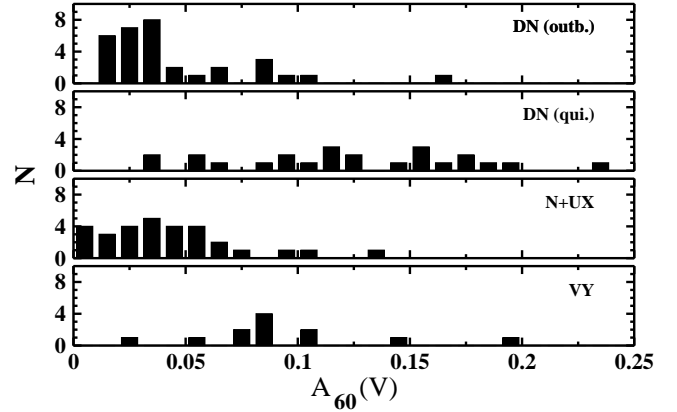}
      \caption[]{Distribution functions of the average values
                 of $A_{60}$ of CVs for
                 various subtypes.}
\label{dist-functions}
\end{figure}

In the previous section the existence of different populations of CV with
respect to the flickering strength already became obvious. This can be
substantiated by investigating the distribution function of $A_{60}$. In
Fig.~\ref{dist-functions} histograms of the average values of $A_{60}$ in
individual systems are shown (from top to bottom) for dwarf novae in outburst 
and in quiescence, novae plus UX~UMa stars, and VY~Scl stars. 

On the statistical
mean the flickering strength is quite similar in outbursting dwarf novae and
in novae and UX~UMa systems on the one side, and quiescent dwarf novae and
VY~Scl stars on the other. Moreover, the distribution functions of the 
latter are much broader than those of the former. This cannot, however, be 
explained assuming that VY~Scl stars are just like quiescent dwarf novae but 
for some reason never experience outbursts because, as Fig.~\ref{amp-mag-dep}
shows, their absolute magnitude is systematically higher than that of the
dwarf novae. This holds even true when limiting this comparison to dwarf novae 
above the period gap (considering that short period dwarf novae are 
intrinsically much fainter than their long period brethren). In this case the
present sample of dwarf novae has an average quiescent magnitude of 
$M_{V,{\rm pr}} = 7.19 \pm 1.23$ versus $M_{V,{\rm pr}} = 5.18 \pm 0.77$ for
the VY~Scl stars. 

\subsection{Flickering evolution around the dwarf nova outburst 
cycle}
\label{Flickering evolution around the dwarf nova outburst cycle}

\begin{figure}
	\includegraphics[width=\columnwidth]{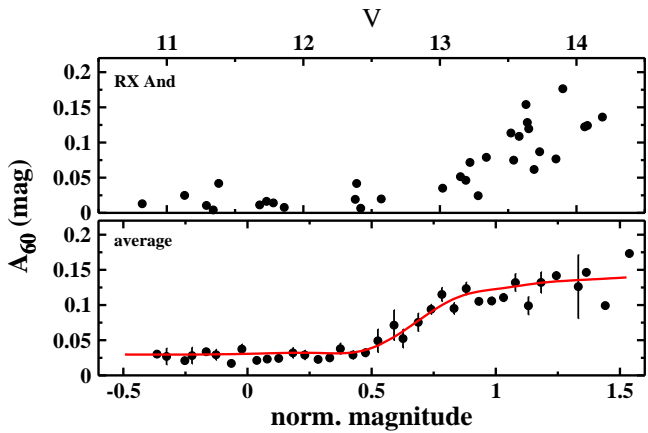}
      \caption[]{{\it Top:} Flickering amplitude of the dwarf nova RX~And as
                 a function of $V$ around the outburst cycle.
                 {\it Bottom:} Average dependence of the flickering amplitude 
                 on the magnitude during the outburst cycle of 26 dwarf novae.
                 The magnitude is normalized such that the mean quiescent 
                 and maximum magnitudes are 1 and 0, respectively (see text
                 for details). The
                 data points are binned in magnitude intervals of width 0.05.
                 The error bars represent the standard error of the mean of all
                 points in each interval. The solid line is a spline fit to
                 the data.}
\label{outburst-cycle}
\end{figure}

In most dwarf novae a tendency is observed in the sense that the flickering 
amplitude decreases when the brightness of the system in the outburst cycle
increases above the quiescent magnitude, but attains a plateau when
it approaches the maximum. An example is shown in the upper frame of
Fig.~\ref{outburst-cycle} where, for RX~And, $A_{60}$ is shown as a function 
of apparent
magnitude in the $V$ band. The same is seen in other dwarf novae, although 
uncertainties in the measurement of $A_{60}$ and an insufficient number of
available light curves makes it often more difficult to recognize this 
tendency. Therefore, it is worthwhile to characterize this behaviour by
averaging the $A_{60} - V$ relation over the ensemble of all dwarf novae.

This is not quite straight forward because the outburst amplitudes differ
greatly from one system to the other, as do, of course, their observed
magnitudes. The latter problem is not resolved by transforming the 
apparent into absolute magnitudes because the systematic brightness
differences as a function of orbital period would still not allow to
determine the average $A_{60} - V$  dependence. In order to overcome
these difficulties I introduce normalized magnitudes $m_{\rm norm}$ such that 
the average magnitudes in quiescence and at outburst maximum are 1 and 0, 
respectively.

This requires the specification of quiescent ($m_{\rm q}$) and maximum 
($m_{\rm m}$) magnitudes for all dwarf novae included in the average. Both 
quantities are not well defined, introducing some uncertainty which, 
however, cannot mask the main findings. In most cases $m_{\rm q}$ and
$m_{\rm m}$ could be estimated with sufficient accuracy from the distribution
of data points in the AAVSO long term light curves. In some cases (in
particular eclipsing systems) the distribution of the out-of-eclipse
magnitudes of the light curves used in this study and information taken
from \citet{Ritter03} was also considered. In the case of SU~UMa stars
$m_{\rm m}$ is taken to be the average maximum magnitude during normal
outbursts.

Having thus transformed the observed into normalized magnitudes the
$A_{60} - V$ relations of all dwarf novae can be averaged. The
result is shown in the lower frame of Fig.~\ref{outburst-cycle} 
where $A_{60}$ is plotted against $m_{\rm norm}$. The data points 
have been binned into intervals of 0.05 in $m_{\rm norm}$. The error bars are 
the standard errors of the mean of all points within an interval. The solid red
line represents a spline fit to the data.

The graph confirms with much more clarity the trend seen in individual systems.
Whenever the dwarf nova is brighter than halfway (in magnitudes)
between quiescence and outburst maximum, $A_{60}$ remains on a constant low 
level. A steep increase occurs between $0.5 < m_{\rm norm} < 0.8$, and at
still fainter magnitudes the amplitude continues to increase at a much lower 
rate (or remains constant on a high level, considering the scatter of the 
data). The exact values of the
break points may be somewhat uncertain in view of the scatter in the data
and the vagaries involved in the determination of the $m_{\rm norm}$. However,
the general shape of the $A_{60}$ -- $m_{\rm norm}$ relation is clear. 

\subsection{The wavelength dependence of the flickering}
\label{The wavelength dependence of the flickering}

\begin{figure}
	\includegraphics[width=\columnwidth]{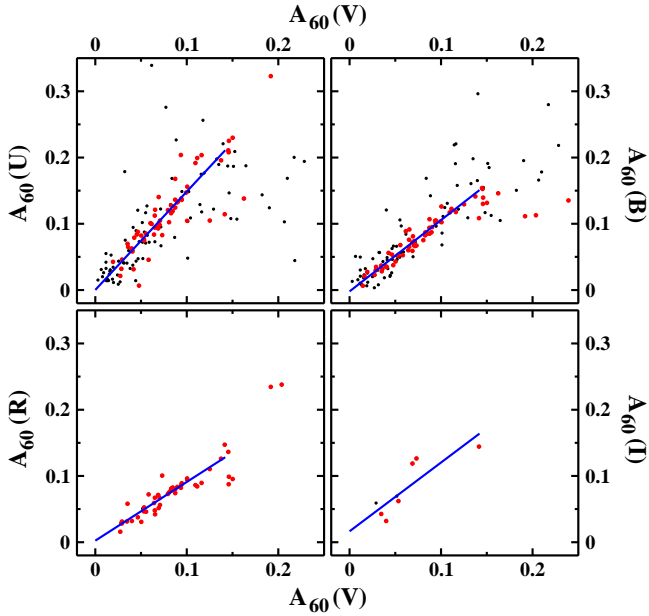}
      \caption[]{Relation between $A_{60}$ measured in $V$ and $A_{60}$
                 measured in other bands ($U$, $B$, $R$ and $I$) of all
                 light curves with simultaneous measurements in different
                 passbands. In contrast to the (smaller) back dots, the red 
                 dots represent light curves of superior quality in which 
                 the measured noise level and consequently the flickering 
                 amplitude no not represent upper limits (see 
                 Sect.~\ref{Data noise correction}). The
                 blue lines are linear least squares fits to the red points
                 with $A_{60} (V) < 0.14$. Note that some data points with
                 high $A_{60}$ fall beyond the limits of the diagrams.}
\label{colors}
\end{figure}

A part of the light curves were observed simultaneously in different passbands.
This opens the path for an assessment of the wavelength dependence of the
flickering amplitude. For this purpose, the
dependence of $A_{60}$ in the $U$, $B$, $R$ and $I$ band is plotted in
Fig.~\ref{colors} as a 
function of $A_{60} (V)$. Red dots refer to light curves for which the noise 
level could be determined, smaller black ones to those where only an upper
limit was measured. The latter are generally based on light curves of inferior 
quality, leading thus to less reliable flickering amplitudes.
Note that some data points representing high flickering amplitudes fall beyond
the limits of the diagrams.  

Up to a certain limit of $A_{60}$ a rather well defined linear relationship
between the amplitude in the $V$ band and in all other bands is observed.
This is particularly so if only the more reliable measurements (red dots)
are regarded. At higher $A_{60}$ the scatter of the data points increases.
A linear fit restricted to the better data (red dots) and to 
$A_{60} (V) < 0.14$ (blue lines in Fig.~\ref{colors}) yields the average 
amplitude ratios $\gamma$ quoted in the second column of 
Table~2. The results confirm the canonical wisdom that 
the flickering increases in strength towards the blue. It may, however, come 
as a surprise that the flickering in the $I$ band is stronger than in $V$ 
and $R$. Although there are only a few data points in $I$ there is no strong 
reason to mistrust their reliability. But, as will be shown below, this does 
not mean that the flickering spectrum rises in the $I$ band.

\begin{table}
\label{Table: gamma beta}
	\centering
	\caption{Ratio of $\gamma$ (the flickering amplitude $A_{60}$),
                 and of $\beta$ (the average flux of a steady state accretion 
                 disk) in different passbands, with respect to the 
                 corresponding values in the $V$ band.} 

\begin{tabular}{lcc}
\hline
%
%

Passband & $\gamma$ & $\beta$ \\
\hline
$U$ & 1.49 $\pm$ 0.12 & 2.66 $\pm$ 0.29  \\ 
$B$ & 1.08 $\pm$ 0.04 & 1.69 $\pm$ 0.10  \\ 
$R$ & 0.89 $\pm$ 0.05 & 0.63 $\pm$ 0.03  \\ 
$I$ & 1.04 $\pm$ 0.25 & 0.36 $\pm$ 0.42  \\ 
\hline
\end{tabular}
\end{table}

In order to estimate the broad band spectrum of the flickering light
source it is assumed that the total flux of a CV can be thought to
be composed of two components (neglecting the contribution of the white
dwarf and remembering that the light of the secondary has already been 
removed when the corresponding correction to the flickering amplitude was 
made). The first one, $F_{\rm c}$, is constant, being emitted by the 
quiet (i.e., non flickering) part of the accretion disk. The second part, 
$F_{\rm fl}$, represents the flickering light source. Since, in general, it 
cannot be assumed that the latter is 100\% modulated (i.e., that at the 
minimum of the flickering activity $F_{\rm fl} = 0$) it can be expressed as 
consisting of a base level, $F_{{\rm fl},0}$, and a part $\Delta F_{\rm fl}$ 
(the latter being that part which is really seen as flickering). Let
$\alpha = \alpha(\lambda)$ describe the spectrum of the flickering light
source, normalized at the wavelength of a reference passband R. Then
\begin{equation}
\label{Equation wavelength dependence of the flickering, Eq. 6}
\alpha = \frac{\beta \left( 10^{0.4 \gamma \Delta {\rm R}} -1 \right)}
              {\Delta F_{\rm fl,R} + F_{{\rm fl,R},0} 
               \left( 1 - 10^{0.4 \Delta {\rm R}} \right)}
\end{equation}
where $\Delta {\rm R}$ is the magnitude difference due to flickering at two
instances of time. Without loss of generality $\Delta {\rm R}$ can be 
identified with $A_{60}$. $\beta$ and $\gamma$ are the spectra of the quiet 
accretion disk and the flickering light source, respectively, normalized at 
the wavelength of R. For the $UBRI$ passbands, $\gamma$ is given in
Table~2. For the derivation of 
Eq.~\ref{Equation wavelength dependence of the flickering, Eq. 6} and
$\beta$, see Appendix~C.

Calculating $\alpha$ for different comparison passbands yields the broad band
spectrum of the flickering light source. However,
Eq.~\ref{Equation wavelength dependence of the flickering, Eq. 6} still 
contains
$F_{{\rm fl,R},0}$ as an unknown quantity. Thus, the flickering spectrum can only
be determined as a function of the not modulated fraction of the flickering
light source. 
Table~3 lists $\alpha$ obtained using
$U$, $B$, $R$, and $I$ as comparison passband for various values of 
$F_{{\rm fl,R},0}$ (expressed in units of the flux of the quiet accretion
disk in the $V$ band). The errors were propagated from the formal errors of 
$\beta$ and $\gamma$.

\begin{table*}
\label{Table: alpha}
	\centering
	\caption{Ratio $\alpha$ (the flux of the flickering light source)
                 in different passbands, with respect to 
                 the $V$ band for different values $F_{{\rm fl,R},0}$ 
                 (in units of the flux of the quiet accretion disk in the $V$ 
                 band) of the unmodulated part of the flickering light source.}

\begin{tabular}{lccccc}
\hline
%
%
%
%
%

Passband               & $F_{{\rm fl,R},0} = 0.00$ & $F_{{\rm fl,R},0} = 0.25$ &
$F_{{\rm fl,R},0} = 0.50$ & $F_{{\rm fl,R},0} = 0.75$ & $F_{{\rm fl,R},0} = 1.00$ \\
\hline
$U$ & 4.04 $\pm$ 0.55 & 4.64 $\pm$ 0.78 & 5.45 $\pm$ 1.12 & 
      6.61 $\pm$ 1.78 & 8.39 $\pm$ 3.13 \\ 
$B$ & 1.82 $\pm$ 0.13 & 1.86 $\pm$ 0.14 & 1.90 $\pm$ 0.15 & 
      1.94 $\pm$ 0.17 & 1.99 $\pm$ 0.19 \\ 
$R$ & 0.57 $\pm$ 0.04 & 0.54 $\pm$ 0.05 & 0.53 $\pm$ 0.05 & 
      0.51 $\pm$ 0.06 & 0.50 $\pm$ 0.06 \\ 
$I$ & 0.38 $\pm$ 0.10 & 0.38 $\pm$ 0.13 & 0.38 $\pm$ 0.15 & 
      0.39 $\pm$ 0.18 & 0.39 $\pm$ 0.21 \\ 
\hline
\end{tabular}
\end{table*}

\begin{figure}
	\includegraphics[width=\columnwidth]{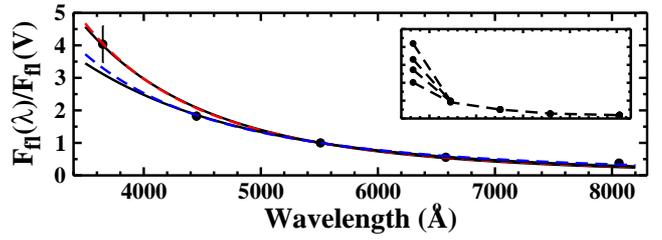}
      \caption[]{Spectrum of the average flickering light source in CVs
                 (black dots), assumed to be fully modulated, compared to 
                 simple models of a power law
                 with a spectral index $a=-2.9$ (blue, dashed) and $a=-3.4$
                 (red, dashed), and
                 to black bodies (black curves) with $T=17\,000$~K 
                 and $T=30\,000$~K (indistinguishable from the red curve).
                 The insert contains the spectrum assuming different
                 degrees of modulation of the flickering light source
                 (from bottom to top: $F_{{\rm fl,R},0} = 0.0, 0.5, 0.75, 1.0$)}
\label{flick-spectrum}
\end{figure}

The main frame of Fig.~\ref{flick-spectrum} shows $\alpha$ (black 
dots) as a function of wavelength for a fully modulated flickering light 
source. Error bars are also drawn, 
but only in the $U$ band they are larger than the plot symbol. 
By definition, in the $V$ band ($\lambda$~ 5510~\AA) $\alpha = 1$. Thus, 
this is the spectrum of the flickering light source. 
The insert of the figure contains the flickering spectra for different
degrees of modulation. Only in the $U$ band the differences with respect to
the fully modulated case surpass the size of the plot symbols. From bottom 
to top $F_{{\rm fl,R},0} = 0.0$, 0.5, 0.75 and 1.0. Concentrating on
$F_{{\rm fl,R},0} = 0$, the spectrum
can be equally well described by a black body (black graph) and 
by a power law of the form $F(\lambda) \propto \lambda^{a}$. The latter is 
shown as a dashed blue graph in the figure for a power law 
index $a=-2.9$. It 
is indistinguishable within the resolution of the figure from the spectrum of
a black body with a temperature of $T=17\,000$~K. 
In both cases only the point representing the $U$ band falls slightly 
above the curves. Using different parameters 
($a=-3.4$ and $T=30\,000$~K) the 
spectra (dashed red and lower black graphs) fit well 
the $U$ point but fall short in $B$.
It is now also seen that the flickering spectrum does not rise in 
the near infrared in spite of $\gamma > 1$ in the $I$ band.

Of course, in the real world a simple black body or power law can well be
an oversimplification for the flickering spectrum. Qualitatively, the
rise being steeper between $B$ and $U$ than can be modelled with one of
these laws (and which becomes more acute if the flickering light source is
not fully modulated), may be explained by an enhanced contribution in $U$ of 
emission from higher Balmer lines and/or continuum emission at wavelengths 
shorter than the Balmer limit.  

\subsection{Relative flux of the flickering light source}
\label{Relative flux of the flickering light source}

The flickering amplitude is a measure of the flux of the flickering light
source relative to the flux of the other components of the system contributing
to the total light. However, since $A_{60}$, taken to be measured at the
wavelength $\lambda_{\rm R}$, does not measure the full amplitude
of the flickering but rather the FWHM of the distribution of data points in
a flickering light curve, it will be explored here how the flux ratio
$F_{\rm fl} (\lambda_{\rm R})/F_{\rm c}(\lambda_{\rm R})$ can be estimated 
from $A_{60}$. Here $F_{\rm fl}$ is taken to be
the average flickering flux in a light curves and $F_{\rm c}$ is the 
non-flickering flux. I will assume the flickering light source to 
be 100\% modulated, leading to a lower limit for 
$F_{\rm fl} (\lambda_{\rm R})/F_{\rm c}(\lambda_{\rm R})$ 
if this is not the case.

\begin{figure}
	\includegraphics[width=\columnwidth]{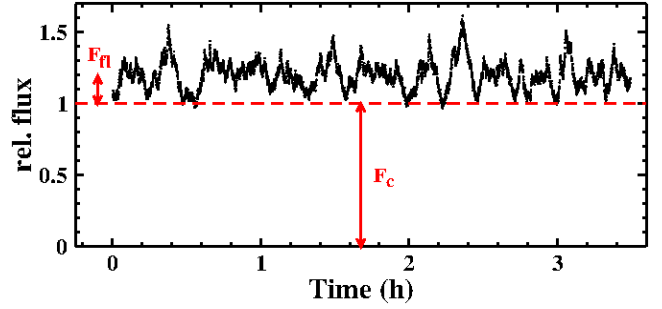}
      \caption[]{Light curve of KR~Aur of 1977, November 14, expressed in
                 fluxes, normalized to the flux at the bottom of the 
                 flickering activity, marked by the horizontal broken
                 line. The arrows indicate the relative flux $F_{\rm c}$
                 (equal to 1 by definition) of the non-flickering light 
                 source and $F_{\rm fl}$ as derived from $A_{60}$. The latter is
                 very similar to the average flux of the flickering light
                 source.}
\label{rel-flux-schema}
\end{figure}

As an example, Fig.~\ref{rel-flux-schema} shows a light curve of KR~Aur
(see Fig.~\ref{example-lc}), transformed from magnitudes into relative 
fluxes. The slight linear trend has 
been subtracted before the transformation. The flux scale has been defined such
that the broken horizontal line, which approximately delimits the lower limit
of the flickering, corresponds to a flux level of 1. Thus, 
$F_{\rm c} (\lambda_{\rm R}) = 1$. The average flux above the level of
the non-flickering light, i.e., the average flux of the flickering light
source, is then $F_{\rm fl} (\lambda_{\rm R}) = 0.120$ in this example, 
leading to a magnitude 
difference between the constant and the average flickering light of
\begin{equation}
\label{Equation relative flux}
\Delta m = -2.5 \log\left( \frac{F_{\rm c} + F_{\rm fl}}{F_{\rm c}} \right) 
         = -0.198
\end{equation}
In the same light curve $A_{60} (\lambda_{\rm Ref})$ was measured to be 0.194.
Thus, substituting $\Delta m$ by $-A_{60}$ in Eq.~\ref{Equation relative flux}
and solving for $F_{\rm fl}$ permits to calculate a satisfactory approximation 
of the relative contribution of the flickering flux to the total light of the
system at the wavelength at which $A_{60}$ was measured. Since no assumptions
about the specific example light curve have been made, this result holds
for all light curves.
Note that it refers to the light of only the primary
component because the contribution of the secondary star is implicitely
taken into account by the corresponding correction to $A_{60}$.

%

Moving from the monochromatic to the polychromatic case requires knowledge
of the spectral energy distribution of both, the non-flickering and the
flickering light source. Here, I draw on the approximations derived in 
Sect.~\ref{The wavelength dependence of the flickering}. Specifically,
I adopt a power law spectrum with index $-2.4$ (from a
fit to the values of $\beta$ in 
Table~2) for the non-flickering, and
a black body spectrum with a temperature of 17\,000~K for the flickering
light source. Scaling both spectra such that their ratio at the wavelength
$\lambda_{\rm R}$ is equal to 
$F_{\rm fl} (\lambda_{\rm R})/F_{\rm c}(\lambda_{\rm R})$, the average of
their ratio over the optical range (which is here taken to be
$3600 \AA \le \lambda \le 8000 \AA$) yields the ratio of the total
optical flux of the flickering light source in units of the flux of the
non-flickering light. 

\begin{figure}
	\includegraphics[width=\columnwidth]{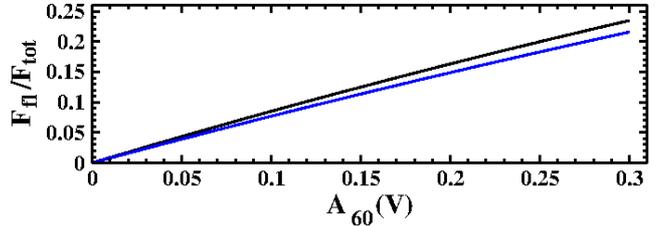}
      \caption[]{Average flux of the flickering light source in units of
                 the total flux of the primary component of a CV
                 as a function of $A_{60}$ measured in the $V$ (black) and
                 $B$ (blue) bands.}
\label{rel-flux}
\end{figure}

Since in the optical range both spectra are quite similar
$F_{\rm fl} (\lambda)/F_{\rm c}(\lambda)$ does not 
change much with $\lambda$. Therefore the ratio of the fluxes integrated
over the entire optical range is similar to the monochromatic flux
ratio. Fig.~\ref{rel-flux} show the contribution of the flickering flux to
the total flux $F_{\rm tot} = F_{\rm fl} + F_{\rm c}$ as a function of $A_{60}$, 
measured in the $V$ (black) and $B$ (blue) band. Thus, for a system with
strong flickering (say, $A_{60}=0.15$), the flickering light source can easily 
contribute on average 15\% of the total flux and twice that during particular 
strong flares. These contributions can even be higher if the flickering
light source is less than 100\% modulated.

Obviously, any physical model for the origin of the flickering must be
able to explain the magnitude of the contribution of the flickering flux
to the total light of a given system. Excepting the paper of
\citet{Dobrotka20} (see Sect.~\ref{Assessment of the results})
I am not aware of a more thorough
assessment of this question in any of the publications which put forward
such models.

\subsection{The phase dependence of the flickering}
\label{The phase dependence of the flickering}

In this section I investigate if the strength of the flickering depends on
the binary phase or, in some cases, the superhump phase. I restrict this
study to a limited number of systems for which high quality, high time 
resolution light curves are available.

In particular, three eclipsing CVs (the novalike variable UX~UMa, and the
dwarf novae U~Gem and V893~Sco in quiescence) are regarded in order to 
verify if the variable aspect of the system around the orbit results in 
changes of the observed flickering activity. Moreover, in two CVs above the 
period gap which exhibit permanent superhumps (V795~Her and TT~Ari), a possible 
dependence of flickering on the superhump phase is studied. Finally the polar 
V834~Cen is regarded where the varying aspect of the accretion site on the 
white dwarf may suggests a change of the observed flickering activity 
depending on the rotational phase of the white dwarf if its source is 
associated to the accretion site on the stellar surface. 

As usual, the long term variations are removed by subtracting a Savitzky-Golay
filtered version from the original data, here using a cut-off time scale of 30
and 60~min for systems below and above the period gap, respectively.
The variance of all data points within a sliding window of width 
$\Delta t$=10~min (below the period gap) or 20~min (above the gap) is then 
calculated\footnote{The variance is used here because the number of data 
points in this window is in general not sufficient to permit a reliable 
determination of the FWHM of a Gaussian fit to the distribution.}, 
using a step width of
1.5~min. After a correction for the average data noise in each light curve
the resulting relationship between variance and time
is folded on the (orbital or superhump) period, yielding the variance vs.\
phase relationship. Combining and binning the result from all examined light 
curves into suitable phase bins of width 0.02 yields the final result 
(hereafter denominated ``variance curve''). Since in the present context
only the phase dependence of the variance matters while the absolute values
are irrelevant, and in order to facilitate a comparison between the considered
systems, the mean of the variance over all phases is first subtracted and it 
is then normalized to its maximum. In the case of the eclipsing 
systems a phase range corresponding to $\Delta t$ before the start of the
eclipse ingress and after the end of egress is ignored in order to
avoid any bias caused by the eclipse. The results are shown in 
Fig~\ref{phase-dep}. The upper frame of each row shows a representative 
phase folded light curve of the considered object as black dots and the
average of all light curves in red (shifted vertically for clarity). The 
lower frame contains the respective variance curves where the error bars 
represent the standard error of the mean in each phase bin. Note that structures
smaller than the width of the sliding window (in all cases of the order of
0.1 in phase) are not independent. In order to verify if the structure of the
variance curves may be dominated by individual light curves with peculiar 
flickering behaviour, the variance curve was re-calculated many times, 
selecting at random only half of the contributing light curves. In all 
cases, the overall shape of the variance curve was recovered.

\subsubsection{The eclipsing systems}
\label{The eclipsing systems}

The left column of Fig.~\ref{phase-dep} contains the results for the 
eclipsing systems. No coherent picture arises, the variance curves being
quite dissimilar for the individual systems.

{\it UX~UMa:} The average out of eclipse light curve is quite flat with at
most a slight indication of a hot spot just before the eclipse.
The variance curve (based on 245 cycles) has a quasi sinusoidal shape 
with a broad maximum centred on
the eclipse and a pronounce minimum at phase 0.5, clearly indicating a non
azimuthally symmetric distribution of the flickering light source which 
appears to be the concentrated on the side of the accretion disk facing the 
secondary star.

{\it U~Gem} is well known to contain a strong orbital hump spanning over almost
half the orbital cycle, as seen here in
both, the individual and the average light curves. Particularly strong 
flickering seen during the hump phase in two light curves by
\citet{Warner71} made these authors to suspect that unstable mass transfer from
the secondary star and, in consequence, variable release of kinetic energy
at the hot spot is the mechanism responsible for flickering; a notion 
that prevailed during the subsequent two decades but which was overcome later. 
Here, the
shape of the variance curve (based on 65 cycles) shows that on average 
flickering does not increase at the hump phase. Instead, it is flat just 
after eclipse, rises quite suddenly to a maximum at phase $\sim$0.4 and
then steadily declines until the start of the subsequent eclipse. This rather 
surprising behaviour is a challenge for any interpretation which I will no 
endeavour to attempt.

{\it V893~Sco} exhibits by far the strongest flickering activity of the three
eclipsing systems regarded here. In spite of considerable variations seen in
individual light curves at any phase, on average, the light curve is flat 
during the first two thirds of the orbital cycle and then shows a hump
similar to, but not as strongly expressed as in U~Gem. The variance curve
(based on 120 cycles) starts high just after eclipse and then declines slowly
but accelerates the decline when the orbital hump becomes visible. As in
UX~UMa there appears to be a dependence on azimuth of the visibility of the
flickering light source, however, in a different sense.

\subsubsection{The permanent superhump systems}
\label{The permanent superhump systems}

\begin{figure}
	\includegraphics[width=\columnwidth]{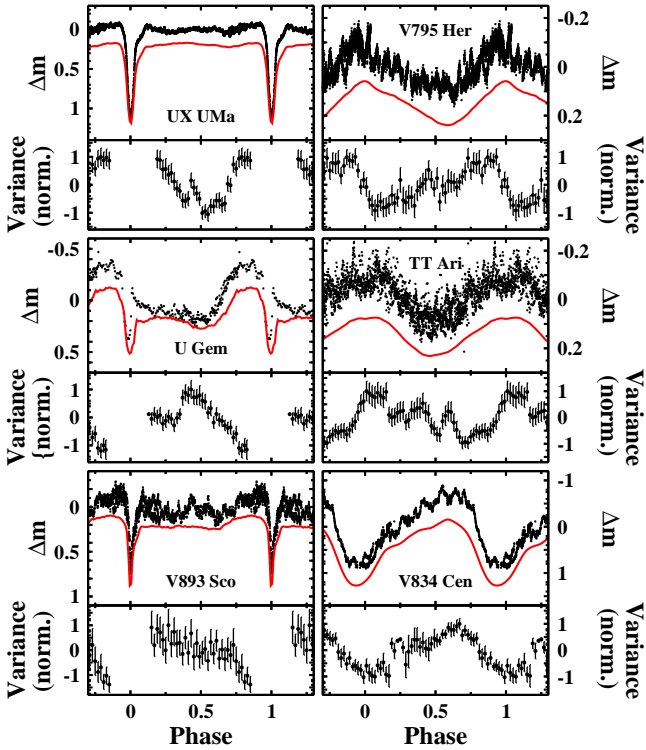}
      \caption[]{Dependence of the flickering activity on the orbital phase
                 in four CVs (left column and lower right frames) and the 
                 superhump phase in two systems above the period gap exhibiting
                 permanent superhumps (right column, top and middle frames). 
                 In each case a representative (black) and the average of all 
                 light curves (red; shifted vertically for clarity) is shown 
                 in the upper frame, while the lower frames contain the 
                 variance curves.}
\label{phase-dep}
\end{figure}

Superhumps in cataclysmic variables are thought to arise as the consequence
of variable tidal stress in the outer parts of an elliptical accretion disk.
The stress and thus the brightness attains a maximum when the elongated part
of the disk extends towards the secondary star. This effect may or may not 
enhance or diminish the flickering activity. This issue is investigated in the
upper and middle frames of the right column of Fig.~\ref{phase-dep}.

{\it V795~Her:}
The superhump properties of V795~Her have been investigated by various authors
\citep[e.g.][]{Kaluzny89, Patterson94, Papadaki06, Simon12}. 37 suitable light
curves (spanning 57 cycles) were used here. They yield a decidedly saw tooth 
shaped average 
superhump waveform with a steep rise, a pointed maximum (arbitrarily chosen 
to represent phase 0) and a more gradual decline. The variance curve does not 
follow the light curve shape. It rather appears as a mirror image of the 
latter, mirrored at the vertical axis. While the maxima approximately 
coincide, the variance drops rapidly after the maximum and then recovers 
slowly. This is just the opposite of the phase dependence of the magnitude. 
There appears to be a secondary maximum centred on phase 0.45, i.e. shortly
before the end of the decline from superhump maximum. Based on a smaller
light curve sample, \citet{Simon12} came to a similar conclusion. As is the 
case here, they see an enhancement during superhump maxima of a parameter 
which measures the relative intensity scatter. They draw particular attention 
to a maximum observed on 2009, April 28 at superhump phase 0.4. This enhances
the credibility that the feature seen here at phase 0.45 is real, although it 
is by no means obvious what is so special at this phase to lead to a 
strengthening of the flickering.

{\it TT~Ari:}
The behaviour of the permanent superhump in this system has been 
studied many times in the past, most recently by \citet{Bruch19} who 
characterized the photometric behaviour of TT~Ari over about 40 years.
The (positive) superhump was always observed, except for a couple of years 
around the turn of the millennium when it was replaced by a negative
superhump and during two VY~Scl type low states. Of the vast quantity of 
available photometry of TT~Ari I selected 184 suitable light curves, 
spanning 247 cycles. The average superhump wave form is very nearly 
sinusoidal. The variance curve has a maximum coinciding with the light
curve maximum. The secondary maxima, in particular the one close to phase
0.6, may not be permanent features because they are not always seen when 
subsets of all light curves are used to calculate the variance curve.

These two examples appear to point at a general correlation between 
the superhump phenomenon and the flickering activity in the sense that 
flickering grows in strength when the superhump attains its maximum. However,
the detailed picture seems to be somewhat more complicated. It may be 
noteworthy that \citet{Bruch96} observed during superoutbursts of the 
SU~UMa type dwarf nova Z~Cha that flickering originates in an extended
region of the accretion disk when the superhump light source is on, but
is restricted to the inner disk (and to an extended range along
the disk rim possibly associated to the hot spot) when it is off.

\subsubsection{The AM~Her stars}
\label{The AM Her stars}

Much of the optical light of polars arises as reprocessed X-rays or as
cyclotron radiation at the bottom of the accretion column close to the
magnetic pole(s) of the white dwarf primary. Thus, if this region is the 
source of flickering in these systems, its strength is expected to be 
correlated with the magnitude which changes as the visibility of the accretion
region changes during the white dwarf rotation cycle.  In polars this is, of
course, equal to the orbital cycle.
  
{\it V834~Cen:}
The data available here only permit the study of the flickering strength
on orbital phase for a single polar, namely V834~Cen. 
Phase resolved light curves of this system are presented by various authors
\citep[e.g.][]{Cropper86, Middleditch91, Imamura00}. In all cases 
flickering is quite strong, the amplitude being higher during the maxima 
than during the minima, as also seen in Fig.~\ref{phase-dep} (lower right
frames). For the present exercise 11 light curves were used. Phases were 
calculated using the ephemeris of \citet{Schwope93}. Thus, phase 0
corresponds to the blue-to-red crossing of the radial velocity of V834~Cen.
The light curves are characterized by a hump with a gradual rise and a 
more rapid decline. The variance curve indeed confirms that flickering is 
stronger during the bright phase of V834~Cen.
The geometry (orbital inclination, co-latitude of accretion region) of 
V834~Cen is such that the accretion column is always in the hemisphere
facing the observer and that at the phase of optical minimum the line
of sight is almost parallel to the magnetic field lines \citep{Bailey83,
Potter04, Costa09}. The dominant light source is cyclotron
radiation from the post shock region, while the light variations are due
to the combined effect of the projected area of the emitting region and
cycloton beaming \citep{Ferrario90}. The variance being maximal during the
phases of best visibility of the dominant light source clearly indicates
that the cyclotron light source is also responsible for the flickering.
Its shape follows rather faithfully the brightness of the system.

\subsection{Long term trends}
\label{Long term trends}

In order to investigate whether the flickering activity suffers systematic
variations over time scales of years a decent coverage with data over a long
time is required. Moreover, short term brightness variations such as outbursts
of dwarf novae and accompanying systematic variations of $A_{60}$ interfere
with any reliable detection of such long term modulations. Thus, only a few
CVs remain for which a respective study is viable. In order to improve
statistics, $A_{60}$ measured in the $V$ and $B$ band are lumped together,
after having applied a correction factor to the $B$ band data in order to
take into account their systematically higher flickering amplitude
(Sect.~\ref{The wavelength dependence of the flickering}). 

\begin{figure}
	\includegraphics[width=\columnwidth]{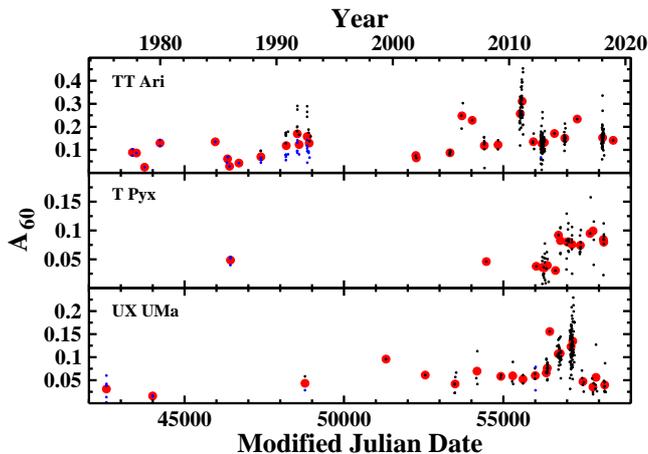}
      \caption[]{Long term evolution of $A_{60}$ for three CVs. Small dots
                 refer to individual light curves [black: $V$ band; blue:
                 $B$ band, divided by the average $A_{60}(B)/A_{60}(V)$ ratio
                 to correct for the systematically higher flickering
                 amplitude in the $B$ band], the big ones are averages
                 within intervals of 100 days.}
\label{longterm}
\end{figure}

In DQ~Her, MV~Lyr, GK~Per, CP~Pup, LX~Ser, RW Tri and IX~Vel no evidence for 
long term variations of the flickering activity was detected. In V603~Aql
and EX~Hya it cannot be excluded that some existing indications for such 
variations are merely the reflection of systematic effects in the data.
They are therefore not considered real. The results for three remaining
system, TT~Ari, T Pyx and UX~UMa, are shown in Fig.~\ref{longterm}. Here,
$A_{60}$ is plotted as a function of time. The small black and blue dots
represent $A_{60}$ as measured in individual light curves in the $V$ and $B$
band, respectively, while the larger red dots are values averaged in time
bins of 100 days. 
 
{\it TT~Ari:} The long term flickering behaviour of TT~Ari was already
discussed by \citet{Bruch19}. An epoch of enhanced activity is observed in 
2010, coincident with the very last phases of the emergence from the 
2009--2010 low state when the system was, on average, still about one
magnitude below its high state magnitude. Note, however, that there is
no linear relationship between $A_{60}$ and magnitude within the
corresponding time interval \citep[see fig.~10 of][]{Bruch19}.

{\it T~Pyx:} With only two exceptions all light curves were
observed after the 2011 outburst of this recurrent nova. $A_{60}$ increases
systematically during this period while the brightness of the system still
declined towards quiescence. Transforming $A_{60}$ from magnitude into flux
units (Sect.~\ref{Flickering evolution around the dwarf nova outburst cycle})
reveals that the flux of the flickering light source did not undergo systematic
changes during this period. Although possibly not statistically significant,
it is interesting to note that the two observations before the outburst yield
values for $A_{60}$ which are similar to those seen just after the outburst
when the magnitude was still $\sim$1~mag brighter than after returning to
quiescent. Considering that the AAVSO long term light curve shows that the
quiescent light of T~Pyx before the 2011 outburst was about 0.7~mag brighter
than afterwards, this means that the flux of the flickering light curve
was about the same long before and just after the outburst.

{\it UX~UMa:} After having remained on an approximately constant level for
many years, the flickering activity of this prototypical novalike variable
started to increase significantly around 2013, culminated in 2015
and then dropped to the previous level one year later. While otherwise the
photometric behaviour of the star remained inconspicuous \citep{Bruch20},
the maximum in 2015 coincided with an unprecedented change: UX~UMa exhibited
a well developed negative superhump \citep{deMiguel16, Bruch20} which was 
not seen in previous or subsequent observing seasons. Since the average
out-of-eclipse brightness of the system did not change appreciably the
rise of $A_{60}$ really reflects an increase of the flux of the flickering
light source. While it is close at hand to associate the high values of 
$A_{60}$ in 2015 to the emergence of the negative superhump, it is not as
obvious why the rise already started in previous years without a change of
the overall photometric behaviour of UX~UMa.
 
\section{Assessment of the results}
\label{Assessment of the results} 

The phenomenology of flickering is quite broad, and the focus of this study, 
its strength, is certainly not the only
characteristic property. Few attempts have been performed in the past to
systematically study the properties of flickering in a larger ensemble of
CVs. When it comes to the flickering strength, only the investigation of
\citet{Fritz98} (hereafter referred to as FB98), although based on different
techniques and significantly fewer data, has a similar breadth. Based on
scalegrams derived from a wavelet analysis of CV light curves they define a 
parameter $\Sigma$ as a proxy for the flickering strength. Thus, whenever
suitable, I will subsequently compare the systematic properties of $A_{60}$
found here with those of $\Sigma$ derived by FB98.
 
When initiating the study I hoped 
that a comparison of the strength of the flickering activity, based on a 
great number of light curves in many CVs would shed light on the system 
parameters which define some basic characteristics of this phenomenon in 
these interacting binary stars. While some systematic trends could be found,
and while it was possible to better quantify some properties of the flickering
in CVs than has been done before, to a large degree the initial expectations 
were frustrated. There is no simple correlation of the flickering strength with
parameters such as component masses, orbital inclination, or the like. If
any such correlations exist, they are masked by other and stronger 
dependences which could not clearly be identified. 

FB98 found a loose correlation between their 
parameter $\Sigma$ and
the orbital periods in systems with accretion disks in the bright state 
(i.e., old nova, novalike variables and dwarf novae in outburst). They
interpret this as an indication of a decrease of the flickering strength with 
the mass accretion rate which, while being difficult to measure individually,
is expected to increase on average with the orbital period. Here, I find that
the relationship between $A_{60}$ and period is even less well defined.
However, this changes if the period is substituted by the absolute magnitude
of the primary component (a quantity not well known for most systems in the
pre-Gaia era), recognizing that, again on average, systems with longer periods
are brighter and their mass accretion rate grows with the orbital period.
Fig.~\ref{amp-mag-dep} shows that this dependence is a function of the specific
CV type, indicating that the mass transfer rate cannot be the only parameter
determining the $M_{V} - A_{60}$ relation. Thus, relating the flickering 
strength to the absolute magnitude, made possible by the accurate distances
provided by the Gaia mission, enables a much more destinct view of the
flickering strength in different types of CVs than was hitherto possible.

Most obvious is the striking difference
of the flickering strength in UX~UMa type novalike variables (and quiescent
novae), and in VY~Scl stars, as already pointed out by FB98. 
While in the former flickering occurs on a low
level and increases slightly with decreasing system brightness, it is much
stonger in the latter (in this case the range of magnitudes covered by the
data is not sufficient to reveal a dependence on the system brightness). 
Since it is not securely known what makes VY~Scl exhibit occasional low states
and what inhibits low states in UX~UMa stars, one can only speculate that the
different flickering behaviour is somehow related to the reason for the 
different long-term photometric properties of the two classes. 

Another dependence of
flickering strength on magnitude concerns dwarf novae in outburst and 
quiescence. This is not a new insight, but is better documented here than
ever before. While during outburst a dependence of the flickering strength
on the magnitude is indicated, such a correlation is not obvious in 
quiescence, considering the large observed scatter of $A_{60}$. Moreover,
there is no continuous transition of the average values of $A_{60}$ between
outburst and quiescence, but rather a dichotomy see 
Fig.~\ref{amp-mag-dep}.

Not only is the dependence the flickering strength on the magnitude of the 
primary component different for different CV subtypes, but there are also
striking exceptions. These refer in particular to CVs with exceptionally
bright primaries (and long orbital periods) such as GK~Per and RS~Oph. For 
no obvious reason flickering in these systems is much stronger than in 
intrinsically fainter novae and UX~UMa stars.

\begin{figure}
	\includegraphics[width=\columnwidth]{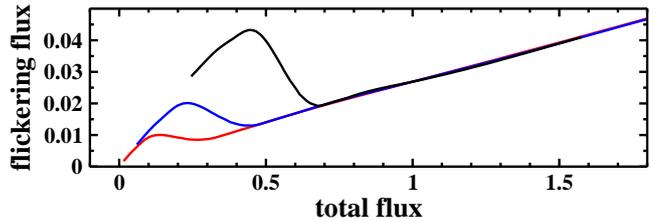}
      \caption[]{Average flux of the flickering light source during the dwarf
                 nova outburst cycle as a function of the total system flux
                 for an assumed outbust amplitude of 1 (black), 2 (blue) and
                 3~mag (red). Flux units are arbitrary.}
\label{fl-fl-vs-fl-total}
\end{figure}

The evolution of the flickering strength through the dwarf nova outburst cycle
suggests a dependence on the state of the accretion disk. During outburst
phases $A_{60}$, expressed in magnitudes, remains constant over a wide 
brightness range. At a higher level, the same is as least approximately true
during quiescence. Independency of $A_{60}$ on magnitude means, of course,
that the flux of the flickering light source increases or decreases in the same
proportion as the total light of the system. Thus, adopting the disk 
instability model for
dwarf nova outbursts, the relative contribution of the flickering light source 
is much reduced when the accretion disk is on the upper branch of the S~curve. 
If it is also reduced on an absolute scale depends on the outburst amplitude.
To show this, in Fig.~\ref{fl-fl-vs-fl-total} (black graph) the spline fit to 
the $A_{60}$ vs.\ $m_{\rm norm}$ relation (Fig.~\ref{outburst-cycle}) has 
been transformed into a relation between the total $V$ band flux and the 
integrated optical flux of the flickering light source, using the relation 
shown in Fig.~\ref{rel-flux} (here it is assumed that the $V$ band 
flux is proportional to the total optical flux). It is characterized
by a linear rise of the flickering flux at high total brightness (outburst)
and another (almost) linear rise at low brightness (quiescence), connected
by a transition phase. During both, quiescence and outburst, the flickering
flux is of the same order of magnitude. The definition of $m_{\rm norm}$ 
implies an outburst amplitude of 1~mag. Of course, in real dwarf 
nova systems the amplitudes are in general larger. To simulated this, the
magnitude axis in Fig.~\ref{outburst-cycle} was stretched by a factor of 2
and 3, corresponding thus to outburst amplitudes of 2 and 3~mag. In
Fig.~\ref{fl-fl-vs-fl-total} this translates into the blue and red curves,
respectively. It is then seen that in a realistic case (amplitude of 3~mag or 
more) the total optical flux of the flickering light source is smaller
during quiescence than during outburst, even if $A_{60}$ is much larger.
Thus, not surprisingly, the changing disk structure during the dwarf nova
outburst cycle is not only reflected in the total brightness of the system,
but also affects the flickering light source. 

\begin{figure}
	\includegraphics[width=\columnwidth]{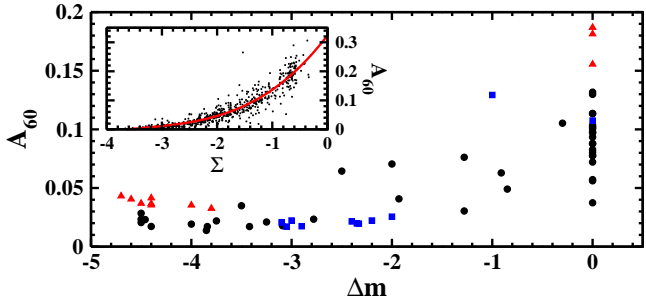}
      \caption[]{$A_{60}$ as calculated using the $\Sigma$ values of FB98
                 for three dwarf novae during their outburst cycle as a
                 function of the magnitude above average quiescence. Circles
                 refer to VW~Hyi, triangles to V436~Cen and squares to 
                 V442~Cen. In insert shows the relationship between $\Sigma$
                 (taken from FB98) and $A_{60}$ based on all light curves in
                 which both quantities have been measured, together with a 
                 least squares fit of a third order polynomial.} 
\label{sigma-a60}
\end{figure}

At first glance, the evolution of the flickering strength in dwarf novae
as a function of outburst phase found here appears to contradict the earlier
results of FB98 which are based on data restricted to only 5 systems (not
counting Z~Cha which FB98 identify as an exceptional case). They concluded
that during the outburst cycle the intrinsic flickering amplitude changes as
the square root of the mean light of the system, but also raise doubts about
the universal validity of such a simple law. However, this contradiction may
merely be a matter of interpretation of their Fig.~16 which is, to a certain
degree, the equivalent of Fig.~\ref{outburst-cycle} of the present paper.
FB98 fit the relation between their $\Delta$m and $\Delta$$\Sigma$, i.e, 
the variation of magnitude and $\Sigma$ over the outburst cycle, by a straight
line. In view of the scatter of data points, this is justified. However, a
close look at their figure (disregarding the exceptional behaviour of Z~Cha)
shows that the distribution of $\Sigma$ far above quiescence can equally
well be considered as being independent from $\Delta$m. This is particularly
clear regarding only V436~Cen, V442~Cen and VW~Hyi which have similar outburst
amplitudes such that a transformation to normalized magniudes is not necessary.
This becomes even clearer regarding Fig.~\ref{sigma-a60}. Here, he insert 
shows the
relationship between $\Sigma$ and $A_{60}$ for all light curves for which both
quantities have been measured. The red curve is a third order least squares
fit to these data. This has been used to transform the $\Sigma$ values of the
three CVs mentioned above into $A_{60}$ which are plotted in the main frame 
of Fig.~\ref{sigma-a60} as a function of the magnitude above the average 
quiescence. It is quite similar to Fig.~\ref{outburst-cycle}. Thus, no 
contradiction between FB98 and the present results exists.

The wavelength dependence of the flickering strength or its colours have 
occasionally been measured in individual systems \citep[e.g.,][]{Middleditch82,
Horne85, Hoard97, Zamanov15} and, more rarely, in small ensembles of 
CVs \citep[e.g.][]{Bruch92}. It is always found that at optical
wavelengths the flickering amplitude increases with decreasing wavelength.
Moreover, FB98 found some evidence for shorter flickering flares to be
at least on the average somewhat bluer than longer ones. FB98 also pointed
out that -- as has not generally been realized -- the wavelength dependence
of the flickering does not directly reflect the spectrum of the flickering 
light source. In 
order to quantify the latter, it is necessary to correct for the wavelength
dependent contribution(s) of the non-flickering sources(s) to the total light.
While this may be difficult to do for individual systems, a statistical 
approach is adopted here. It is then found that on the mean over most of the 
optical range the flickering spectrum can equally well be approximated by a 
power law with a spectral index of $-2.9$ and a back body with a temperature of 
17\,000~K. Only in the $U$ band an excess over these simple
laws is observed which increases if the flickering light source is less than
100\% modulated. Qualitatively, this excess can be explained by enhanced
emission in the Balmer continuum and/or from higher Balmer lines. Confirmation
for this idea must await more detailed model calculations for the spectrum, 
or -- observationally -- detailed and statistically sound spectrophotometry 
of the flickering light source.

The range of brightness modulations of flickering in many CV light curves 
immediately suggests that an appreciable fraction of the total optical
radiation is due to the flickering light source. This is quantified here.
In particular in quiescent dwarf novae and VY~Scl stars the elevated values
of $A_{60}$ imply that on the temporal average easily 15~\% or more of the
total energy emitted in the optical range is due to flickering. This 
reinforces my opinion, expressed many years ago \citep{Bruch95}, that we
cannot claim to really understand CVs as long as the origin of flickering
remains unknown. The only attempt that I am aware of 
to reconcile the amount of energy radiated by the flickering light source 
in a disk dominated system with predictions derived from models was recently
performed by \citet{Dobrotka20}. They try to explain optical flickering in
MV~Lyr as reprocessed unstable X-ray radiation from a corona above the inner 
disk and find that the available X-ray energy is only sufficient to explain
a small fraction of the optical flickering. 

It is not surprising that flickering in polars is stronger when the accretion
foot point on the surface of the white draws is viewed favourably, as evidenced
in the case of V834~Cen (Sect.~\ref{The AM Her stars}), suggesting that this
is its place of origin. 
More surprising is a dependence of the flickering strength on the superhump
phase in the permanent superhump systems TT~Ari and V795~Her as indicated
by the respective variance curves. 
But these only reflect the shape of the light curve in 
general terms, not in detail. Remembering that the superhump phenomenon is
thought to arise due to the extra stress exerted on an elliptically deformed
accretion disk when the secondary star passes close to the elongated part
of the disk, it might be conjectured that this additional pertubation 
leads to enhanced disturbances and, in consequence, to stronger flickering.
Being aware that this is only a speculative explanation, I will refrain from 
pondering further about details of the structure of the variance curves.

A dependence of the visibility of structures in the accretion disk on
orbital phase in high inclination CVs may also lead to a phase dependence 
of flickering if its light source is distributed asymmetrically. Indeed,
in the three systems investigated in this respect the variance due to 
flickering is clearly a function of phase. But this finding does not lead
to a systematic picture, since the phase dependence is neither the same nor
does the variance seem to follow systematic brightness variations around
the orbit (e.g., hot spots).

Even for a given object in a well defined photometric state $A_{60}$,
measured in different light curves, can assume significantly different
values, as evidenced by the standard deviations quoted in Table~1 which 
sometimes assume the same order of magnitude as the mean values themselves.
This scatter is certainly in part due to the inherent uncertainties in the
determination of $A_{60}$ in particular in light curves which are not of 
the best quality (S/R, time resolution). In the face of difficulties to 
quantify these uncertainties,
assessing if the short term variations of the flickering 
strength are at least partly real, is not easily possible. Regarding 
dwarf novae in quiescence, the mean values quoted in Table~1 are, in general,
based on light curves encompassing a range of average magnitudes. In
these cases variations of $A_{60}$ are not surprising and are likely real
to a certain degree. During outburst, when the ensemble of all present
data indicates no dependence of the flickering strength on magnitude, the 
numerical values of $A_{60}$ are small, enhancing the contribution of 
measurement errors to their scatter. But also in non-outbursting systems
the standard deviation of the average of $A_{60}$ can be quite high, suggesting
real changes of the flickering strength over time.

As detailed in Sect.~\ref{Long term trends} the available data permit only
in a small number of CVs an assessment of systematic long term variations 
of $A_{60}$. In only 3 out 
of 12 systems apparently systematic deviations of $A_{60}$ from an otherwise 
more or less constant value have been seen. The few identified events may
all be associated to changes in the structure of the systems. This 
confirms the findings of FB98, but extends the investigated time base by 
more than a factor of 2.  

\section{Summary and outlook}
\label{Summary} 

Flickering is an elusive phenomenon. In spite of its general presence in
accreting systems I am not aware of any model which can really explain the
underlying physics. The models promoted in the literature (see references
in Sect.~\ref{Introduction}), referring to disk dominated  
CVs, mostly invoke turbulences and the ensuing propagation 
of mass transfer fluctuations. Others \citep[e.g.][]{Scaringi14a, Dobrotka17}
additionally conjecture reprocessing of variable X-ray emission to explain
optical flickering. X-ray reprocessing is also invoked as the origin of 
optical flickering in magnetic systems \citep[e.g.][]{Semena14}. 

Concentrating on disk dominated CVs, 
the models are mostly heuristic, without a solid phyiscal base.
They have deficiencies at both ends: At the beginning, since they
do not specify the physical origins of such turbulences. Consequently, any
assumptions about the frequency, amplitude or place of origin of such 
turbulences
are {\it ad hoc} and not based on physical principles. At the other end, 
quantities such a the power spectra of the propagated turbulences (density 
fluctuations) are derived from the models and compared to flickering power 
spectra. But it remains an open question 
how the presumed density fluctuations are
transformed into brightness variations.
Few attempts to solve this question have come to my knowledge (and 
they are not necessarily related to
models of propagating turbulences). The theoretical study of \citet{Kley97},
investigating instabilities in the boundary layer may be seen as a step in 
this direction. They find burst-like variations of the luminosity which, as
they say, could have important observational consequences. In a different
approach \citet{Pearson05} model the time dependent emission of fireballs.
The results are compatible with the properties of individual flares in 
flickering light curves \citep[see also][]{Bruch15}. I also mention
\citet{Williams02} who carried out hydrodynamical simulations of flares in
CV disks. Moreover, it has never
been shown that any of the models is able to explain the observed flickering
amplitude, i.e., the often considerable fractional contribution of the 
flickering light source to the total optical luminosity of the systems.
While I recognize that the idea of propagating disturbances in accretion
disks may well be part of the solution of the flickering enigma, I am
afraid that the undoubted progress achieved during the past two or three
decades only refers to the simpler part of the problem, disregarding the
physics at the two ends mentioned above. 

The situation appears to be somewhat more favourable when it comes
to magnetic CVs. Here, more detailed models of the time dependent structure 
and emission from accretion shocks and the post-shock region above the white
dwarf surface have been
developed. For a brief overview and further references, see the introductory
section of \citet{Semena14}.

Having said this, I emphasize 
that it has not been the purpose of this study
to address the issues raised above. Instead, investigating a single property
of flickering, namely its strength relative to the total brightness of the
underlying object, parameterized by $A_{60}$ as a proxy of the average 
amplitude, it was attempted to find systematic features in its behaviour, 
such as correlations with other properties, which may shed light on its
origins, and in order to identify boundary conditions for physical models.

To do so, many more data of numerous CVs have been investigated than ever
before in any single observational study of flickering. Tools and reduction 
procedures for a rigorous determination of the flickering strength, 
rendering possible a comparison of $A_{60}$ measured in light curves of 
diverse characteristics, have been developed. This enabled to
(i) identify patterns in the flickering properties depending on the
absolute magnitude of the primary component, (ii) the quantification 
of the average broad band spectrum of the flickering light source and
its contribution to the overall optical flux, (iii) a dependence of the
flickering strength on the orbital phase in some systems and (iv) a
remarkable systematic behaviour of the flickering around the outburst 
cycle of dwarf novae.

Concluding, and as a personal remark, I wonder if, in spite of all the 
detailed knowledge about flickering gained during the past and possibly in 
the future, we will ever really understand it. Or will the focus of science 
shift, and flickering be left as an interesting phenomenon the understanding 
of which is not essential for further progress in astrophysics?

\section*{Acknowledgements}

This work is partly based on observations taken at the Observat\'orio do
Pico dos Diaz operated by the Laborat\'orio Nacional de Astrof\'{\i}sica,
Brazil. Many observers worldwide have generously put their data at my disposal.
A great amount of archival data were downloaded from the AAVSO International 
Database and other data banks. I thank the numerous dedicated AAVSO and other 
observers for their contributions, and the AAVSO staff for maintaining their 
valuable database. Without their efforts this work would not have been 
possible. 
I am also grateful to Claudia V.\ Rodrigues for a critical reading of the
manuscript and for the helpful comments on an anonymous referee. This 
research made use of the VizieR catalogue access 
tool, CDS, France (DOI: 10.26093/cds/vizier). 

\section*{Data availability}

The majority of the data used in this article are available in the AAVSO 
International Database (https://www.aavso.org), the OPD databank 
(https://www.gov.br/mcti/pt-br/rede-mcti/lna/composicao/coast/obs/opd) 
and the MEDUZA archive
(http://var2.astro.cz/EN/meduza/index.php). Data not to be found in these
archives will be shared on reasonable request to the author.







\appendix

\section{The secondary star contribution}
\label{The secondary star contribution}

The contribution of the secondary star to the light of a CV can have a 
substantial bearing on the observed flickering amplitude. In order to
apply a correction, it must be known how much light is due to the secondary
at the observed wavelength. Numerical values for this contribution taken from
the literature often refer to a different wavelength. Then, a reduction to
the observed wavelength is required, involving numerous assumptions
which will be outlined here together with an assessment of the resulting
uncertainties. 

The approach used
here assumes that the secondary star radiates like a black body. Sometimes
the temperature is quoted in the literature. However, more often, information
on its spectral type is available. Then, the corresponding temperature is
interpolated in the relevant tables \citep[e.g.][]{Schmidt-Kaler82B}. 
The primary
spectrum is assumed to be that of a standard steady state accretion disk
\citep[][being aware that his is only a rough 
approximation in the case of quiescent dwarf novae]{Lynden-Bell74B}.
It is defined by the inner and outer disk radii, the mass of
the central star, and the mass transfer rate in the disk. If available,
the white dwarf mass is taken from the literature. Otherwise the average
mass \citep{Zorotovic11B} is adopted. Its radius, calculated using 
\citet{Nauenberg72B} mass-radius relation for white dwarfs, is taken to
be the inner disk radius. The outer radius is assumed to be equal to 40\%
of the component separation, calculated from Kepler's third law, using
literature values for the mass of the secondary star, or an estimate
based on its spectral type (this is not a critical parameter), and the
orbital period. Unless more specific values are found in the literature for 
the mass accretion rate, a standard value of 
$10^{-8} M_\odot/{\rm y}$ is used if the accretion disk is in a bright
state (e.g., in novalike variables or old novae), and $10^{-9} M_\odot/{\rm y}$
if it is in a faint state. Any effect that emission lines may have on the
spectrum are not taken into account. Knowing thus the spectra of the two
components the contribution of the secondary star to the total light
at a wavelength $\lambda_2$ can be calculated if it is known at another
wavelength $\lambda_1$.

The procedure involves 5 independent parameters, namely the primary star
mass $M_1$ (which also determines the inner disk radius via the mass-radius
relation for white dwarfs), the secondary star mass $M_2$ (or, equivalently, 
the mass ratio $q$, if $M_1$ is given), the orbital period which together with
the component masses determines the component separation, the outer
disk radius $R_{\rm out}$ expressed in units of the component separation $a$, 
and the mass transfer rate $\dot{M}$ in the disk.

\begin{figure*}
	\includegraphics[width=\textwidth]{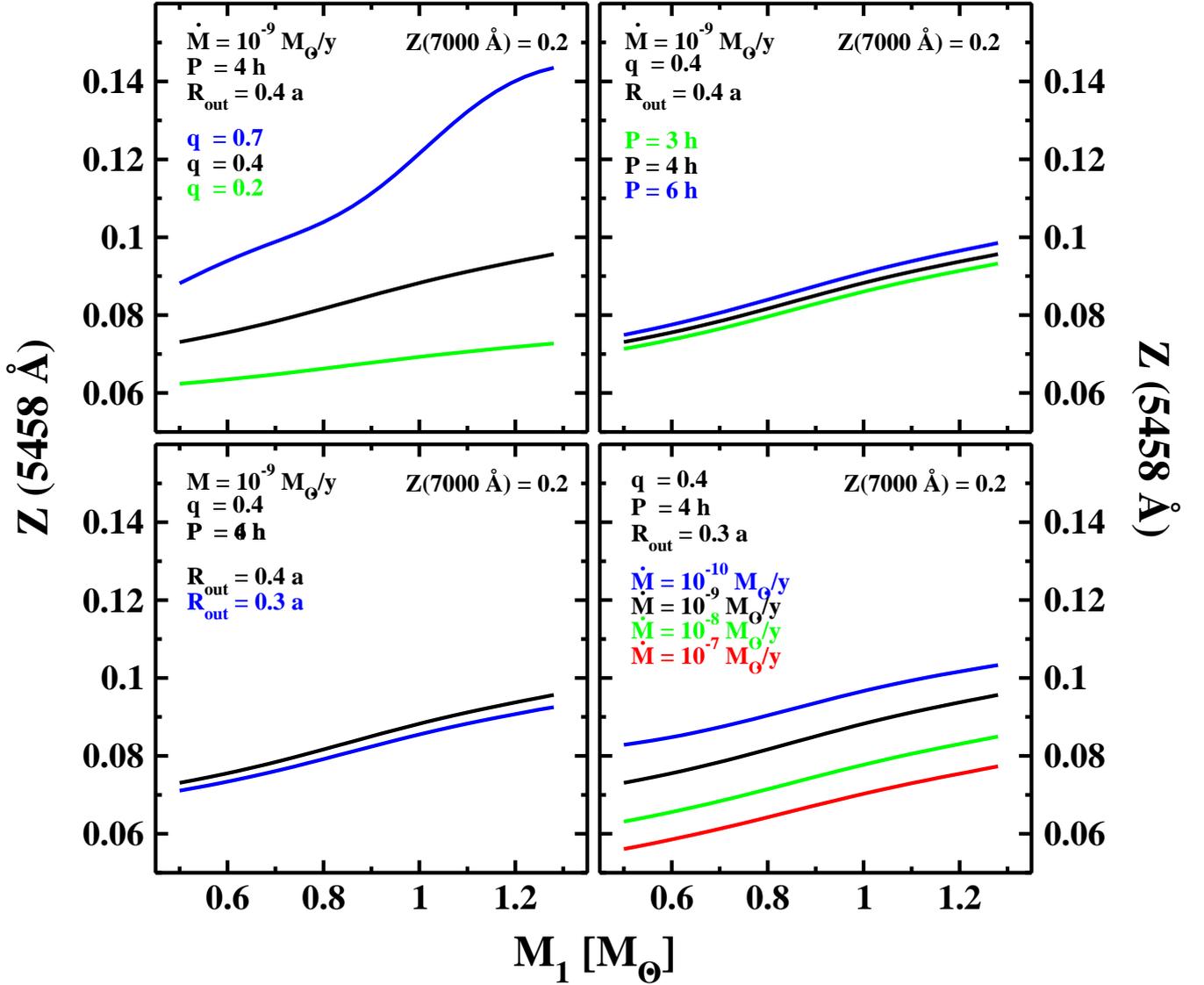}
      \caption[]{The fractional secondary star contribution to the total
                 light of a CV at the isophotal wavelength of the $V$ band
                 (5468~\AA), as a function of the primary star mass, assuming 
                 a contribution of 20\% at 7000~\AA. In
                 all frames the black graph refers to the standard case,
                 while the coloured graphs illustrate the situation when
                 one variable parameter is changed while all others
                 are fixed to those of the standard case. For details, see
                 text.}
\label{second-contrib}
\end{figure*}

In order to get a feeling for the dependence of the secondary star 
contribution $Z(\lambda)$ at a given wavelength $\lambda$ on these
parameters, I investigate a typical case of a CVs contributing 20\% to
the total light at 7000~\AA. What is its contribution at the isophotal
wavelength of the $V$ band (5468~\AA)? The
answer is given in Fig.~\ref{second-contrib} where $Z (5468 {\rm \AA})$
is shown as a function of $M_1$. In all four frames the black graph refers
to what may be considered a standard case: $q = 0.4$, $P = 4 {\rm h}$, 
$R_{\rm out} = 0.4 a$, and $\dot{M} = 10^{-9} M_\odot/{\rm y}$. The coloured
graphs in the different frames then explore the situation when one of the
parameters is modified, fixing all the others. Note that not all
parameter combinations may be realized in nature.

In the upper left frame,
the mass ratio is changed. Of all parameters this has the greatest bearing
on $Z(5468 {\rm \AA})$, in particular at high primary star masses: A higher 
mass ratio implies a larger mass for the secondary, and thus a higher
temperature and a bluer spectrum. It will thus contribute relatively more
at short wavelengths. The upper right frame shows that the influence of the
orbital period is quite small. The same is true for the outer disk radius
(lower left frame). Finally, the lower right frame shows that the mass
transfer rate is of intermediate importance. But note, that at very high
mass transfer rates a secondary star contribution of as much as 20\% at 
7000~\AA\ is not very likely.

This little exercise shows that the secondary star contribution in the $V$
band can be estimated within tolerable errors (considering the numerous other
error sources involved in determining the flickering amplitude) if it is 
known at some other wavelength (noting that any such estimate has, in 
general, also considerable uncertainties). At shorter wavelengths the
contribution will be smaller, meaning that errors in its determination
will have a smaller impact on the corrected amplitude. 

The secondary star contribution to the total light was determined for
the 25 CVs listed in 
In Table\ref{Table: Relative contribution of the secondary star}.
The specific information required to 
derive these numbers comes from numerous literature sources. The fractional
contribution is given only for those passbands for which light curves of the
respective system are available. $m_{\rm ref}$ is the reference magnitude to
which these values refer.

\begin{table}
\label{Table: Relative contribution of the secondary star}
	\centering
	\caption{Estimate of the relative contribution of the secondary star 
                 in CVs at the wavelengths of the $UBVRI$ passbands and at the 
                 reference magnitude $m_{\rm ref}$.}

\begin{tabular}{lcccccc}
\hline
Name & $m_{\rm ref}$ & $U$ & $B$ & $V$ & $R$ & $I$ \\
\hline
RX And    & 14.00 & 0.04 & 0.25 & 0.30 & 0.42 &      \\
DX And    & 15.05 &      &      & 0.86 &      &      \\
AT Ara    & 14.90 &      &      & 0.58 & 0.66 &      \\
V363 Aur  & 14.50 &      & 0.19 & 0.35 &      &      \\
SY Cnc    & 12.20 &      & 0.24 & 0.35 &      &      \\ [1ex]
AC Cnc    & 14.30 &      & 0.19 & 0.30 & 0.40 &      \\ 
BV Cen    & 13.05 & 0.23 & 0.35 & 0.50 &      &      \\
MU Cen    & 14.95 &      &      & 0.73 &      &      \\
V442 Cen  & 15.00 &      & 0.35 &      &      &      \\ 
ST Cha    & 16.46 &      &      & 0.30 &      &      \\ [1ex]
T CrB     & 11.65 & 0.10 &      &      &      &      \\
SS Cyg    & 12.10 & 0.17 & 0.35 & 0.43 &      &      \\
EM Cyg    & 12.90 & 0.04 & 0.09 & 0.19 &      &      \\
DO Dra    & 15.90 & 0.02 & 0.08 & 0.24 & 0.38 &      \\
U Gem     & 14.25 &      & 0.13 & 0.23 &      &      \\ [1ex]
AH Her    & 14.30 &      & 0.06 & 0.13 & 0.19 &      \\
BT Mon    & 15.50 &      &      & 0.02 &      &      \\
RS Oph    & 11.50 & 0.07 & 0.25 & 0.55 & 0.72 & 0.85 \\
V426 Oph  & 12.80 & 0.07 & 0.16 & 0.33 &      &      \\
V345 Pav  & 13.71 &      & 0.23 & 0.36 &      &      \\ [1ex]
IP Peg    & 15.25 & 0.04 & 0.11 & 0.28 & 0.44 &      \\
GK Per    & 13.10 & 0.27 & 0.34 & 0.38 & 0.40 & 0.40 \\
RW Sex    & 11.66 & 0.01 & 0.02 & 0.04 & 0.07 & 0.11 \\
RW Tri    & 13.23 &      & 0.01 & 0.03 & 0.05 &      \\
UX UMa    & 12.90 &      & 0.01 & 0.03 & 0.06 &      \\
\hline
\end{tabular}
\end{table}

Hereafter, a brief discussion follows of those systems of this study with a 
noticeable contribution of the secondary star to the total light. 

{\it RX~And:} \citet{Hutchings82B} observed absorption
lines in the blue spectrum of RX~And and remark that {\it ``compared with 
SS~Cyg, the late-type spectrum appears similar, but less intense by a factor of 
about two''}. In the same spectral range \citet{Stover79B} estimate the late 
type star contribution in SS~Cyg to be 45\%. 
\citet{Bitner07B} come to a similar conclusion.
Thus, it should be of the order 
of 20\% -- 30\% in RX~And. To be definite, I will adopt a contribution of 25\% 
in $B$, assuming that this holds for the average quiescent magnitude of 14~mag. 
With a primary star mass of $0.8 M_\odot$ \citep{Sion01B} and a secondary 
corresponding to a main sequence star with a spectral type K4.75 
\citep{Knigge06B} the contribution in other band as listed in 
Table~\ref{Table: Relative contribution of the secondary star} results.

{\it DX~And:} \citet{Bruch97B} determined the veiling factor due
to the dilution of the primary star light by the dominating secondary. 
This translates into a fractional contribution of the secondary to the 
total light of 86\% in the $V$ band. I assume that this value refers to 
the average quiescent magnitude of DX~And, i.e., 15.05~mag according to 
the AAVSO long term light curve. Although the veiling is quite strong,
considering that the data used in the present study were all obtained 
close to outburst maximum, the contribution of the secondary star is 
very much reduced and the corresponding correction is small. 

{\it AT~Ara:} The secondary contributes a significant fraction of the total 
light of AT~Ara. The veiling factor derived by \citet{Bruch03B} yields a 
secondary contribution of 65\% in the wavelength range centred on 
$\lambda$ 6300~\AA. Using the standard assumptions, this leads to a 
contribution of 58\% in the $V$ band and 66\% in the $R$ band. The exact 
magnitude of AT~Ara at the time of the spectroscopic observations is not 
known. I take it to be the average quiescent magnitude 
of 14.9~mag \citep{Bruch94B}.

{V363~Aur:} \citet{Thoroughgood04B} estimate that
$19\% \pm 4\%$ of the blue light and $35\% \pm 7\%$ of the red light originates
from the secondary. I assume that this hold for the average out-of-eclipse
magnitude of 14.5~mag.

{\it SY~Cnc:} According to \citet{Casares09B} the mass ratio of SY~Cnc is
$1.18 \pm 0.14$ and the spectral type of the secondary
is G8$\pm$2~V. They estimate its contribution to the total 
light of the system to be $42\% \pm 2\%$ at H$\alpha$, while 
\citet{VandePutte03B} find a smaller contribution of only 20\% 
at $\lambda$~8000~\AA. This apparent contradiction is resolved considering 
that according to the long-term AAVSO light curve the stronger contribution 
was measured in spectroscopic data when SZ~Cnc was at the very end of the 
decline from an outburst at an average visual magnitude of 13.2~mag, while 
the system just started the decline from maximum at an average magnitude of 
11.8~mag in the other case. Then, the secondary contribution should be 
$\sim$12\% at H$\alpha$ in the brighter state and somewhat more than that at
$\lambda$~8000~\AA. This is of the observed order of magnitude.
Assuming the secondary to contribute 42\% of the
light at H$\alpha$, it is supposed to contribute 24\% and 35\%, respectively,
at the isophotal wavelength of the $B$ and $V$ bands at 13.2~mag.

{\it AC~Cnc:} \citet{Thoroughgood04B} estimate that
$19\% \pm 2\%$ of the blue light and $40\% \pm 11\%$ of the red light originates
from the secondary. Its contribution in the $V$ band should then be 
approximately 30\%. I assume that this holds for the average out-of-eclipse
magnitude of 14.3~mag.

{\it BV~Cen:} \citet{Vogt80B} classify the secondary star as G5-8~IV-V and
estimate its contribution to the total light in the $V$ band as about 50\%.
A similar value can be derived from fig.~4 of \citet{Gilliland82B}. This holds
for quiescence at a magnitude of 13.05~mag averaged over the orbital 
cycle. Using a mass transfer rate of 
$\dot{M} = 3 \times 10^{-9} M_\odot/{\rm y}$ \citep{Sion07B} then yields a
contribution of 35\% at $B$ and 23\% at $U$.

{\it MU~Cen:} The light curves of MU~Cen are dominated by ellipsoidal 
variation of the secondary star. Model fits of \citet{Bruch16B} indicate 
that it contributes 73\% in the $V$ band at the average magnitude of 
14.95~mag. 

{\it V442 Cen:} There is
no information in the literature about the contribution of the scondary
star to the total optical light of the system. However, at the long orbital
period of about 11~h \citep{Marino84B} and the early spectral type of the 
secondary \citep[G6$\pm$2;][]{Harrison04B} it is expected to
contribute a considerable fraction. The spectral type being similar to
that of the BV~Cen secondary and the orbital period also being comparable,
I will assume -- somewhat arbitrarily -- that the secondary contribution
to the total light in the $B$ band in V442~Cen is the same as in BV~Cen,
i.e., 35\%. This holds true for the quiescent 
magnitude which I take to be 15.0~mag.

{\it ST~Cha:} At the orbital period of about 5.5~mag \citep{Bruch17B}
it should be expected that the secondary
star contributes non-negligibly to the total system light. However, no
specific information relative to this question is available in the 
literature. Therefore, I (rather arbitrarily) assume a contribution of
30\% (at $V = 16.46$, the average magnitude of the observed light curve), 
slightly less than the contribution of 42.5\% derived for SS~Cyg
which has a one hour longer period.

{\it T~CrB:} The secondary star strongly dominates the total light of T~CrB
in $V$. \citet{Zamanov98B} estimate that it can contribute 90\%. However, 
this number is uncertain. Correcting the amplitude of the flickering in the 
visual for the contribution of the secondary star may therefore lead to huge 
errors. Therefore, I prefer to use only data obtained in the $U$ band for 
this study. Here, the secondary contributes much less and any results are 
therefore more reliable. \citet{Zamanov98B} calculated that 10\% of the total
light comes from the red component at these wavelengths. This hold for
the average $U$ magnitude of their sample, which is 11.65~mag.

{\it SS Cyg:} Based on 
spectroscopic observations at an epoch when SS~Cyg had a visual
brightness of $V = 12.10 \pm 0.18$~mag according to the 
AAVSO long-term light curve, \citet{Bitner07B} measured a contribution
of the primary component to the total light at 5500~{\AA} (i.e., very close
to the isophotal wavelength of the $V$ band) of $0.575 \pm 0.075$.
The secondary is then responsible for 42.5\%. With the temperature appropriate
for its spectral type of K4V -- K5V \citep{Stover80B, Cowley80B, Bailey81B,
Friend90B, Harrison00B, North02B} the contribution in other bands
can be calculated.  

{\it EM~Cyg:} The analysis of EM~Cyg is complicated by an unrelated star
of spectral type K2-5~V along the line of sight, discovered by
\citet{North00B} who find that it contributes 16\% to the observed light.
Moreover, the secondary star, which is of a very similar spectral type of K3,
contributes $23.1 \pm 0.5$~\% to the remaining light. Then, the combined
light of the late type stars constitutes 35\% of the total. 
This holds true at an average wavelength of 6450~\AA\ and at an
epoch, when EM~Cyg was at an approximate visual magnitude of 12.9~mag
according the the AAVSO long time light curve. The usual assumptions then
lead to a combined contribution of the K stars
to the total light of about 4\% in the $U$ band, 9\%
in $B$ and 19\% in $V$.

{\it DO~Dra:} According to \citet{Mateo91B} the secondary star contributes 
55\% of the total light of DO~Dra at 8200~\AA. A mass transfer rate of the 
order of $10^{-10} M_\odot/{\rm y}$ \citep{Patterson92B} together with the
usual assumptions lead to contributions of 2\% in $U$, 8\% in $B$, 24\% in $V$
and 38\% in $R$. The flux level of the spectra used by \citet{Mateo91B}
correspond roughly to an average magnitude of $I = 15.56$.
Again using the assumption about the spectral behaviour of the components
defined above, this translates into a magnitude in the $V$ band of
15.90~mag.

{\it U~Gem:} The model calculations of \citet{Zhang87B} yield an average 
contribution of the secondary of 23\% to the total light in the $V$ band
at a visual magnitude of $14.25 \pm 0.29$~mag. \citet{Wade79B} also observed 
an appreciable secondary star contribution, but in this case the error margins
are too wide to reliably infer a numerical value. Using the contribution of
23\% derived by \citet{Zhang87B}, the usual assumptions yield
a contribution of 13\% in the $B$ band.

{\it AH~Her:} Absorption features of the secondary star were seen in the 
optical spectrum of AH~Her by \citet{Horne86B} and \citet{Bruch87B}.
The latter author restricted the spectral type to be in the range 
${\rm K2} \le {\rm SP} \le {\rm M0}$.
He also determined the distance as a function of the spectral
type. It is consistent with the GAIA distance of $324.3 \pm 3.2$~pc if 
${\rm Sp} \approx {\rm K8}$. Then, the contribution of the secondary star
to the total light at $\sim$5900~\AA\ is about 15\% at a
visual magnitude of 14.3~mag. This agrees well with a contribution
of 10\% -- 15\% at 5200~\AA\ (and 30\% -- 40\% at 6500~\AA) found by 
\citet{Horne86B}. Using the usual assumptions yields a secondary contribution
of 6\% in $B$, 13\% in $V$ and 19\% in $R$.

{\it BT~Mon:} \citet{Smith98B} detected weak late type absorption lines in 
the spectrum of BT~Mon. They classified the secondary as
being of type G8 and found that while its contribution out-of-eclipse is of the
order of 8\% in the red, it is virtually zero in the blue (4570 -- 5370~\AA;
see their Fig.~14). Interpolation in the same figure indicates that the 
contribution in the $V$ band should be of the order of 2\%. This is assumed
to hold for the average out-of-eclipse magnitude of 15.5~mag.

{\it RS~Oph:} The secondary star being a red giant, its contribution to the 
total optical light of RS~Oph is expected to be substantial.
\citet{Kelly14B} estimated that the giant star should be 0.65~mag 
fainter in $V$ than the total magnitude of 11.5~mag of the binary 
system. This corresponds to a secondary star contribution of 55\%.
\citet{Anupama99B} classify the cool component of RS~Oph
as M0~III and find that the hot component mimics a B9-A4 star. Adopting A0 
for the latter, and using the colours derived from synthetic 
photometry listed by \citet{Pickles98B}, I find
secondary star contributions of 7\% in $U$, 25\% in $B$,
72\% in $R$ and 85\% in $I$. 

{\it V426~Oph:} \citet{Hessman88B} estimates that the secondary star of
V426~Oph contributes 1/3 of total light in the $V$ band when the system 
had a $B$ band brightness of 13.2~mag. With the
average quiescent $B-V$ colours taken from \citet{Bruch94B} this
corresponds to $V = 12.8$. Adopting a spectral type of K3 for the 
secondary \citep{Hessman88B} the usual assumptions leads to contribution
of 7\% in U, 16\% in $B$, 33\% and $V$. 

{\it V345~Pav:} The spectrum of V345~Pav
is peculiar in the sense that it exhibits metallic absorption lines typical 
of a G8 dwarf star which cannot be attributed to the secondary of the CV
system. Instead, \citet{Buckley92B} conjectured the presence of a third
star which contributes 23$\pm$5\% to the $B$ band light. The usual assumptions
then lead to a contribution of 36\% in $V$. I take this to hold true for
the average magnitude of 13.71~mag of the present data.

{\it IP~Peg:} The secondary star of IP~Peg contributes appreciably to the
optical light. However, it is only quantified in the near infrared passbands
\citep{Ribeiro07B, Froning99B, Szkody86B, Littlefair01B}. The usual assumptions
then yield to a rough estimate of this contribution of 4\% in $U$, 11\% in $B$,
28\% in $V$ and 44\% in $R$. I assume that this refers of an average quiescent
magnitude of 15.25~mag.

{\it GK~Per:} The optical spectrum is strongly affected by absorption lines 
of a slightly evolved late type star of type $\sim$K2 \citep{Kraft64B, 
Gallagher74B, Crampton86B, Garlick94B, Reinsch94B}.
\citet{Gallagher74B} estimate the contribution of the secondary star to 
the total light in September 1973 to be approximately one-third at the 
wavelength of the Ca~II $\lambda$ 4227~\AA\ absorption line. At the time of 
their observations GK~Per had a visual magnitude of $\sim$13.1~mag
\citep{Bianchini83B}. Similarly, \citet{Crampton86B} estimate a contribution 
of 33\% at $\sim$4200~\AA\ during their 1984 observations. They
do not specify the magnitude of the system, but from the AAVSO long term light
curve an average magnitude of $13.1 \pm 0.2$~mag is
derived. Finally, \citet{Reinsch94B} also quotes a contribution of 1/3 
in the spectral range between 4200~\AA\ and 5000~\AA, 
when the average magnitude of the system according to the AAVSO long term 
light curve was at the same level. The disk spectrum resulting from the
usual assumptions depends most strongly on the unknown inner disk radius 
which is expected to be truncated in this intermediate polar. I rather 
arbitrarily take it to be the co-rotation radius of the white dwarf. 
I then find fractional contributions
of the secondary star in the $U$, $B$, $V$, $R$ and $I$ bands of 27\%,
34\%, 38\%, 40\% and 40\%, respectively, at a visual magnitude of 
13.1~mag.

{\it RW~Sex:} \citet{Beuermann92B} suggest that the late type
star, being of spectral type $\sim$M0, contributes about 10\% to the light
at $\lambda$~8000~\AA. With the usual assumptions and adopting a mass 
accretion rate of $\approx 4\, 10^{-9} M_\odot/{\rm y}$ \citep{Bolick87B} the
fractional contributions of the secondary star to the total light in the
$U$, $B$, $V$, $R$ and $I$ bands are, respectively, 0.6\%, 1.6\%; 4.1\%,
6.7\% and 11.1\%.
This holds for the average magnitude of $V = 10.66$ \citep{Bolick87B}.

{\it RW Tri:} \citet{Poole03B} estimate the contribution of the secondary
star of RW~Tri to be $\sim$10\% in the $I$ band. Adopting a spectral type
of M0 \citep{Dhillon00B} and the usual assumptions it is found to be
negligible in the $U$ band (0.2\%), and 1\%, 3\% and 5\% in $B$, $V$ and $R$, 
respectively. These numbers are taken to hold for the average magnitude 
of 13.23~mag.

{\it UX~UMa:} \citet{VandePutte03B} estimate that the secondary star of UX~UMa 
contributes about 20\% to the total light in the region of 8000~\AA.
Similarly, \citet{Rutten94B} estimate a contribution of 
6\% at 6000~\AA; and 15\% at 9000\%. The exact system magnitude at the 
respective observing epochs is not known. However, according to the AAVSO
long term light curve, in a broad interval around these dates UX~UMa
remained at a mean magnitude of 12.9~mag. With the spectral type of M4+ 
\citep{VandePutte03B} and the usual assumptions the secondary star
contributions are negligible in $U$ and amount to 1\%, 3\% and 6\% in
the $B$, $V$ and $R$ bands, respectively.


\section{System parameters}
\label{System parameters} 

The search for correlations between the flickering amplitude and other
characteristics of the investigated objects makes it necessary to define some 
important parameters of these systems. This requires often a decision between
sometimes contradicting numerical values found in the literature. In many
cases there is no objective way to do this, making subjective and even
arbitrary choices unavoidable. While this undoubtedly introduces uncertainties
into the results, it may be hoped that these will statistically cancel each
other to a certain degree. The parameters finally used in this study are
summarized in Table~\ref{Table: Final object parameters}.

\begin{table*}
\label{Table: Final object parameters}
	\centering
{\bf Table B1:} Object parameters adopted in the flickering analysis
\vspace{1em}

\begin{tabular}{lll@{\hspace{1ex}}lrr@{\hspace{1ex}}lr@{\hspace{1ex}}lr@{\hspace{1ex}}l
                 r@{\hspace{1ex}}l}
\hline
Star  & Type  
      & \multicolumn{2}{c}{$P_{\rm orb}$ (Ref)}   &
        \multicolumn{1}{c}{d}           & \multicolumn{2}{c}{$E_{B-V}$ (Ref)}   &
        \multicolumn{2}{c}{$M_1$ (Ref)} & \multicolumn{2}{c}{$q$      (Ref)}   &
        \multicolumn{2}{c}{$i$   (Ref)} \\
      &      
      & \multicolumn{2}{c}{(h)}   &
        \multicolumn{1}{c}{(pc)}        & \multicolumn{2}{c}{ }                &
        \multicolumn{2}{c}{($M_\odot$)}  & \multicolumn{2}{c}{ }                &
        \multicolumn{2}{c}{($^o$)}      \\
\hline
RX And          & DN\hspace{1em}Z  & \phantom{0}5.037   & (81)  & 
198 $\pm$ \phantom{00}2 & 0.00 & 
(84,92,196,      & 1.14 & (133)              & 0.42 & (133)   & 51 & (133)\\
                &         &                    &       &        & 
& 197,204)         &      &        &      &         &    &      \\

AR And          & DN\hspace{1em}UG & \phantom{0}3.91248 & (181) & 
436 $\pm$ \phantom{0}33 & 0.02 & 
(84,92,175)      &      &        &      &         &    &      \\

DX And          & DN\hspace{1em}UG &          10.57205 & (27)  & 
589 $\pm$ \phantom{00}9 & 0.20 & 
(44)           &      &         & 0.66 & (27)    & 45 & (44) \\

V455 And        & DN\hspace{1em}IP &\phantom{0}1.35142 & (24)  & 
75.4 $\pm$ \phantom{}0.3& 0.01 & 
(145)          &      &         &      &         & 75 & (1)  \\

V704 And        & NL\hspace{1em}VY &\phantom{0}3.69898 & (211) & 
401 $\pm$ \phantom{0}10 & 0.09 & 
(145)          &      &         &      &         &    &      \\

UU Aqr          & NL\hspace{1em}UX &\phantom{0}3.97930 & (5)   & 
254 $\pm$ \phantom{00}5 & 0.06 & 
(145)          & 0.67 & (5)     & 0.30 & (5)     & 78 & (5)  \\

VY Aqr          & DN\hspace{1em}SU &\phantom{0}1.5142  & (187) & 
138 $\pm$ \phantom{00}3 & 0.00 & 
(92)           & 0.54 &(109,190)& 0.11 & (109)   & 35 & (3)  \\

HL Aqr          & NL\hspace{1em}UX &\phantom{0}3.25368 & (138) & 
585 $\pm$ \phantom{0}17 & 0.00 & 
(92)           &      &         &      &         & 23 & (138)\\

CZ Aql          & DN?              &\phantom{0}4.812   & (160) & 
612 $\pm$ \phantom{0}17 & 0.09 & 
(145)          &      &         &      &         &    &      \\

DH Aql          & DN\hspace{1em}SU &\phantom{0}1.855   & (54)  & 
281 $\pm$ \phantom{0}16 & 0.15 & 
(145)          &      &         &      &         &    &      \\

V603 Aql        & N                &\phantom{0}3.31682 & (127) & 
311 $\pm$ \phantom{00}7 & 0.07 & 
(47,52,55,88,  & 1.20 & (2)     & 0.24 & (2)     & 13 & (2)  \\
                &                  &                   &       &
        &      & 
106,132,151,   &      &         &      &         &    &      \\
                &                  &                   &       &
        &      & 
152,189,210)   &      &         &      &         &    &      \\

V725 Aql        & DN\hspace{1em}SU &\phantom{0}2.276   & (189) &
        & 0.27 & 
(145)          &      &         &      &         &    &      \\

V794 Aql        & NL\hspace{1em}VY &\phantom{0}3.672   & (73)  & 
641 $\pm$ \phantom{0}21 & 0.00 & 
(92,196)       & 0.90 & (59)    &      &         & 60 & (59) \\

V1315 Aql       & NL\hspace{1em}UX &\phantom{0}3.35256 & (124) & 
443 $\pm$ \phantom{00}7 &      &
                & 0.73 & (40)    & 0.41 & (40)    & 82 & (40) \\

AT Ara          & DN\hspace{1em}UG &\phantom{0}8.9976  & (19)  & 
958 $\pm$ \phantom{0}49  & 0.13 & 
(145)          & 0.53 & (19)    & 0.79 & (19)    & 38 & (19) \\

TT Ari          & NL\hspace{1em}VY &\phantom{0}3.30121 & (218) & 
256 $\pm$ \phantom{00}5 & 0.04 &
(88,89,92,197   & 1.24 & (218)   & 0.19 & (218)   & 29 & (218)\\
                &                  &                   &       &
         &      & 
203,205)       &      &         &      &         &    &      \\

T Aur           & N                &\phantom{0}4.90508 & (35)  & 
711 $\pm$ \phantom{0}26 & 0.39 & 
(106,151,152,  & 0.68 & (13)    & 0.93 & (13)    & 57 & (13) \\
                &                  &                   &       &
         &      & 
204,210)       &      &         &      &         &    &      \\

KR Aur          & NL\hspace{1em}VY &\phantom{0}3.9058  & (79)  & 
451 $\pm$ \phantom{0}93  & 0.05 & 
(92,196)       & 0.70 & (154)   & 0.66 & (154)   &    &      \\

V363 Aur        & NL\hspace{1em}UX &\phantom{0}7.70980 & (182) & 
858 $\pm$ \phantom{0}23 & 0.30 & 
(176,204)      & 0.90 & (182)   & 1.17 & (182)   & 70 & (182)\\

SY Cnc          & DN\hspace{1em}Z  &\phantom{0}9.17701 & (29)  & 
443 $\pm$ \phantom{00}9 & 0.00 & 
(92,196)       &      &         & 1.18 & (29)    & 26 & (153)\\

YZ Cnc          & DN\hspace{1em}SU &\phantom{0}2.08618 & (193) & 
238 $\pm$ \phantom{00}3 & 0.00 & 
(92,196)       & 0.80 & (156)   & 0.22 & (86,156)& 41 & (156)\\

AC Cnc          & NL               &\phantom{0}7.21146 & (182) & 
660 $\pm$ \phantom{0}17 & 0.00 & 
(92)           & 0.76 & (182)   & 1.02 & (182)   & 76 & (182)\\

OY Car          & DN\hspace{1em}SU &\phantom{0}1.51490 & (216) & 
90.6 $\pm$ \phantom{}0.2 & 0.00 & 
(92)           & 0.88 & (105)   & 0.11 & (105)   & 83 & (105)\\

QU Car          & NL\hspace{1em}UX &          10.896   & (120) & 
580 $\pm$ \phantom{0}13 & 0.12 & 
(144,196,203,  &      &         &      &         & 50 & (97) \\
                &                  &                   &       &
        &      & 
204)           &      &         &      &         &    &      \\

HT Cas          & DN\hspace{1em}SU &\phantom{0}1.76753 & (51)  & 
141 $\pm$ \phantom{00}1 & 0.03 & 
(204)          & 0.61 & (97)    & 0.15 & (97)    & 81 & (75) \\

WX Cen          & NL\hspace{1em}UX &          10.00704 & (128) & 
2699 $\pm$ \phantom{0}11 & 0.05 & 
(92)           & 0.90 & (119)   &      &         &    &      \\

BV Cen          & DN\hspace{1em}UG &          14.64278 & (58)  & 
363 $\pm$ \phantom{00}5 & 0.15 & 
(213)          & 1.18 & (208)   & 0.89 & (208)   & 53 & (208)\\

MU Cen          & DN\hspace{1em}UG &\phantom{0}8.20512 & (21)  & 
513 $\pm$ \phantom{00}9 & 0.07 & 
(145)          &      &         &      &         & 58 & (21) \\

V436 Cen        & DN\hspace{1em}SU &\phantom{0}1.50    & (57)  & 
157 $\pm$ \phantom{00}2 & 0.13 & 
(203,204)      & 0.80 & (115)   &      &         & 75 & (115)\\

V442 Cen        & DN\hspace{1em}UG &          11.0     & (101) & 
336 $\pm$ \phantom{00}5  & 0.15 & 
(203)          &      &         &      &         &    &      \\

V504 Cen        & NL\hspace{1em}VY &\phantom{0}4.21360 & (62)  & 
600 $\pm$ \phantom{}120 & 0.05 & 
(145)          &      &         &      &         &    &      \\

V834 Cen        & AM               &\phantom{0}1.69194 & (150) & 
112 $\pm$ \phantom{00}3 & 0.00 & 
(92)           & 0.66 & (150)   & 0.24 & (141)   & 44 & (33) \\

V1033 Cen       & AM               &\phantom{0}3.15643 & (135) & 
244 $\pm$ \phantom{00}9 & 0.25 & 
(145)          &      &         &      &         & 70 & (28) \\

WW Cet          & DN\hspace{1em}Z  &\phantom{0}4.21941 & (180) & 
217 $\pm$ \phantom{00}2 & 0.04 & 
(85,92,196,    & 1.05 & (180)   & 0.37 & (180)   & 48 & (180)\\
                &                  &                   &       &
        &      & 
203,204)       &      &         &      &         &    &      \\

WX Cet          & DN\hspace{1em}SU &\phantom{0}1.39824 & (172) & 
262 $\pm$ \phantom{0}11 & 0.00 & 
(43)           & 0.55 & (139)   & 0.09 & (139)   & 40 & (139)\\

BO Cet          & NL\hspace{1em}UX &\phantom{0}3.3343  & (138) & 
513 $\pm$ \phantom{0}10 & 0.03 & 
(145)          &      &         &      &         &    &      \\

Z Cha           & DN\hspace{1em}SU &\phantom{0}1.78793 & (37)  & 
115 $\pm$ \phantom{00}2 & 0.00 & 
(65,92,204)    & 0.80 & (105)   & 0.19 & (105)   & 80 & (105)\\

ST Cha          & DN\hspace{1em}Z  &\phantom{0}5.496   & (25)  & 
701 $\pm$ \phantom{0}12 & 0.21 & 
(145)          &      &         &      &         &    &      \\

TV Col          & NL\hspace{1em}IP &\phantom{0}5.48641 & (68)  & 
505 $\pm$ \phantom{00}5 & 0.02 &
(92,104,113,196)&      &         &      &         & 70 & (69) \\

T CrB           & NR               &\phantom{0}5464    & (95)  & 
806 $\pm$ \phantom{0}32 & 0.15 & 
(30,210)       & 1.37 & (171)   & 0.82 & (171)   & 67 & (171)\\

SS Cyg          & DN\hspace{1em}UG &\phantom{0}6.03114 & (71)  & 
114 $\pm$ \phantom{00}1 & 0.03 & 
(92,196,203)   & 0.94 & (72)    & 0.63 & (72)    & 45 & (72) \\

EM Cyg          & DN\hspace{1em}Z  &\phantom{0}6.98182 & (36)  & 
564 $\pm$ \phantom{}113  & 0.04 & 
(92,174,196)   & 1.00 & (212)   & 0.77 & (212)   & 67 & (116)\\

V751 Cyg        & NL\hspace{1em}VY &\phantom{0}3.46992 & (48)  & 
758 $\pm$ \phantom{0}13 & 0.25 & 
(61)           &      &         &      &         & 30 & (61) \\

HR Del          & N                &\phantom{0}5.13996 & (90)  & 
932 $\pm$ \phantom{0}31 & 0.17 &
(47,132,151,152,& 0.75 & 
(90)    & 0.77 & (90)    & 41 & (17) \\
                &                  &                   &       &
        &      & 
196,204,210)   &      &         &      &         &    &      \\

DO Dra          & DN\hspace{1em}IP &\phantom{0}3.96898 & (67)  & 
197 $\pm$ \phantom{00}1 & 0.02 & 
(145)          & 0.83 & (67)    & 0.45 & (67)    & 45 & (67) \\

AQ Eri          & DN\hspace{1em}SU &\phantom{0}1.4626  & (185) & 
373 $\pm$ \phantom{0}17 & 0.03 & 
(145)          &      &         &      &         &    &      \\

KT Eri          & N                &\phantom{0}4.685   & (23)  & 
3688 $\pm$ \phantom{}476& 0.04 & 
(145)          &      &         &      &         &    &      \\

U Gem           & DN\hspace{1em}UG &\phantom{0}4.24575 & (102) & 
93.1 $\pm$ \phantom{}0.3 & 0.03 & 
(92,121,196,204)& 1.12 & (220)  & 0.47 & (220)   & 70 & (220)\\

DM Gem          & N                &                  &        & 
3097 $\pm$ \phantom{}903 & 0.11 & 
(145)         &        &      &         &    &      \\

IR Gem          & DN\hspace{1em}SU &\phantom{0}1.64    & (179) & 
264 $\pm$ \phantom{00}6 & 0.00 & 
(92,196,204)   &      &         &      &         &    &      \\

RZ Gru          & NL\hspace{1em}UX &                   &       & 
535 $\pm$ \phantom{0}17 & 0.00 & 
(92)           &      &         &      &         &    &      \\

AH Her          & DN\hspace{1em}Z  &\phantom{0}6.19478 & (74)  & 
321 $\pm$ \phantom{00}3 & 0.03 & 
(66,92,174,196,& 0.95 & (74)    & 0.80 & (74)    & 46 & (74) \\
                &                  &                   &       &
        &      & 
197,204)       &      &         &      &         &    &      \\
\hline
\end{tabular}
\end{table*}

\begin{table*}
\centering
{\bf Table B1} (continued)
\vspace{1em}

\begin{tabular}{lll@{\hspace{1ex}}lrr@{\hspace{1ex}}lr@{\hspace{1ex}}lr@{\hspace{1ex}}l
                 r@{\hspace{1ex}}l}
\hline
Star  & Type  
      & \multicolumn{2}{c}{$P_{\rm orb}$ (Ref)}   &
        \multicolumn{1}{c}{d}           & \multicolumn{2}{c}{$E_{B-V}$ (Ref)}   &
        \multicolumn{2}{c}{$M_1$ (Ref)} & \multicolumn{2}{c}{$q$      (Ref)}   &
        \multicolumn{2}{c}{$i$   (Ref)} \\
      &      
      & \multicolumn{2}{c}{(h)}   &
        \multicolumn{1}{c}{(pc)}        & \multicolumn{2}{c}{ }                &
        \multicolumn{2}{c}{($M_\odot$)}  & \multicolumn{2}{c}{ }                &
        \multicolumn{2}{c}{($^o$)}      \\
\hline
AM Her          & AM               &\phantom{0}3.09425 & (38)  & 
87.5 $\pm$ \phantom{}0.1  & 0.00 & 
(92)           & 0.58 & (39)    & 0.50 &(169,207)& 60 & (207)\\

DQ Her          & N\phantom{L}\hspace{1em}IP  &\phantom{0}4.64690 & (34)  & 
494 $\pm$ \phantom{00}6 & 0.07 & 
(92,106,132,152,& 0.60 & (76)   & 0.66 & (76)    & 87 & (76) \\
                &                  &                   &       &
        &      & 
196,294,210)   &      &         &      &         &    &      \\

V533 Her        & N                &\phantom{0}5.328   & (183) & 
1165 $\pm$ \phantom{0}44 & 0.03 &
(110,151,152,196)& 0.95 & (136)  & 0.35 & (136)   & 62 & (136)\\

V795 Her        & NL\hspace{1em}UX &\phantom{0}2.59830 & (158) & 
580 $\pm$ \phantom{0}13 & 0.03 & 
(145)          & 0.80 & (111)   &      &         & 41 & (136)\\

EX Hya          & DN\hspace{1em}IP &\phantom{0}1.63761 & (50)  & 
56.9 $\pm$ \phantom{}0.1   & 0.00 & 
(92,196,204)   & 0.78 & (50)    & 0.13 & (50)    & 78 & (205)\\

VW Hyi          & DN\hspace{1em}SU &\phantom{0}1.78251 & (83)  & 
53.9 $\pm$ \phantom{}0.1   & 0.00 & 
(92,196,204)   & 0.71 & (167)   & 0.15 & (167)   & 60 & (149)\\

WX Hyi          & DN\hspace{1em}SU &\phantom{0}1.79552 & (149) & 
232 $\pm$ \phantom{00}2 & 0.00 & 
(92,196)       & 0.90 & (149)   & 0.18 & (149)   & 40 & (149)\\

BL Hyi          & AM               &\phantom{0}1.89396 & (108) & 
130 $\pm$ \phantom{00}1 & 0.00 & 
(92)           & 0.60 & (215)   & 0.50 & (80)    & 40 & (10) \\

DI Lac          & N                &          13.0505  & (133) & 
1570 $\pm$ \phantom{0}51 & 0.35 & 
(45,132,152,   & 0.70 & (114)   &      &         & 18 & (114)\\
                &                  &                   &       &
        &      & 
152,210)       &      &         &      &         &    &      \\

X Leo           & DN\hspace{1em}UG &\phantom{0}3.9456  & (155) & 
393 $\pm$ \phantom{0}14 & 0.00 & 
(66,92,204)    & 1.03 & (191)   &      &         & 41 & (191)\\

GW Lib          & DN\hspace{1em}WZ &\phantom{0}1.2768  & (186) & 
112 $\pm$ \phantom{00}1 & 0.02 & 
(145)          & 0.84 & (194)   & 0.06 & (194)   & 11 & (194)\\

BR Lup          & DN\hspace{1em}SU &\phantom{0}1.908   & (107) & 
627 $\pm$ \phantom{0}76 & 0.11 & 
(145)          & 0.98 & (107)   & 0.15 & (107)   & 48 & (107)\\

AY Lyr          & DN\hspace{1em}SU &\phantom{0}1.769   & (188) & 
448 $\pm$ \phantom{0}27 & 0.00 & 
(92,196)       &      &         &      &         & 18 & (200)\\

MV Lyr          & NL\hspace{1em}VY &\phantom{0}3.190   & (164) & 
493 $\pm$ \phantom{0}12  & 0.00 & 
(196)          & 0.88 & (199)   & 0.42 &(148,164)& 12 & (164)\\

AQ Men          &                  &                   &       & 
552 $\pm$ \phantom{00}8 & 0.08 & 
(145)          &      &         &      &         &    &      \\

BT Mon          & N                &\phantom{0}8.01153 & (166) & 
1413 $\pm$ \phantom{0}97 & 0.21 & 
(151,152,204)  & 1.04 & (166)   & 0.84 & (166)   & 82 & (166)\\

KQ Mon          & NL               &\phantom{0}3.08    & (146) & 
618 $\pm$ \phantom{0}11 & 0.00 & 
(92)           & 0.60 & (214)   &      &         & 60 & (214)\\

RS Oph          & RN               &\phantom{0}10886   & (16)  & 
2137 $\pm$ \phantom{}252 & 0.76 & 
(31,46,168,173,& 1.35 & (64)    &      &         & 51 & (16) \\
                &                  &                   &       &
        &      & 
204,209,210)   &      &         &      &         &    &      \\

V380 Oph        & NL\hspace{1em}VY &\phantom{0}3.69857 & (138) & 
667 $\pm$ \phantom{0}33 & 0.18 & 
(145)          &      &         &      &         &    &      \\

V426 Oph        & DN\hspace{1em}Z  &\phantom{0}6.84754 & (70)  & 
191 $\pm$ \phantom{00}2 & 0.05 & 
(92,177,204)   & 0.90 & (70)    & 0.79 & (70)    & 59 & (70) \\

V442 Oph        & NL\hspace{1em}VY &\phantom{0}2.98392 & (41)  & 
492 $\pm$ \phantom{00}7 & 0.01 & 
(92,178)       &      &         &      &         &    &      \\

V841 Oph        & N                &          14.43130 & (127) & 
805 $\pm$ \phantom{0}18 & 0.45 & 
(45,132,151,152,& 0.90& (127)   & 1.00 & (42)    & 30 & (127)\\
                &                  &                   &       &
        &      & 
201,204)       &      &         &      &         &    &      \\

V2051 Oph       & DN\hspace{1em}SU &\phantom{0}1.49827 & (8)   & 
112 $\pm$ \phantom{00}1 & 0.00 & 
(92,196)       & 0.78 & (7)     & 0.19 & (7)     & 83 & (7)  \\

CN Ori          & DN\hspace{1em}UG &\phantom{0}3.91656 & (9)   & 
348 $\pm$ \phantom{00}5 & 0.00 & 
(92,196)       & 0.74 & (53)    & 0.49 & (53)    & 67 & (99) \\

CZ Ori          & DN\hspace{1em}UG &\phantom{0}5.2536  & (170) & 
490 $\pm$ \phantom{0}27 & 0.00 & 
(92)           & 0.67 & (170)   & 0.79 & (170)   & 24 & (170)\\

V1193 Ori       & NL\hspace{1em}UX &\phantom{0}3.96002 & (124) & 
797 $\pm$ \phantom{0}26 & 0.07 & 
(145)          &      &         &      &         & 79 & (63) \\

GS Pav          & NL\hspace{1em}UX &\phantom{0}3.72648 & (22)  & 
544 $\pm$ \phantom{0}16 & 0.04 & 
(145)          &      &         &      &         &    &      \\

V345 Pav        & NL\hspace{1em}UX &\phantom{0}4.75431 & (22)  & 
600 $\pm$ \phantom{}120 & 0.05 & 
(145)          &      &         &      &         &    &      \\

IP Peg          & DN\hspace{1em}UG &\phantom{0}3.79695 & (32)  & 
141 $\pm$ \phantom{00}1 & 0.00 & 
(92,204)       & 1.16 & (32)    & 0.48 & (32)    & 84 & (32) \\

LQ Peg          & NL\hspace{1em}VY &\phantom{0}2.99393 & (122) & 
1052 $\pm$ \phantom{0}60 & 0.05 & 
(145)          &      &         &      &         &    &      \\

TZ Per          & DN\hspace{1em}Z  &\phantom{0}6.30975 & (49)  & 
472 $\pm$ \phantom{00}6 & 0.27 & 
(66,85,92,93)  &      &         &      &         &    &      \\

GK Per          & N\phantom{L}\hspace{1em}IP  &          47.923   & (112) & 
437 $\pm$ \phantom{00}8 & 0.29 &
(132,152,204,217)& 0.87& (201)   & 0.55 & (112)   & 68 & (201)\\

KT Per          & DN\hspace{1em}UG &\phantom{0}3.90388 & (49)  & 
241 $\pm$ \phantom{00}3 & 0.18 & 
(66,92)        &      &         &      &         &    &      \\

RR Pic          & N                &\phantom{0}3.48062 & (198) & 
504 $\pm$ \phantom{00}8 & 0.01 & 
(55,151,152,   & 1.00 & (130)   & 0.10 & (130)   & 65 & (130)\\
                &                  &                   &       &
        &      & 
196,204)       &      &         &      &         &    &      \\

TY PsA          & DN\hspace{1em}SU &\phantom{0}2.02    & (206) & 
183 $\pm$ \phantom{00}2 & 0.00 & 
(175)          &      &         &      &         & 63 & (117)\\

VV Pup          & AM\hspace{1em}VY &\phantom{0}1.67390 & (202) & 
137 $\pm$ \phantom{00}1 & 0.00 & 
(92)           & 0.73 & (78)    & 0.14 & (78)    & 73 & (162)\\

CP Pup          & N                &\phantom{0}1.47034 & (14)  & 
795 $\pm$ \phantom{0}13 & 0.23 & 
(151,15,204)   &      &         &      &         &    &      \\

V348 Pup        & NL\hspace{1em}UX &\phantom{0}2.44413 & (140) & 
790 $\pm$ \phantom{0}21 & 0.12 & 
(145)          & 0.75 & (137)   & 0.31 & (140)   & 81 & (137)\\

T Pyx           & RN               &\phantom{0}1.82945 & (192) & 
2926 $\pm$ \phantom{}364 & 0.32 & 
(26,60,152)    & 1.25 &(143,159)& 0.10 & (87,192)& 15 & (125)\\

WZ Sge          & DN\hspace{1em}WZ &\phantom{0}1.36050 & (126) & 
45.1 $\pm$ \phantom{}0.1 & 0.00 
& (92,196,209)   &      &         &      &         & 76 & (163)\\

V3885 Sgr       & NL\hspace{1em}UX &\phantom{0}4.97186 & (131) & 
132 $\pm$ \phantom{00}1 & 0.02 & 
(92,195,203,204)& 0.70& (98)    & 0.68 & (98)    & 65 & (98) \\

V4140 Sgr       & DN\hspace{1em}SU &\phantom{0}1.47431 & (8)   & 
599 $\pm$ \phantom{0}62 & 0.11 & 
(145)          & 0.73 & (15)    & 0.13 & (15)    & 80 & (15) \\

V893 Sco        & DN\hspace{1em}SU &\phantom{0}1.82398 & (20)  & 
124 $\pm$ \phantom{00}1 & 0.02 & 
(145)          & 0.89 & (103)   & 0.20 & (103)   & 73 & (103)\\

VY Scl          & NL\hspace{1em}VY &\phantom{0}3.84    & (147) & 
630 $\pm$ \phantom{0}24 & 0.04 & 
(92,203,204)   &      &         &      &         & 15 & (147)\\

VZ Scl          & NL\hspace{1em}VY &\phantom{0}3.47    & (91)  & 
552 $\pm$ \phantom{0}18 & 0.00 & 
(92)           &      &         &      &         & 87 & (118,161)\\

LX Ser          & NL\hspace{1em}VY &\phantom{0}3.80238 & (94)  & 
486 $\pm$ \phantom{0}10 & 0.00 & 
(92,142)       & 0.41 & (100)   & 0.50 & (219)   & 79 & (100)\\

RW Sex          & NL\hspace{1em}UX &\phantom{0}5.8810  & (12)  & 
235 $\pm$ \phantom{00}5 & 0.00 & 
(92,196)       &      &         & 0.74 & (12)    & 34 & (12) \\

KK Tel          & DN\hspace{1em}SU &\phantom{0}2.02    & (77)  & 
611 $\pm$ \phantom{}180 & 0.03 & 
(145)          &      &         &      &         &    &      \\

RW Tri          & NL\hspace{1em}UX &\phantom{0}5.56520 & (134) & 
312 $\pm$ \phantom{00}5 & 0.10 & 
(92,142,196)   & 0.70 & (142)   & 0.86 & (142)   & 75 & (142)\\

EF Tuc          & DN\hspace{1em}UG &                   &       & 
1336 $\pm$ \phantom{0}41 & 0.02 & 
(145)          &      &         &      &         &    &      \\

SU UMa          & DN\hspace{1em}SU &\phantom{0}1.8327  & (184) & 
219 $\pm$ \phantom{00}1 & 0.00 & 
(92,174,196,204)&     &         &      &         &    &      \\

SW UMa          & DN\hspace{1em}SU &\phantom{0}1.3634  & (157) & 
162 $\pm$ \phantom{00}2 & 0.00 & 
(92)           &      &         &      &         &    &      \\

UX UMa          & NL\hspace{1em}UX &\phantom{0}4.72011 & (6)   & 
295 $\pm$ \phantom{00}2 & 0.00 & 
(92,142,196)   & 0.78 & (195)   & 0.60 & (195)   & 73 & (165)\\

IX Vel          & NL\hspace{1em}UX &\phantom{0}4.65420 & (11)  & 
90.3 $\pm$ \phantom{}0.2  & 0.01 & 
(56,92,196,204)& 0.80 & (11)    & 0.65 & (11)    & 57 & (96) \\
\hline
\end{tabular}
\end{table*}

\begin{table*}
\centering
{\bf Table B1} (continued)
\vspace{1em}

\begin{tabular}{lll@{\hspace{1ex}}lrr@{\hspace{1ex}}lr@{\hspace{1ex}}lr@{\hspace{1ex}}l
                 r@{\hspace{1ex}}l}
\hline
Star  & Type  
      & \multicolumn{2}{c}{$P_{\rm orb}$ (Ref)}   &
        \multicolumn{1}{c}{d}           & \multicolumn{2}{c}{$E_{B-V}$ (Ref)}   &
        \multicolumn{2}{c}{$M_1$ (Ref)} & \multicolumn{2}{c}{$q$      (Ref)}   &
        \multicolumn{2}{c}{$i$   (Ref)} \\
      &      
      & \multicolumn{2}{c}{(h)}   &
        \multicolumn{1}{c}{(pc)}        & \multicolumn{2}{c}{ }                &
        \multicolumn{2}{c}{($M_\odot$)}  & \multicolumn{2}{c}{ }                &
        \multicolumn{2}{c}{($^o$)}      \\
\hline
HV Vir          & DN\hspace{1em}WZ &\phantom{0}1.37004 & (82)  & 
351 $\pm$ \phantom{0}57 & 0.02 & 
(145)          &      &         &      &         &    &      \\
CTCV J2056-3014 & NL\hspace{1em}IP &\phantom{0}1.75    & (4)   & 
261 $\pm$ \phantom{00}7 & 0.01 & 
(145)          &      &         &      &         &    &      \\

EC 21178-5417   & NL               &\phantom{0}3.70262 & (22)  & 
529 $\pm$ \phantom{00}9 & 0.02 & 
(145)          &      &         &      &         & 83 & (221)\\

LS IV -08 3     & NL\hspace{1em}UX &\phantom{0}4.68699 & (25)  & 
210 $\pm$ \phantom{00}3 & 0.30 & 
(145)          &      &         &      &         &    &      \\
\hline									
\end{tabular}								
\par
{\it References:} 
(1)   \citet{Araujo-Betancor05C};                             
(2)   \citet{Arenas00C};                                      
(3)   \citet{Augusteijn94C};                                          
(4)   \citet{Augusteijn10C};                                  
(5)   \citet{Baptista94C};                                    
(6)   \citet{Baptista95C};                                    
(7)   \citet{Baptista98C};                                    
(8)   \citet{Baptista03C};                                    
(9)   \citet{Barrera89bC};                                    
(10)  \citet{Beuermann89C};                                
(11)  \citet{Beuermann90C};                                 
(12)  \citet{Beuermann92C};                                   
(13)  \citet{Bianchini80C};                                           
(14)  \citet{Bianchini12C};                                   
(15)  \citet{Borges05C};                                  
(16)  \citet{Brandi09C};                                      
(17)  \citet{Bruch82bC};                                              
(18)  \citet{Bruch00C};                                               
(19)  \citet{Bruch03C};                                               
(20)  \citet{Bruch14C};                                               
(21)  \citet{Bruch16C};                                               
(22)  \citet{Bruch17bC};                                              
(23)  \citet{Bruch18aC};                                              
(24)  \citet{Bruch20C};                                               
(25)  \citet{Bruch17dC};                                       
(26)  \citet{Bruch81C};                                       
(27)  \citet{Bruch97C};                                       
(28)  \citet{Buckley00C};                                     
(29)  \citet{Casares09C};                                     
(30)  \citet{Cassatella82C};                                  
(31)  \citet{Cassatella85C};                                  
(32)  \citet{Copperwheat10C};                                 
(33)  \citet{Costa09C};                
(34)  \citet{Dai09C};                                         
(35)  \citet{Dai10aC};                                        
(36)  \citet{Dai10bC};                                        
(37)  \citet{Dai09aC};    
(38)  \citet{Dai13C};                                         
(39)  \citet{Davey96C};                                      
(40)  \citet{Dhillon91C};                                     
(41)  \citet{Diaz01C};                                                
(42)  \citet{Diaz03C};                                     
(43)  \citet{Downes90C};                                              
(44)  \citet{Drew93C};                                        
(45)  \citet{Duerbeck80C};                                            
(46)  \citet{Duerbeck81C};                                            
(47)  \citet{Dulcin-Hacyan80C};                               
(48)  \citet{Echevarria02C};                                     
(49)  \citet{Echevarria99C};                             
(50)  \citet{Echevarria16C};                             
(51)  \citet{Feline05C};                                      
(52)  \citet{Ferland82C};                                     
(53)  \citet{Friend90C};                                      
(54)  \citet{Gaensicke09C};                                  
(55)  \citet{Gallagher74C};                                   
(56)  \citet{Garrison84C};                                    
(57)  \citet{Gilliland82aC};                                          
(58)  \citet{Gilliland82bC});                                          
(59)  \citet{Godon07C};                                       
(60)  \citet{Godon14C};                                       
(61)  \citet{Greiner99C};                                     
(62)  \citet{Greiner10C};                                     
(63)  \citet{Groot98C};                                       
(64)  \citet{Hachisu18C};                                     
(65)  \citet{Harlaftis92C};                                   
(66)  \citet{Hassall85C};                                             
(67)  \citet{Haswell97C};                                     
(68)  \citet{Hellier93C};                                             
(69)  \citet{Hellier91C};                                     
(70)  \citet{Hessman88C};                                             
(71)  \citet{Hessman84C};                                     
(72)  \citet{Hill17C};                                        
(73)  \citet{Honeycutt98aC};                              
(74)  \citet{Horne86C};                                       
(75)  \citet{Horne91C};                                       
(76)  \citet{Horne93C};                                       
(77)  \citet{Howell91C};                                      
(78)  \citet{Howell06C};                                      
(79)  \citet{Hutchings83C};                                   
(80)  \citet{Hutchings85C};                                   
(81)  \citet{Kaitchuck89C};                                            
(82)  \citet{Kato01C};                                        
(83)  \citet{Kato14C};                                        
(84)  \citet{Klare82C};                                       
(85)  \citet{Klare82C};                                       
(86)  \citet{Knigge06C};                                              
(87)  \citet{Knigge00C};                                      
(88)  \citet{Krautter81aC};                                   
(89)  \citet{Krautter81bC};                                   
(90)  \citet{Kuerster88C};                                 
(91)  \citet{Krzeminski66C};                                          
(92)  \citet{laDous91C};                                             
(93)  \citet{laDous85C};                                     
(94)  \citet{Li17C};                                          
(95)  \citet{Lines87C};                                       
(96)  \citet{Linnell07C};                                     
(97)  \citet{Linnell08bC};                                     
(98)  \citet{Linnell09C};                                     
(99)  \citet{Mantel87C};                                      
(100) \citet{Marin07C};                                       
(101) \citet{Marino84C};                                    
(102) \citet{Marsh90C};                                       
(103) \citet{Mason01C};                                       
(104) \citet{Mateo85C};                                       
(105) \citet{McAllister19C};                                  
(106) \citet{McLaughlin60C};                                          
(107) \citet{Mennickent98C};                               
(108) \citet{Mennickent99C};                                  
(109) \citet{Mennickent06C};                                  
(110) \citet{Miroshnichenko88C};                                      
(111) \citet{Mizusawa10C};                                    
(112) \citet{Morales-Rueda02C};                               
(113) \citet{Mouchet81C};                                     
(114) \citet{Moyer03C};                                       
(115) \citet{Nadalin01C};                                     
(116) \citet{North00C};                                       
(117) \citet{ODonoghue92C};                             
(118) \citet{ODonoghue87C};                                  
(119) \citet{Oliveira04C};                                 
(120) \citet{Oliveira14C};                                    
(121) \citet{Panek84C};                                       
(122) \citet{Papadaki06C};                                    
(123) \citet{Papadaki06C};                                   
(124) \citet{Papadaki09C};                                    
(125) \citet{Patterson98C};                                   
(126) \citet{Patterson18C};                                   
(127) \citet{Peters06C};                               
(128) \citet{Qian13C};                                        
(129) \citet{Rayne81C};                                     
(130) \citet{Ribeiro06C};                                     
(131) \citet{Ribeiro07C};                                     
(132) \citet{Ringwald96aC};                                    
(133) \citet{Ritter03C};                                      
(134) \citet{Robinson91C};                                    
(135) \citet{Rodrigues98C};                                   
(136) \citet{Rodriguez-Gil02C};            
(137) \citet{Rodriguez-Gil01C};                          
(138) \citet{Rodriguez-Gil07C};                         
(139) \citet{Rogoziecki01C};                  
(140) \citet{Rolfe00C};                                       
(141) \citet{Rosen87C};                                       
(142) \citet{Rutten92C};                                      
(143) \citet{Schaefer10aC};                                    
(144) \citet{Schild69C};                                              
(145) \citet{Capitano17C};                              
(146) \citet{Schmidtobreick05C};                              
(147) \citet{Schmidtobreick18C};                              
(148) \citet{Schneider81C};                                   
(149) \citet{Schoembs81bC};                                    
(150) \citet{Schwope93C};                                     
(151) \citet{Selvelli04C};                                            
(152) \citet{Selvelli13C};                                
(153) \citet{Shafter83aC};                                            
(154) \citet{Shafter83bC};                                            
(155) \citet{Shafter86aC};                                 
(156) \citet{Shafter88C};                                  
(157) \citet{Shafter86bC};                                     
(158) \citet{Shafter90C};                                     
(159) \citet{Shara18C};                                       
(160) \citet{Sheets07C};                                      
(161) \citet{Sherrington84C};                                
(162) \citet{Sirk98C};                                      
(163) \citet{Skidmore02C};                                    
(164) \citet{Skillman95C};                                    
(165) \citet{Smak94C};                                                
(166) \citet{Smith98C};                                       
(167) \citet{Smith06C};                                       
(168) \citet{Snijders87C};                                           
(169) \citet{Southwell95C};                                   
(170) \citet{Spogli94C};                                    
(171) \citet{Stanishev04C};                                   
(172) \citet{Sterken07C};                                     
(173) \citet{Svolopoulos66C};                                         
(174) \citet{Szkody81aC};                                             
(175) \citet{Szkody85C};                                              
(176) \citet{Szkody81bC};                                     
(177) \citet{Szkody88C};                                     
(178) \citet{Szkody83C};                                   
(179) \citet{Szkody84C};                                      
(180) \citet{Tappert97C};                                     
(181) \citet{Taylor96C};                               
(182) \citet{Thoroughgood04C};                                
(183) \citet{Thorstensen00C};                               
(184) \citet{Thorstensen86C};                                 
(185) \citet{Thorstensen96C};                                 
(186) \citet{Thorstensen02C};                                 
(187) \citet{Thorstensen97C};                                
(188) \citet{Udalski88C};                                
(189) \citet{Uemura01C};                                      
(190) \citet{Urban06C};                                       
(191) \citet{Urban06C};                                       
(192) \citet{Uthas10C};                                       
(193) \citet{vanParadijs94C};                                 
(194) \citet{vanSpaandonk10C};                               
(195) \citet{VandePutte03C};                                 
(196) \citet{Verbunt87C};                                             
(197) \citet{Verbunt84C};                                     
(198) \citet{Vogt17C};                                        
(199) \citet{Voikhanskaya88C};                                        
(200) \citet{Voikhanskaya96C};                                        
(201) \citet{Wada18C};                                        
(202) \citet{Walker65C};                                              
(203) \citet{Warner76aC};                                              
(204) \citet{Warner87C};                                              
(205) \citet{Warner95C};                                              
(206) \citet{Warner89C};                                      
(207) \citet{Watson03C};                                      
(208) \citet{Watson07C};                                      
(209) \citet{Webbink78C};                                             
(210) \citet{Weight94C};                                      
(211) \citet{Weil18C};                                        
(212) \citet{Welsh07C};                                       
(213) \citet{Williger88C};                                    
(214) \citet{Wolfe13C};                                       
(215) \citet{Wolff99C};                                       
(216) \citet{Wood89C};                                        
(217) \citet{Wu89C};                                          
(218) \citet{Wu02C};                                           
(219) \citet{Young81bC};                                      
(220) \citet{Zhang87C};                                 
(221) \citet{Khangale20C}.
\end{table*}                      

The classification of a CV into types and subtypes is sometimes ambiguous.
Here I restrict myself to assign a principal type and at most one subtype to 
each object, using the usual nomenclature: N = nova, NR = recurrent nova, NL =
novalike variable (subtypes: UX = UX~UMa star, VY = VY~Scl star), DN =
dwarf nova (subtypes: UG = U~Gem star, Z = Z~Cam star, SU = SU~UMa star,
WZ = WZ~Sge star) and the magnetic types AM = AM~Her star (or polar) and
IP = intermediate polar.

The orbital period is often known to a high accuracy. Here, I restrict
myself to a maximum of five decimals. More accurate values can be found
in the quoted references. But there are still systems for which this
most fundamental of all binary parameters is only known with a much
lower precision or not at all. In two cases (V725~Aql and AY~Lyr) the
period given here is implied from the measured superhump period, using
the Schoembs-Stolz relation \citep{Schoembs81aC} as updated by
\citet{Gaensicke09C}, noting that for the purpose of the present
study the period needs not be known with a high accuracy. 

One of the greatest problems when it comes to system parameters, namely the 
distance of the objects which is essential to calculate properties such as 
absolute magnitudes and luminosities, is largely resolved since the 2$^{nd}$ 
Gaia data release became available. The distances quoted in 
Table~\ref{Table: Final object parameters} 
are in most cases taken from \citet{Bailer-Jones18C}, based on the 
Gaia parallaxes.
Three objects have no Gaia parallax: V725~Aql, V504~Cen and V345~Pav. For
V725~Aql no independent distance estimate is available. \citet{Greiner10C}
quote a distance of 600~pc for V504~Cen. The same distance is estimated for
V345~Pav by \citet{Buckley92C}.
A Gaia parallax is available for T~Aur, but this
star is not in the list of \citet{Bailer-Jones18C}. For simplicity, the
inverse of the parallax is adopted here for its distance. As pointed out
by \citet{Schaefer18C} the Gaia parallaxes of recurrent novae with a giant
companion are unreliable because the wobble induced by the motion of the
stellar components around their centre of mass cannot correctly be accounted
for in Gaia DR2. Here, this effects T~CrB and RS~Oph. Even so, the Gaia
distances are quoted in 
Table~\ref{Table: Final object parameters}, noting
that these values are not used in the analysis of flickering in these systems.

The interstellar extinction, parameterized by the reddening parameter
$E_{B-V}$ has been determined by various authors for many of the systems 
investigated here. The adopted values are listed in 
Table~B1 were taken from the references
quoted in parenthesis. If more than one extinction determination is available
the average is taken\footnote{Note that the values used to calculate the
average may not always constitute independent measurements.}, rejecting,
however, strongly deviating values (and thus possibly introducing a personal
bias). For a considerable number of objects no specific extinction measurements
could be found. In those cases it was estimated using the 
three-dimensional extinction maps of \citet{Capitano17C}, adopting the
distances as derived from the Gaia parallaxes.

Mass estimates for CVs are difficult, and frequently contradicting values are 
published in the literature. The same is true for the orbital inclination. 
This makes a choice among the quoted values often ambiguous and subjective. 
The primary star mass $M_1$, the mass ratio $q$ and the
inclination $i$ listed in the table are therefore not free from a
certain personal bias of the present author. Moreover, they are 
sometimes not taken directly form the mentioned references but
inferred from other information taken from these.

Only for the distance errors are given in the table, defined here as 
the average of 
the differences between the nominal distance value and the upper and lower 
limits of the confidence intervals quoted by \citet{Bailer-Jones18C}. 
For the orbital period the errors are in most 
cases much too small to be of any consequence in the 
context of this study (and often smaller than the limited accuracy of the
table entries). The errors of the dynamical and geometrical system parameters
are probably only to a small degree random, but are dominated by systematic 
effects caused by the multiple assumption normally involved in their derivation.
Therefore, quoting formal errors may be misleading. The reader should
exercise the necessary caution when using the table values.

Subsequently, a brief description of each of the stars is given, focussing 
on the observational history in the optical and providing references in
particular on basic dynamical and geometrical system parameters (orbital
period, component masses and orbital inclination) without trying to discuss 
in depth conflicting values quoted in the literature. Whenever flickering
properties of a particular system have been investigated in the past, these
are also mentioned [refraining, however, from mentioning explicitely the
work of \citet{Fritz98C} on wavelet properties of the flickering in a large
ensemble of CVs, most of which are also studied here]. Finally, the 
observations used in the present study are also briefly summarized for each
star.

{\it RX~And} 
is a Z~Cam type dwarf nova. The orbital period has been measured by
\citet{Kaitchuck89C}, while \citet{Shafter83aC} derived the inclination.
The system is very active, and never remains for a long time in quiescence 
\citep[][see also the AAVSO long term light curve]{Glasby70C}. 
Consequently, the available light curves cover apart from 
quiescence a wide range of outburst phases, including maxima, rise 
and decline phases, and a standstill. They consist of two high time 
resolution white light curves, one run in $UBV$ \citep[also used in the
flickering study of][]{Bruch92}, and a number of 
lower time resolution light curves found in the AAVSO International
Database\footnote{https://www.aavso.org}. The white 
light data were already analysed with respect to rapid oscillations by 
\citet{Patterson81C}.

{\it AR~And} is a little studied dwarf nova with an orbital period 
measured spectroscopically by \citet{Taylor96C}.
That paper, together with the spectroscopic study of \citet{Shafter95C}
appears to be the only somewhat detailed optical investigations of this
system. Thus, not much is known about AR~And. In particular, the component 
masses and the orbital inclination remain unknown. Since no light curves 
spanning the entire orbital period are available, it cannot even be said if 
AR~And is eclipsing or not. However, as the AAVSO long term light curve reveals,
it is a very active systems with frequent outbursts in fairly regular 
intervals of approximately 20 days. 
Not many data are available for the present study. They consist of a
single high time resolution unfiltered light curve provided
by R.E.\ Nather, complemented by some $V$ band light curves retrieved from the
AAVSO International Database.

{\it DX~And}, discovered by \citet{Romano58C}, is another little studied dwarf 
nova. The long term light curve was investigated by \citet{Simon00C} who noted 
a long recurrence time between the outburst. The first optical spectrum was 
observed by \citet{Bruch89C}. The red continuum and traces of absorption lines
suggested a strong contribution of the secondary star and a long orbital 
period. This was confirmed by two more detailed studies of the optical
spectrum by \citet{Drew93C} and \citet{Bruch97C}. The former authors
measured the mass ratio, restricted the range of possible orbital inclinations
and found an upper limit for the the secondary star mass. \citet{Bruch97C} 
derived the orbital period and quantified the significant veiling factor
due to the secondary star (see Appendix~B).
The data of DX~And available for the current study are quite scarce. They
consist of only three $V$ band light curves retrieved from the MEDUZA archive
\footnote{http://var2.astro.cz/EN/meduza/index.php}.

{\it V455~And} was discovered in the context of the Hamburg Quasar Survey 
\citep{Hagen95C} as HS2331+3905. It is a DQ~Her type intermediate polar 
with a short orbital period 
\citep{Araujo-Betancor05C, Bruch20C}. It is also a dwarf nova of
WZ~Sge subtype with only one outburst having been observed so far 
\citep{Nogami09C, Matsui09C}. The orbital light curve is double
humped, reflecting probably the presence of two accreting magnetic poles
on the white dwarf and contains a shallow grazing eclipse. Moreover, 
V455~And exhibits a permanent negative superhump 
\citep{Kozhevnikov15C, Bruch20C}. A puzzling
feature is a 3.5~h spectroscopic period detected by \citet{Araujo-Betancor05C}
which is in no way related to the orbital period. Dynamical and 
geometrical parameters such as the component masses and the orbital 
inclinations have not yet been determined.
In the high frequency regime several periods were first observed
by \citet{Araujo-Betancor05C} and later studied in more detail by other 
authors. Signals on time scales of 5 -- 6~min which change slightly from 
night to night are interpreted as non-radial pulsations of the white 
dwarf. A coherent 1.12~min signal is thought to be the rotation period of
the primary star. A precise spin period 
has been measured by \citet{Mukadam16C}. Closely related is another, 
slightly shorter and drifting period, first mentioned by \citet{Gaensicke07C}
who interpreted it as being due to the illumination by the rotating white
dwarf of a warped inner accretion disk which precesses retrogradely. 
The observational data used for this study were all retrieved from the 
MEDUZA archive and the AAVSO International Database.

{\it V704~And} (= LD~317) was discovered as a blue variable star by 
\citet{Dahlmark99C}. It appears in the \citet{Downes01C} catalogue as
a possible CV. The first more detailed observations were published by
\citet{Papadaki06C} who, based on different brightness levels seen
in different epochs, classified the star as a novalike variable
of the VY~Scl type. This was confirmed by \citet{Weil18C}
spectroscopically and through an analysis of the long term light curve.
\citet{Weil18C} also measured the orbital period.
For this study I used the (unfiltered) light curves of \citet{Papadaki06C}.
There are complemented by many $V$ band light curves retrieve from the AAVSO 
International Database. All these data refer to the high state or at most 
slightly below.

{\it UU~Aqr} is an eclipsing novalike variable. Although known as a variable
star for a long time \citep{Beljawsky26C} it was identified as a CV 
only in 1986 by \citet{Volkov86C}. The most accurate
values for the orbital period was measured by \citet{Baptista94C}. Using
measurements of the eclipse contact phases these authors also derived
geometrical and dynamical system parameters.
Alternative values, based on spectroscopy, were derived by \citet{Diaz91aC}.
Other detailed spectroscopic studies include \citet{Kaitchuck98C}
[who found a significantly different mass ratio compared to that of 
\citet{Baptista94C}; a discrepancy explained by \citet{Baptista08C}],
\citet{Hoard98C} and \citet{Baptista00C}.
The photometric variability of UU~Aqr is not restricted to eclipses. On the
time scale of years variations of some tenths of a magnitudes are observed
\citep{Baptista94C}, interpreted as being caused by variations of the
mass transfer rate from the secondary. Short outbursts with an amplitude of
$\sim$1~mag (``stunted'' dwarf nova outbursts) occur on time scales of 
days \citep{Honeycutt98bC}. Superhumps were 
observed by \citet{Patterson05C} but were absent in extensive
photometry of \citet{Bruch19aC}.
Flickering is quite strong in UU~Aqr and has been the subject of dedicated
studies. \citet{Dobrotka12C} performed
a comparison of the observed power spectrum with the predictions of a 
statistical model simulating flickering caused by turbulent angular
momentum transport in the accretion disk \citep{Dobrotka10C}.
While the flickering observed in KR~Aur is consistent with this model
this is not so for UU~Aqr. The authors speculate that spiral waves in 
the accretion disk \citep{Baptista08C} are the culprit.
The present study is based on light curves retrieved from the
OPD data bank\footnote{https://lnapadrao.lna.br/OPD/databank/databank} 
and the AAVSO International Database, complemented by
the white light light curves already used by \citet{Bruch19aC} in a different
context. 

{\it VY~Aqr} is a dwarf nova of SU~UMa type and has thus a short orbital period,
measured spectroscopically by \citet{Thorstensen97C}. It may have evolved beyond
the orbital period limit \citep[``period bouncer'',][]{Mennickent02C}. 
The early history of the
system is summarized by \citet{Patterson93C}. 
The system parameters are quite uncertain. The values listed in Table~C1 
were taken from 
\citet{Augusteijn94C}, \citet{Mennickent06C} and \citet{Urban06C}.
Properties of the secondary star are discussed by \citet{Mennickent02C},
\citet{Harrison09C} and \citet{Littlefair00C}.
The amount of data available for the present study is quite limited and not
of high quality. They were all retrieved from the OPD data bank, the AAVSO
International Database and the MEDUZA archive.

{\it HL~Aqr} (= PHL~227) was identified as a novalike variable by 
\citet{Hunger85C}. Despite its high apparent magnitude of $\sim$13.5~mag
it remained comparatively little studied. The orbital period, measured
spectroscopically and originally determined by \citet{Haefner87C}
was later refined by \citet{Rodriguez-Gil07C}. It lies in the range of 
the SW~Sex stars and HL~Aqr is indeed identified as such. The system is
seen almost face-on. \citet{Rodriguez-Gil07C} determined an orbital
inclination in the range $19^{\rm o} < i < 27^{\rm o}$.
Component masses are not known. 
The observational data available for this study consist of 5 Walraven
light curves obtained by \citet{Hollander93C}.

{\it CZ Aql} was discovered as a 
variable star by \citet{Reinmuth25C} who did not provide a classification. 
Based on spectroscopic evidence \citet{Cieslinski98C} considered the star 
to be a dwarf nova. \citet{Sheets07C} measured a spectroscopic orbital 
period of 4.812~h but cannot exclude the true period to be an alias of 
this value. They also speculate about a magnetic nature of the system. 
\citet{Bruch17aC} discusses photometry of CZ~Aql and found in the 
power spectra of his light curves indications of a photometric
period $\sim$8\% longer than the orbital 
period. This might be due to a superhump in the system. Indeed, choosing
a close-by alias (nearly as strong in the power spectrum), the Schoembs-Stolz 
relation between orbital and superhump periods as revised by 
\citet{Gaensicke09C} is obeyed to within $1.2 \sigma$. No information 
about geometrical and dynamical system parameters exists.
The data available for the present study consist of the unfiltered high
time resolution light curves taken from \citet{Bruch17aC}
and several more data sets with lower time resolution retrieved from 
the OPD data bank.

{\it DH~Aql} was identified as a dwarf nova by \citet{Tsessevich69C} and 
found to be of SU~UMa subtype by \citet{Nogami95aC}. Its nature was confirmed
spectoscopically by \citet{Mason03C}. Apart from long-term monitoring
by \citet{Bateson82aC} it remains little studied. The orbital period used
here is deduced from the superhump period \citep{Kato09C} together with the
revised Schoembs-Stolz relation of \citet{Gaensicke09C}. System masses are
unknown, as is the orbital inclination. The absence of eclipses in the
light curve excludes a high inclination. 
All data used in the present study were taken during the July-August 2002
superoutburst of DH~Aql and were retrieve from the
AAVSO International Database.

{\it V603~Aql:} Among all known classical novae V603~Aql has the highest 
apparent magnitude in quiescence. As such, it is very well studied and the 
amount of literature on the object is vast. V603~Aql erupted in 1918 and 
reached quiescence in 1937 \citep{Strope10C}. From then on it exhibited 
a slight secular fading to a mean quiescent magnitude of 
$\sim$11.8~mag in recent years \citep{Johnson14C}.
The orbital period was first determined spectroscopically by \citet{Kraft64aC}
and later verified by other authors. \citet{Peters06C} were able to combine
their own radial velocity measurements with results from previous studies
to derive long term ephemeries. \citet{Arenas00C} measured the component 
masses and the orbital inclination.
V603~Aql exhibits a puzzling multitude of photometric periods none of which
is identical to the orbital period \citep{Haefner85C, Drechsel83C,
Udalski89C, Schwarzenberg-Czerny92C, Bruch91aC}.
The only consistent modulation present in the light curve 
has a slightly variable period close to 3.5~h, about
6\% longer than the orbital period. It has
first been detected by \citet{Haefner81C} and then be confirmed by many 
authors \citep[see][]{Bruch18bC}. Additionally, \citet{Patterson97C}
observed a modulation with a period being about 3\% less than the
orbital period. Today, the longer of these  variations is interpreted as a
permanent superhump in V603~Aql, while the shorter one is considered to be
an occasionally visible negative superhump. 
An analysis of the flickering in a very limited sample of light curves of 
V603~Aql has been performed by \citet{Bruch92C}. His data are part of the 
vastly larger sample data investigated here which, apart from many light
curves retrieved from the AAVSO International Database include multicolour
light curves obtained by the author and his collaborators, Walraven light 
curves observed by \citet{Hollander93C}, and extremely high time resolution
UBVR* light curves provided by E.\ Robinson. The latter were used
by \citet{Bruch15C} in a study of time lags of the flickering as a function of
wavelength. Flickering flares in V603~Aql clearly occur slightly earlier at
shorter optical wavelength compared to longer wavelength. This time lag
amounts to $(5.0 \pm 0.3) \times 10^{-3}$ sec/{\AA}. 

{\it V725~Aql:} After some confusion about the identity of V725~Aql, 
\citet{Nogami95bC} identified the system unambiguously and classified 
it as a dwarf nova. Based on superoutburst observations, \citet{Uemura01C} 
revealed its nature as a SU~UMa star. 
The superhump period suggests it to be located within the 2 -- 3 h period gap
of CVs. In fact, the superhump period implies an orbital period of 2.2757~h,
applying the Schoembs-Stolz relation revised by \citet{Gaensicke09C}.
Nothing is known about the component masses and the orbital inclination.
The absence of eclipses in the light curve indicates that the latter is 
not high. The distance of V725~Aql is also unknown. The system has no
parallax measured by Gaia DR2.
The data used in the present study were retrieved from the
AAVSO International Database.

{\it V794~Aql} is a novalike variable of the VY~Scl subtype. The long term
light curve has been monitored for many years \citep{Honeycutt85C,
Honeycutt94C, Honeycutt98aC, Honeycutt04C}. The orbital period was 
determined spectroscopically by \citet{Honeycutt98aC}. 
Dynamical and geometrical system parameters are only known
from modelling of the far UV spectrum by \citet{Godon07C}.
The system was included in a specific study of the flickering in some
accreting white dwarf systems by \citet{Zamanov16C}. They found that
the amplitude (measured as the flux difference between the maximum and the
minimum in a light curve) as well as the rms of all investigated systems 
follows a common linear
relationship with the average system brightness, when plotted on a logarithmic
scale. The present study includes only one unfiltered light curve, observed 
with high time resolution by the author, plus two light curves
retrieved from the AAVSO International Database.

{\it V1315~Aql} was first detected as a suspected variable star 
by \citet{Metik61C} and classified as an eclipsing CV by \citet{Downes86C}. 
It is now considered to be a novalike variable of the SW~Sex subtype 
\citep{Szkody90C}. Detailed studies have also been performed by 
\citet{Smith93C}, \citet{Dhillon91C} and \citet{Hellier96C}. Recently, 
\citet{Sahman18C} discovered a shell around the system, pointing at a 
nova outburst roughly 1000 years ago.
The orbital period has last been updated by \citet{Papadaki09C}. 
The few data available for the present study were retrieve from the
OPD data bank. 

{\it AT~Ara:} This dwarf nova was
originally suspected to be a SU~UMa star \citep{Warner76bC} and as such
would have a short orbital period. However, the long-term photometric
behaviour studied by \citet{Bateson78C} does not support this hypothesis
since the system does not show the typical dichotomy of normal and 
superoutburst which characterize the SU~UMa stars. Moreover, the meager
spectroscopic information available about AT~Ara points at a long orbital
period: \citet{Vogt76C} mentions possible G-band absorption, and a
low resolution spectrum of \citet{Bruch82aC} clearly shows the 
Mg~I $\lambda$ 5167--5184~{\AA} triplet and Ca~I $\lambda$ 5262--5270~{\AA}
in absorption, indicating a considerable contribution of the secondary
star to the optical light which is further detailed in Appendix~B.
This is confirmed by the so far only detailed study
of AT~Ara, presented by \citet{Bruch03C}.
It clearly shows the system to be a double lined spectroscopic
binary. It has a long orbital period of 8.9976~h. \citet{Bruch03C} also
determined dynamcial and geometrical system parameters.
The data available for this study consist of the light curves taken by
\citet{Bruch03C}, complemented by unpublished data.

{\it TT~Ari} is a well known VY~Scl-type novalike variable and one of the 
brightest CVs in the sky. As such, it has been extensively
studied spectroscopically and photometrically from infra-red to X-ray
wavelengths. The orbital period was
originally measured by \citet{Cowley75C} and later refined, last by
\citet{Wu02C}, who also determined the dynamical and geometrical
properties used here.
Many photometric investigations have been published about TT~Ari. The
light curve shows modulations which, however, are notoriously unstable.
They may be somewhat longer or shorter than the orbital period and can be 
interpreted as positive an negative superhumps. Additionally, numerous 
authors report variations on the time scales of some tens of minutes
\citep[see][for a thorough discussion and further references]{Bruch19bC}. 
Some papers dealing with flickering properties of TT~Ari are 
relevant in the present context. \citet{Kraicheva99bC} observed 
variations in the flickering activity on the time scale of a year. They 
also relate the $\approx$20~min variations consistently present in the
light curves to flickering and perform an
investigation of the flickering power spectrum. This was also done by
\citet{Belova13C}. Just as V603~Aql, TT~Ari exhibits a clear time lag
of the flickering activity as a function of wavelength, albeit at a smaller
level of $(1.5 \pm 0.1) \times 10^{-3}$ sec/{\AA} \citep{Bruch15C}.
For the current study I use a vast number of light curves from many different
sources. They were already investigated in another context by \citet{Bruch19bC}.
I restrict myself to observations performed during the high state of TT~Ari.

{\it T~Aur} is a classical nova the outburst of which ocurred in 1891. 
\citet{Walker63C} detected shallow partial eclipses in its light curve and 
first measured the orbital period, which was last refined by 
\citet{Dai10aC}. Surprisingly
little work has been published on the detailed optical characteristics of
the central binary star in the system. The only attempt to determine masses
and the orbital inclination was performed by \citet{Bianchini80C}.
The presently available data consist of some high time resolution
light curves provided by R.E.\ Nather, supplemented by several light
curves retrieved from the AAVSO International Database.

{\it KR~Aur} is a well known novalike variable of the VY~Scl subtype. The long
term behaviour has been extensively monitored in the literature 
\citep[see, e.g.,][]{Honeycutt04C}. The early history of the system was 
summarized by \citet{Kato02C}. The orbital period has been measured 
spectroscopically by \citet{Hutchings83C} and \citet{Shafter83bC} and
photometrically by \citet{Rodriguez-Gil20C}. The latter authors also 
provide a geometrical and dynamical model of the system, quantifying
respective parameters.
Apart from the frequent low states which characterize KR~Aur as a VY~Scl
star, the system exhibits strong variability also on short time scales.
While no variability on the orbital period has been reported, 
\citet{Kozhevnikov07C} claims the presence of negative superhumps which,
however, were not detected by \citet{Kato02C}. Significant
signals with unstable periods on the time scale of several hundred seconds
have been seen by \citet{Singh93C} et al.\ (1993) and \citet{Kato02C}.
\citet{Biryukov90C} claim the presence of 25~min variations, but these 
are quite unstable and can at most be classified as QPOs (Quasi Periodic
Oscillations).
Flickering is strong in KR~Aur and has been subject of several studies.
\citet{Kato02C} investigate the power spectrum of their light curves.
A more detailed study has been performed by \citet{Zamanov16C}
who included KR~Aur in their study of the amplitude-flux relation (see
V794~Aql). \citet{Dobrotka12C}, performed
a comparison of the observed power spectrum with the predictions of a 
statistical model simulating flickering caused by turbulent angular
momentum transport in the accretion disk \citep{Dobrotka10C}
and found agreement. They concluded that flickering is concentrated in the
inner disk regions.
The present study includes only two unfiltered high state light curves of 
KR~Aur, provided by R.E.\ Nather, which are, however, of very high quality. 

{\it V363~Aur} was detected as Lanning~10 in a survey of UV bright objects
\citep{Lanning73C}. It was suggested to be a CV by \citet{Margon81C},
and eclipses were detected by \citet{Lanning81C}. Today, it is
classified as a novalike variable of the SW~Sex subclass. The orbital period,
last updated by \citet{Thoroughgood04C}, is comparatively long 
for a CV. It is therefore not surprising that the spectral features of the 
secondary star are readily visible in the optical spectrum and V363~Aur thus 
constitutes one of the few eclipsing double-lined variables among CVs. This 
enable to derive masses with a minimum of assumptions. These, and the orbital 
inclination were measured by \citet{Schlegel86C} and \citet{Thoroughgood04C}.
Flickering in V363~Aur has been commented on by \citet{Horne82C}.
The present study draws on light curves provided by R.E.\ Nather. 

{\it SY~Cnc} is a very active dwarf nova which hardly remains in quiescence for
more than a handful of days. It is also classified by \citet{Simonsen14bC}
as a bona-fide Z~Cam star. Its orbital period, based on spectroscopic 
observations taken over several years \citep{Casares09C}, is quite long 
compared to most CVs. Dynamical and geometric
system parameters are discussed by \citet{VandePutte03C}, \citet{Smith05C}
and \citet{Shafter83aC}. At the long orbital period the secondary star 
leaves considerable traces in the optical spectrum (for more details, see
Appendix~B).
Flickering in SY~Cnc was studied by \citet{Middleditch82C} during
outburst in three photometric bands between the UV and the red spectral
range. They find that the spectrum of the flickering rises too rapidly
towards shorter wavelength to be consistent with any simple thermal model.
Moreover, they did not find a time lag larger than a fraction of a second 
between flickering in different passbands.
The data used for the present study consist of 14 unfiltered high time
resolution light curves provided by R.E.\ Nather, encompassing
all outburst stages. They were already analysed in a different context
by \citet{Patterson81C}. These data are complemented by further light curves 
retrieved from the AAVSO International Database.

{\it YZ~Cnc} is a SU~UMa type dwarf nova with an orbital period of just over
2~h, first measured by \citet{Shafter88C} and later refined
by \citet{vanParadijs94C}. Component masses were discussed by
\citet{Shafter88C} and \citet{Knigge06C}.
Early on it was perceived that YZ~Cnc exhibits very strong flickering during
quiescence. \citet{Moffett74C} observed a typical peak-to-peak amplitude
of 0.75~mag, sometimes even reaching 1.1~mag. However, this 
is not borne out by the presently studied data, where the peak-to-peak
amplitude is more of the order of 0.5~mag. The system
was one of the first CVs to be subjected to a dedicated study of its
flickering activity. \citet{Elsworth86C} observed the star with
subsecond time resolution and calculated power spectra of the light
curves. Some of these exhibit the typical $P \propto f^{-\alpha}$ behaviour
of red noise, where $P$ is the power, $f$ the frequency and $\alpha$ a
parameter which describes the power distribution between high and low 
frequencies. The majority, however, has a shape which \citet{Elsworth86C}
interpret as being due to diffusion in a scattering cloud of an
immersed flickering light source.
For the present study I used a dozen high time resolution 
light curves provided by R.E.\ Nather, observed in various
outburst and superoutburst stages, and many more light curves
retrieved from the AAVSO data base covering all phases of the outburst cycle.

{\it AC Cnc:} This novalike variable  was detected as a variable star by 
\citet{Kurochkin60C}. \citet{Kurochkin80C}
first saw eclipses in the system and determined the
orbital period, which was last refined by \citet{Thoroughgood04C}. 
The system belongs to the long period CVs. Slight period changes 
have been analysed by \citet{Qian07C}. Soon after the detection of eclipses
the nature of AC~Cnc as a CV was discovered in spectroscopic observations by 
\citet{Okazaki82C}, and in the same year \citet{Downes82C} detected the
secondary star in the spectrum, making AC~Cnc one of the few eclipsing
double-lined variable among CVs and thus enabling to measure the masses
with a minimum of assumptions \citep[see][]{Schlegel86C, Thoroughgood04C}.
The present study draws on light curves of AC~Cnc retrieved from
the AAVSO International Database. 

{\it OY~Car:} This eclipsing system is one of the best studied SU~UMa type 
dwarf novae. Discovered by \citet{Hoffmeister63C}, it has been the object of 
intensived photometric and spectroscopic studies from the infrared to X-rays 
in particular in the last two decades of the 20$^{\rm th}$ century. 
Its orbital period was measured by \citet{Cook85C}, \citet{Schoembs87C} and
\citet{Wood89C}.
\citet{Greenhill06C} found slight deviations from purely linear ephemeris
due either secular or cyclic period variations. Basic dynamical and geometrical
properties were derived on the basis of a detailed analysis of the eclipse 
properties by \citet{Schoembs87C}, \citet{Wood89C} and \citet{McAllister19C}. 
The data analysed in the present study consist of the $UBV$ and white light
observations of \citet{Bruch96aC}, some unpublished $B$ light curves 
(courtesy B.\ Borges), $B$ light curves of \citet{Schlindwein17C} and 
$V$ light curves retrieved from the AAVSO International Database.

{\it QU~Car} was discovered as an irregular variable by \citet{Stephenson68C}
and classified as a cataclysmic variable by \citet{Schild69C}. At an average
magnitude slightly fainter than 11~mag it is one of the brightest
CVs in the sky. Currently considered to be a novalike variable,
\citet{Kafka08C} also proposed a classification as V~Sge star, i.e., 
a system with characteristics similar to supersoft X-ray sources
\citep{Steiner98C}.
The system parameters of QU~Car are not well known. The orbital period
was first determined spectroscopically by \citet{Gilliland82C}
as 10.9~h. This period was later confirmed by \citet{Oliveira14C}
who also find a photometric modulation with the same period, but
only when the flickering activity is reduced and the system hovers
at a slightly lower magnitude level than normal. Modelling the ultraviolet 
spectrum \citet{Linnell08aC} put limits on the mass of the primary star and
the orbital inclination. The failure of \citet{Gilliland82C} to detect 
absorption lines of the secondary star (expected at the long orbital period)
may be due to the exceptional high luminosity of the system \citep{Drew03C}.
The data of QU Car used here consist of light curves taken from 
\citet{Bruch91bC} and others retrieved from the AAVSO International 
Database and the OPD data bank.

{\it HT~Cas} was detected and classified as a dwarf nova by 
\citet{Hoffmeister43C}. The short orbital period, most precisely 
measured by \citet{Feline05C}, places the system below the period 
gap of cataclysmic variables and thus in the realm of the SU~UMa subclass of 
dwarf novae. HT~Cas is an eclipsing system. Basic dynamical parameters have 
be determined by \citet{Horne91C}.
In contrast to many other eclipsing dwarf novae HT~Cas does not exhibit an
orbital hump in its light curve. Thus, the contribution of the hot spot to the
total light appears not the be significant and no hot spot flickering is
expected. This is confirmed by \citet{Bruch00C} who found that the flickering
in HT~Cas is restricted to the region on or in the immediate vicinity of the
white dwarf. Here, I analyse high time resolution light curves provided by 
R.E.\ Nather [some of which have also been investigated by 
\citet{Patterson81C}] observed during quiescence and superoutburst. These 
are complemented by numerous light curves observed during all photometric 
states, retrieved from the AAVSO International Data Base.

{\it WX~Cen} resembles in many aspects the bright novalike variable QU~Car.
It was originally detected as a variable star by
H.\ Leavitt \citep{Pickering06C} and acquired interest when \citet{Eggen68C}
identified it as the optical counterpart of the X-ray source Cen~XR-2.
\citet{Steiner98C} counted WX~Cen among their original list of V~Sge
stars. The orbital period was first determined spectroscopically by 
\citet{Diaz95C} and later confirmed photometrically by \citet{Oliveira04C} 
and \citet{Qian13C}. At just over 10~h it is rather long for a CV.
Even so, just as QU Car, WX~Cen does not exhibit absorption line features in
it optical spectrum \citep{Diaz95C}, in agreement with the high
luminosity found in V~Sge stars. The orbital light curve studied by
\citet{Oliveira04C} exhibits a well defined orbital modulation
which, however, deviates somewhat from a simple sine-wave. \citet{Oliveira04C}
also determined a tentative primary star mass. 
No further information about dynamical and geometrical system parameters is 
available. 
The present study draws on light curves found in the OPD data bank and others
retrieved from the AAVSO International Database. 

{\it BV~Cen} has one of the longest orbital periods of all dwarf novae. First
measured by \citet{Vogt80C}, it was improved by R.F.\ Webbink
\citep[see note added in proof to][]{Gilliland82bC}.
Dynamical and geometrical system parameters have been determined by 
\citet{Gilliland82bC}, \citet{Vogt80C} and \citet{Watson07C}. 
At the long orbital period the secondary star should be somewhat evolved and
contributes a significant fraction to the total optical light of the system
(see Appendix~B).
The data used for the present study consist of some unfiltered light curves
provided by B.\ Warner, complemented by several light curves observed in the
Walraven system by \citet{Bruch92C} (used in his original flickering study)
and \citet{Hollander93C}, and $V$ band light curves
retrieved from the AAVSO International Database.

{\it MU~Cen} is a little studied long period dwarf nova. \citet{Friend90C},
using radial velocity measurements of absorption features of the secondary 
star observed in near infrared spectra, first derived an orbital period 
which is in agreement with the more accurate photometric
period measured by \citet{Bruch16C}. Both, \citet{Friend90C} and
\citet{Bruch16C}, discuss component masses. The latter author also
delimits the orbital inclination. 
At the long orbital period the secondary star contributes
a substantial fraction of the optical light (see Appendix~B for details).
The data used in this study are those already analysed by \citet{Bruch16C}. 

{\it V436~Cen} is a SU~UMa type dwarf nova and as such has a small orbital
period \citep{Gilliland82aC}. Not much is known about the system 
parameters. From model fits to the ultraviolet spectrum 
\citet{Nadalin01C} estimated the primary star mass and the orbital 
inclination. The data used in the present study consist of unfiltered 
light curves provided by B.\ Warner \citep[see also][]{Warner75bC} and of 
$B$ band data taken from \citet{Semeniuk80C}.

{\it V442~Cen:} Very few details are known about this dwarf nova. 
Even the orbital period, determined by \citet{Marino84C} to be 11.0~h
has never been confirmed. However, the period must be
long, considering the early spectral type of the secondary star, classified
by \citet{Harrison04C} as G6$\pm$2, based on infrared spectra (see also
Appendix~B).
Other system parameters have never been measured to any degree of
reliability. The data available for this study were all provided by 
B.\ Warner.

{\it V504~Cen} is classified in the General Catalogue of Variable Stars as a
possible R~CrB star \citep{Kholopov85C}. However, based on its spectrum
\citet{Kilkenny89C} identify it as a cataclysmic variable, and
the photometric behaviour made them suspect that V504~Cen belongs to the
VZ~Scl class. This was confirmed by \citet{Kato03C} and \citet{Greiner10C}. 
The latter authors, observing the star during a low state, also detected 
photometric variations in optical light, as well as in X-rays, and radial 
velocity variations which permitted them to determine the orbital period. 
12 unpublished light curves observed by the author in unfiltered 
light are available for the present study.

{\it V834~Cen} (= E1405-401) is a well observed polar. It was first discovered
as an X-ray source by \citet{Jensen82C}. The optical counterpart was
identified by \citet{Mason83C} who also suggested the system to be of
AM~Her type. This is corroborated by its optical polarization first observed 
by \citet{Tapia82C} and soon thereafter confirmed by \citet{Visvanathan83C}
and \citet{Bailey83C}. Many detailed photometric, polarimetric and 
spectroscopic studies followed \citep[e.g.][]{Rosen87C, Cropper89C, 
Ferrario92C, Potter04C}. An optical modulation in the light of V834~Cen 
was already observed by \citet{Mason83C}. Interpreting it as the rotation 
period of the white dwarf, the orbital period should be the same. It was
later improved spectroscopically by \citet{Schwope93C}.
Dynamical and geometrical system parameters are discussed by 
\citet{Tuohy85C}, \citet{Cropper86C}, \citet{Rosen87C}, \citet{Schwope93C}, 
\citet{Ramsay00C} and \citet{Costa09C}.
Oscillations with periods or quasi-periods in the range of a few seconds
have repeatedly been observed in the light curve of V834~Cen 
\citep{Mason83C, Larsson85C, Middleditch91C, Imamura00C, Mouchet17C}.
However, their period is so small that even at the high time resolution
of the light curves used for the present study they will be smeared out
and thus have no bearing on the flickering properties as measured here.
The available data consist of unpublished unfiltered light curves 
observed by the author, complemented by data provided by L.\ Chiappetti
\citep[see also][]{Sambruna91C}

{\it V1033~Cen} (= RX~J1141.3-6410) was detected as a ROSAT X-ray source by
\citet{Motch96C} and identified in the optical with a 16.6~mag star exhibiting 
properties of a synchronized magnetic system, i.e., a polar. The 
observation of polarization by \citet{Rodrigues98C} and \citet{Buckley00C}
confirmed this classification. Periodic light variations observed by 
\citet{Cieslinski97C} permitted to determine the orbital period. A refined 
value was measured by \citet{Rodrigues98C}. Component masses 
of V1033~Cen remain unknown. \citet{Buckley00C} estimated an orbital 
inclination of $\sim$70$^{\rm o}$. 
Only three light curves retrieved from the OPD data bank are available for 
this study. They were already used by \citet{Cieslinski97C}.

{\it WW~Cet} is normally classified as a U~Gem type dwarf nova. However, a
standstill observed by \citet{Simonsen11C} interrupted the $\sim$40 year 
historical record of normal outburst behaviour and suggests that the system
may belong to the Z~Cam stars. The orbital period was first measured 
spectroscopically by \citet{Thorstensen85C} and then refined by 
\citet{Ringwald96bC} and \citet{Tappert97C}. \citet{Tappert97C} also 
derived dynamical and geometrical system parameters. Several $UBVRI$ 
(unpublished) and Walraven \citep{Bruch92C} light curves observed by the 
author were supplemented for this study by Walraven light curves
taken from \citet{Hollander93C}.

{\it WX~Cet} is a short period dwarf nova of the SU~UMa type, albeit with a
long outburst cycle \citep[880 days for superoutbursts and 200 days for short
eruptions;][]{Sterken07C}. The orbital period has first been measured
spectroscopically by \citet{Mennickent94C} and was then confirmed (modulo a 
cycle count ambiguity) by \citet{Thorstensen96C}. \citet{Rogoziecki01C} 
found the same period photometrically. The most accurate value is provided
by \citet{Sterken07C}. \citet{Rogoziecki03C} are the only authors to 
derive component masses and the orbital inclination for WX~Cet. The data 
used in the present study consist of light curves provided by B.\ Warner 
and were already used by \citet{ODonoghue91C}. They were all observed 
during the maximum and early decline from the June 1989 superoutburst.

{\it BO~Cet} is a little studied novalike CVs. While \citet{Zwitter95C}
present a single spectrum, the only time resolved spectroscopy was published
by \citet{Rodriguez-Gil07C}. They classify the system as being of
SW~Sex type. Apart from this study and some time resolved photometry of
\citet{Bruch17aC} no further detailed optical observations of BO~Cet have been
obtained. A photometric period reported in an informal communication by J.\, 
Patterson\footnote{http://cbastro.org/communications/news/messages/0274.html; 
or http://cbastro.org/pipermail/cba-public/2002-October/000300.html} based on
Center for Backyard Astrophysics (CBA) data has been refined by 
\citet{Bruch17aC} and is compatible with the spectroscopic period measured by
\citet{Rodriguez-Gil07C}. Dynamical and geometrical
system parameters are unknown. 
The data used in the present study are those already investigated by
\citet{Bruch17aC} supplemented by light curves retrieved from the
AAVSO International Database.  

{\it Z Cha:} Like OY~Car and HT~Cas, Z~Cha is another well known deeply 
eclipsing dwarf nova of SU~UMa type. Its orbital period
\citep{Dai09aC} is slightly variable with a period of 32.6 
years \citep{Dai09aC} or 28$\pm$2 years \citep{Baptista02C}.
Dynamical parameters were determined by \citet{Wood86C}, \citet{Wade88C}
and \citet{McAllister19C}.
The location of the flickering light source in Z~Cha has been investigated by
\citet{Bruch96C}. He found that in quiescence and normal maximum flickering
comes from the immediate vicinity of the white dwarf. However during
quiescence a second source, located at the hot spot, also contributes.
During superoutburst the situation is more complex. When the superhump
light source shines flickering originates in an extended region of the 
accretion disk. When it is off, the immediate vicinity of the white dwarf
as well as an extended range along the disk rim (possibly enhanced at the
bright spot location) flickers.
Here, I analyse numerous unfiltered high time resolution
light curves kindly provided by B.\ Warner. 
Some of these have also been used in the studies of \citet{Warner74C},
\citet{Warner88C} and \citet{Wood86C}. They encompass all photometric
states. More light curves were retrieved from the OPD data bank and the
AAVSO International Database, again encompassing all photometric states.

{\it ST Cha:} Very little is know about this star. Originally discovered by 
\citet{Luyten34C}, its classification history was summarized by 
\citet{Simonsen14aC} who concluded that it belongs to the Z~Cam subtype of 
dwarf novae. The tentative orbital period mentioned by \citet{Cieslinski98C}
was recently shown to be erroneous by \citet{Bruch17dC} who measured a
spectroscopic period of 5.50~h during the final rise from a standstill 
to an outburst. Nothing is known about dynamical and geometrical system 
parameters. Only one light curve is used for the present study. It was 
already reproduced by \citet{Bruch17dC}.

{\it TV Col:} This intermediate polar was identified as the optical counterpart
of the hard X-ray source 2H~0526-328 by \citet{Charles79C}. It
exhibits complicated photometric variations which, apart from flickering,
includes 3 periods observed in optical light, in addition to another
period only seen in X-rays. The reader is referred
to the introduction of \citet{Rana04C} for references to observations
of these periods. 
A coherent modulation with a period of 32~s in X-rays was first
detected by \citet{Schrijver85C} in EXOSAT observations, confirmed by 
\citet{Schrijver87C} and \citet{Rana04C}, and analysed in more detail 
by \citet{Norton89C}. It is interpreted as the rotation period of
the white dwarf in TV~Col and is the reason for the classification of
the system as an intermediate polar. However, unlike in most other system
of this kind the white dwarf rotation does not reveal itself in the optical
light \citep{Barrett88C}. Instead, TV~Col exhibits a spectroscopic period, 
a $\sim$5\% shorter photometric modulation, and longer term
variation on their beat period, first mentioned by \citet{Hutchings81C} and 
later confirmed by many other authors. The spectroscopic period is
interpreted to be orbital, while the photometric modulation is thought to
be due to the nodal precession of an inclined accretion disk 
\citep{Barrett88C, Hellier93C}, i.e., it is a negative superhump. The
most precise value of the orbital period has been measured by 
\citet{Hellier93C} who observed shallow 
eclipses, confirming that this period really reflects the binary revolution.
The eclipses are only partial, involving a part of the accretion 
disk but not the primary. They are not very conspicuous in the light curve 
and where therefore not removed before analysing the flickering in this
study. TV~Col exhibits sometimes outbursts with amplitudes of the order of 
2~mag \citep{Szkody84aC, Hellier93aC, Hudec05C} which
last only some hours and are thus quite different from normal dwarf novae
outbursts. \citet{Hellier93aC} and \citet{Hameury17C} interpret
them as events of enhanced mass transfer from the secondary star rather than
in the framework of the thermal-viscous disk instability model.
Most of the observations of TV~Col concentrate on the X-ray and optical
variability. Almost no detailed optical spectroscopy has been performed.
This may be one reason for the absence of mass determinations for the 
components. The geometrical properties have also not been explored. Only 
the orbital inclination has been estimated by \citet{Hellier91C}, based on 
their detection of eclipses.
The data available for this study consist light curves
provided by B.\ Warner and C.\ Hellier. They were discussed previously 
in different contexts by \citet{Barrett88C}, \citet{Hellier91C},
\citet{Hellier93C} and \citet{Hellier93aC}.

{\it T~CrB} is a recurrent nova with outbursts observed in 1866 and 1946.
Unlike most other cataclysmic variables the secondary star is not close
to the main sequence but a giant. Therefore, T~CrB is also classified as a
symbiotic system and a symbiotic nova. As such, its orbital period is
much longer than that of most CVs. It has first been measured 
spectroscopically by \citet{Sanford49C}. The most accurate value of 227.67 d 
was derived by \citet{Lines87C}. Initial determinations of dynamical system
parameters \citep{Kraft58C, Paczynski65C, Kenyon86C} proved to be incompatible
with a CV model. This issue was resolved by \citet{Belczynski98C}, 
\citet{Hric98C} and \citet{Stanishev04C}.
Flickering in T~CrB was relatively well studied in the past. Early 
observations were reported by \citet{Walker54C}, \citet{Ianna64C}, 
\citet{Lawrence67C}, \citet{Walker77C} (who also noted the much higher 
amplitude in the $U$ band compared to the $I$ band), \citet{Bianchini76C}
and \citet{Oskanian83C}. The latter two authors noted the absence of the 
flickering at certain epochs. Negative observations are also reported by
\citet{Dobrzycka96C} and \citet{Mikolajewski97C}.
A strong increase of the flickering amplitude with decreasing wavelength
has been noted by \citet{Raikova86C}, \citet{Bruch92C} and \citet{Hric98C}.
More detailed studies of the flickering in T~CrB start with
\citet{Zamanov98C} who investigated $U$
band light curves and found that -- unlike the behaviour during the
outburst cycles of dwarf novae -- the ratio of the flux of the
flickering light source and the quiet part of the primary remains
constant. They conclude that with the exception of the peculiar, not
well understood feature that the activity can disappear temporarily,
flickering in T~CrB is on the whole indistinguishable from that in 
normal CVs in spite of the vastly larger geometrical dimensions.
The constancy of the flux ratio of the flickering light source and
average flux of the primary component in the presence of considerable
variations of the latter is confirmed by \citet{Zamanov04C}.
They also find that the slope of the high frequency part of the power
spectrum (on the double logarithmic scale) is always the same, independent 
of the brightness of T~CrB in the $U$ band, as is the e-folding time of the
auto-correlation function. \citet{Zamanov16C}
show that the linear amplitude-flux relation in T~CrB repeats itself with the
same proportionality constant in other symbiotic systems just as in several
novalike variables over four orders of magnitude in flux. They conclude that
viscosity in the accretion disk is practically the same in all investigated
systems. \citet{Dobrotka10C} analysed $V$ band flickering in T~CrB. 
They simulated the statistical distribution of flare durations, assuming
that the aperiodic variability is produced by turbulent elements in the 
disk and conclude that the majority of flickering events is concentrated
in the inner disk which is weakly truncated. Finally, \citet{Ilkiewics16C}
also study flickering in T~CrB during active phases, emphasizing flickering
in X-rays, and suggest that it originates in the boundary layer  between the
accretion disk and the white dwarf.
Flickering in T~CrB being strongest at short wavelength is in the first
place due to the diluting effect of the giant secondary star the contribution
of which in the visual range is overwhelming. \citet{Zamanov98C}
estimate that it can be 90\% in $V$. However, this number is uncertain.
Correcting the amplitude of the flickering in the visual for the contribution
of the secondary star can thus lead to huge errors. Therefore, I
prefer to use only data obtained in the $U$ band for this study. Here,
the secondary contributes much less and any results are therefore more
reliable (see Appendix~B).
The data used here consist of suitable light curves of \citet{Zamanov98C}
supplemented by some unpublished light curves of the author.

{\it SS Cyg:} As the prototype dwarf nova and one of the brightest members 
of its class SS~Cyg is arguably the best studied of all cataclysmic 
variables. The number of publications about the system is legion. It has a 
comparatively long orbital period. The (still) most accurate value 
is based on spectroscopic measurements published by
\citet{Hessman84C} more than three decades ago.
Geometrical and dynamical system parameters have been determined by many 
groups, e.g., \citet{Giovannelli83C}, \citet{Bitner07C}, \citet{Voloshina00C},
and \citet{Hill17C}.
Although SS~Cyg is not generally recognized as a magnetic system, claims in 
this sense have been brought forward repeatedly in the past (see 
\citet{Giovannelli12C} and references therein).
In her epic work on SS~Cyg, \citet{Zuckermann61C}
included the first ever systematic study on the flickering in any
cataclysmic variable, although it did not leave a lasting imprint in the
literature. \citet{Bruch92} also included SS~Cyg in his original flickering
study. More recently, \citet{Aranzana18C} investigated simultaneous
high time resolution observations in 3 photometric bands. They found a
soft lag of the flickering events with an amplitude of about 5~s at
a time scale of 250~s. To my knowledge no other paper dealing 
specifically with flickering in this system has be published. The 
contribution of the secondary star to the total system light is
discussed in Appendix~B.
The available data for this study comprise some high time resolution
light curves provided by R.E.\ Nather, several more light curves observed
by the author and his collaborators, and many data sets retrieved from the 
AAVSO International Database.

{\it EM~Cyg} is an eclipsing Z~Cam type dwarf nova. At the comparatively
long orbital period of almost 7~h, originally measured by \citet{Mumford69C}
and last refined by \citet{Dai10bC}, the absorption spectrum of the 
secondary star is visible together with the accretion disk emission 
lines \citep{Kraft64bC},
making the system one of the few eclipsing double lined spectroscopic 
binaries among CVs. However, any determination of dynamical system parameters
is complicated by the pollution of the light of EM~Cyg by an unrelated 
K2-5~V star along the line of sight, discovered by \citet{North00C}.
Component masses were derived by various authors but due to the contaminating 
light of the unrelated K-star (not noted by most of them) the results are not
reliable. It appears that the best values were derived by \citet{Welsh07C}.
The orbital inclination was determined by
\citet{North00C} to be $67^{\rm o} \pm 2^{\rm o}$ which means that
during eclipses the secondary star only covers a part of the accretion
disk but not the white dwarf. This is also immediately suggested by
the shallow profile of the eclipses \citep[e.g.,][]{Mumford69C}.
The contribution of the secondary star (and the unrelated star in the
line of sight) is discussed in Appendix~B.
The data used here consist of some high time resolution unfiltered light 
curves provided by R.E.\ Nather \citep[see also][]{Patterson81C}, supplemented
by light curves retrieved from the AAVSO International Database.

{\it V751~Cyg} is a novalike variable, probably of the VY~Scl subtype. The 
early history is summarized by \citet{Patterson01C}.
According to \citet{Greiner98C} the magnitude
of the star ranges between 13.2 and 17.8~mag. The AAVSO long term light 
curve suggests that the normal brightness of V751~Cyg appears to decrease
slightly in recent years, attaining an average level of 
$\approx$15.1~mag currently. It also exhibits short lived excursions
to faint states of different magnitude levels. 
The orbital period of V751~Cyg was first determined spectroscopically by 
\citet{Patterson01C} and then refined by \citet{Echevarria02C}. 
The latter authors also observed a shorter photometric period
which they interpret as a negative superhump. Its presence was
confirmed by \citet{Papadaki09C}. The masses of the binary
components are not known, while the orbital inclination seems to be low
\citep{Greiner99C}.
No detailed investigation of the flickering in the system has been performed,
but \citet{Patterson01C} measured the power spectrum to follow 
$P \propto \nu^{-2.0 \pm 0.1}$ for $\log{\nu} > 1.8$ ($\nu$ in cycles/day).
The data available for this study consist of unfiltered, high time resolution 
light curves provided by R.E.\ Nather, and light curves taken from 
\citet{Papadaki09C}. All data refer to the normal high state of V751~Cyg.

{\it HR~Del} is one of the brightest old novae. The outburst in 1967 was
discovered by \citet{Alcock67C} and peaked at a maximum brightness of 
3.5~mag \citep{Krempec70C}.
After a long fading it reached is pre-outburst magnitude of 12~mag
\citep{Robinson75C} by the end of the decade of 1970 \citep[][and references
therein]{Bruch82bC} but kept on declining slowly, attaining a mean
magnitude of $\approx$12.3~mag in recent times (see the AAVSO
long term light curve). 
HR~Del has a fairly long orbital period, first determined spectroscopically
by \citet{Bruch82bC} and later refined by \citet{Kuerster88C}.
The latter authors also measured dynamical parameters of the system.
From the structure of the nebula expelled during the nova explosion an orbital
inclination of $i = 40^{\rm o} \pm 2^{\rm o}$ is derived \citep{Solf83C, 
Pilyugin85C, Kuerster88C}, confirming an earlier estimate of \citet{Bruch82bC}. 
The data available here consist of high time resolution  unfiltered light 
curves provided by R.E.\ Nather and E.\ Robinson, complemented by light curves 
of I.\ Voloshina and the author \citep[some of them used in the flickering study
of][]{Bruch92}, and many more retrieved from the AAVSO 
International Database. 

{\it DO Dra:} Due to a confusion concerning the identification of this star 
and another variable the denominations DO~Dra and YY~Dra are both used for 
the same system in the literature [see \citet{Patterson87C} and
\citet{Kholopov88C}]. It is a dwarf nova and at the same time an 
intermediate polar. The first indication for the magnetic nature of the 
system was observed by \citet{Patterson92C} who detected a low amplitude 
($\sim$1\%) variation in the $U$ band with a period of 275 sec. However, the 
presence of sub-harmonics suggests that the true period (interpreted as the 
rotation period of the magnetic white dwarf) is twice this value. This was 
confirmed through X-ray observations first by \citet{Patterson93aC}
and then by other groups. \citet{Patterson92C} also saw double humped 
orbital variations in the IR light curve of DO~Dra, attributed to ellipsoidal 
variations of the secondary star 
\citep[confirming earlier observations of][]{Mateo91C} and 
furthermore conclude that the mass accretion rate is quite low.
The orbital period was first determined spectroscopically in the near 
infrared by \citet{Mateo91C} and later refined by \citet{Haswell97C}.
Both groups also determined dynamical and geometrical system parameters.
In spite of its intermediate polar nature with variations on the time
scale of 10 min caused by the white dwarf rotation I include DO~Dra in
this study because the modulation is of quite low amplitude in the
optical range \citep{Patterson92C} and thus have at most a minimal
impact on the flickering parameters. The secondary star contribution is
discussed in Appendix~B.
The data available here consist of two very high time resolution multicolour
light curves provided by E.\ Robinson, some light curves published by
\citet{Andronov08C}\footnote{avoiding their light curves observed on 
JD~2454162 and JD~2454163
because these are dominated by variations which the authors term
transient periodic oscillations and which would significantly bias the
flickering parameters.} and many more retrieved from the
AAVSO International Database. 

{\it AQ Eri}, discovered by \citet{Morgenroth34C}, is
an SU~UMa type dwarf nova which, except for the study of superhumps during
its superoutbursts \citep[see, e.g.,][who also summarize the observational 
history of the star]{Kato16C} has not been
investigated in much detail. \citet{Thorstensen96C} measured
spectroscopically an orbital period of 1.463~h which makes AQ~Eri one of 
the shortest period SU~UMa stars. Nothing is known about dynamical and 
geometrical system parameters. 
The present study draws exclusively on light curves retrieved from the 
AAVSO International Database, all observed during superoutburst.

{\it KT~Eri} is a classical nova which erupted in 2009 \citep{Itagai09C}. 
While the outburst was well documented, investigations of the quiescent
state are rare. Long term variations occurring on the time scale of
hundreds of days were examined by \citet{Jurdana-Sepic12C} and
\citet{Munari14C}. Using additional data, \citet{Bruch18aC} showed
that the periods suspected by these authors to be present in KT~Eri are not
stable and hence cannot be the orbital period in an scenario which puts
the system in the vicinity of recurrent novae with evolved secondary stars.
The orbital period is uncertain. C.\ Stockdale (Kato, vsnet-alert 11755) 
reports a possible period of 135~min. However, this value is unconfirmed
and the corresponding observations remain undocumented. Moreover, they were
performed soon after the nova outburst when KT~Eri was at least 5 magnitudes
above the average post-nova brightness. It is doubtful if orbital variations
manifest themselves at a light level so much above quiescence. 
\citet{Bruch18aC} detected a modulation in the post-outburst light with a 
period of 4.685~h which may be orbital, although
other origins, such as a permanent superhump as seen in several old novae and
novalike variables, may also be possible. The data used in this study are 
limited to the high time resolution light curves already presented by 
\citet{Bruch18aC}.

{\it U~Gem}, like SS~Cyg, is considered as prototype of dwarf novae. It was
the first of its kind to be discovered \citep{Hind56C}, although at the time
its nature was, of course, not yet recognized. Consequently, it has a
very long history of observations and the list of publications
is extensive. U~Gem has an orbital period of $\sim$4.25~h \citep{Marsh90C} 
which may show cyclic variations \citep{Eason83C}. Dynamical and 
geometrical system parameters have been determined by various authors, e.g.,
\citet{Stover81C}, \citet{Zhang87C}, \citet{Naylor05C} and 
\citet{Unda-Sanzana06C}. As discussed in Appendix~B,
the secondary star contributes significantly to the total light of the system.
The observational material available for U~Gem consists of high time
resolution unfiltered light curves provided by R.E.\ Nather, and numerous
light curves retrieved from the AAVSO International Database.

{\it DM~Gem} is an old nova which erupted in 1903. Very few details of the
underlying binary system are known. A modulation with a period of 
$\sim$2.95~h was reported by \citet{Lipkin00C}
and interpreted as quasi-periodic oscillations. Variations at the
same period were also seen by \citet{Rodriguez-Gil05C}.
They consider them to be probably linked to the orbital period. In the
absence of stronger evidence I will consider the binary period of DM~Gem
to be unknown. \citet{Rodriguez-Gil05C} also claim the presence of 
short-term variations on the time scale of $\sim$20~min.
Only one high time resolution unfiltered light curve, provided
by R.E.\ Nather, is available for this study. Note that it does not contain
variations with a preferred time scale of $\sim$20~m. The respective 
variability seen by \citet{Rodriguez-Gil05C} may therefore
well be caused by the accidental occurrence of strong flickering flares on
that time scale.

{\it IR~Gem} was discovered as a dwarf nova by \citet{Popowa61C}
and classified as being of SU~UMa subtype by \citet{Szkody83C}
who conducted the first detailed study of this star. The spectroscopic
orbital period was measured by \citet{Szkody84C} and \citet{Feinswog88C}. 
Orbital brightness variations during quiescence are also observed, however, 
with strongly variable amplitude \citep{Szkody89C, Feinswog88C, Fu04C}.
Nothing is known about dynamical and geometrical system parameters, execept
that the absence of eclipses precludes a high orbital inclination.
The data available for this study consist of one high time resolution 
unfiltered light curve provided by R.E.\ Nather, plus several light curves 
retrieved from the AAVSO International Database. They refer to different
photometric states between quiescence and superoutburst.

{\it RZ~Gru} is a little studied novalike variable of the UX~UMa subtype.
Originally discovered as a variable star by \citet{Hoffmeister49C}
its nature as a cataclysmic variable was first suspected by
\citet{Kelly81C} and then confirmed by \citet{Stickland84C}.
The orbital period is uncertain. In a short notice \citet{Tappert98C}
quote two possible values: 8.64~h and 9.79~h. Based on the narrowness of
the emission lines in the spectrum of RZ~Gru, \citet{Stickland84C}
suspect an orbital inclination of $<$20$^{\rm o}$. A quite different value of
$i = 60^{\rm o}$ was derived by \citet{Bisol12C} via modelling of the 
ultraviolet spectrum. These authors also found a
primary star mass of $0.55 M_\odot$ and a distance of only 116~pc. The
latter value is in stark contrast to the GAIA distance of $542.8 \pm 17.3$~pc,
casting doubts upon the reliability of the adopted model. Therefore, I will
consider all binary parameters of RZ~Gru (including the unconfirmed orbital
period) to be unknown.
The data available for this study consist of unpublished high time 
resolution unfiltered light curves observed by the author.

{\it AH~Her} is a dwarf nova of the Z~Cam subtype. It oscillates between 
about 14.3~mag during minimum and 11.3~mag in
outburst without ever remaining at a well defined quiescent light level
(see AAVSO long-term light curve). Starting with \citet{Mumford66C}
various authors performed time resolved photometry. \citet{Moffat84C}
found a period in their extensive photometry which was later confirmed 
spectroscopically to be orbital \citep{Horne86C}. The latter 
authors also derived geometrical and dynamical system parameters.
Absorption features of secondary star were seen in the optical spectrum of 
AH~Her by \citet{Horne86C} and \citet{Bruch87C} \citep[who also derived some
flickering properties in][]{Bruch92}. Its contribution to the
total system light is discussed in Appendix~B.
The data available for this study consist of some high time resolution 
unfiltered light curves provided by R.E.\ Nather. Most of these were already 
analyzed in a different context by \citet{Patterson81C}. To these I add
light curves retrieved from the AAVSO International Database.

{\it AM~Her} is the prototype of strongly magnetized CVs, the polars. Their
structure is significantly different from that of non-magnetic or only 
weakly magnetic systems. In particular, mass transfer does not occur
via an accretion disk, but matter is guided by the strong magnetic
field of the primary white dwarf directly from the L$_1$ point to the vicinity
of its magnetic poles where it forms accretion columns. This absence of a disk 
and a classical boundary layer between the disk and the white dwarf
surface, considered to be the most probable site of flickering in 
non-magnetic CVs, obviously means that the flickering mechanism must be 
different in polars. Thus, it may be expected that the properties of 
flickering in these systems are also different.
The magnetic nature of AM~Her was revealed when \citet{Tapia77aC} discovered 
strong optical linear polarization.
As the brightest member of its class it has been extensively observed
in particular at optical, UV and X-ray wavelengths. An extensive, albeit by
far not exhaustive list of references is given in the introduction of the 
paper of \citet{Dai13C} to which the reader is referred.  
The system exhibits distinct photometric states. During the high state it
attains a visual magnitude of the order of 12.3~mag while in the
low state the magnitude averages about 15.3~mag (see the AAVSO
long term light curve). The two states are thought
to be caused by an enhanced and reduced, respectively, mass transfer between 
the components.
The optical light is periodically modulated due to the changing visibility
of the accreting magnetic pole of the rotating white dwarf. Since this
rotation is coupled to the binary orbit, the period of these light changes
is equal to the orbital period, the most precise values of which has been
determined by \citet{Dai13C}, based on an analysis of these variations
over many years. Although
many attempts have been undertaken to determine dynamical and geometrical 
system parameters they remain uncertain 
\citep{Brainerd85C, Cropper98C, Davey96C, Gaensicke98C, Greeley99C, 
Mouchet83C, Mukai87C, Southwell95C, Watson03C, Wickramasinghe91C, Wu95C,
Young81bC}. Some work on the flickering in AM~Her has been done in the past. 
\citet{Greeley99C} observed rapid variability in the far ultraviolet range 
where the brightness changes as much as 50\% on time scales as short as 10~s.
The spectrum of the variable light source is blue. The autocorrelation
function suggests typical flare durations of some tens of seconds. This
is quite similar to flickering observed in the optical light of AM~Her by
\citet{Stockman79C} and \citet{Panek80C}. Simultaneous observations in
hard X-rays and the optical range by \citet{Beardmore97C} revealed
a correlation of the variations in these bands, decreasing towards shorter
optical wavelengths. This is interpreted as a correlation between 
bremsstralhung (X-ray) and cyclotron (optical, decreasing to the blue)
radiation. Such a correlation was, however, not seen in simultaneous
optical and soft xray observations by \citet{Szkody80aC}.
\citet{Szkody80C} observed flickering in the $V$ and $U$ band 
and in a narrow band centred on the He~II $\lambda$~4686 emission line. 
The currently available data consist exclusively of light curves retrieved
from the AAVSO International Database. 

{\it DQ~Her} (= Nova Herculis 1934) is the prototype of the class of 
intermediate
polars. In these CVs the white dwarf has a magnetic field which -- while in
most cases not strong enough to dominate the accretion flow from the secondary 
star from the very beginning as in AM~Her stars (or polars) -- is able to
disrupt the inner accretion disk and guide matter via an accretion curtain
to the magnetic poles. In contrast to polars the white dwarf rotation 
period is not synchronized with the orbital period. The accretion energy 
seen by the observer in the optical range (often as reprocessed X-ray
radiation) is therefore modulated on the white dwarf spin period.
While the denominations DQ~Her star and intermediate polar are often used 
synonymously, sometimes the former term is used for a small subset of
intermediate polars where the white dwarf rotation is particularly rapid
with periods not exceeding $\sim$2~min. In the case of DQ~Her itself
the spin period, first detected by \citet{Walker56C}, is 71~s. There are, 
however, claims that the real period is twice this value. For a discussion of 
this still unsettled question, see \citet{Bloemen10C} and references therein.
This spin period manifests itself in the light curves as a highly coherent
low amplitude oscillation. Component masses of DQ~Her have been determined 
by several authors. Initial studies resulted in either high or low masses. 
This issue was discussed in some detail and resolved by \citet{Horne93C}.
DQ~Her is a deeply eclipsing system. \citet{Horne93C} found a very high
orbital inclination of $i = 86.5^{\rm o} \pm 1.6^{\rm o}$. 
An even higher value of $i = 89.7^{\rm o} \pm 0.1^{\rm o}$
(i.e., very nearly edge-on) was suggested by \citet{Wood05C}.
The orbital period has originally been determined by \citet{Walker56C}. Later
other authors detected period variations. These studies were most recently
summarized by \citet{Dai09C} who discuss possible origins. 
The available data were all retrieved form the AAVSO International Data
base and comprise 206 light curves. 
 
{\it V533~Her} is the remnant of a bright nova which exploded in 1963. The
quiescent system obtained some attention when \citet{Patterson79bC}
observed coherent oscillations with a period of 63.3~s which indicated that 
the system belongs to the rapidly rotating intermediate polars of the DQ~Her 
subclass. However, later observations by \citet{Robinson83C}
showed that these oscillations had disappeared, casting doubts on the
magnetic rotator model for V533~Her.
The spectroscopic orbital period was measured by \citet{Thorstensen00C}.
A slightly shorter photometric period was observed by 
\citet{McQuillin12C} in SuperWASP data. It may be related to a negative 
superhump. Geometrical and dynamical system parameters we determined by
\citet{Rodriguez-Gil02C}.
The data available for this study consist of an extremely high time
resolution multicolour light curve provided by E.\ Robinson, complemented 
by several light curves retrieved from the AAVSO International Database.

{\it V795 Her}, discovered in the Palmar Green survey \citep{Green82C}, was
initially suspected to be an intermediate polar \citep{Zhang91C, 
Thorstensen96aC, Shafter90C}. However, it appears now to be
consensus that it is a SW~Sex type novalike system as first
proposed by \citet{Casares96C}. With an orbital period of about 2.6~h
determined spectroscopically by \citet{Shafter90C}
it lies right within the period gap of cataclysmic variables. 
\citet{Mizusawa10C} estimate the primary star mass and the orbital
inclination.
V795~Her shows photometric variations with a period somewhat longer
than the spectroscopic period \citep[see, e.g.,][]{Kaluzny89C, Patterson94C,
Papadaki06C, Simon12C}, establishing the system as a permanent superhumper. 
Superposed upon the superhumps are rapid flickering variations and
occasional QPOs with periods of 10 -- 20~min \citep{Rosen95C, Patterson94C,
Papadaki06C}. \citet{Simon12C}
investigate the flickering in some detail. In particular, they find a 
correlation between the strength (amplitude) of the flickering on the
phase of the superhump modulations 
(see Sect.~\ref{The permanent superhump systems} of the main paper). 
The presently available data consists of light curves already
discussed by \citet{Kaluzny89C}, \citet{Papadaki06C} and \citet{Simon12C}, 
one very high time resolution multicolour light curve provided by 
E.L.\ Robinson and numerous light curves retrieved from the AAVSO 
International Database.

{\it EX~Hya} is a well-known intermediate polar. It has a short 
orbital period, first measured by W.\ Krzeminski \citep[cited by][]{Mumford64C}
and last revised by \citet{Echevarria16C}. The optical light
is dominated by a stable 67~min variation \citep{Vogt80aC}
which is also seen in X-rays \citep{Kruszewski81C}. It is
interpreted as being due to the rotation of the magnetic white dwarf primary
of EX~Hya \citep{Warner82C}. The ratio between the rotational and orbital 
period of almost exactly 2/3 is unusually large for intermediate polars.
Many authors have tried to measure the component masses of EX~Hya 
\citep[see the list of references in Table~4 of][]{Echevarria16C}. 
EX~Hya is normally encountered in a quiescent state at an average visual 
magnitude of 13.0~mag (see AAVSO long term light curve; however, 
in recent years the quiescent magnitude appears to decrease slightly). 
But occasionally this state is interrupted by short lived dwarf nova-like
outburst which can reach 9 -- 10~mag. \citet{Hellier00C}
estimate the average outburst frequency to be 
once every 1.5 years. They are unable to decide whether the outbursts are
triggered by a disk instability or by an enhanced mass transfer from the
secondary.
EX~Hyi is an eclipsing system. However, flickering is particularly strong
to the extend that, depending on the phase of the 67~min variation, it may
be difficult to distinguish the eclipses from minima between flickering flares. 
To my knowledge, the only study \citep[apart from][]{Bruch92} dealing in 
some detail with flickering in this 
star was performed by \citet{Semena14C}. They interpret flickering 
in this system as reprocessed X-ray radiation and determine properties of 
the accretion footprints on the white dwarf. 
The observational data used in the present study consist of many high time
resolution quiescent light curves provided by B.\ Warner, R.E.\ Nather and 
C.\ Sterken, supplemented by light curves taken from \citet{Reinsch90C} and 
others retrieved from the AAVSO International Database.

{\it VW~Hyi}, discovered by \citet{Luyten32C}, is a 
well known SU~UMa type dwarf nova and probably one of the
best studied objects of the kind. In particular during the last two
decades of the past century many publications were dedicated
to the system. A prominent hump in its light curve permitted \citet{Vogt74C}
to measure the orbital period. After subsequent refinements over time
the most accurate value modern value is quoted by \citet{Kato14C}.
Many authors have tried to determine the component masses of VW~Hyi. A
critical assessment of the various applied methods and their results 
has been performed by \citet{Smith06C}. The orbital inclination remains 
rather uncertain. There appears to be no more accurate estimate than that of 
\citet{Schoembs81bC} of $i = 60^{\rm o} \pm 10^{\rm o}$. VW~Hyi is a rich source 
of DNOs (dwarf nova oscillations) and QPOs \citep[see, e.g.,][]{Warner06C,
Woudt10C, Blackman10C}. However, no specific studies of the substantial 
flickering exhibited by the system have come to my knowledge. 
The data available for this study consist of numerous high time resolution 
light curves provided by B.\ Warner, used previously by \citet{Warner75aC}
\citet{Warner78C} and \citet{Robinson84C}. These data are complemented by 
many more light curves retrieved from the AAVSO International Database.

{\it WX~Hyi} is another SU~UMa type dwarf nova, but by far not as well studied
as more famous members of the group. While the long term visual behaviour was
documented by \citet{Bateson86C} and \citet{Bateson90C}, detailed
accounts of the optical behaviour are mostly restricted to the discussion
of light curves \citep[e.g.,][]{Walker76C, Bailey79C, Kuulkers91C}. The
orbital period was spectroscopically determined by 
\citet{Schoembs81bC}. These authors provide also the only estimate of 
dynamical and geometrical system parameters which, however, remain quite 
uncertain. The data available
for this study consist of high quality light curves provided by B.\ Warner, 
observed in all photometric states. These are complemented by Walraven light 
curves, some observed by the present author [also used by \citet{Bruch92} 
to characterize flickering in WX~Hyi], but
mostly taken form \citet{Kuulkers91C}, and two unpublished $UBVRI$
light curves.      

{\it BL~Hyi} (= H0538+608) was discovered and classified as a polar by
\citet{Agrawal83C}.
It belongs to the better studied CVs with numerous dedicated publications.
Many of these deal with the complicated magnetic field
structure and the X-ray properties. Even so, not much reliable information
exists on some basic binary parameters. An exception is the orbital period
which has been determined at similar levels of accuracy by various groups 
using different techniques \citep{Pickles83C, Cropper87C, Beuermann89C,
Wolff99C, Mennickent99C}. It lies below
the CV period gap, as is the case for most polars. The orbital inclination
and the masses are not well known \citep{Beuermann89C, Wolff99C}.
Just as many other polars, BL~Hyi exhibits low and high states in its
long term light curve \citep[see, e.g.,][]{Gerke06C}.
During high states, to which all observations used in the present study
refer, strong orbital variations reaching almost 1.5~mag are caused
by the various aspect of the active magnetic (accretion) regions on the white
dwarf. The available data consist of 11 unpublished light 
curves observed by the author at a high time resolution.

{\it DI~Lac} is an old novae which erupted in 1910. Although at an average
magnitude of $\sim$15.6~mag \citep{Hoard00C} it belongs to the brighter nova
systems, details about the structure of the system are scarce. Even the
orbital period is not well documented, only being quoted by 
\citet{Ritter03C} as a privat communication from R.F.\ Webbink. From model fits 
limits on the primary star mass and the orbital inclination are given
by \citet{Gill00C} and \citet{Moyer03C}.
The data available for this study consist of a single unfiltered light curve
with a high time resolution provided by R.E.\ Nather.  

{\it X~Leo:} This dwarf nova has not received much attention in the past. The
AAVSO long term light curve reveals that the system suffers frequent 
outbursts and remains in quiescence only for short periods of time.
The orbital period has been determined spectroscopically by 
\citet{Shafter86aC}.
Based on a synthetic spectral analysis of IUE spectra \citet{Urban06C}
estimated the primary star mass and the orbital inclination.
The data available for this study consist of only one light curve
observed close to outburst maximum retrieved from the AAVSO International 
Database and one quiescent high time resolution light curve provided by 
R.E.\ Nather. 
 
{\it GW~Lib:} \citet{Gonzales83C}
detected this dwarf nova during an outburst. A second large amplitude
(super-) outburst was observed and extensively documented in 2007, 
establishing the system as a WZ~Sge star. GW~Lib became famous as the
first CV to display WD oscillations of the ZZ~Cet type in its light curves
\citep{vanZyl00C} which have been extensively investigated thereafter 
\citep[e.g.,][]{vanZyl04C, Townsley04C, Szkody16C, Chote16C, Toloza16C}. 
The oscillations subsided during the 2007 superoutburst and in subsequent 
years  \citep{Copperwheat09C, Vican11C, Bullock11C}
probably because the white dwarf left the instability strip due to
heating during the outburst.
Apart from the short period WD oscillations a modulation with a period of
2.1~h was repeatedly observed during quiescence \citep{Woudt02C, 
Copperwheat09C, Vican11C}. The latter authors suspect
that the true period is twice this value, i.e, 4.2~h \citep[see also][]
{Bullock11C}. It is of unknown origin, unrelated to the much shorter
orbital period, first measured spectroscopically by \citet{Szkody00C}
and then rectified by \citet{Thorstensen02C}. The
most complete geometrical and dynamical model was published by
\citet{vanSpaandonk10C}. The data available for the present study 
consist of light curves retrieved from the OPD data bank and the AAVSO 
International Database. Most of the observations were taken during the
plateau phase of the 2007 outburst while some were acquired
after the steep decent at the end of the superoutburst during the gradual
decline to quiescence, but still considerably above the quiescent magnitude
of $\sim$17~mag \citep{Thorstensen02C}. As mentioned above, during 
these phases the white dwarf oscillations had ceased and
therefore do not bias the determination of the flickering parameters.

{\it BR~Lup:} Not many dedicated studies of this star exist. The detection 
of superhumps by \citet{ODonoghue87C}
during an outburst in 1986 permitted a classification as a dwarf nova of
SU~UMa subtype. The orbital period was measured by 
\citet{Mennickent98C} in photometric observations during quiescence. 
These authors also quote dynamical and geometrical system parameters.
The data available for the present study consist of high time resolution
light curves provided by B.\ Warner. These are the
same data already used by \citet{ODonoghue87C}. They are supplemented by
observations retrieved from the AAVSO International Database.

{\it AY~Lyr} is a little studied cataclysmic variable. Discovered as a 
dwarf nova by \citet{Hoffmeister28C}, it was recognized as a SU~UMa star by 
\citet{Patterson79aC}.
Not much is known about the system parameters. Not even the orbital period
has been measured directly. The period quoted by \citet{Ritter03C} has been 
inferred from the observed superhump period \citep{Udalski88C}. The system 
appears to be seen almost pole-on. \citet{Voikhanskaya96C} estimated 
$i \le 18^{\rm o}$.
The data available for the present study consist of some light curves
provided by R.E.\ Nather \citep[already having been used by][]{Patterson79aC}. 
They are complemented by a few data retrieved from the AAVSO International 
Database.

{\it MV~Lyr} is a well studied novalike variable of the VY~Scl subtype. The
long term light curves was characterized by many authors, e.g.,
\citet{Wenzel80C}, \citet{Rosino93C}, \citet{Pavlenko98C}, or
\citet{Honeycutt04C}. During high states MV~Lyr has a mean magnitude 
around $V = 12.5$. During low states it can become as faint as 18~mag.
The orbital period has been determined spectroscopically by 
\citet{Voikhanskaya88C}, \citet{Borisov92C} and last by \citet{Skillman95C}.
MV~Lyr also exhibits a slightly longer photometric period
\citep{Borisov92C, Skillman95C} which may be interpreted as being due to a 
superhump. Dynamical and geometrical system parameter were investigated by
\citet{Schneider81C}, \citet{Voikhanskaya88C} and \citet{Skillman95C}. 
Of all individual CVs, MV~Lyr is arguably the system where flickering has 
been studied most intensively. These investigations started with
\citet{Pavlenko98C}, who fitted a shot noise model to the 
observed flickering, and \citet{Kraicheva99aC}, who studied the power 
spectrum of the flickering and constructed autocorrelation functions to 
learn about the involved time scales. However, the long term, high cadence 
monitoring by the Kepler spacecraft provided an unprecedented opportunity 
to explore flickering in MV~Lyr. \citet{Scaringi12aC}
detected a linear relationship between the rms-variations in the light
curve and the flux of the system; a feature previously observed in black
hole and neutron star binaries and in AGNs, suggesting a common physical 
origin for broad band variability in all compact accretors. The same
group also investigated the power density spectrum of the flickering light
source and 
were able to decompose it into a number of Lorentzian shaped components
\citep{Scaringi12bC}. On the basis of these findings \citet{Scaringi14C}
developed a model to explain flickering as fluctuation of the mass accretion
rate at different disk radii which propagate inwards and couple together.
Using the same Kepler data of MV~Lyr, \citet{Dobrotka15C} modelled the 
flickering variability in the framework of a statistical
model based on disk angular momentum transport via discrete turbulent
eddies with an exponential distribution of the dimension scale. The 
simulated light curves exhibit the typical linear rms-flux relationship
and the observed log-normal flux distribution. The authors can also explain
at least some of the breaks in the power spectra 
\citep[see also][]{Dobrotka17C}. Finally, I mention the work of 
\citet{Zamanov16C}
who showed that many accreting systems (CVs and symbiotic stars), 
including MV~Lyr, obey the same rms-flux relation over four orders of 
magnitude in flux.
The data available for the present study consist of light curves provided
by R.E.\ Nather and S.\ Shugarov, as well as data retrieved from the
AAVSO International Database. All light curves refer to the high state of 
MV~Lyr.

{\it AQ Men:} Only three studies of the optical properties of AQ~Men have 
been published. The star was detected in the Edinburgh-Cape Blue Object 
Survey as EC~05114-7955 by \citet{Chen01C}
and classified as a cataclysmic variable. They suspected a dwarf nova
nature but did not exclude the possibility of AQ~Men being a novalike
variable. \citet{Armstrong13C}
prefer the lattere classification based on the absence of 
observed outbursts even many years after discovery and the compatible 
spectrum. From radial velocity variations covering only 1.5 cycles 
\citet{Chen01C} derived a period of 3.12~h. In extensive photometric 
observations \citet{Armstrong13C} found a similar but more precise period 
and suspect the
presence of grazing eclipses. They also found indications for negative
superhumps and a super-orbital signal with a period of 3.78~d. However,
light curves observed by \citet{Bruch20C} contain a strong modulation with
a period of 1.95~h. It is not clear if any of the claimed periods is
really due to the orbital motion of AQ~Men.
The present study draws on the light curves already used by \citet{Bruch20C}.

{\it BT~Mon} is an old nova which suffered an outburst in 1939. Deep eclipses
in the light curve were detected by \citet{Robinson82C} allowing to measure 
the orbital period which was later refined by \citet{Smith98C}.
Based on spectroscopic data the latter authors also derived a complete
geometrical and dynamical model for the system. 
The present study is based on light curves with high time resolutions
provided by R.E.\ Nather \citep[see also][]{Robinson82C}, supplemented
by light curves retrieved from the AAVSO International Database.

{\it KQ~Mon:} Detected by \citet{Hoffmeister43C}, KQ~Mon was originally 
classfied as an irregluar variable and only in 1982 identified as a 
UX~UMa type novalike variable by
\citet{Sion82C} \citep[later confirmed by][]{Zwitter94C}. 
Not much is know about the object. \citet{Schmidtobreick05C} measured a 
spectroscopic orbital period. Modelling UV spectra, \citet{Wolfe13C} obtained a 
primary star mass of roughly $0.6 M_\odot$, and an orbital inclination
of no more than 60$^{\rm o}$. However, their model may be questionable because
it implies a distance of only 144 -- 165~pc, while the GAIA distance to KQ~Mon
is $628.7 \pm 11.5$~pc. 
Only three unpublished light curves by the author are available for this
study.

{\it RS~Oph} is well known as one of the most active recurrent novae. Since
1998 six outbursts have been observed and two more are suspected to have
occurred undetected \citep{Schaefer10C}.
In contrast to most other cataclysmic variables, the secondary star is not
on or close to the main sequence, but a M type giant. Thus, RS~Oph also
ranks among the symbiotic stars. It is one of the few symbiotic stars
exhibiting flickering which may be seen as a clear indication for the
presence of an accretion disk. \citet{Schaefer10C} presents a comprehensive
overview of the photometric history of the system in the context of 
recurrent novae in general.
As a system containing a giant component, the orbital period is much longer
than that of the bulk of CVs. The most precise measurement is that or
\citet{Brandi09C} (who also discuss earlier period measurements): 
$453.6 \pm 0.05$~d. As most
recurrent novae, RS~Oph contains a white dwarf with a high mass close to the
Chandrasekhar limit. Dynamical and geometrical system parameters are
discussed by \citet{Hachisu06C}, \citet{Hachisu18C}, \citet{Brandi09C} and 
\citet{Somero13C}. Beginning with the early observations of \citet{Walker57C} 
and \citet{Walker77C}, reports on rapid variations in RS~Oph abound in the 
literature. Starting with \citet{Bruch92} some authors
performed detailed studies of the flickering. Observing $UBVRI$ light curves,
\citet{Zamanov10C}
determined the colours of the flickering light source and estimated its
temperature to be $9500 \pm 500$~K, its radius to be $3.5 \pm 0.5 R_\odot$
and its luminosity to be $50 - 150 L_\odot$ (assuming a distance of 1.6~kpc).
\citet{Zamanov15C} found a significant correlation between the flickering 
amplitude and the average flux of the hot component. This was already 
noticed by \citet{Simon04C}. \citet{Zamanov16C} 
showed that many accreting systems (CVs and symbiotic 
stars), including RS~Oph obey the same rms-flux relation over four orders of 
magnitude in flux. Finally, in a series of papers, \citet{Georgiev19C}
determined statistical parameters of intra-night variations in numerous
light curves of RS~Oph and concluded that flickering occurs on several
preferred time scales (modes) which follow a power function.
The present study draws on multicolour light curves observed by the author
\citep[some of them published in][]{Bruch86C}, and numerous light curves
retrieved from the MEDUZA archive and the AAVSO International Database.

{\it V380~Oph} was identified as a variable star by \citet{Hoffmeister29C}. 
Its spectrum is considered a ``textbook example'' for a CV by 
\citet{Liu99C}.
The orbital period was spectroscopically determined by 
\citet{Shafter83dC} and later refined by \citet{Rodriguez-Gil07C} 
who also classified V380~Oph as a SW~Sex star. A 
long-term light curve with rather smooth variations between 
$\sim$14.3 and 16~mag is shown by \citet{Kafka04C}, but an excursion 
to a low state of $\sim$17.5~mag observed by 
\citet{Shugarov05C} and in particular light curves generated from more 
recent data provided by the AAVSO, the AFOEV (Association Fran\c{c}aise
des Observatuers d'\'Etoiles Variables) and the BAAVSS (British Astronomicial
Association Variable Star Section) also justify to 
count V380~Oph among the VY~Scl stars. Nothing is known about geometrical and
dynamical system parameters. 
The data available for this study consist high time 
resolution light curves taken from \citet{Bruch17aC}. They were
obtained at a brightness slightly below the normal high state magnitude. 
These data are complemented by light curves retrieved from the 
AAVSO International Database.

{\it V426~Oph} is a comparatively bright dwarf nova. A complete dynamical model
based on spectroscopic observations has been presented by \citet{Hessman88C}
who also summarizes all previous observations of the system.
At the long orbital period of almost 7~h
the secondary spectrum is visible in the optical range, making
the system a double lined spectroscopic cataclysmic binaries and
thus permitting a direct measurement of the mass ratio.
Based on apparently periodic modulation in X-rays and in optical light,
V426~Oph was repeatedly suspected to contain a magnetic white dwarf and
thus to belong to the intermediate polars \citep[see, e.g., the discussion 
in][]{Homer04C}. However, none of them has survived the test of time. 
Previous studies of flickering in this system are restricted to \citet{Bruch92}.
The data available for the present study consist of light curves
observed during outburst by \citet{Shugarov83C}, Walraven light 
curves observed by the author (quiescence) and taken from \citet{Hollander93C}
(outburst), and many light curves retrieved from the AAVSO International 
Database, referring to different photometric states.

{\it V442~Oph:} This novalike variable was identified as a cataclysmic 
variable by \citet{Szkody80bC}. It has a spectroscopic orbital period
\citep{Diaz01C} which
places the system close to the upper edge of the CV period gap. Like most
novalike variables in this period range, V442~Oph shares many spectroscopic
charateristics with the SW~Sex stars \citep{Hoard00C}. The occasional
ocurrence of low states in the light curve \citep{Garnavich88C} also 
makes it a VY~Scl star. The presence of superhumps (negative and positive) 
was reported by \citet{Patterson02C}.
Nothing is known about dynamical and geometrical system parameters
except that the absence of eclipses in the light curves precludes a high
orbital inclination. 
Strong continuum flickering has been seen in V442~Oph by various observers
\citep{Szkody89C, Szkody83C, Patterson02C} but has not been investigated in 
detail. \citet{Diaz01C} studied the H$\alpha$ emission line flickering. The 
data available for this study consist of light curves provided by B.\ Warner,
supplemented by by data retrieved from the OPD data bank and the AAVSO 
International Database.

{\it V841~Oph} erupted as a nova in 1848 at the dawn of the epoch of 
astrophysics \citep{Hind48C} and is thus one of the oldest old nova known.
Slight semi-regular brightness variations with an amplitude of the order of 
0.4~mag appear to occur on time scales of about 27 days \citep{Turner21C} or 
50 days \citep{Barnard21C, Steavenson23C, DellaValle87C, Hoard00C}.
The latter authors also report possibly periodic variations on a 
time scale fo 3.5 -- 5~y. It took a long time before the orbital period
could be determined. The first reliable value was measured spectroscopically,
using both, emission lines and late type absorption features, by 
\citet{Diaz03C} and was soon afterwards improved by \citet{Peters06C}.
At more than 14~h it is
among the longest of all CVs. Consequently, the secondary star has a fairly
early spectral type of K3$\pm$2 \citep{Peters06C}. Dynamical and geometrical
system parameters were derived by \citet{Diaz03C}.
The data available for this study consist of light curves taken from 
\citet{Warner95aC} and unpublished data observed by the author.

{\it V2051~Oph:} This deeply eclipsing SU~UMa type dwarf nova is one of the 
better studied CVs with numerous dedicated publications dealing in particular 
with the optical properties of the system. It was discovered as a variable 
star by \citet{Sanduleak72C} who already suspected its dwarf nova nature. 
Eclipses were first seen by \citet{Grauer81C}. The orbital period is slightly 
variable as shown by \citet{Echevarria93C} and \citet{Baptista03C}
\citep[see also][]{Qian15C}. 
Geometrical and dynamical parameters are quoted by \citet{Baptista98C}
and \citet{Longa-Pena15C}.
The location of the flickering light source in V2051~Oph was determined by
\citet{Bruch00C}. It resides very
close to the white dwarf while the outer parts of the accretion disk do
not take part in the flickering to a perceptible degree. Additionally,
flickering occurs also to a certain degree at the impact region of the 
transferred matter onto the accretion disk. This holds true for short 
period time scales of $\le$1~min. Using a different approach 
\citet{Baptista04C} were able to investigate
flickering at both, low and high frequencies. They found the low frequency
flickering to be associated mainly to the gas stream from the secondary
which overflows the accretion disk. The maximum occurs at the location of
the closest approach of the gas stream to the white dwarf. In contrast,
the high frequency flickering originates in the accretion disk and has
a distribution similar to the steady disk light. There is no evidence for
emission from the hot spot, gas stream or white dwarf. 
The observational data used here consist light curves provided by
R.E.\ Nather and B.\ Warner, all referring to quiescence.
In part, they were already used by \citet{Warner83C}, \citet{Warner87aC}
and \citet{Bruch00C}. Additional quiescent data sets were
taken from \citet{Hollander93C}. Further light curves obtained during
quiescence and outburst were retrieved from the OPD data bank and the AAVSO
International Database.

{\it CN~Ori:} The long-term light curve of the dwarf nova CN~Ori in unusual 
in the sense that in contrast to most other systems it does not contain well
defined quiescent phases but rather a successive series of outbursts
with a mean cycle length of only 16 days \citep{Bateson79C}.
In spite of being a comparatively bright object, few specific studies
of CN~Ori have been published. Extensive photometry was performed by
\citet{Schoembs82C} and \citet{Mantel88C}. Based on spectroscopic and
photometric data \citet{Barrera89aC} measured the orbital period.
Geometrical and dynamical system parameters are not well known. The
most reliable values quoted in the literature appear to be those dirived by
\citet{Mantel87C}.
The data sets used here consist of light curves provided by
R.E.\ Nather, B.\ Warner and E.\ Robinson. To these I add many light curves 
retrieved from the AAVSO International Database. The entire data set 
comprises all photometric stages between quiescence and outburst maximum.

{\it CZ Ori:} This dwarf nova was discovered by \citet{Hoffmeister28C}. 
\citet{Szkody84bC} studied the long term light curve, showing that the system 
oscillates between an outburst magnitude of 11.8~mag and 16.3~mag in
quiescence. However, more recent observations (see
the long-term AAVSO light curve) reveal that in quiescence CZ~Ori can
be even fainter than 17~mag. Based on spectroscopic observations
\citet{Spogli94C} and \citet{Ringwald94C} derived slightly different values
for the orbital period. Component masses and orbital inclination were
estimated by \citet{Spogli94C}.
Only one light curve provided by R.E.\ Nather, observed during early decline 
from an outburst is available for this study.

{\it V1193~Ori} was discovered by \citet{Hamuy86C} who noted its similarity to 
cataclysmic variables. This classification was confirmed by \citet{Bond87C} 
who showed that the system is a UX~UMa-type novalike variable. A spectroscopic
period measured by \citet{Ringwald94C} is consistent with the photometric 
period derived by \citet{Papadaki09C}. However, while the latter
authors did not see any other period in their data, \citet{Ak05C}
claim the presence of three different periods which they interpret as (i)
orbital [claiming that the periods of \citet{Ringwald94C} and 
\citet{Papadaki09C} are 1/day aliases), (ii) a negative superhump period,
and (iii) the beat between the other two periods. Nothing is known about 
geometrical
and dynamical system parameters. Flickering in V1193~Ori is rather strong. 
\citet{Warner88aC} draw attention to the fact
that its amplitude is significantly higher than that of the flickering in
other {\it bona fide} UX~UMa type novalike variables. \citet{Papadaki09C}
investigate the power spectrum of the flickering and mention the dificulties
to interpret the power law index. 
The data available for the present study consist of light curves taken
from \citet{Papadaki09C}, complemented by light curves retrieved from 
the AAVSO International Database.

{\it GS~Pav} was discovered as a variable star by \citet{Hoffmeister63C}.
\citet{Zwitter95C} published a spectrum. Only the Balmer emission 
lines can reliably be identified upon a blue continuum. The photometric 
measurements of \citet{Groot98C} revealed the star to be deeply 
eclipsing and thus permitted to derive the orbital period which was
later refined by \citet{Bruch17bC}. \citet{Groot98C} restricted the
orbital inclination to $74^{\rm o} < i < 83^{\rm o}$, but their mass
estimates depend on unreliable assumptions and will therefore not be used
here. While they classified the system to be a novalike variable of the 
RW~Tri subclass, it is listed as being of VY~Scl subtype in the
\citet{Ritter03C} catalogue. However, as \citet{Bruch17bC} argues,
there is no reason for the latter classification. Therefore I consider
GS~Pav as a normal UX~UMa type novalike variable. 
The data used in this study were all taken from \citet{Bruch17bC}.

{\it V345~Pav:} Most of what is known about V345~Pav (= EC~19314-5915) 
comes from a study of
\citet{Buckley92C}. They presented a detailed spectroscopic analysis,
supplemented by some photometry. Based on eclipses observed in the latter
they derived the orbital period which was later refined by \citet{Bruch17bC}. 
While \citet{Buckley92C} tentatively consider the star to be of Z~Cam subtype 
the absence of any outburst activity
in later data and the constancy of the out-of-eclipse light curve as
evidenced, e.g., by the Catalina Sky Survey (2005 -- 2013) renders a 
classification as an UX~UMa novalike variable likely, as quoted in the
\citet{Ritter03C} catalogue. The spectrum of V345~Pav
is peculiar in the sense that it exhibits metallic absorption lines typical 
of a G8 dwarf star which cannot be attributed to the secondary of the CV
system. Instead, \citet{Buckley92C} conjectured the presence of a third
star which contributes 23$\pm$5\% to the $B$ band light (see also Appendix~B). 
Nothing is known
about the component masses. The system being eclipsing, the orbital
inclination must evidently be high, but no specific values is quoted in
the literature. The light curves used in this study were all taken from 
\citet{Bruch17bC}.

{\it IP~Peg} is an eclipsing dwarf nova with a period originally measured by 
\citet{Goranskij85C} and most recently updated by 
\citet{Copperwheat10C}. The light curve in quiescence 
is dominated by a prominent orbital hump caused by the phase dependent 
visibility of the hot spot. 
IP~Peg belongs to the better studied CVs with numerous papers dedicated to it.
Geometrical and dynamical system parameters were derived by various authors
\citep[e.g.,][]{Marsh88C, Martin87C, Beekman00C, Smak02C, Ribeiro07aC,
Copperwheat10C}. 
The contribution of the cool companion to the total light of IP~Peg cannot be 
neglected. It is discussed in more detail in Appendix~B.
Specific statements on the flickering in IP~Peg are rare in the literature.
Using UV spectra observed with the Hubble Space Telescope
\citet{Hoard97C} investigated the spectrum of the flickering component
in the wavelength range $\lambda \lambda$ 1300 -- 2200~\AA. They find an 
excess between $\approx$1300 and $\approx$1500~\AA\ which they interpret
as an indication for a dominant contribution of the bright spot to the
flickering. This is in agreement with \citet{Bruch00C} who,
using quite different techniques, also concludes that most of the
flickering originates in the bright spot in this system.
The data used here consist of a light curve taken from \citet{Goranskij85C},
multicolour light curves provided by 
E.\ Robinson, and one light curve downloaded from 
the OPD data bank. The bulk of the data, however, was retrieved from AAVSO 
International Database. The observations encompass all photometric states.

{\it LQ~Peg} (= PG~2133+115) was discovered in the Palomar-Green survey
\citep{Green86C} and classified as a novalike variable by
\citet{Ferguson84C}. Most of the time hovering at a magnitude of
about 14.8~mag, occasional low
states, first seen by \citet{Sokolov96C} and then by
\citet{Watanabe99C}, \citet{Kato99C}, \citet{Schmidke02C} and \citet{Kafka05C},
testify to a VY~Scl nature of LQ~Peg. 
Not many details about the structure of the system are known. In particular,
no time resolved spectroscopy has been performed. Based on regular brightness
variations \citet{Papadaki06C} derived an orbital period of 2.99~h. However,
even this is contested by \citet{Rude12C} who argue that this period is 
rather due to a negative superhump. They consider the orbital period (not
observed directly, but inferred from indirect clues) to be 
3.22~h. The data available for the present study are those published by
\citet{Papadaki06C}, supplemented by light curves retrieved from the
AAVSO Internacional Database.

{\it TZ~Per:} This Z~Cam type dwarf nova was discovered as a variable star by
\citet{dEsterre12C}. The early observational history of the star is summarized
by \citet{Bruch92cC}. Based on radial velocity measurements \citet{Ringwald95C}
first measured the orbital period which was later refined by 
\citet{Echevarria99C}. Nothing
is known about geometrical and dynamical system parameters. 
Only two light curves retrieved from the AAVSO International Database 
are available for this study. They refer to outburst maximum and standstill, 
respectively.

{\it GK~Per} erupted in 1901 as the first bright nova in the 20$^{\rm th}$ 
century. After the outburst irregular fluctuations developed over
time into a well defined alternation between quiescence and dwarf nova 
outbursts recurring in intervals of several hundred days \citep{Bianchini83C}. 
Thus, today the system is established as one of a few old novae which became 
dwarf novae decades after the outburst. At very nearly 2 days the
orbital period of GK~Per is unusually long. The most precise measurement
was published by \citet{Morales-Rueda02C}.
The almost exact commensurability of the period with the solar day severely 
hampers observations destined to cover all orbital phases of the system.
The long period may explain some peculiarities of the outburst light curve
such as the long outburst intervals, the long duration and the symmetrical
shape of outburst rise and decline \citep{Evans09C}; properties which GK~Per 
shares with other long period dwarf novae such as BV~Cen \citep{Bateson74C}, 
V630~Cas \citep{Shears09C} and possibly V1129~Cen \citep{Bruch17cC}.
Many authors tried to determine geometrical and dynamical parameters of 
GK~Per \citep[see][for a summary]{Wada18C}.
\citet{Watson85C} detected a 351~s oscillation in X-rays, 
establishing GK~Per as an intermediate polar. In the optical range sometimes
QPOs with a period at or close to the X-ray period are seen.
\citep{Patterson81C, Mazeh85C}. The former author observed the optical 
counterpart of the 351~s X-ray pulsation in the $U$ band at a small and 
variable amplitude of $<$0.002 -- 0.016~mag. He did not see these
variations in blue light but detected instead QPOs at a similar period. 
In high speed photometric observations  \citet{Pezzuto96C}
did not detect a corresponding signal. If present, these modulations remain
on a small magnitude scale and occur predominantly in the ultraviolet. 
Therefore, they are not expected to have a significant impact on the
flickering amplitudes determined here. Some properties of the flickering in
GK~Per were investigated by \citet{Bruch92}. The optical light contains a 
significant contribution of the secondary star which is detailed in Appendix~B.
The data used for the present study refer to all photometric states of the
GK~Per dwarf nova cycle and consists of unfiltered light curves
provided by R.E.\ Nather and multicolour light curves observed by E.\ Robinson
and the author, respectively 
\citep[the later were already used by][]{Mazeh85C}. These data are 
suplemented by light curves retrieved from the MEDUSA archive and the AAVSO 
International Database. 

{\it KT Per:} Various publications count this dwarf nova among the Z~Cam star.
This is apparently based on a tentative classification by \citet{Loechel65C}.
However, as \citet{Simonsen11aC} points out, the long term AAVSO light curve, 
covering more than 40 years, does not contain evidence of standstills. The 
long-term light curve has also been studied by \citet{Szkody84C}.
The system is very active, remaining never in quiescence for an extended
period of time.
The first time resolved spectroscopy was published by \citet{Ratering93C}
who measured radial velocity variations and determined the orbital period.
Combing these data with data of \citet{Thorstensen97aC} and their own 
spectroscopy, \citet{Echevarria99C} refined the period.
Nothing is known on dynamical and geometrical system parameters. However,
the lack of eclipses in the light curve excludes a high inclination.
The data available for this study consist of light curves provided by 
R.E.\ Nather \citep[see][]{Robinson79C, Patterson81C}.
These data are supplemented light curves retrieved from the
AAVSO International Database. All light curves were observed in different
outburst states, well above the quiescent magnitude.

{\it RR Pic} (Nova Pic 1925), in contrast to other similarly bright novae, 
has attracted relatively little attention. While a number of photometric
studies have been published in the last decades of the past century
\citep[e.g.,][]{Warner81C, Warner86C, Kubiak84C, Haefner91C}
few detailed optical spectroscopic investigations have been performed
\citep{Schmidtobreick03C, Ribeiro06C}. In consequence
not much is known about the geometrical and dynamical system parameters.
The values cited in Table~C1 are
nothing more that the centres of wide ranges of permitted values
quoted by \citet{Ribeiro06C}. The orbital period of RR~Pic reveals itself 
through a periodic hump in the light curve. It was first determined by 
\citet{Vogt75C} and last updated by \citet{Vogt17C}.
The long term behaviour of the hump was studied by \citet{Fuentes-Morales18C}.
\citet{Schmidtobreick08C} and \citet{Fuentes-Morales18C}
claim the detection of superhumps in the light curve of RR~Pic. 
The former authors also characterize the power spectrum of the 
fickering activity, however, without drawing specific conclusions. They 
confirm the presence of rapid oscillation, already previously seen by, e.g.,
\citet{Warner81C} and \citet{Schoembs81aC}.
The present study is based on white light observations published by 
\citet{Warner86C}, complemented by light curves 
retrieved from the AAVSO International Database.

{\it TY~PsA:} The SU~UMa type nature of this star, originally identified as 
the high galactic latitude early type star PS~74 by \citet{Philip72C},
was discovered by \citet{Barwig82C}. It remains a little studied dwarf 
nova with few dedicated publications. \citet{Warner89C} performed
photometric and spectroscopic observations in quiescence and different
outburst states. They measured a photometric period
but could not detect this 
period in radial velocity measurements. Only later, \citet{ODonoghue92C}
found a spectroscopic period compatible with the photometric one.
\citet{Warner89C} also suspect the presence of shallow eclipses
seen during the decline from a superoutburst but not in quiescence.
Similar features were already mentioned by \citet{Barwig82C}. A range of 
possible component masses is mentioned by \citet{ODonoghue92C},
but it is too wide to be useful in the present context. They also mention
a possible range for the orbital inclination .
The data available for the present study consist of light curves already 
presented by \citet{Warner89C}. 

{\it VV~Pup} is the third among all detected polars and one of the best 
studied systems of its kind. Its nature as a magnetic cataclysmic variable 
was noticed by \citet{Tapia77bC}
on behalf of its strong circular polarization. Even before its identification
as an AM~Her star strong photometric modulations permitted \citet{Walker65C}
to measure a period which was interpreted as being the orbital period, and 
which was later confirmed spectroscopically by \citet{Schneider80C} and 
\citet{Cowley82C}. Component masses were determined by \citet{Howell06C}. 
The orbital inclination was measured by various authors 
\citep{Cropper88C, Vennes95C, Sirk98C}.
The data available for the present study consist light curves provided
by R.E.\ Nather and observed by the author.

{\it CP~Pup:} (Nova Puppis 1942) was one of the brightest novae ever 
observed. At a quiescent magnitude of $\sim$15~mag \citet{Bruch94C} it
is bright enough for detailed observations. Yet the nature of this object
is still shrouded in mysteries. Not only are the spectroscopic and 
photometric periods different from each other, but various authors 
found discrepant values at different epochs \citep[see][and references 
therein for a more detailed discussion]{Bianchini12C}. \citet{Mason13C}
even raise doubts whether the
radial velocity variations reflect the binary period at all or whether
they might be caused by the rotation of a magnetic white dwarf (see below)
in a system with a longer orbital period.
Another mystery concerns the dynamical properties of the system which lead 
to considerable uncertainties about the component masses and the orbital 
inclination. \citet{ODonoghue89C}
point out the problems concerning a dynamical solution based on radial
velocity measurements. They lead to low white dwarf masses 
\citep{Duerbeck87C, Barrera89bC, White93C}. This 
is in contradiction with the high outburst amplitude and the rapid decline rate
of $t_3 = 8$ days \citep{Duerbeck81C} which according to our current 
understanding of nova outbursts requires a massive white dwarf. 
Not the least of the enigmas about the CP~Pup concern its possible nature as a
magnetic CV. Several groups suggested that the unstable photometric period
may be due to a rotation of the white dwarf with a period slightly out of
synchronization with the orbital period \citep{Warner85C, Diaz91bC, White93C}.
Based on their spectroscopic observations, \citet{Bianchini12C} come to a 
similar conclusion. The X-ray properties of the star 
make \citet{Balman95C} also suggest an intermediate polar nature for 
CP~Pup. \citet{Mason13C} thoroughly discuss this issue.
The data available for the present study consist of 
light curves provided by B.\ Warner. They were already published by 
\citet{Warner85C} and \citet{ODonoghue89C}. These data are 
complemented by numerous light curves retrieved from the AAVSO International
Database.

{\it V348~Pup:} \citet{Tuohy90C}
identified the faint HEAO X-ray source 1H0709-360 as an eclipsing novalike
variable in the CV period gap. They also measured the orbital period, later
confirmed by \citet{Baptista96C}, \citet{Rolfe00C} and \citet{Dai10cC}. The 
latter authors report a small period increase over time.  Photometric 
variations with a period slightly different from the orbital period made 
\citet{Tuohy90C} suspect an intermediate polar
nature of the star; a notion which could neither be confirmed no rejected
in pointed X-ray observations by \citet{Rosen94C}, while \citet{Froning03aC}
found no evidence for a magnetic nature. Instead of intermediate polar
type variations, \citet{Rolfe00C} reported superhumps in the system which, 
however, were not seen when \citet{Saito16C} observed the V348~Pup at 
a later epoch. 
Dynamical and geometrical system parameters were determined by
\citet{Rolfe00C}, \citet{Rodriguez-Gil01C} and \citet{Saito16C}.
The observational data used in this study consist of light curves
retrieved from the OPD data bank.

{\it T~Pyx} is one of the best known and thoroughly observed recurrent novae 
which suffered its first observed outburst in 1890. Thereafter it erupted on 
average every 20 years \citep{Schaefer10C} until 1967, then requiring 
another 44 year until its (so far) last eruption in 2011. \citet{Schaefer10aC}
argue that these recurrent nova type eruptions were preceded by an
(unobserved) classical nova outburst in 1866. \citet{Godon18C} summarize the 
extensive literature dealing with T~Pyx and discuss in some detail the system 
parameters which have a bearing on the current work. The system 
belongs to the small group of recurrent novae with short (time scale of
hours) orbital periods \citep{Schaefer10C}. It was photometrically
measured first by \citet{Schaefer92C} and then seen and found to be slighly 
increasing over time by various authors \citep{Patterson98C, Uthas10C,
Patterson17C}.
All model calculations of recurrent novae outbursts indicate that the mass
of the white dwarf is high \citep[close to the Chandrasekhar limit; see, 
e.g.,][and later references]{Starrfield85C}. This is in line with a primary 
star mass for T~Pyx derived by \citet{Schaefer10aC} and \citet{Shara18C}, 
respectively, based on nova outburst models. However, a much smaller mass  
is advocated by \citet{Uthas10C} on grounds of their radial velocity study. 
Conflicting secondary star masses have been derived by \citet{Patterson17C}
and \citet{Selvelli95C}. There is also no agreement concerning the orbital\
inclination of the system \citep{Patterson98C, Uthas10C, Shore13C, 
Patterson17C}.
The observational data available for this study consist of light curves 
provided by B.\ Warner, supplemented by numerous data sets retrieved from 
the AAVSO International Database. The latter were all observed during the late
decline from the 2011 outburst and the subsequent quiescence between 2012
and 2018. 

{\it WZ~Sge} is the prototype of a small class of dwarf novae which do not
exhibit normal outbursts but only rare high amplitude superoutburst
with intervals which can reach decades. Initially, it was thought to
be a nova when it was first observed in outburst in 1913 by J.C.\ Mackie,
as reported by \citet{Leavitt19C}. A second outburst observed by 
K.\ Himpel and reported by \citet{Mayall46C} led to a classification as 
recurrent nova. Only much later \citet{Warner76aC}
suggested the system to be dwarf nova. WZ~Sge is doubtlessly one of the
best studied CVs ever. 
Periodic photometric variations with a primary minimum which is now known 
to be a shallow eclipse of the bright spot in the system first permitted 
\citet{Krzeminski62C} to measure the orbital period. Refinements have been 
performed various times afterwards, most recently by \citet{Patterson18C} 
who also discuss period variations. 
Even ignoring discrepant early mass estimates (performed before the 
canonical CV model was established) mass determinations of WZ~Sge are 
controversial \citep{Gilliland86C, Smak93C, Spruit98C, Skidmore00C, Steeghs07C,
Harrison16C}
In contrast to the masses, there is not much disagreement concerning the
binary inclination which cannot be small because the system exhibits
eclipses and not very high because the eclipses are only grazing
\citep{Smak93C, Spruit98C, Skidmore02C}. 
The available data consist of light curves provided by R.E.\ Nather. They
were already used by \citet{Patterson80C} and \citet{Patterson81aC}.
All data refer to quiescence. There
are many more light curves in the AAVSO International Database. However,
these do not have a suitable quality to be of use here.

{\it V3885~Sgr:} At an average magnitude of $B = 10.3$ the novalike variable 
V3885~Sgr (= CD~-42$^{\rm o}$~14462) is not only one of the brightest CVs in 
the sky, but it is also quite stable, exhibiting only slight variations since 
the first photometric observations in 1899 \citep{Bond78C}.
The orbital period was measured by \citet{Hartley05C} and \citet{Ribeiro07C} 
with consistent results. Geometrical and dynamical system 
parameters were determined by \citet{Linnell09C} by modelling the 
ultraviolet and far ultraviolet spectral energy distribution. Their best 
model parameters are within the limits derived by \citet{Hartley05C}.
Flickering in V3885~Sgr, although remaining on a low level, has been
observed for a long time, often together with short period oscillations
\citep{Warner73C, Hesser74C, Cowley77bC}. It has been subjected
to a thorough analysis by \citet{Ribeiro07C} who studied also
the relationship between the continuum flickering and rapid variations in
the H$\alpha$ emission line. They found that the source of the line flickering
may be associated with the secondary star and explain it as  being due to
disk flickering in the UV reprocessed on the illuminated face
of the secondary. A cross-correlation study between the continuum and line
flickering confirms a correlation between both. 
The data available for the current study consist of light curves provided 
by B.\ Warner, and multicolour light curves observed by \citet{Hollander93C}
and the author.

{\it V4140~Sgr:} Discovered as a variable star by 
\citet{Hoffmeister63C} V4140~Sgr (= NSV~12615) was identified as an 
eclipsing SU~UMa type dwarf nova by \citet{Jablonski87C}.
The orbital period was originally measured by the latter authors
and later discussed more thoroughly by \citet{Baptista92C} and 
\citet{Baptista03C}. While the mean period
identifies the system as being close to the lower CV period limit, 
it exhibits slight variations which may either be secular or periodic. 
Geometrical and dynamical system parameters have been determined by
\citet{Borges05C}. Several
properties of the system are quite different from those of normal SU~UMa
stars. As pointed out by \citet{Borges05C},
the normal outbursts, which occur in intervals of 80 -- 90 days, have
an amplitude of only about 1~mag, and the superoutbursts are 1~mag
brighter. These are remarkably low values. Moreover, the disk temperature,
both, in quiescence and in outburst, always remains lower than the critical 
temperature for outbursts to occur according to the disk instability model 
\citep{Lasota01C}. This is confirmed by \citet{Baptista16C} who also consider 
the quiescent disk to be in a high-viscosity steady state regime. Therefore 
\citet{Borges05C} and \citet{Baptista16C} conclude that the outbursts of 
V4140~Sgr are not due to a disk instability but instead are powered by 
bursts of enhanced mass transfer from the secondary star.  
The location of the flickering in V4140~Sgr was investigated by 
\citet{Baptista16C} via eclipse tomography. Their maps show flickering sources
at an azimuthally extended stream-disk impact region and in the inner disk.
Both are responsible for flickering on longer time scales ($>$500~s). Instead,
flickering on shorter time scales (which is dominant in optical light)
has its origin in an extended disk region. \citet{Baptista16C} conclude
that -- if flickering is caused according to the MHD turbulence model of 
\citet{Geertsema92C} -- the quiescent viscosity parameter is as large as 
$\alpha \sim 0.2 - 0.4$. 
All data used for the current study were either published by 
\citet{Baptista89C} or retrieved from the OPD data bank.

{\it V893~Sco:} \citet{Satyvoldiev72C} identified this dwarf nova 
as a variable star,
but it got lost thereafter. Only much later \citet{Kato98C} re-identified
the system. Soon thereafter \citet{Bruch00aC} discovered grazing eclipses
which were also seen in X-rays by \citet{Mukai09C}. The orbital period 
was first measured by \citet{Bruch00aC} and later refined by \citet{Bruch14C}
who found a low amplitude cyclic variations which may be attributed to a
light travelling effect due to the presence of a giant planet orbiting
V893~Sco. Time resolved spectroscopic observations were published by
\citet{Matsumoto00C} and \citet{Mason01C}. The latter authors also derived
component masses and the orbital inclination.
The orbital period being below the CV period gap, V893~Sco is expected to
be of SU~UMa type. However, it took a long time to detect the defining 
properties of these stars, i.e., superoutbursts in addition to normal 
outbursts. To my knowledge, in spite of a dense coverage of the long term 
light curve since 1999, the first superoutburst was only observed in 2016 as
reported by \citet{Kato17C}.
Apart from strong flickering V893~Sco exhibits QPOs \citep{Warner03C}
and possibly DNOs \citep{Pretorius06C}. \citet{Bruch14C} also saw transient
semi-periodic variations on time scales of several minutes which may be 
identified as QPOs. However, he pointed out the difficulty to draw a
dividing line between QPOs and flickering. The bulk of the data used in 
this study are  observations already used by \citet{Bruch00aC} and
\citet{Bruch14C}. They are complemented by light curves retrieved from
the AAVSO International Database.

{\it VY~Scl} is the prototype of a class of novalike variables, most of 
them with orbital periods in the range of 3 -- 4~h, which normally are 
found in a high state but occasionally exhibit low states up to a couple 
of magnitudes fainter than usual. VY~Scl was discovered as a blue object 
with large amplitude variations (i.e., high and low states) by 
\citet{Luyten59C}.
In spite of its relatively high magnitude during the high state the system
has not been as extensively observed as may be expected considering its
status of prototype of a whole class. There is even some controversy
concerning basic parameters such as the orbital period. It was first
measured spectroscopically by \citet{Hutchings84C} at a value which is shorter 
than the period advocated by \citet{Martinez-Pais00C}. Only recently, using 
additional spectroscopic observations performed during a low state
\citet{Schmidtobreick18C} claim to have resolved this discrepancy, 
finding a period which is compatible with the earlier data of
\citet{Hutchings84C}.
Although \citet{Martinez-Pais00C} and \citet{Schmidtobreick18C}
quote dynamical system parameters, the allowed masses encompass a large 
range. Therefore I will assume them to be unknown. The orbital inclination
found by these authors is also uncertain, but it is definitely quite low.
The data available for the present study consist of light curves observed
by \citet{Hollander93C} in the Walraven system.

{\it VZ~Scl} was discovered by \citet{Krzeminski66C} as a deeply eclipsing 
novalike variable with the a period of 3.47~h. Although photometric and 
spectroscopic observations have thus a high potential to
reveal many details about the properties and the structure of the system,
surprisingly little work has been dedicated to the star. The photometric
history as detailed by \citet{Sherrington84C}. \citet{ODonoghue87aC}
revealed low states in VZ~Scl leading to its classification as a VY~Scl star.
No reliable mass determinations for the system components exist. As is evident
from the deep eclipses the orbital inclination must be high
\citep{Sherrington84C, ODonoghue87aC}.
While no specific study of the flickering in VZ~Scl has been performed
before \citep[except for the determination of wavelet parameters by][] 
{Fritz98}, \citet{ODonoghue87aC} mention that the source of the 
flickering appears to be centred on the white dwarf primary and is not
associated with the bright spot.
The observational data available for this study consists of light curves
already published by \citet{ODonoghue87aC}, complemented by light curves
retrieved from the OPD data bank and AAVSO International Database.

{\it LX~Ser}, also known as Stepanyan's star, was first noticed by 
\citet{Stepanyan79C}. Is is a deeply eclipsing novalike variable. The 
orbital period 
has been steadily refined by many authors over the years and was last
updated by \citet{Li17C}. The long-term light curve shows a low
state \citep{Liller80C}. The system may therefore be considered to be
of VY~Scl subtype \citep{Leach99C}. Dynamical and geometrical
system parameters have been studied by \citet{Young81aC} and \citet{Marin07C}.
All light curves used in the present study were retrieved from the AAVSO 
International Database.

{\it RW~Sex} (= BD -7$^{\rm o}$ 3007) is one of the brightest novalike variables.
\citet{Hesser72C} performed early photometric 
observations and found low amplitude flickering. \citet{Cowley77aC}
were the first to suggest an orbital period (albeit with
possible aliases) based on spectroscopy, which was then confirmed and
improved by \citet{Bolick87C} and \citet{Beuermann92C}. While
\citet{Bolick87C} derive tentative component masses, \citet{Beuermann92C} 
argue that the uncertainty of the orbital inclination renders individual 
masses very uncertainty. Therefore, I only quote the mass ratio in Table~C1.
The data available for this study consist of light curves provided by 
R.E.\ Nather, and multicolour data sets 
observed by the author and his collaborators \citep[see also][]{Bruch91bC}.

{\it KK~Tel:} Discovered by \citet{Hoffmeister62C}
as a  variable blue object and suspected by him to be a dwarf nova,
KK~Tel remains a little studied object. The dwarf nova classification was
confirmed photometrically by \citet{Bateson82aC} and \citet{McIntosh89C},
and spectroscopically by \citet{Zwitter95C}. A photometric modulation of 
2.02~h, interpreted as the orbital period, observed by \citet{Howell91C}
during minimum, together with the large magnitude difference between minimum 
and maximum of 6.2~mag suggested a nature as SU~UMa star. In fact,
\citet{Kato03aC} and \citet{Patterson03C} observed superhumps during a 
bright state. Their period is slightly longer that the photometric period 
reported by \citet{Howell91C}, confirming
the latter to be orbital in origin. Nothing is known about dynamical and
geometrical system parameters. 
Only one light curve, observed during superoutburst and retrieved 
from the AAVSO International Database, is available for this study.

{\it RW~Tri} is a deeply eclipsing novalike variable of the UX~UMa type. 
The orbital period was determined by \citet{Robinson91C}
and appears not to have been refined since then. 
Low amplitude ($\approx 0.5$~mag) oscillations on time scales of some 
tens of days have been observed by \citet{Honeycutt94C}, \citet{Honeycutt01C} 
and even more clearly by \citet{Bruch20C}. Apart from this the system is
relatively stable as is corroborated by the long 
term AAVSO light curve which shows the out-of-eclipse magnitude in general 
to be between 12.5 and 13~mag. However, there are exceptions.
An isolated measurement in the AAVSO light curve on 2011, April 14 shows 
the system at 8.8~mag\footnote{No observations are available
in the interval of 7 days before and 10 days after this event.}. Another
bright state was seen in spectrophotometric observations performed by
\citet{Still95C} in 1988, October, when RW~Tri was 3.5~mag 
brighter than normal. \citet{Smak19C} reports negative superhumps in
his observations obtained in 1984 and possibly 1957. They were, however,
not present in 2015 -- 2016 \citep{Bruch20C}.
Component masses have repeatedly been determined in the past, but
with conflicting results \citep{Kaitchuck83C, Shafter83aC, Rutten92C, Smak95C, 
VandePutte03C, Poole03C, Mizusawa10C}. Orbital inclinations between 
67$^{\rm o}$ and 80$^{\rm o}$ are reported in the literature
\citep{Kaitchuck83C, Rutten92C, Smak95C, Mason97C}. The contribution of the 
secondary star to the total system light is discussed in Appendix~B.
The data used here consist light curves provided by R.E.\ Nather, multicolour
light curves observed by E.\ Robinson and the author, and numerous data sets
retrieved from the AAVSO International Database.

{\it EF~Tuc} (= EC~23593-6724) was discovered as a dwarf nova in the 
Edinburgh-Cape Blue Object Survey. The spectrum shown by \citet{Stobie95C}
is dominated by strong hydrogen emission lines. \citet{Chen01C}
investigated the spectrum in somewhat more detail and suspected EF~Tuc to be 
of SU~UMa type. A photometric study of the system was presented by 
\citet{Bruch17aC} who -- while not confirming the dwarf nova nature of the 
system -- argues against the classification as an SU~UMa star. Indeed,
the average quiescent magnitude of roughly 14.5~mag as judged from 
the AAVSO long term light curve together with the distance and interstellar
reddening as quoted in Table~C1
leads to an absolute magnitude of $\approx$3.8~mag, way too bright for a 
quiescent SU~UMa type star. Such a 
classification is also not possible if the orbital period of 3.6 hours 
quoted by the \citet{Ritter03C} is correct. This value is based on an 
informal communications by J.\ Patterson in 2003 and 
2006\footnote{http://cbastro.org/communications/news/messages/0350.html;
http://cbastro.org/communications/news/messages/0487.html} where it is
considered a candidate period. It has never been confirmed and it is
not seen in the data of \citet{Bruch17aC}. Therefore, I consider the orbital
period of EF~Tuc unknown. The same is true for dynamical and geometrical
system parameter. 
As already mentioned by \citet{Chen01C} and confirmed by \citet{Bruch17aC},
EF~Tuc exhibits strong flickering. \citet{Bruch17aC} also provides some
quantitative information on the flickering parameters which are re-investigated
with more rigour here. The data used in the present study are all taken
from \citet{Bruch17aC}, including the AAVSO data mentioned in that publication.

{\it SU~UMa} is the prototype of a class of short period dwarf novae which
apart from normal outburst exhibit from time to time so-called superoutbursts
which last longer and are brighter at maximum than these. While the
outburst properties have extensively been investigated \citep[e.g.,][]
{Szkody84C, Udalski90C,  vanParadijs94C, Rosenzweig00C, Imada13C} 
much less is known about the struture of the binary system. 
\citet{Voikhanskaya83C} quote two possible values for the orbital period, 
one of which was later confirmed with higher precision by 
\citet{Thorstensen86C}. The component masses remain unknown just 
as the orbital inclination. The latter is probably moderately low 
\citep{Thorstensen86C}.
The data available for this study consist of light curves provided by 
R.E.\ Nather, complemented light curves retrieved from the
AAVSO International Database. The data comprise all photometric states
between quiescence and superoutburst.

{\it SW~UMa} is a SU~UMa type dwarf nova. While variations during outbursts
have been extensively studied \citep[e.g.][]{Robinson87C, Kato92C,
Semeniuk97C, Nogami98C, Pavlenko00C, Soejima09C} 
the basic dynamical and geometrical system parameters remain to a large degree
unknown. In particular the component masses and the orbital inclination have
not yet been determined. The orbital period was first measured spectroscopically
by \citet{Shafter83dC} and then slightly revised by \citet{Shafter86bC}.
SW~UMa has been suspected to be an intermediate polar because 
\citet{Shafter86bC} saw a modulation with a period of 15.9~min in optical 
light during quiescence and in X-rays. However, these variations were never 
seen again and the intermediate polar nature remains unconfirmed.
The data available for this study consist of light curves provided by 
R.E.\ Nather and are supplemented by data retrieved from the AAVSO 
International Database. All observations refer to superoutburst or to the 
early decline.

{\it UX~UMa} is the prototype of novalike stars, in particular of those system 
which, in contrast to VY~Scl star, have never been observed to go into a low 
state. It is an eclipsing system which in many aspects is similar to RW~Tri
(see above). As the prototype of its class and the
brightest eclipsing novalike variable, UX~UMa has been extensively studied 
in the past \citep[see][for a short summary of previous observations]
{Neustroev11C}. The orbital period was last refined by
\citet{Baptista95C}. Dynamical and geometrical system parameters have been
discussed by many authors \citep{Baptista95C, Froning03bC, Smak94C, 
VandePutte03C}. In extensive photometric observations during the 2015
observing season \citet{deMiguel16C} found a modulation with a period of 
3.680~d in the light curve of UX~UMa which they
interpret as being due to a retrograde precession of the accretion disk.
An associated negative superhump at the beat period of the precession and
the orbit is also seen. \citet{Bruch20C} confirmed this behaviour but also
noted that it was restricted to that particular season and did not repeat
itself in previous or following years.  
Rapid oscillations with a period of 29~s and a low amplitude 
were first detected by \citet{Warner72C} in optical observations, but
were not seen by \citet{Froning03bC} in the far UV. Instead, the latter
authors detected quite strong (flux variations of more than a factor of two 
within 5 minutes) flickering in this wavelength range. It is restricted to 
the continuum with no flickering in the emission lines. The optical 
flickering was studied by \citet{Bruch00C} who concluded that at
least its high frequency part occurs very close to the white dwarf, while
the hot spot may also contribute. The contribution of the secondary star
to the total light is discussed in Appendix~B.
Of all systems investigated in this study the number of light curves of
UX~UMa is the second largest (after TT~Ari). 
The bulk of them were retrieved from the AAVSO International Database.
To these data I add light curves provided by R.E.\ Nather and E.\ Robinson. 

{\it IX~Vel:} This novalike variable is the brightest CV in the sky. It may 
therefore be surprising that -- when compared to other system -- relatively
few dedicated studies have been devoted to this star. \citet{Linnell07C}
extensively review the previous literature about IX~Vel. With $\sim$4.65~h 
\citep{Beuermann90C}
it has a similar orbital period as other famous novalikes like UX~UMa
and TW~Tri, but in contrast to these systems IX~Vel is not eclipsing.
Dynamical and geometrical system parameters have been determined by
\citet{Beuermann90C}, \citet{Linnell07C} and \citet{Hoard14C}.
Occasionally, IX~Vel exhibits coherent oscillations with 
periods in the range 24 -- 29~s and with amplitudes of
$\sim$1~mmag \citep{Warner85aC}. Flickering occurs on a rather low
scale. A study of autocorrelations functions of its light curves make
\citet{Williams84C} suggest that two time scales exist in the
flickering activity, namely 77 and 500~s. 
Most of the data used here were provided by B.\ Warner.
Some the them were already used in the study of \citet{Warner85aC}.
These data are supplemented by unpublished light curves observed by the 
author and data sets retrieved from the OPD data bank.

{\it HV~Vir} belongs to the class of WZ~Sge stars, i.e., short period
dwarf novae which exhibit rare large amplitude superoutbursts, without
interspersed normal outbursts (as seen in SU~UMa type stars). The star
was discovered as a variable by \citet{Schneller31C} and for a long time 
thought to be a nova. Only when \citet{Schmeer92C} detected a second outburst 
in 1992 and \citet{DellaValle92C} verfied that the spectrum was that of a 
dwarf nova in outburst instead of a (recurrent) nova, the classification was 
rectified. The faintness of HV~Vir during quiescence 
\citep[19.2~mag,][]{Howell92C} explains the almost complete absence 
\citep[mentioning the notable exception of][]{Szkody02C} of detailed optical 
studies of the system in this phase. In contrast, the 1992 outburst was well 
observed \citep{Ingram92C, Barwig92C, Leibowitz94C, Kato01C}. Another 
outburst in 2002 was covered by \citet{Ishioka03C}.
Details about the structure of the system are rare. The orbital period
has never been measured directly. Assuming that early superhumps in 
WZ~Sge stars occur with the same period, \citet{Kato01C}, using
data from various previous publications, suggested the orbital period to
be 1.3700~h. Component masses and orbital inclination remain unknown. 
The data available for this study consist of light curves already included 
in the study of \citet{Leibowitz94C}, complemented by light curves retrieved 
from the AAVSO International Database. All observations were taken during 
superoutburst.

{\it CTCV~J2056-3014:} Little is known about this star. \citet{Augusteijn10C}
observed the star as part of follow-up observations of CV candidates
identified in the Cal\'an-Tololo Survey \citep[see][and references therein]
{Maza89C}. They measured a spectroscopic period of 1.757~h and
suspect the system to be an intermediate polar, based upon apparently periodic
variations of 15.4~min observed in one light curve. While the arguments which
lead \citet{Augusteijn10C} to this classification were refuted by 
\citet{Bruch18aC}, more recently \citet{Lopes20C} revived this notion
because they found a coherent 29.6~s oscillation in X-ray observations
and an X-ray spectrum compatible with that of the recently identified class
of low luminosity IPs. A re-analysis of the data of \citet{Bruch18aC} 
revealed the presence of oscillation with the same period also in the optical. 
Associating it to the rotation of the white dwarf means that CTCV~J2056-3014
harbours the white dwarf with the shortest securely known spin period. 
The data available for this study consist of the light curves already
used by \citet{Bruch18aC}.

{\it EC~21178-5417:} Unitil quite recently, not much was known about this CV. 
Identified as a blue object in the Edinburgh Cape survey \citep{Stobie97C},
it was first mentioned as a cataclysmic variable by \citet{Warner03C} who 
found it to be an eclipsing novalike system. Time resolved photometry
was published by \citet{Bruch17bC} who measured the orbital period. 
It was refined by \citet{Khangale20C} who published a spectroscopic
analysis of the system. Component masses remain unknown. The orbital 
inclination must be high in view of the deep eclipses in the 
light curves. This is confirmed by \citet{Khangale20C} who estimate
$i = 83^o \pm 7^o$.  The data available for this study consist of light curves
already presented by \citet{Bruch17bC}.

{\it LS~IV~--08$^o$~3} was originally classified as an OB star 
\citet{Nassau63C}. At $V \approx 11.5$~mag \citep{Hog00C}
it is a fairly bright star. The first detailed investigation
of the star was performed by \citet{Stark08C}, who also give an overview
of the history of our knowledge about this object. They performed a detailed 
spectroscopic study, found LS~IV~--08$^{\rm o}$~3 to be a binary, and 
measured the orbital period. The characteristics of the system 
suggested a reclassification as a nova-like variable of the UX~UMa subtype. 
\citet{Stark08C} also present some time resolved photometry which 
reveals low scale (a few hundredths of a magnitude) apparently stochastic 
variations superposed on slight orbital variations. These findings were
confirmed and enhanced by \citet{Bruch17dC}. They improved the precision
of the orbital period. Component masses and orbital 
inclination of LS~IV~--08$^{\rm o}$~3 remain unknown.
The observational data used here are the light curves already discussed by 
\citet{Bruch17dC}, supplemented by light curves retrieved from the AAVSO 
International Database.


\section{Derivation of the flickering spectrum}
\label{Flickering spectrum} 

Neglecting the contribution of the white
dwarf and remembering that the light of the secondary has already been 
removed when the corresponding correction to the flickering amplitude was 
made, the total flux of a CV can be thought to be composed of two components. 
The first one, $F_{\rm c}$, is constant, being emitted by the 
quiet (i.e., non flickering) part of the accretion disk. The second part, 
$F_{\rm fl}$, represents the flickering light source. Since, in general, it 
cannot be assumed that the latter is 100\% modulated (i.e., that at the 
minimum of the flickering activity $F_{\rm fl} = 0$) it can be expressed as 
consisting of a base level, $F_{{\rm fl},0}$, and a part $\Delta F_{\rm fl}$ 
(the latter being that part which is really seen as flickering). Let 
${\rm R}$ and ${\rm C}$ denominate a reference and a comparison passband, 
respectively, and $\Delta {\rm R}$ and $\Delta {\rm C}$ the flickering 
amplitude observed in the two passbands. If $m_0$ and $m_1$ are 
the magnitudes of the system at the minimum of the flickering activity and
at some other brightness level, respectively, the difference $m_0 - m_1$ in
the reference passband can be written as: 
\begin{eqnarray}
\label{Equation wavelength dependence of the flickering, Eq. 1}
\Delta {\rm R} & \equiv & m_{{\rm R},0} - m_{{\rm R},1} \nonumber \\
               & = &
-2.5 \log \left( \frac{F_{\rm c,R} + F_{{\rm fl,R},0}}
                      {F_{\rm c,R} + F_{{\rm fl,R},0} + \Delta F_{\rm fl,R}}
          \right) 
\end{eqnarray}
Similarly, the magnitude difference in the comparison passband is:
\begin{eqnarray}
\label{Equation wavelength dependence of the flickering, Eq. 2}
\Delta {\rm C} & \equiv & m_{{\rm C},0} - m_{{\rm C},1} \nonumber \\
               &  = &
-2.5 \log \left( \frac{F_{\rm c,C} + F_{{\rm fl,C},0}}
                      {F_{\rm c,C} + F_{{\rm fl,C},0} + \Delta F_{\rm fl,C}}
          \right)
\end{eqnarray}
The flux unit is arbitrary, permitting to define the flux of the constant
light source in the reference passbands to be $F_{{\rm c,R}} = 1$. Its 
flux in the comparison passband can then be expresses as 
$F_{\rm c,C} = \beta F_{{\rm c,R}} = \beta$. 
If the spectrum of the constant source can be specified, $\beta$ is known. 
Let $\alpha = \alpha (\lambda)$ describe the spectrum of the flickering
light source, normalized at the wavelength of the reference passband.
Assuming that $\alpha (\lambda)$ does not change when the flickering light
brightens or weakens, $F_{{\rm fl,C},0} = \alpha F_{{\rm fl,R},0}$, and 
$\Delta F_{\rm fl,C} = \alpha \Delta F_{\rm fl,R}$. Finally, the ratio 
$\Delta {\rm C}/\Delta {\rm R} \equiv \gamma$ is known from 
Table~2. Thus,
Eqs.~\ref{Equation wavelength dependence of the flickering, Eq. 1} and
\ref{Equation wavelength dependence of the flickering, Eq. 2} can be 
expressed as
\begin{equation}
\label{Equation wavelength dependence of the flickering, Eq. 3}
\Delta {\rm R} = -2.5 \log \left( \frac{1 + F_{{\rm fl,R},0}}
                                {1 + F_{{\rm fl,R},0} + \Delta F_{\rm fl,R}}
                            \right)
\end{equation}
and
\begin{equation}
\label{Equation wavelength dependence of the flickering, Eq. 4}
\Delta {\rm C} = \gamma \Delta {\rm R} =
-2.5 \log \left( \frac{\beta + \alpha F_{{\rm fl,R},0}}
                 {\beta + \alpha F_{{\rm fl,R},0} + \alpha \Delta F_{\rm fl,R}}
          \right)
\end{equation}

Before proceeding it is now necessary to quantify the parameter $\beta$,
i.e., the spectrum of the non-flickering light. Since the uncertainties in
the current approach do not warrant a more sophisticated treatment, I will 
assume it to radiate like a standard steady-state accretion disk 
\citep[e.g.,][]{Shakura73D, Frank02D} in the black body approximation. 
The spectrum is then a function of the
mass $M_1$ of the central star, the mass transfer rate $\dot{M}$ through
the disk, and the inner ($r_{\rm in}$) and outer ($r_{\rm out}$) disk radii. Here,
$r_{\rm in}$ is taken to be equal to the radius of the central star which
is assumed to be that of a white dwarf of mass $M_1$ according to
\citet{Nauenberg72D}. $r_{\rm out}$ is taken to be 40\% of the distance
between the binary components, calculated from Kepler's third law. This
implies the knowledge of the secondary star mass (or the mass ratio
$q = M_2/M_1$) and the binary period. I calculate the disk spectrum for a 
statistical sample of 10\,000 hypothetical systems, choosing random values 
from a uniform distribution of $0.5 M_\odot \le M_1 \le 1.43 M_\odot$, 
$0.05 \le q \le 1.0$ and $10^{-10} M_\odot/{\rm y} \le
\dot{M} \le 10^{-8} M_\odot/{\rm y}$. The period is taken to increase
linearly with the mass ratio from 80~min at $q=0.05$ to 8~h at $q=1$.
The 10\,000 trials yield fluxes at the effective wavelength of
the $U$, $B$, $R$ and $I$ bands relative to the flux at $V$ (i.e., the
parameter $\beta$) as listed in the last column of 
Table~\ref{Table: gamma beta}. 

From Eq.~\ref{Equation wavelength dependence of the flickering, Eq. 3} we 
have
\begin{equation}
\label{Equation wavelength dependence of the flickering, Eq. 5}
\Delta F_{\rm fl,R} = \left( 1 + F_{{\rm fl,R},0} \right) 
                    \left( 10^{0.4 \Delta {\rm R}} - 1 \right)
\end{equation}
which can then be used to solve 
Eq.~\ref{Equation wavelength dependence of the flickering, Eq. 4} 
for $\alpha$:
\begin{equation}
\label{Equation wavelength dependence of the flickering, Eq. 6a}
\alpha = \frac{\beta \left( 10^{0.4 \gamma \Delta {\rm R}} -1 \right)}
              {\Delta F_{\rm fl,R} + F_{{\rm fl,R},0} 
               \left( 1 - 10^{0.4 \Delta {\rm R}} \right)}
\end{equation}

Due to the non-linear relation between the flux and the magnitude
scale there is a slight dependence of $\alpha$ on $\Delta {\rm R}$. In other
words, if $\alpha$ is supposed to be constant, $\gamma$ is a weak
function of $\Delta {\rm R}$. Taking the $V$ band to be the reference and $U$ 
($B$, $R$, $I$) as the comparison band, $\alpha$ varies by 6.9\% (1.1\%, 
1.5\%, 0.6\%) of its average value in the range 
$0.01 \le \Delta {\rm R} \le 0.3$ 
in the case of a fully modulated flickering light source 
($F_{{\rm fl,R},0} = 0$). Neglecting these small variations, and to be 
definite, I assume $\Delta {\rm R} = 0.1$.


\section{An atlas of flickering light curves}
\label{atlas}

The characterization of flickering in CVs and the comparison of its
properties between different systems is complicated by several factors.
The most obvious of these are:

{\it The time resolution of the light curves:} The visual properties of the
flickering, as apparent to the eye of an investigator, depend strongly on
the time resolution at which the data points in a light curve were taken.
If the sampling time is long, details of the rapid variations which give 
rise to flickering are washed out. The light curve may then even appear
to consist of pure noise. The ideal situation would be a time
resolution which permits to resolve every individual flare. However, so far
there is no indication of a minimum time scale on which these occur. Therefore,
whatever time resolution is chosen for the observations, it is too long to
permit to see all the details of the flickering. As an additional problem
the noise level will rise when decreasing the sampling time, and this depends
also on the apparent brightness of the observed system and the aperture of the
telescope used. According to my experience, in order to get a good visual
impression of the flickering properties, a 5 second cadence of the 
observations is appropriate in a high signal to noise light curve. Of course, 
the acceptable time resolution also depends on the tools which are applied 
to characterize the flickering.

{\it The noise level of the light curves:} The acceptable noise level depends
strongly on the amplitude of the flickering signal. For a light curve to be
useful for flickering studies the signal to noise ratio must remain high enough 
to permit the secure detection of variations occurring on time scales no 
longer than a couple (say, 4 or 5) integrations. Thus, a system with a strong
flickering signal requires less stringent limits on the noise than a system
where the flickering is weak. 

{\it Variations other than flickering:} Many systems show variations which 
apparently are not related to flickering. The most obvious of these are 
orbital variations or (negative or positive) superhumps. These occur on time
scales significantly longer than the flickering and can thus be
subtracted with relative ease (if it is possible to determine their wave 
form reliably). More disturbing are variations which occur in magnetic 
systems such as polars and in particular intermediate polars. Here, rapid 
variations related to the changing aspect of the system as a function of the
rotational phase of the white dwarf can occur on much shorter 
time scales, and a correction is not always easy. It is often not obvious
if variations on comparatively long time scales (say, some 10 minutes), if
they are not strictly periodic and transient (often termed quasi-periodic
oscillations), should be regarded as flickering, or if they 
are due to unrelated mechanisms. If variations other than 
flickering occur on time scales equal to those normally associated with 
flickering, these have often small amplitudes and therefore - even if they 
cannot readily be recognized - they will have only a limited influence on 
any properties measured with tools applied to study flickering (unless they 
are - strictly - period, in which case it is, however, much easier to 
recognize them).

\begin{figure*}
	\includegraphics[width=\textwidth]{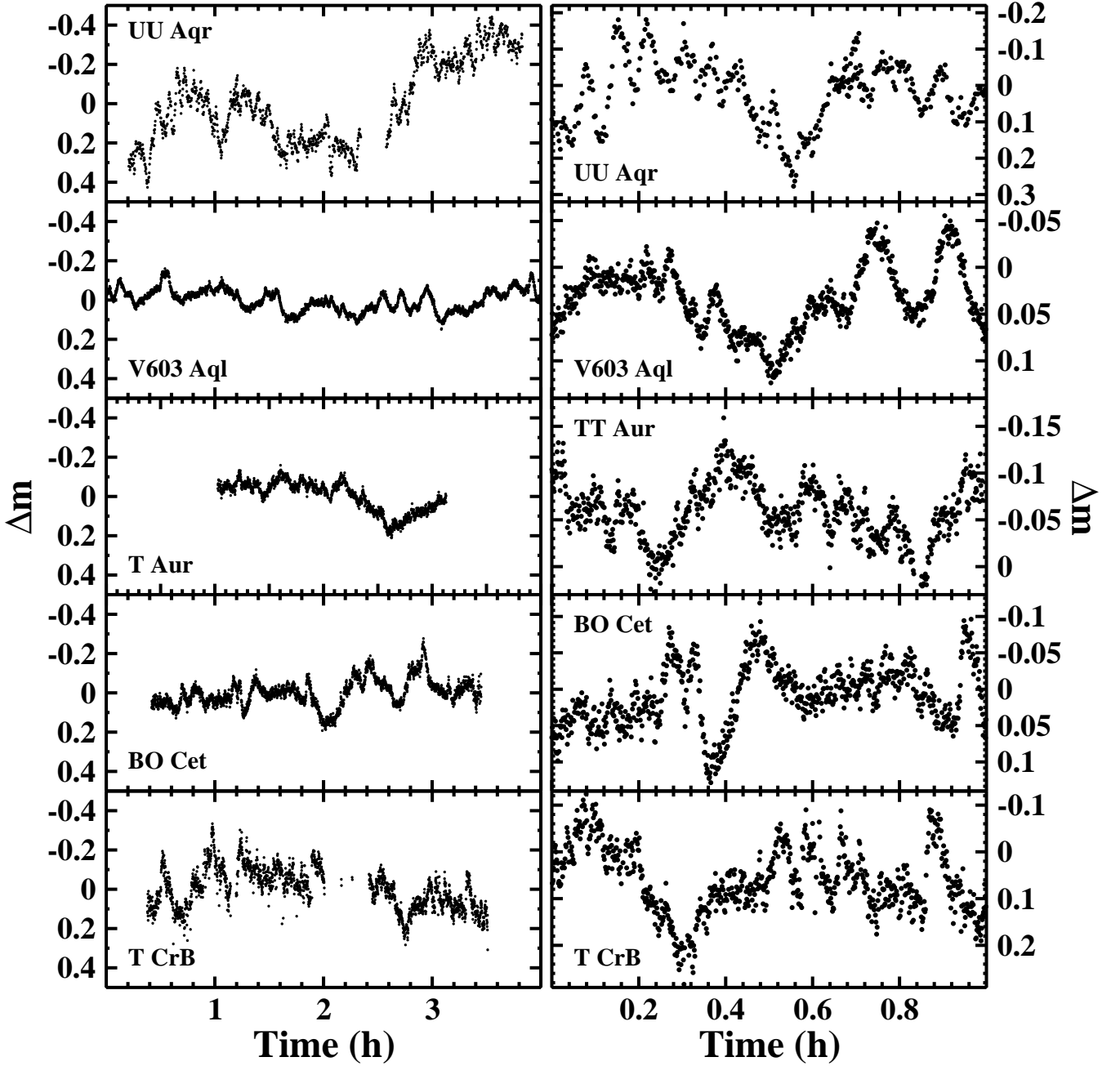}
      \caption[]{Light curves of novae during quiescence and
                 UX~UMa type nova-like variables.
                 On the left a section of four hours duration (or less, if
                 the total duration of the data set is inferior) is shown.
                 All light curves are on the same magnitude scale. Each frame
                 comprises one magnitude in the vertical direction. On the
                 right one hour sections of the same light curves are shown with
                 the magnitude scale adapted to the total amplitude of the
                 variations in the light curve.}
\label{atlas-N-UX-1}
\end{figure*}

\begin{figure*}
	\includegraphics[width=\textwidth]{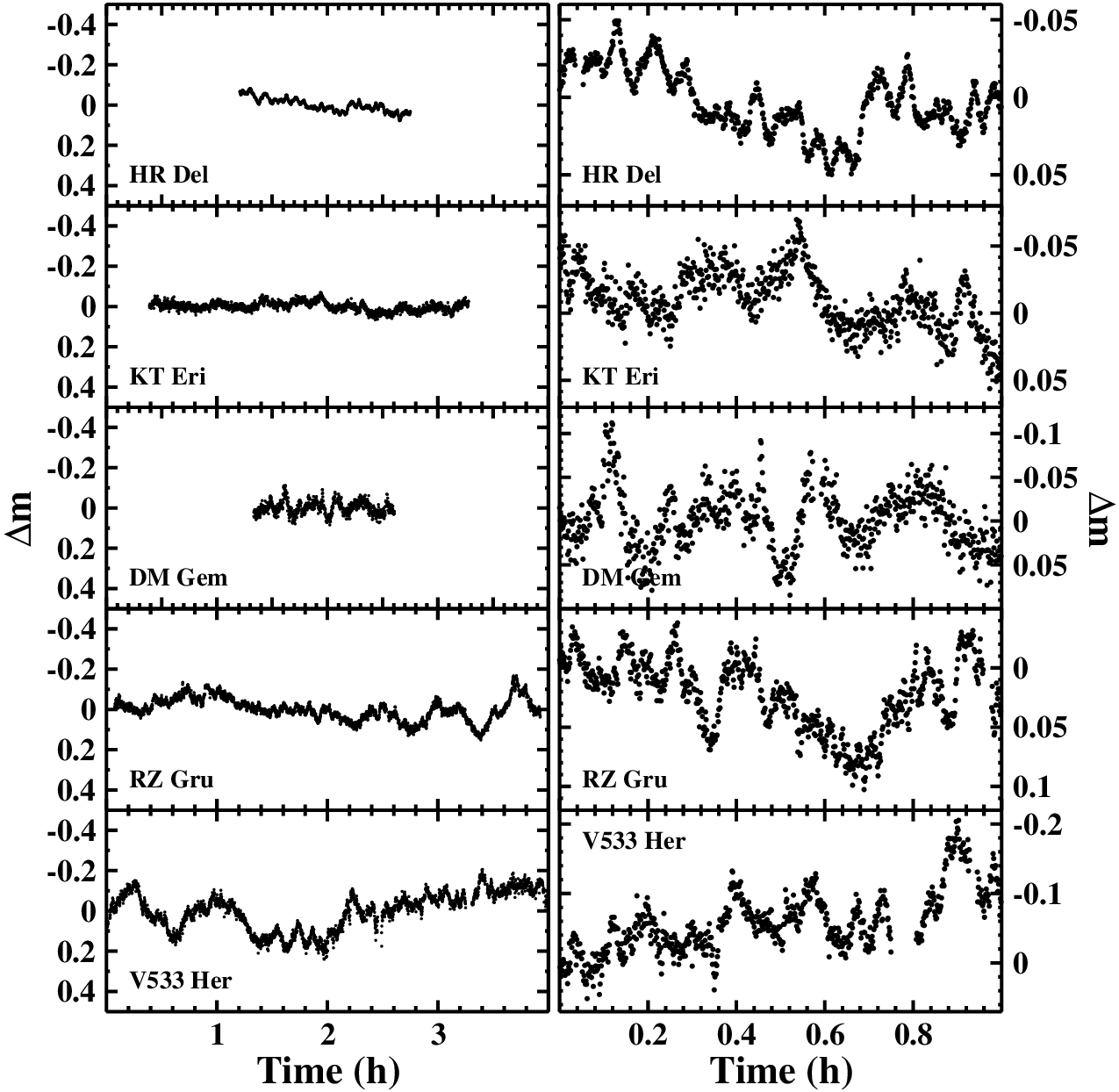}
      \caption[]{Light curves of novae and UX~UMa type nova-like variables
                 (continued).}
\label{atlas-N-UX-2}
\end{figure*}

\begin{figure*}
	\includegraphics[width=\textwidth]{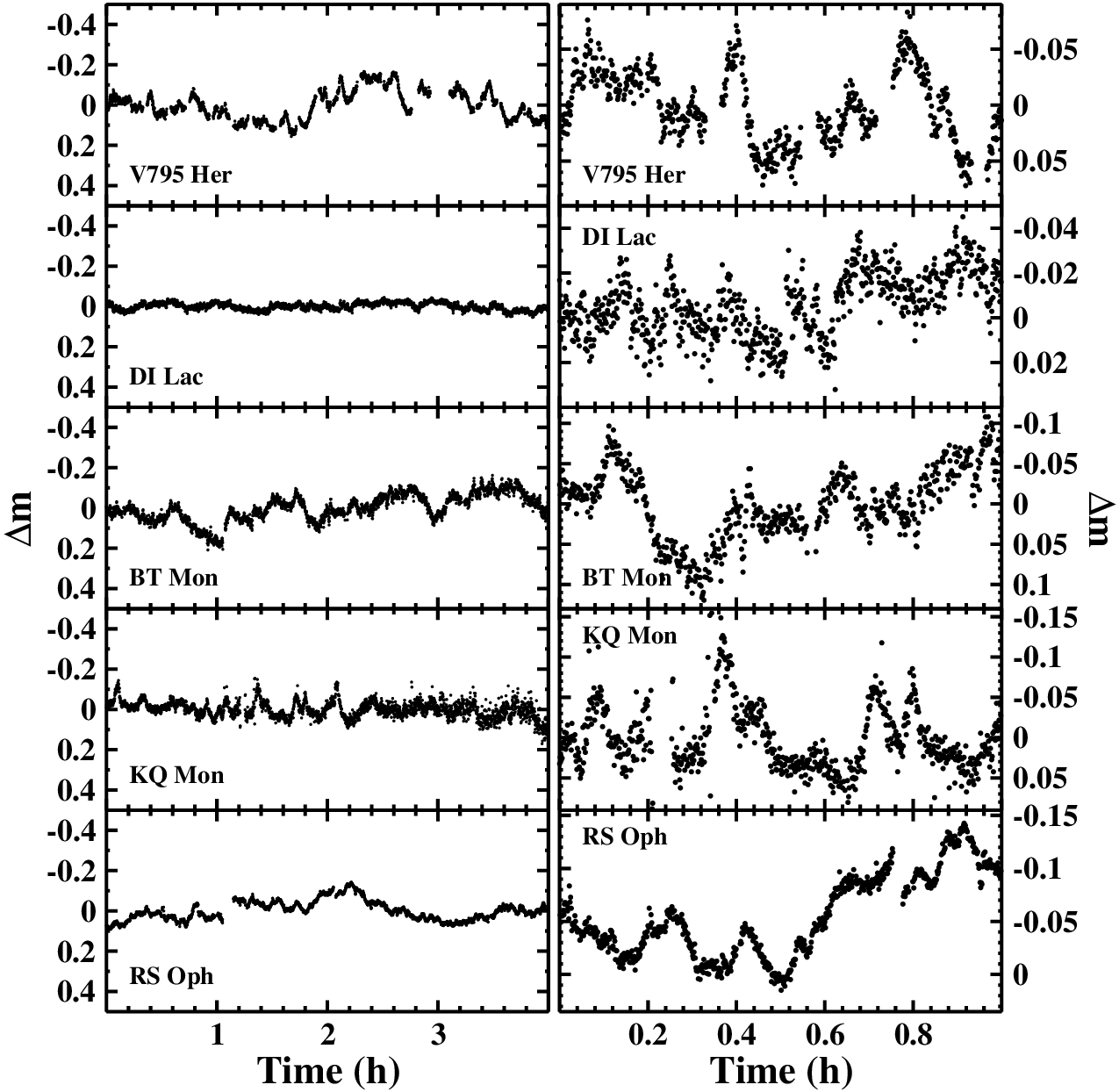}
      \caption[]{Light curves of novae and UX~UMa type nova-like variables
                 (continued).}
\label{atlas-N-UX-3}
\end{figure*}

\begin{figure*}
	\includegraphics[width=\textwidth]{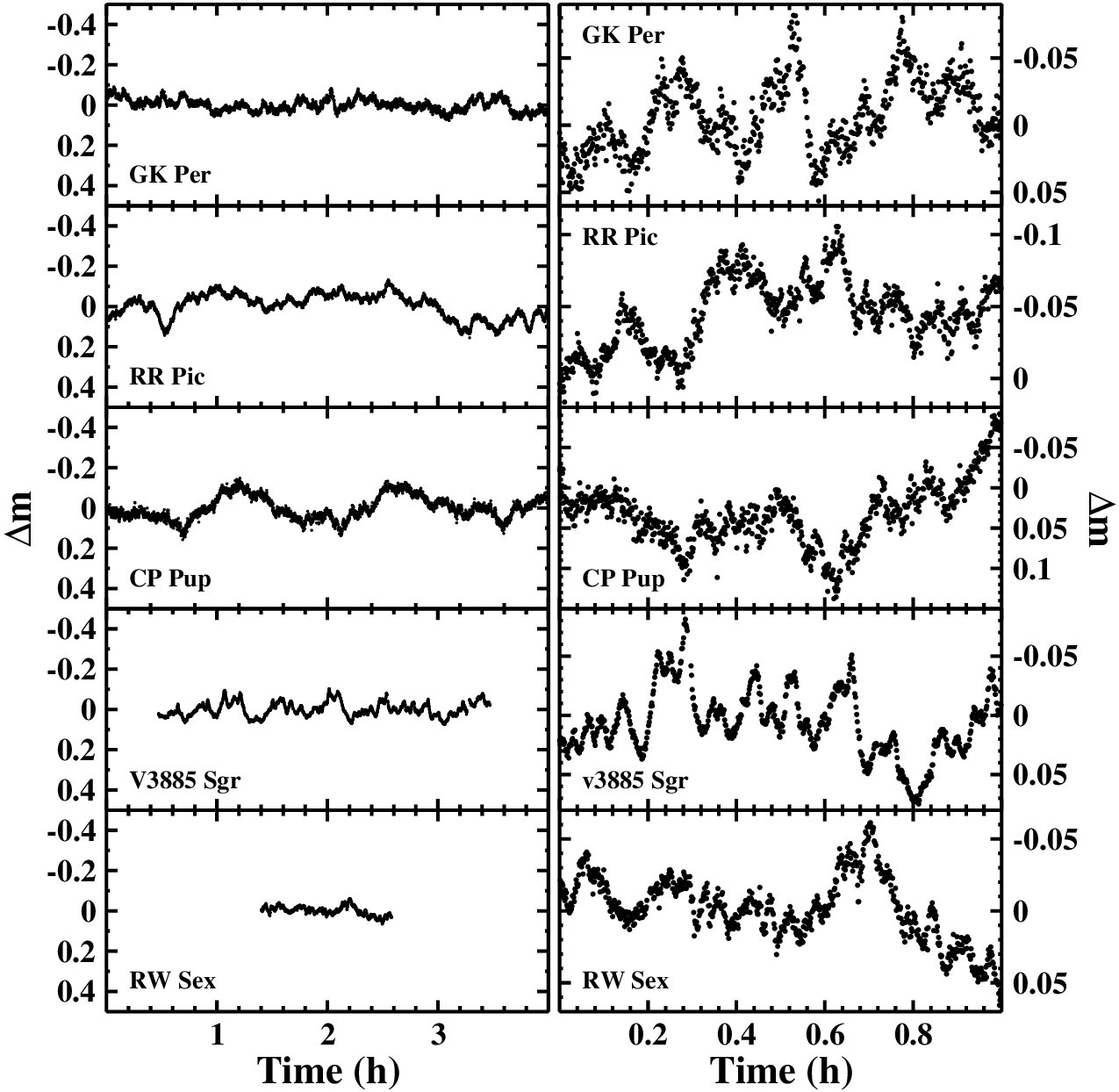}
      \caption[]{Light curves of novae and UX~UMa type nova-like variables
                 (continued).}
\label{atlas-N-UX-4}
\end{figure*}

In this situation a synoptic view of light curves of many CVs of different 
types and photometric states is useful in order to enable a comparison of 
their visual flickering properties. Such an atlas is presented here in 
Figs.~\ref{atlas-N-UX-1} -- \ref{atlas-AM}. To compile it,
suitable light curves from the light curve collection of the author
were chosed and are presented in a way which facilitates their comparison.
In Table~A1 the light curves used for the 
atlas are listed. It is structured as follows:

{\parindent0em {\it Column 1:} The name of the star in lexigraphical order}

{\parindent0em {\it Column 2:} The type of the variable. The following 
abbreviations are used:} 
   \begin{description}
   \item[] N = (classical) nova
   \item[] RN = recurrent nova
   \item[] UX = novalike variable of UX~UMa subtype
   \item[] VY = novalike variable of VY~Scl subtype
   \item[] DN = dwarf nova; may be of any of the various subtypes
   \item[] IP = intermediate polar (DQ Her star)
   \item[] AM = polar (AM Her star)
   \end{description}

{\parindent0em {\it Column 3:} Date (start of observations)}

{\parindent0em {\it Column 4:} Passband (w stands for white light)}

{\parindent0em {\it Column 5:} Photometric state. This is only given for 
dwarf novae and normally is based on information provided by the original 
observer in the respective observing logs. The light curves of all other
systems were taken during their normal (high) state (quiescence for novae
and recurrent novae). The following abbreviations are used:}
   \begin{description}
   \item[] Q = quiescence
   \item[] O = outburst; not further specified; not always maximum
   \item[] M = outburst maximum
   \item[] SM = superoutburst (in SU UMa type dwarf novae)
   \end{description}

{\parindent0em {\it Column 6:} Number of the figure in which the light 
curve is reproduced}

{\parindent0em {\it Column 7:} Reference to a paper where the light curve 
is published}

\begin{table*}
\label{Table: List of atlas objects}
	\centering
	\caption{List of light curves reproduced in 
                 Figs.~\ref{atlas-N-UX-1} -- \ref{atlas-AM}}

\begin{tabular}{lcccccl}
\hline
Name & Type & Obs. & Pass- & photom. & Figure & Reference \\
     &      & Date & band  & state   &        &           \\
\hline
RX And    & DN & 1976, Aug 20 & w &  O  & A12 & \citet{Patterson81A} \\
UU Aqr    & UX & 2016, Aug 28 & B &     & A1  &     \\  
AE Aqr    & IP & 1992, Aug 05 & B &     & A17 &     \\
CZ Aql    & UG & 2014, Jun 17 & w &  Q  & A8  & \citet{Bruch17A} \\
V603 Aql  & N  & 1991, Jul 10 & B &     & A1  &     \\ [1ex]
AT Ara    & UG & 2001, Jul 01 & w &  Q  & A8  &     \\ 
TT Ari    & VY & 1979, Dec 24 & w &     & A6  &     \\ 
T Aur     & N  & 1977, Nov 10 & w &     & A1  &     \\
SS Aur    & DN & 1977, Set 09 & w &  O  & A12 & \citet{Patterson81A} \\
KR Aur    & VY & 1977, Nov 14 & w &     & A6  &     \\ [1ex]
SY Cnc    & DN & 1977, Jan 24 & w &  Q  & A8  & \citet{Patterson81A} \\ 
YZ Cnc    & DN & 1978, Jan 05 & w &  SM & A15 & \citet{Patterson79A} \\ 
BG CMi    & IP & 1982, Feb 21 & w &     & A17 &     \\
OY Car    & UG & 2014, Fev 12 & B &  Q  & A8  &     \\
HT Cas    & DN & 1978, Set 10 & w &  Q  & A8  & \citet{Patterson81A} \\ [1ex]
BV Cen    & DN & 1987, May 26 & w &  O  & A12 &     \\ 
MU Cen    & DN & 2015, Jun 10 & w &  Q  & A9  & \citet{Bruch16A} \\
V436 Cen  & DN & 1973, Apr 10 & w &  Q  & A9  &     \\
V436 Cen  & DN & 1984, Jan 24 & w &  SM & A15 &     \\ 
V442 Cen  & DN & 1982, May 25 & w &  M  & A12 &     \\ [1ex]
V504 Cen  & VY & 2014, Apr 29 & w &     & A6  & \citet{Bruch18A} \\ 
V834 Cen  & AM & 1984, Mar 24 & V &     & A19 & \citet{Chiapetti89A} \\
WW Cet    & DN & 1976, Aug 20 & w &  Q  & A9  & \citet{Patterson81A} \\
BO Cet    & UX & 2016, Aug 11 & w &     & A1  & \citet{Bruch17A} \\
Z Cha     & DN & 1972, Dec 08 & w &  Q  & A9  & \citet{Warner74A} \\ [1ex]
Z Cha     & DN & 1973, Jan 08 & w &  O  & A12 & \citet{Warner74A}, 
                                        \citet{Warner88A} \\
Z Cha     & DN & 1980, Feb 19 & w &  SM & A15 & \citet{Warner88A} \\
TV Col    & IP & 1986, Jan 01 & w &     & A17 &     \\
T CrB     & RN & 1996, Feb 28 & U &     & A1  & \citet{Zamanov98A} \\
EM Cyg    & DN & 1978, Apr 12 & w &  O  & A13 & \citet{Patterson81A} \\ [1ex]
V751 Cyg  & VY & 1977, Jul 18 & w &     & A6  &     \\ 
HR Del    & N  & 1977, Sep 13 & w &     & A2  &     \\
DO Dra    & DN & 1992, Jan 31 & B &     & A9  &     \\ 
KT Eri    & N  & 2014, Oct 23 & w &     & A2  &     \\
U Gem     & UG & 1983, Apr 12 & w &  Q  & A10 &     \\ [1ex]
DM Gem    & N  & 1977, Nov 11 & w &     & A2  &     \\ 
RZ Gru    & UX & 2017, Aug 23 & w &     & A2  &     \\
V533 Her  & N  & 1992, Jun 02 & B &     & A2  &     \\
V795 Her  & UX & 1987, May 28 & B &     & A3  & \citet{Kaluzny89A} \\
EX Hya    & IP & 1982, Feb 82 & w &     & A17 &     \\ [1ex]
VW Hyi    & DN & 1972, Set 12 & w &  Q  & A10 & \citet{Warner75A} \\ 
BL Hyi    & AM & 2001, Jun 30 & w &     & A19 &     \\
DI Lac    & N  & 1976, Oct 22 & w &     & A3  &     \\
MV Lyr    & VY & 1978, Jun 08 & w &     & A6  &     \\
BT Mon    & N  & 1977, Dec 14 & w &     & A3  &     \\ [1ex]
KQ Mon    & UX & 2001, Mar 01 & w &     & A3  &     \\ 
RS Oph    & RN & 1993, Aug 05 & w &     & A3  &     \\
V380 Oph  & VY & 2014, Jun 18 & w &     & A7  & \citet{Bruch17A} \\
V442 Oph  & VY & 1983, Jul 11 & w &     & A7  &     \\
V2051 Oph & DN & 1985, Aug 10 & w &  Q  & A10 & \citet{Warner87A} \\ [1ex]
CN Ori    & DN & 1976, Oct 25 & w &  O  & A13 & \citet{Patterson81A} \\ 
RU Peg    & DN & 1975, Nov 08 & w &  M  & A13 &     \\
IP Peg    & DN & 1990, Oct 26 & B &  Q  & A10 &     \\
GK Per    & N  & 1978, Dec 04 & w &     & A4  &     \\
KT Per    & DN & 1977, Dec 12 & w &  M  & A13 & \citet{Patterson81A} \\ [1ex]
RR Pic    & N  & 1984, Dec 29 & w &     & A4  & Warner (1986) \\ 
\hline
\end{tabular}
\end{table*}

\begin{table*}
{\bf Table A1.} List of light curves reproduced in Figs.~\ref{atlas-N-UX-1} --
                \ref{atlas-AM} (continued)
\vspace{1em}

\begin{tabular}{lcccccl}
\hline
Name & Type & Obs. & Pass-  & photom. & Figure & Reference \\
     &      & Date & band   & state   &        &           \\
\hline
AO Psc    & IP & 1980, Aug 29 & w &     & A17 & \citet{Warner81A} \\
TY PsA    & DN & 1987, Jul 29 & w &  Q  & A10 & \citet{Warner89A} \\
TY PsA    & DN & 1984, Sep 23 & w &  O  & A13 & \citet{Warner89A} \\
TY PsA    & DN & 1984, Jun 22 & w &  SM & A15 & \citet{Warner89A} \\
CP Pup    & N  & 1985, Feb 15 & w &     & A4  & \citet{Warner85A} \\ [1ex]
V1223 Sgr & IP & 2000, Mai 06 & w &     & A18 &     \\
V3885 Sgr & UX & 1982, Jul 26 & w &     & A4  &     \\
VZ Scl    & VY & 1985, Aug 11 & w &     & A7  & \citet{ODonoghue87A} \\
WZ Sge    & DN & 1977, Sep 10 & w &  Q  & A11 & \citet{Patterson80A} \\
WZ Sge    & DN & 1978, Dez 02 & w &  SM & A15 & \citet{Patterson81Ab} \\ [1ex]
V893 Sco  & DN & 2000, May 23 & w &  Q  & A11 & \citet{Bruch14A}    \\ 
RW Sex    & UX & 1989, May 09 & w &     & A4  &     \\
RW Tri    & UX & 1991, Jan 16 & B &     & A5  &     \\
EF Tuc    & DN & 2014, Sep 22 & w &  Q  & A11 & \citet{Bruch17A} \\ 
SU UMa    & DN & 1978, Dec 23 & w &  Q  & A11 &     \\ [1ex]
SW UMa    & DN & 1986, Mar 21 & w &  SM & A16 &     \\
UX UMa    & UX & 1979, May 05 & w &     & A5  &     \\
IX Vel    & UX & 1983, Dec 18 & w &     & A5  &     \\
TW Vir    & DN & 1975, May 17 & w &  O  & A14 &     \\
HV Vir    & DN & 1992, Apr 23 & w &  SM & A16 & \citet{Leibowitz94A} \\

\hline
\end{tabular}
\end{table*}

The individual light curves are displayed in Figs.~\ref{atlas-N-UX-1} -- 
\ref{atlas-N-UX-5} for classical and recurrent novae, and novalike variables
of the UX~UMa type subtype, Figs.~\ref{atlas-VY-1} -- \ref{atlas-VY-2} for
novalike variable of the VY~Scl subtype, Figs.~\ref{atlas-DNqui-1} -- 
\ref{atlas-DNqui-4} for quiescent dwarf novae, Figs.~\ref{atlas-DNoutb-1} -- 
\ref{atlas-DNoutb-3} for dwarf novae in outburst, Figs.~\ref{atlas-DNsuper-1} 
-- \ref{atlas-DNsuper-2} for SU~UMa type dwarf novae during supermaximum, 
Figs.~\ref{atlas-IP-1} -- \ref{atlas-IP-2} for intermediate polars,
and Fig.~\ref{atlas-AM} for AM~Her stars.

The left hand column of each figure is structured such as to permit a direct
comparison between different light curves. The scale in x (time) and y
(magnitude) is the same for all. Thus, a section of 4 hours (or less, if the
entire duration of the data set is less) of a light curve is shown. The
magnitude scale comprises always $1^{\rm m}$, independent of the amplitude of
the variations in the light curve. This enables a visualization of the 
relative strength of the flickering in different systems.
In the right hand part of each figure a 1 hour subset of the same light
curves is shown. Now, the magnitude scale is adapted to the total range of
variability of the star within the chosen time frame. This enables a better
comparison of the structure of the flickering (e.g., if the variations
occur more rapidly or more slowly). 

In some light curves
the variations are strongly dominated by mechanisms other than flickering.
In these cases in right hand part of the figures such variations have been
removed by subtracting a smoothed version of the same data.
This has been done for the dwarf novae WW~Cet, U~Gem and IP~Peg in quiescence,
YZ~Cnc, V436~Cen, Z~Cha and TY~PsA in superoutburst, and
the polars V834~Cen, BL~Hyi.

This is not the place for a detailed assessment of the light curves shown in
the atas. However, some quick remarks are in order. 

It is obvious that the strength of the flickering, expressed as the total
amplitude of the variations, varies significantly from one system to the
next even in systems which are expected to be rather similar in structure.
Even ignoring variations on longer time-scales (which may not be 
related to flickering) the total amplitude can reach 0.5 magnitudes or more;
see e.g.\ KR~Aur (Fig.~\ref{atlas-VY-1}) or SU~UMa in quiescence 
(Fig.~\ref{atlas-DNqui-4}). In other systems, such as IX~Vel 
(Fig.~\ref{atlas-N-UX-5}), the flickering only
occurs on a very small scale of not more than a few hundredths of a 
magnitude. But on that small scale flickering is definitly present and
at first glance has a temporal structure which does not differ drastically
from that of systems with much stronger flickering (e.g.\ V751~Cyg,
Fig.~\ref{atlas-VY-1}). Dwarf novae in outburst also often have a small 
flickering amplitude (see RU~Peg and KT~Per, Fig.~\ref{atlas-DNoutb-2}). 
But this is not always the case, as testified by the outburst light curves
of  Z~Cha (Fig.~\ref{atlas-DNoutb-1}), TY~PsA (Fig.~\ref{atlas-DNoutb-2}) 
and TW~Vir Fig.~\ref{atlas-DNoutb-3}). 

Although complicated by differences in the signal-to-noise level of the
present light curves, it can also be said that the temporal structure of
the flickering, as seen on the right hand
side of the figures, can differ significantly, even for systems of the same
type and in the same state. Variations of V442~Cen in outburst
(Fig.~\ref{atlas-DNoutb-1}), for instance, appear to occur on a quite small
magnitude scale but on comparatively long time scales. Variations of
KT~Per in outburst (Fig.~\ref{atlas-DNoutb-2}) occur on a similar magnitude
scale but much more rapidly.

A quantitative assessment of the flickering variations in different system 
and light curves is subject of the main part of this paper.

\begin{figure*}
	\includegraphics[width=\textwidth]{atlas-N-UX-5.eps}
      \caption[]{Light curves of novae and UX~UMa type nova-like variables
                 (continued).}
\label{atlas-N-UX-5}
\end{figure*}

\clearpage

\begin{figure*}
	\includegraphics[width=\textwidth]{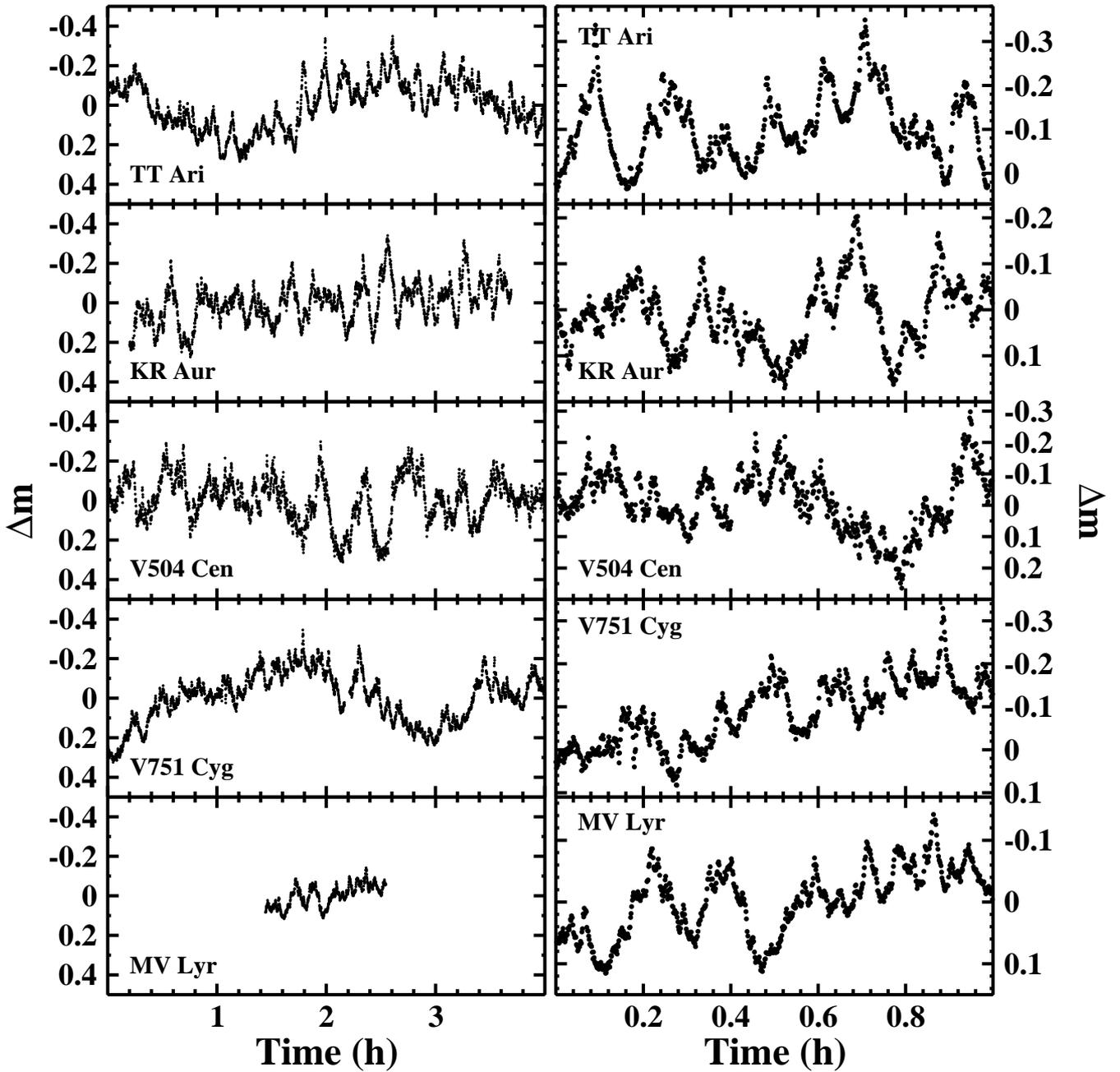}
      \caption[]{Light curves of VY~Scl type nova-like variables.}
\label{atlas-VY-1}
\end{figure*}

\begin{figure*}
	\includegraphics[width=\textwidth]{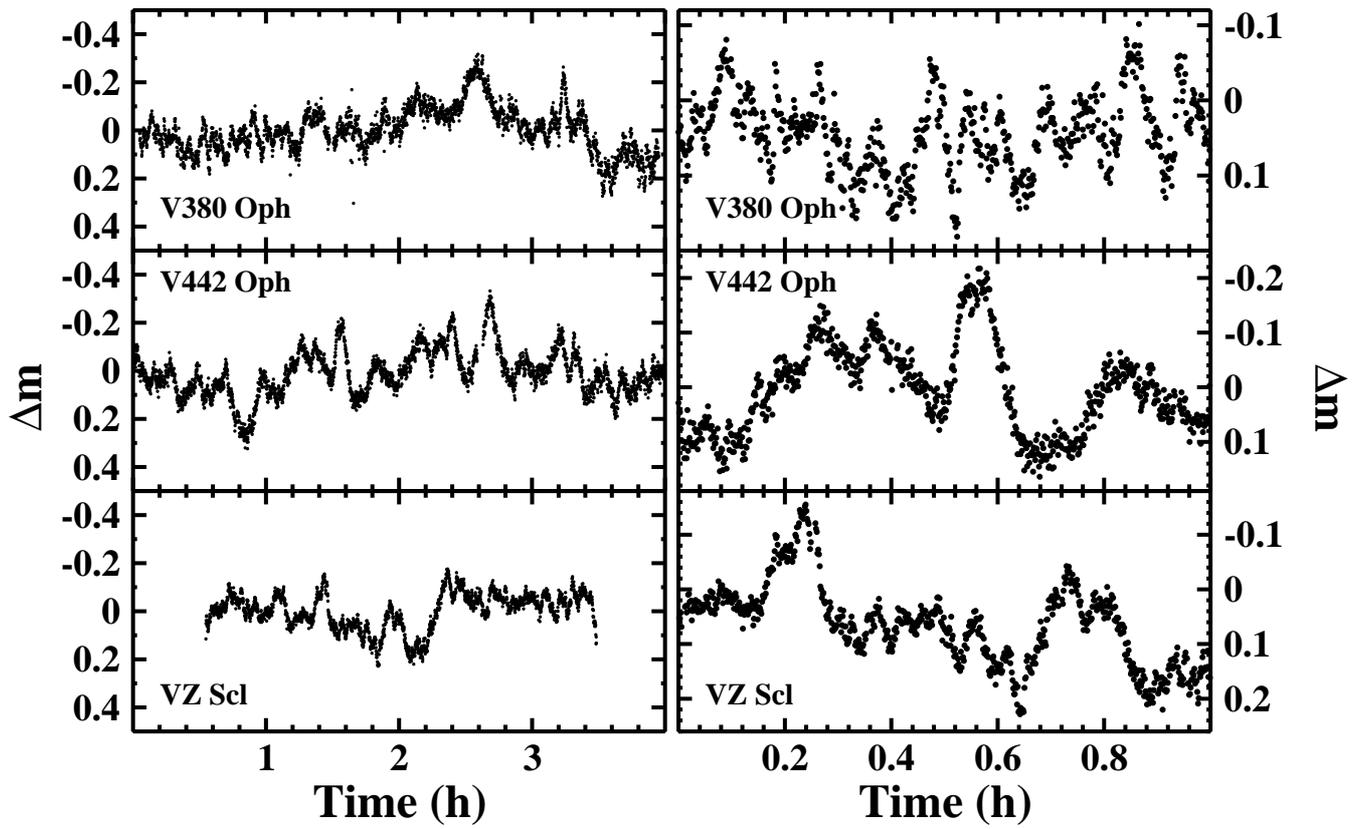}
      \caption[]{Light curves of VY~Scl type nova-like variables
                 (continued).}
\label{atlas-VY-2}
\end{figure*}

\clearpage

\begin{figure*}
	\includegraphics[width=\textwidth]{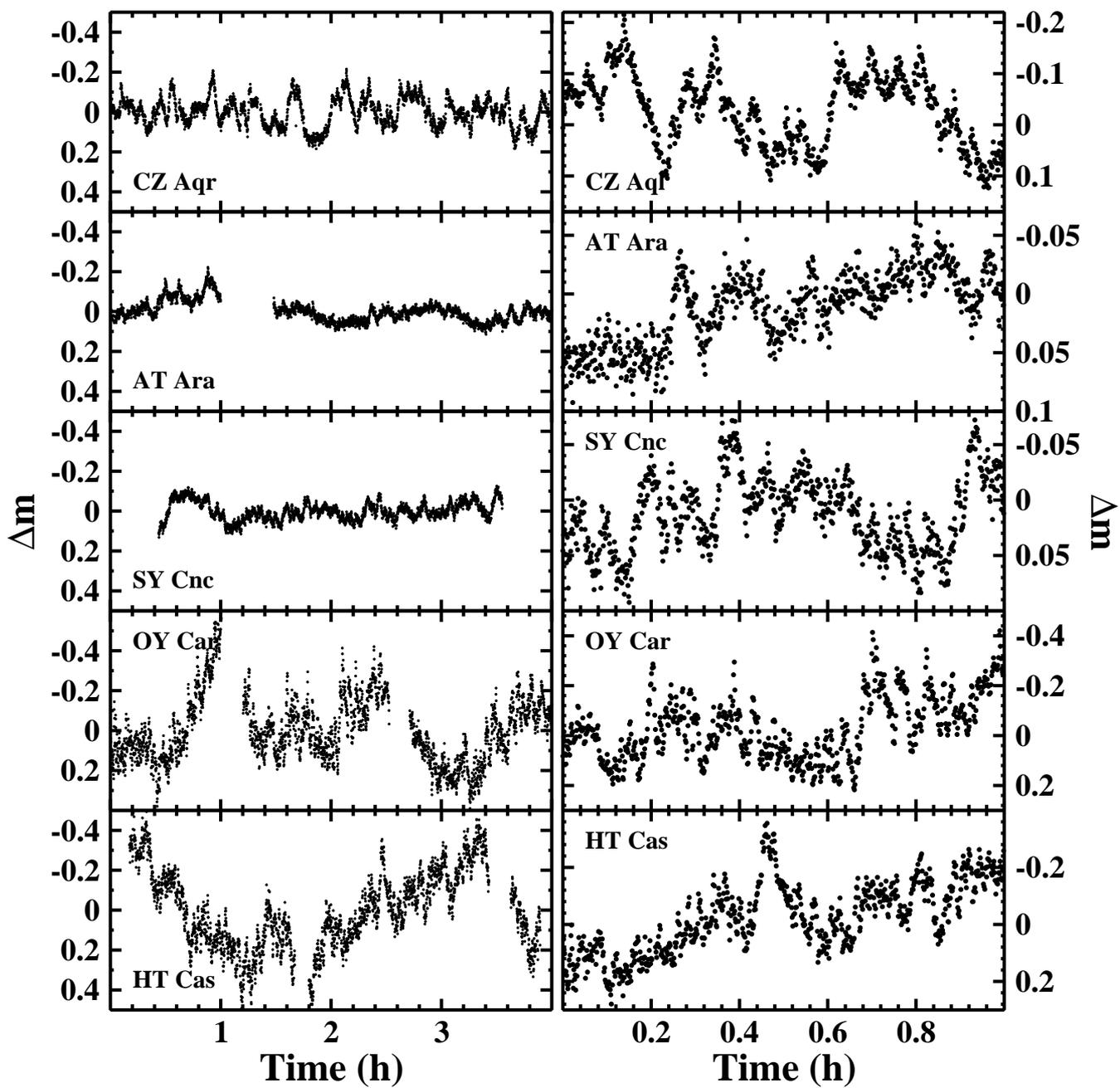}
      \caption[]{Light curves of dwarf novae in quiescence.}
\label{atlas-DNqui-1}
\end{figure*}

\begin{figure*}
	\includegraphics[width=\textwidth]{atlas-DNqui-2.eps}
      \caption[]{Light curves of dwarf novae in quiescence
                 (continued).}
\label{atlas-DNqui-2}
\end{figure*}

\begin{figure*}
	\includegraphics[width=\textwidth]{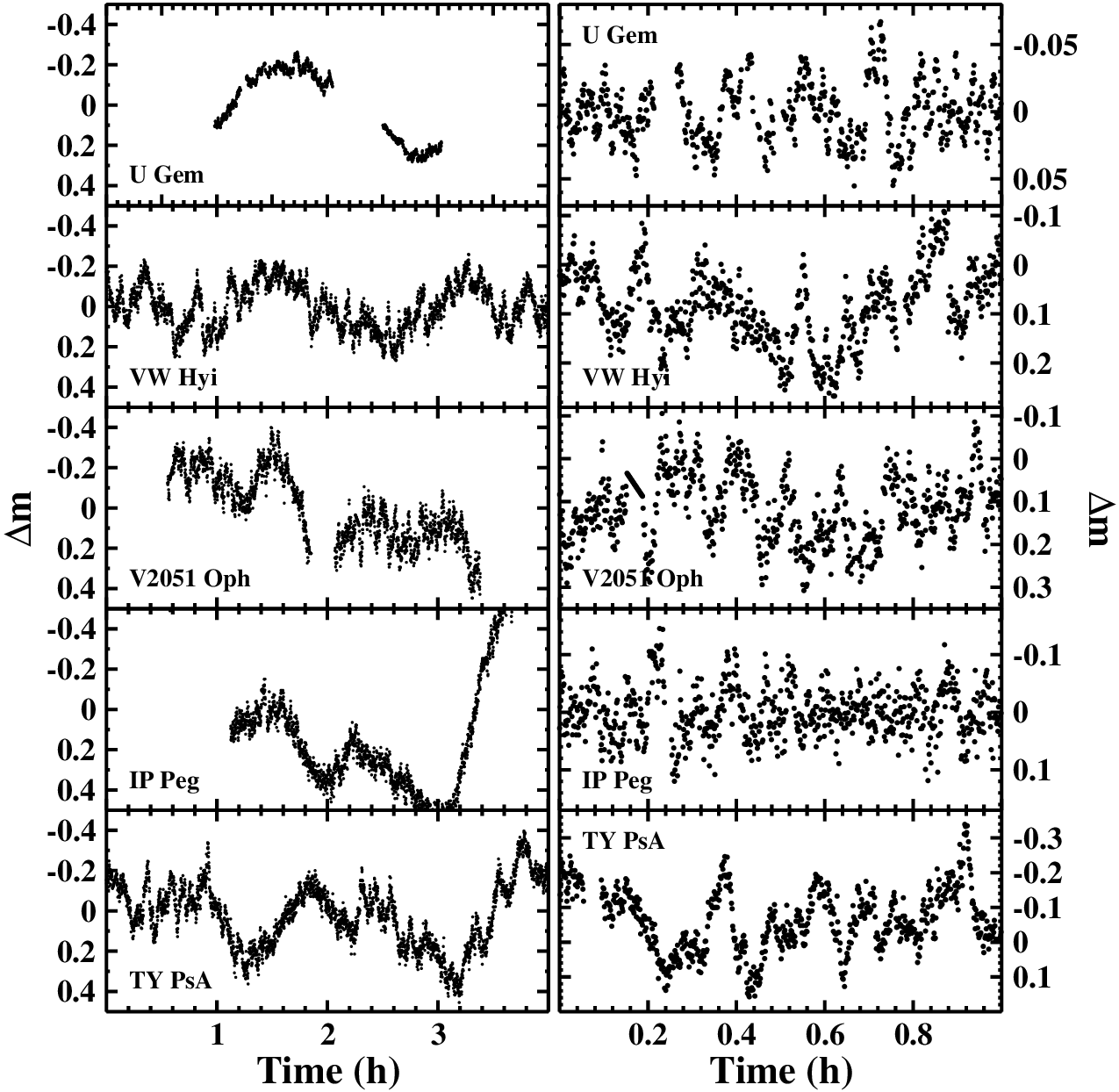}
      \caption[]{Light curves of dwarf novae in quiescence
                 (continued).}
\label{atlas-DNqui-3}
\end{figure*}

\begin{figure*}
	\includegraphics[width=\textwidth]{atlas-DNqui-4.eps}
      \caption[]{Light curves of dwarf novae in quiescence
                 (continued).}
\label{atlas-DNqui-4}
\end{figure*}

\clearpage

\begin{figure*}
	\includegraphics[width=\textwidth]{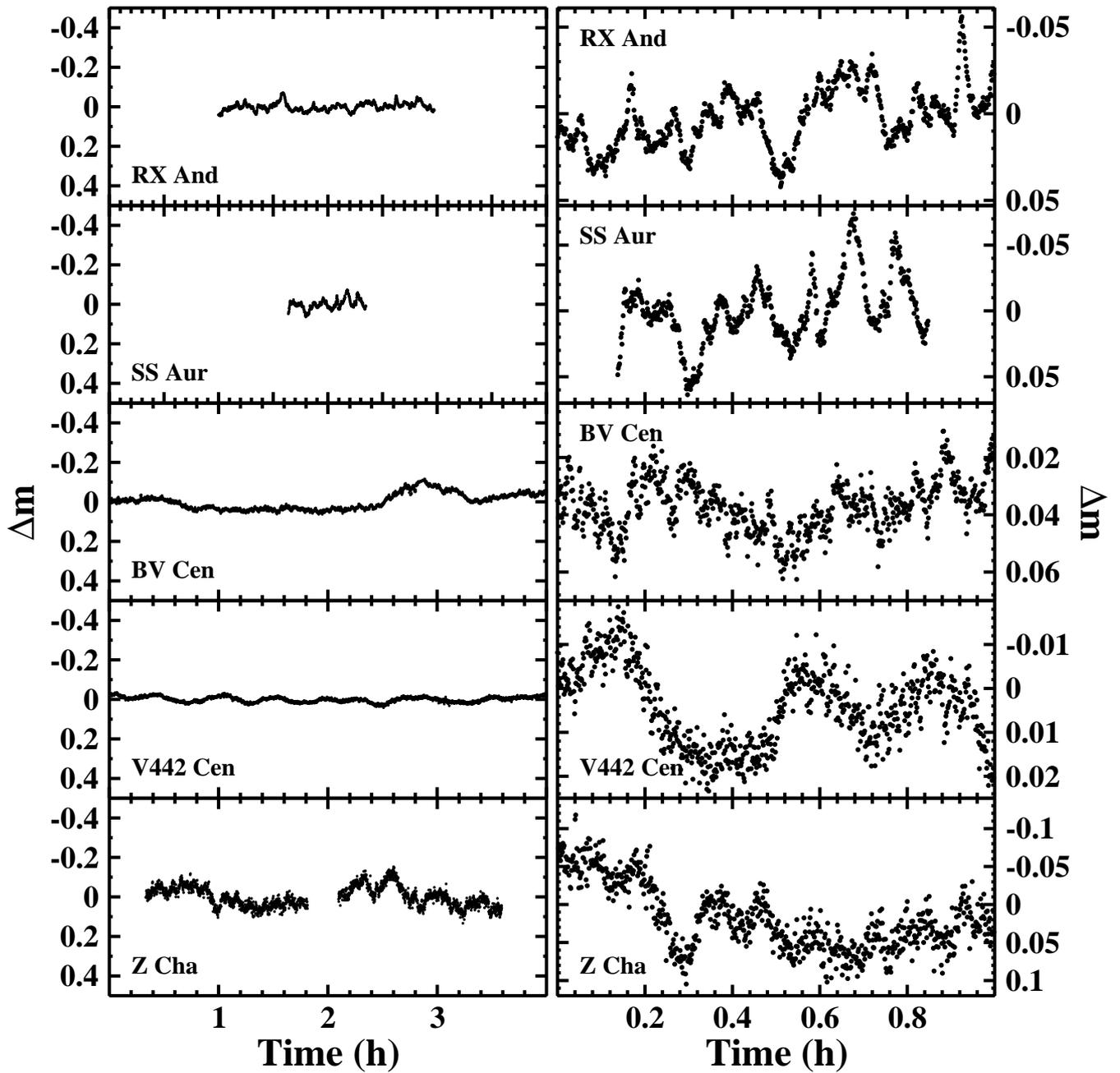}
      \caption[]{Light curves of dwarf novae in outburst.}
\label{atlas-DNoutb-1}
\end{figure*}

\begin{figure*}
	\includegraphics[width=\textwidth]{atlas-DNoutb-2.eps}
      \caption[]{Light curves of dwarf novae in outburst
                 (continued).}
\label{atlas-DNoutb-2}
\end{figure*}

\begin{figure*}
	\includegraphics[width=\textwidth]{atlas-DNoutb-3.eps}
      \caption[]{Light curves of dwarf novae in outburst
                 (continued).}
\label{atlas-DNoutb-3}
\end{figure*}

\clearpage

\begin{figure*}
	\includegraphics[width=\textwidth]{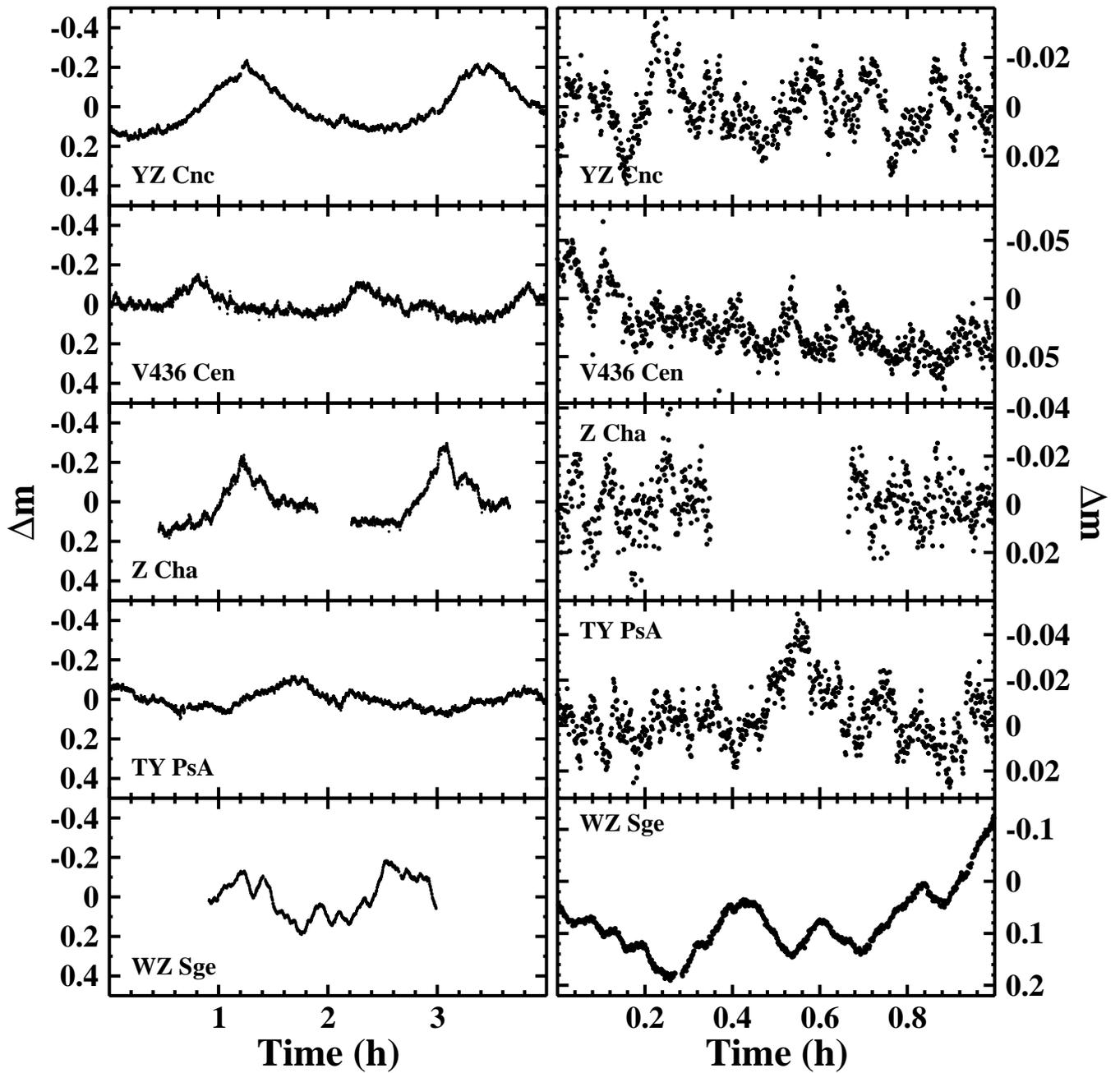}
      \caption[]{Light curves of SU~UMa type dwarf novae in superoutburst.}
\label{atlas-DNsuper-1}
\end{figure*}

\begin{figure*}
	\includegraphics[width=\textwidth]{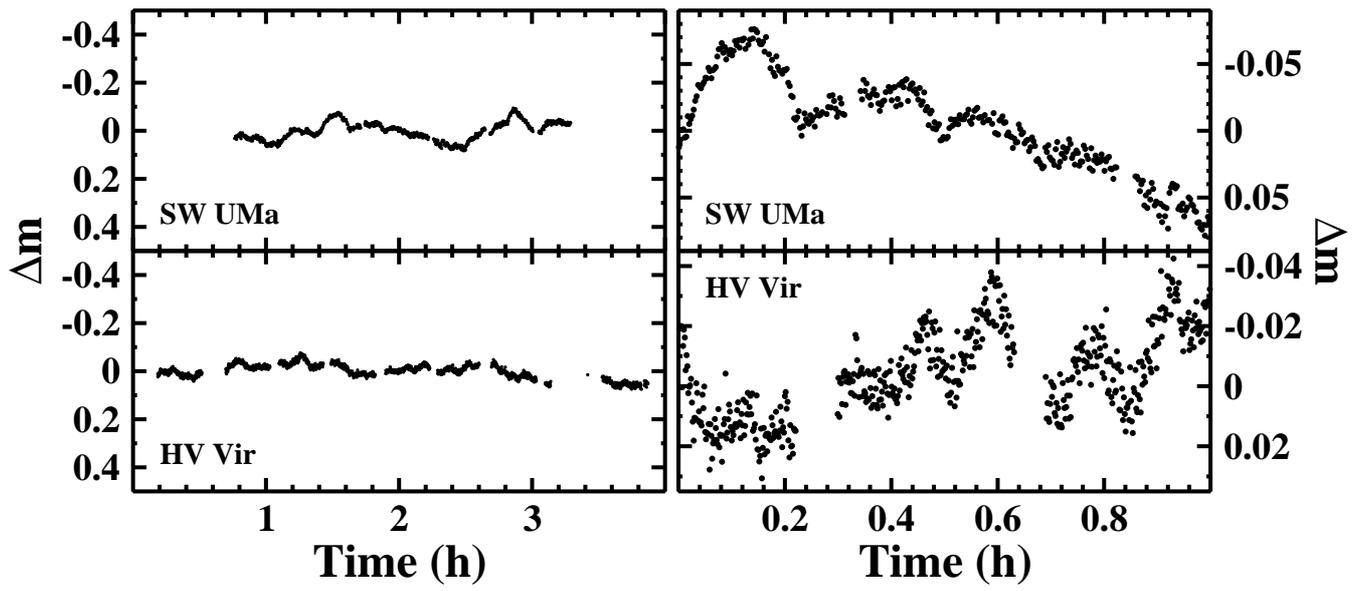}
      \caption[]{Light curves of SU~UMa type dwarf novae in superoutburst
                 (continued).}
\label{atlas-DNsuper-2}
\end{figure*}

\clearpage

\begin{figure*}
	\includegraphics[width=\textwidth]{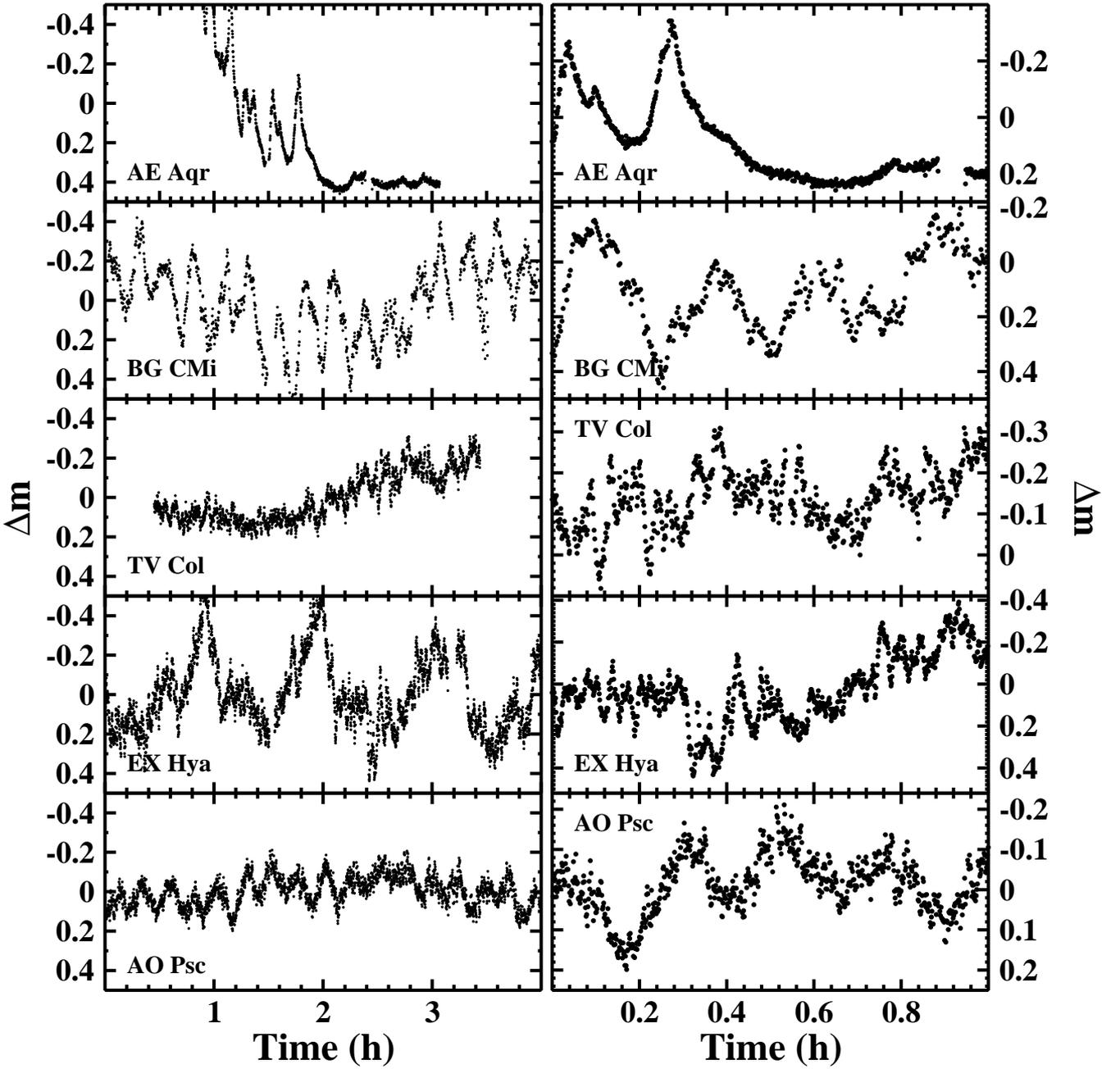}
      \caption[]{Light curves of intermediate polars.}
\label{atlas-IP-1}
\end{figure*}

\begin{figure*}
	\includegraphics[width=\textwidth]{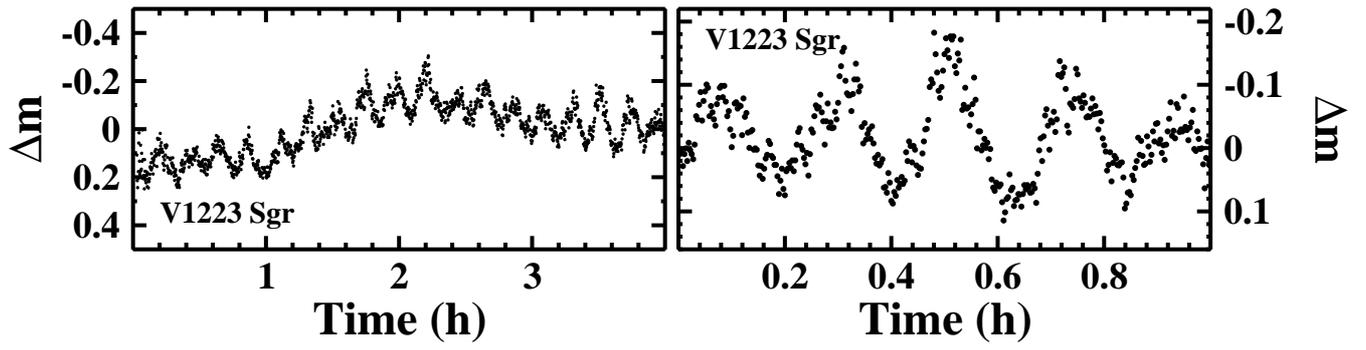}
      \caption[]{Light curves of intermediate polars (continued).}
\label{atlas-IP-2}
\end{figure*}

\clearpage

\begin{figure*}
	\includegraphics[width=\textwidth]{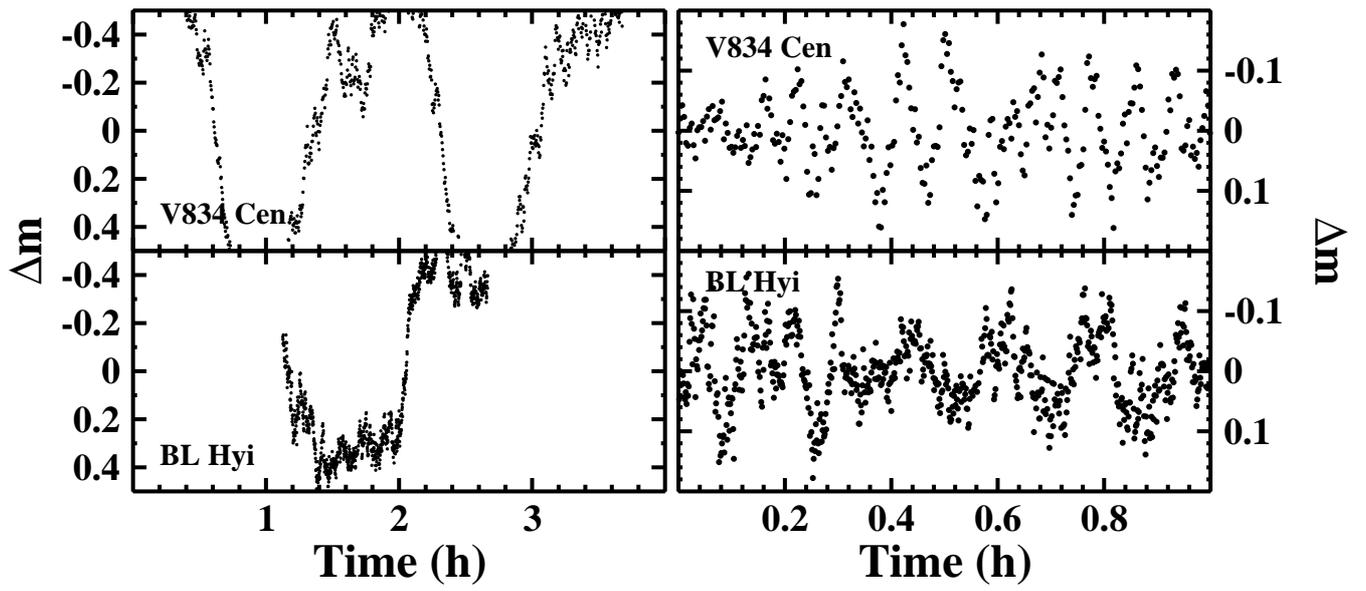}
      \caption[]{Light curves of polars.}
\label{atlas-AM}
\end{figure*}

\clearpage

\bsp	
\label{lastpage}

\begin{thebibliography}{}
\makeatletter
\relax
\def\mn@urlcharsother{\let\do\@makeother \do\$\do\&\do\#\do\^\do\_\do\%\do\~}
\def\mn@doi{\begingroup\mn@urlcharsother \@ifnextchar [ {\mn@doi@}
  {\mn@doi@[]}}
\def\mn@doi@[#1]#2{\def\@tempa{#1}\ifx\@tempa\@empty \href
  {http://dx.doi.org/#2} {doi:#2}\else \href {http://dx.doi.org/#2} {#1}\fi
  \endgroup}
\def\mn@eprint#1#2{\mn@eprint@#1:#2::\@nil}
\def\mn@eprint@arXiv#1{\href {http://arxiv.org/abs/#1} {{\tt arXiv:#1}}}
\def\mn@eprint@dblp#1{\href {http://dblp.uni-trier.de/rec/bibtex/#1.xml}
  {dblp:#1}}
\def\mn@eprint@#1:#2:#3:#4\@nil{\def\@tempa {#1}\def\@tempb {#2}\def\@tempc
  {#3}\ifx \@tempc \@empty \let \@tempc \@tempb \let \@tempb \@tempa \fi \ifx
  \@tempb \@empty \def\@tempb {arXiv}\fi \@ifundefined
  {mn@eprint@\@tempb}{\@tempb:\@tempc}{\expandafter \expandafter \csname
  mn@eprint@\@tempb\endcsname \expandafter{\@tempc}}}

\bibitem[\protect\citeauthoryear{{Andronov}, {Kolosov}, {Movchan}  \&
  {Rudenko}}{{Andronov} et~al.}{1992}]{Andronov92}
{Andronov} I.~L.,  {Kolosov} D.~E.,  {Movchan} A.~I.,   {Rudenko} A.~N.,  1992,
  Soobshcheniya Spetsial'noj Astrofizicheskoj Observatorii, \href
  {https://ui.adsabs.harvard.edu/abs/1992SoSAO..69...79A} {69, 79}

\bibitem[\protect\citeauthoryear{{Andronov} et~al.,}{{Andronov}
  et~al.}{1999}]{Andronov99}
{Andronov} I.~L.,  et~al., 1999, \mn@doi [\aj] {10.1086/300665}, \href
  {https://ui.adsabs.harvard.edu/abs/1999AJ....117..574A} {117, 574}

\bibitem[\protect\citeauthoryear{{Andronov} et~al.,}{{Andronov}
  et~al.}{2005}]{Andronov05}
{Andronov} I.~L.,  et~al., 2005, Information Bulletin on Variable Stars, \href
  {https://ui.adsabs.harvard.edu/abs/2005IBVS.5664....1A} {5664, 1}

\bibitem[\protect\citeauthoryear{{Bailer-Jones}, {Rybizki}, {Fouesneau},
  {Mantelet}  \& {Andrae}}{{Bailer-Jones} et~al.}{2018}]{Bailer-Jones18}
{Bailer-Jones} C.~A.~L.,  {Rybizki} J.,  {Fouesneau} M.,  {Mantelet} G.,
  {Andrae} R.,  2018, \mn@doi [\aj] {10.3847/1538-3881/aacb21}, \href
  {https://ui.adsabs.harvard.edu/abs/2018AJ....156...58B} {156, 58}

\bibitem[\protect\citeauthoryear{{Belova}, {Suleimanov}, {Bikmaev}, {Khamitov},
  {Zhukov}, {Senio}, {Belov}  \& {Sakhibullin}}{{Belova}
  et~al.}{2013}]{Belova13}
{Belova} A.~I.,  {Suleimanov} V.~F.,  {Bikmaev} I.~F.,  {Khamitov} I.~M.,
  {Zhukov} G.~V.,  {Senio} D.~S.,  {Belov} I.~Y.,   {Sakhibullin} N.~A.,  2013,
  \mn@doi [Astronomy Letters] {10.1134/S1063773713020011}, \href
  {https://ui.adsabs.harvard.edu/abs/2013AstL...39..111B} {39, 111}

\bibitem[\protect\citeauthoryear{{Bessell}}{{Bessell}}{2005}]{Bessell05}
{Bessell} M.~S.,  2005, \mn@doi [\araa]
  {10.1146/annurev.astro.41.082801.100251}, \href
  {https://ui.adsabs.harvard.edu/abs/2005ARA&A..43..293B} {43, 293}

\bibitem[\protect\citeauthoryear{{Bruch}}{{Bruch}}{2014}]{Bruch14}
{Bruch} A.,  2014, \mn@doi [\aap] {10.1051/0004-6361/201423576}, \href
  {https://ui.adsabs.harvard.edu/abs/2014A&A...566A.101B} {566, A101}

\bibitem[\protect\citeauthoryear{{Bruch}}{{Bruch}}{2016}]{Bruch16}
{Bruch} A.,  2016, \mn@doi [\na] {10.1016/j.newast.2015.12.005}, \href
  {https://ui.adsabs.harvard.edu/abs/2016NewA...46...60B} {46, 60}

\bibitem[\protect\citeauthoryear{{Bruch} \& {Cook}}{{Bruch} \&
  {Cook}}{2018}]{Bruch18}
{Bruch} A.,  {Cook} L.~M.,  2018, \mn@doi [\na] {10.1016/j.newast.2018.02.002},
  \href {https://ui.adsabs.harvard.edu/abs/2018NewA...63....1B} {63, 1}

\bibitem[\protect\citeauthoryear{{Cowley}, {Crampton}, {Hutchings}  \&
  {Marlborough}}{{Cowley} et~al.}{1975}]{Cowley75}
{Cowley} A.~P.,  {Crampton} D.,  {Hutchings} J.~B.,   {Marlborough} J.~M.,
  1975, \mn@doi [\apj] {10.1086/153341}, \href
  {https://ui.adsabs.harvard.edu/abs/1975ApJ...195..413C} {195, 413}

\bibitem[\protect\citeauthoryear{{Deeming}}{{Deeming}}{1975}]{Deeming75}
{Deeming} T.~J.,  1975, \mn@doi [\apss] {10.1007/BF00681947}, \href
  {https://ui.adsabs.harvard.edu/abs/1975Ap&SS..36..137D} {36, 137}

\bibitem[\protect\citeauthoryear{{Dobrotka}, {Mineshige}  \& {Ness}}{{Dobrotka}
  et~al.}{2015}]{Dobrotka15}
{Dobrotka} A.,  {Mineshige} S.,   {Ness} J.~U.,  2015, \mn@doi [\mnras]
  {10.1093/mnras/stu2631}, \href
  {https://ui.adsabs.harvard.edu/abs/2015MNRAS.447.3162D} {447, 3162}

\bibitem[\protect\citeauthoryear{{Dobrzycka}, {Kenyon}  \&
  {Milone}}{{Dobrzycka} et~al.}{1996}]{Dobrzycka96}
{Dobrzycka} D.,  {Kenyon} S.~J.,   {Milone} A. A.~E.,  1996, \mn@doi [\aj]
  {10.1086/117794}, \href
  {https://ui.adsabs.harvard.edu/abs/1996AJ....111..414D} {111, 414}

\bibitem[\protect\citeauthoryear{{Eastman}, {Siverd}  \& {Gaudi}}{{Eastman}
  et~al.}{2010}]{Eastman10}
{Eastman} J.,  {Siverd} R.,   {Gaudi} B.~S.,  2010, \mn@doi [\pasp]
  {10.1086/655938}, \href
  {https://ui.adsabs.harvard.edu/abs/2010PASP..122..935E} {122, 935}

\bibitem[\protect\citeauthoryear{{G{\"a}nsicke}, {Sion}, {Beuermann}, {Fabian},
  {Cheng}  \& {Krautter}}{{G{\"a}nsicke} et~al.}{1999}]{Gaensicke99}
{G{\"a}nsicke} B.~T.,  {Sion} E.~M.,  {Beuermann} K.,  {Fabian} D.,  {Cheng}
  F.~H.,   {Krautter} J.,  1999, \aap, \href
  {https://ui.adsabs.harvard.edu/abs/1999A&A...347..178G} {347, 178}

\bibitem[\protect\citeauthoryear{{Gianninas}, {Bergeron}  \&
  {Ruiz}}{{Gianninas} et~al.}{2011}]{Gianninas11}
{Gianninas} A.,  {Bergeron} P.,   {Ruiz} M.~T.,  2011, \mn@doi [\apj]
  {10.1088/0004-637X/743/2/138}, \href
  {https://ui.adsabs.harvard.edu/abs/2011ApJ...743..138G} {743, 138}

\bibitem[\protect\citeauthoryear{{Hessman}, {G{\"a}nsicke}  \&
  {Mattei}}{{Hessman} et~al.}{2000}]{Hessman00}
{Hessman} F.~V.,  {G{\"a}nsicke} B.~T.,   {Mattei} J.~A.,  2000, \aap, \href
  {https://ui.adsabs.harvard.edu/abs/2000A&A...361..952H} {361, 952}

\bibitem[\protect\citeauthoryear{{Hirose} \& {Osaki}}{{Hirose} \&
  {Osaki}}{1990}]{Hirose90}
{Hirose} M.,  {Osaki} Y.,  1990, \pasj, \href
  {https://ui.adsabs.harvard.edu/abs/1990PASJ...42..135H} {42, 135}

\bibitem[\protect\citeauthoryear{{Hollander} \& {van Paradijs}}{{Hollander} \&
  {van Paradijs}}{1992}]{Hollander92}
{Hollander} A.,  {van Paradijs} J.,  1992, \aap, \href
  {https://ui.adsabs.harvard.edu/abs/1992A&A...265...77H} {265, 77}

\bibitem[\protect\citeauthoryear{{Honeycutt} \& {Kafka}}{{Honeycutt} \&
  {Kafka}}{2004}]{Honeycutt04}
{Honeycutt} R.~K.,  {Kafka} S.,  2004, \mn@doi [\aj] {10.1086/422737}, \href
  {https://ui.adsabs.harvard.edu/abs/2004AJ....128.1279H} {128, 1279}

\bibitem[\protect\citeauthoryear{{Horne} \& {Stiening}}{{Horne} \&
  {Stiening}}{1985}]{Horne85}
{Horne} K.,  {Stiening} R.~F.,  1985, \mn@doi [\mnras]
  {10.1093/mnras/216.4.933}, \href
  {https://ui.adsabs.harvard.edu/abs/1985MNRAS.216..933H} {216, 933}

\bibitem[\protect\citeauthoryear{{Hutchings} \& {Cote}}{{Hutchings} \&
  {Cote}}{1985}]{Hutchings85}
{Hutchings} J.~B.,  {Cote} T.~J.,  1985, \mn@doi [\pasp] {10.1086/131613},
  \href {https://ui.adsabs.harvard.edu/abs/1985PASP...97..847H} {97, 847}

\bibitem[\protect\citeauthoryear{{Kato} et~al.,}{{Kato} et~al.}{2009}]{Kato09}
{Kato} T.,  et~al., 2009, \mn@doi [\pasj] {10.1093/pasj/61.sp2.S395}, \href
  {https://ui.adsabs.harvard.edu/abs/2009PASJ...61S.395K} {61, S395}

\bibitem[\protect\citeauthoryear{{Kim}, {Andronov}, {Cha}, {Chinarova}  \&
  {Yoon}}{{Kim} et~al.}{2009}]{Kim09}
{Kim} Y.,  {Andronov} I.~L.,  {Cha} S.~M.,  {Chinarova} L.~L.,   {Yoon} J.~N.,
  2009, \mn@doi [\aap] {10.1051/0004-6361:200810005}, \href
  {https://ui.adsabs.harvard.edu/abs/2009A&A...496..765K} {496, 765}

\bibitem[\protect\citeauthoryear{{Kozhevnikov}}{{Kozhevnikov}}{2007}]{Kozhevnikov07}
{Kozhevnikov} V.~P.,  2007, \mn@doi [\mnras]
  {10.1111/j.1365-2966.2007.11819.x}, \href
  {https://ui.adsabs.harvard.edu/abs/2007MNRAS.378..955K} {378, 955}

\bibitem[\protect\citeauthoryear{{Kozhevnikov}}{{Kozhevnikov}}{2012}]{Kozhevnikov12}
{Kozhevnikov} V.~P.,  2012, \mn@doi [\na] {10.1016/j.newast.2011.06.004}, \href
  {https://ui.adsabs.harvard.edu/abs/2012NewA...17...38K} {17, 38}

\bibitem[\protect\citeauthoryear{{Kraicheva}, {Stanishev}, {Iliev}, {Antov}  \&
  {Genkov}}{{Kraicheva} et~al.}{1997}]{Kraicheva97}
{Kraicheva} Z.,  {Stanishev} V.,  {Iliev} L.,  {Antov} A.,   {Genkov} V.,
  1997, \mn@doi [\aaps] {10.1051/aas:1997286}, \href
  {https://ui.adsabs.harvard.edu/abs/1997A&AS..122..123K} {122, 123}

\bibitem[\protect\citeauthoryear{{Kraicheva}, {Stanishev}, {Genkov}  \&
  {Iliev}}{{Kraicheva} et~al.}{1999}]{Kraicheva99}
{Kraicheva} Z.,  {Stanishev} V.,  {Genkov} V.,   {Iliev} L.,  1999, \aap, \href
  {https://ui.adsabs.harvard.edu/abs/1999A&A...351..607K} {351, 607}

\bibitem[\protect\citeauthoryear{{Larwood}}{{Larwood}}{1998}]{Larwood98}
{Larwood} J.,  1998, \mn@doi [\mnras] {10.1046/j.1365-8711.1998.01978.x}, \href
  {https://ui.adsabs.harvard.edu/abs/1998MNRAS.299L..32L} {299, L32}

\bibitem[\protect\citeauthoryear{{Lasota}}{{Lasota}}{2001}]{Lasota01}
{Lasota} J.-P.,  2001, \mn@doi [\nar] {10.1016/S1387-6473(01)00112-9}, \href
  {https://ui.adsabs.harvard.edu/abs/2001NewAR..45..449L} {45, 449}

\bibitem[\protect\citeauthoryear{{Livio} \& {Pringle}}{{Livio} \&
  {Pringle}}{1994}]{Livio94}
{Livio} M.,  {Pringle} J.~E.,  1994, \mn@doi [\apj] {10.1086/174202}, \href
  {https://ui.adsabs.harvard.edu/abs/1994ApJ...427..956L} {427, 956}

\bibitem[\protect\citeauthoryear{{Lomb}}{{Lomb}}{1976}]{Lomb76}
{Lomb} N.~R.,  1976, \mn@doi [\apss] {10.1007/BF00648343}, \href
  {https://ui.adsabs.harvard.edu/abs/1976Ap&SS..39..447L} {39, 447}

\bibitem[\protect\citeauthoryear{{Melikian}, {Tamazian}, {Docobo},
  {Karapetian}, {Kostandian}  \& {Henden}}{{Melikian}
  et~al.}{2010}]{Melikian10}
{Melikian} N.~D.,  {Tamazian} V.~S.,  {Docobo} J.~A.,  {Karapetian} A.~A.,
  {Kostandian} G.~R.,   {Henden} A.~A.,  2010, \mn@doi [Astrophysics]
  {10.1007/s10511-010-9129-6}, \href
  {https://ui.adsabs.harvard.edu/abs/2010Ap.....53..373M} {53, 373}

\bibitem[\protect\citeauthoryear{{Papadaki}, {Boffin}, {Sterken}, {Stanishev},
  {Cuypers}, {Boumis}, {Akras}  \& {Alikakos}}{{Papadaki}
  et~al.}{2006}]{Papadaki06}
{Papadaki} C.,  {Boffin} H.~M.~J.,  {Sterken} C.,  {Stanishev} V.,  {Cuypers}
  J.,  {Boumis} P.,  {Akras} S.,   {Alikakos} J.,  2006, \mn@doi [\aap]
  {10.1051/0004-6361:20054679}, \href
  {https://ui.adsabs.harvard.edu/abs/2006A&A...456..599P} {456, 599}

\bibitem[\protect\citeauthoryear{{Papadaki}, {Boffin}, {Stanishev}, {Boumis},
  {Akras}  \& {Sterken}}{{Papadaki} et~al.}{2009}]{Papadaki09}
{Papadaki} C.,  {Boffin} H.~M.~J.,  {Stanishev} V.,  {Boumis} P.,  {Akras} S.,
   {Sterken} C.,  2009, Journal of Astronomical Data, \href
  {https://ui.adsabs.harvard.edu/abs/2009JAD....15....1P} {15, 1}

\bibitem[\protect\citeauthoryear{{Patterson} \& {Skillman}}{{Patterson} \&
  {Skillman}}{1994}]{Patterson94}
{Patterson} J.,  {Skillman} D.~R.,  1994, \mn@doi [\pasp] {10.1086/133491},
  \href {https://ui.adsabs.harvard.edu/abs/1994PASP..106.1141P} {106, 1141}

\bibitem[\protect\citeauthoryear{{Patterson}, {Thorstensen}, {Fried},
  {Skillman}, {Cook}  \& {Jensen}}{{Patterson} et~al.}{2001}]{Patterson01}
{Patterson} J.,  {Thorstensen} J.~R.,  {Fried} R.,  {Skillman} D.~R.,  {Cook}
  L.~M.,   {Jensen} L.,  2001, \mn@doi [\pasp] {10.1086/317973}, \href
  {https://ui.adsabs.harvard.edu/abs/2001PASP..113...72P} {113, 72}

\bibitem[\protect\citeauthoryear{{Patterson} et~al.,}{{Patterson}
  et~al.}{2005}]{Patterson05}
{Patterson} J.,  et~al., 2005, \mn@doi [\pasp] {10.1086/447771}, \href
  {https://ui.adsabs.harvard.edu/abs/2005PASP..117.1204P} {117, 1204}

\bibitem[\protect\citeauthoryear{{Retter}, {Hellier}, {Augusteijn}, {Naylor},
  {Bedding}, {Bembrick}, {McCormick}  \& {Velthuis}}{{Retter}
  et~al.}{2003}]{Retter03}
{Retter} A.,  {Hellier} C.,  {Augusteijn} T.,  {Naylor} T.,  {Bedding} T.~R.,
  {Bembrick} C.,  {McCormick} J.,   {Velthuis} F.,  2003, \mn@doi [\mnras]
  {10.1046/j.1365-8711.2003.06331.x}, \href
  {https://ui.adsabs.harvard.edu/abs/2003MNRAS.340..679R} {340, 679}

\bibitem[\protect\citeauthoryear{{Rijf}, {Tinbergen}  \& {Walraven}}{{Rijf}
  et~al.}{1969}]{Rijf69}
{Rijf} R.,  {Tinbergen} J.,   {Walraven} T.,  1969, \bain, \href
  {https://ui.adsabs.harvard.edu/abs/1969BAN....20..279R} {20, 279}

\bibitem[\protect\citeauthoryear{{R\"o{\ss}iger}}{{R\"o{\ss}iger}}{1987}]{Roessiger87}
{R\"o{\ss}iger} S.,  1987, Information Bulletin on Variable Stars, \href
  {https://ui.adsabs.harvard.edu/abs/1987IBVS.3007....1R} {3007, 1}

\bibitem[\protect\citeauthoryear{{R{\"o}{\ss}iger}}{{R{\"o}{\ss}iger}}{1988}]{Roessiger88}
{R{\"o}{\ss}iger} S.,  1988, Zentralinstitut fuer Astrophysik Sternwarte
  Sonneberg Mitteilungen ueber Veraenderliche Sterne, \href
  {https://ui.adsabs.harvard.edu/abs/1988MitVS..11..112R} {11, 112}

\bibitem[\protect\citeauthoryear{{Savitzky} \& {Golay}}{{Savitzky} \&
  {Golay}}{1964}]{Savitzky64}
{Savitzky} A.,  {Golay} M.~J.~E.,  1964, Analytical Chemistry, \href
  {https://ui.adsabs.harvard.edu/abs/1964AnaCh..36.1627S} {36, 1627}

\bibitem[\protect\citeauthoryear{{Scargle}}{{Scargle}}{1982}]{Scargle82}
{Scargle} J.~D.,  1982, \mn@doi [\apj] {10.1086/160554}, \href
  {https://ui.adsabs.harvard.edu/abs/1982ApJ...263..835S} {263, 835}

\bibitem[\protect\citeauthoryear{{Scaringi}, {Maccarone}, {D'Angelo}, {Knigge}
  \& {Groot}}{{Scaringi} et~al.}{2017}]{Scaringi17}
{Scaringi} S.,  {Maccarone} T.~J.,  {D'Angelo} C.,  {Knigge} C.,   {Groot}
  P.~J.,  2017, \mn@doi [\nat] {10.1038/nature24653}, \href
  {https://ui.adsabs.harvard.edu/abs/2017Natur.552..210S} {552, 210}

\bibitem[\protect\citeauthoryear{{Schwarzenberg-Czerny}, {Semeniuk}, {Tremko},
  {Urban}  \& {Zboril}}{{Schwarzenberg-Czerny}
  et~al.}{1988}]{Schwarzenberg-Czerny88}
{Schwarzenberg-Czerny} A.,  {Semeniuk} I.,  {Tremko} J.,  {Urban} Z.,
  {Zboril} M.,  1988, Contributions of the Astronomical Observatory Skalnate
  Pleso, \href {https://ui.adsabs.harvard.edu/abs/1988CoSka..17...49S} {17, 49}

\bibitem[\protect\citeauthoryear{{Semeniuk}, {Schwarzenberg-Czerny},
  {Duerbeck}, {Hoffmann}, {Smak}, {Stepien}  \& {Tremko}}{{Semeniuk}
  et~al.}{1987}]{Semeniuk87}
{Semeniuk} I.,  {Schwarzenberg-Czerny} A.,  {Duerbeck} H.,  {Hoffmann} M.,
  {Smak} J.,  {Stepien} K.,   {Tremko} J.,  1987, \mn@doi [\apss]
  {10.1007/BF00654990}, \href
  {https://ui.adsabs.harvard.edu/abs/1987Ap&SS.130..167S} {130, 167}

\bibitem[\protect\citeauthoryear{{Shafter}, {Szkody}, {Liebert}, {Penning},
  {Bond}  \& {Grauer}}{{Shafter} et~al.}{1985}]{Shafter85}
{Shafter} A.~W.,  {Szkody} P.,  {Liebert} J.,  {Penning} W.~R.,  {Bond} H.~E.,
   {Grauer} A.~D.,  1985, \mn@doi [\apj] {10.1086/163028}, \href
  {https://ui.adsabs.harvard.edu/abs/1985ApJ...290..707S} {290, 707}

\bibitem[\protect\citeauthoryear{{Skillman} et~al.,}{{Skillman}
  et~al.}{1998}]{Skillman98}
{Skillman} D.~R.,  et~al., 1998, \mn@doi [\apjl] {10.1086/311534}, \href
  {https://ui.adsabs.harvard.edu/abs/1998ApJ...503L..67S} {503, L67}

\bibitem[\protect\citeauthoryear{{Smak} \& {St\c{e}pie\'n}}{{Smak} \&
  {St\c{e}pie\'n}}{1969}]{Smak69}
{Smak} J.,  {St\c{e}pie\'n} K.,  1969, Commmunications of the Konkoly
  Observatory Hungary, \href
  {https://ui.adsabs.harvard.edu/abs/1969CoKon..65..355S} {65, 355}

\bibitem[\protect\citeauthoryear{{Smak} \& {St\c{e}pie\'n}}{{Smak} \&
  {St\c{e}pie\'n}}{1975}]{Smak75}
{Smak} J.,  {St\c{e}pie\'n} K.,  1975, \actaa, \href
  {https://ui.adsabs.harvard.edu/abs/1975AcA....25..379S} {25, 379}

\bibitem[\protect\citeauthoryear{{Stanishev}, {Kraicheva}  \&
  {Genkov}}{{Stanishev} et~al.}{2001}]{Stanishev01}
{Stanishev} V.,  {Kraicheva} Z.,   {Genkov} V.,  2001, \mn@doi [\aap]
  {10.1051/0004-6361:20011312}, \href
  {https://ui.adsabs.harvard.edu/abs/2001A&A...379..185S} {379, 185}

\bibitem[\protect\citeauthoryear{{Strohmeier}, {Kippenhahn}  \&
  {Geyer}}{{Strohmeier} et~al.}{1957}]{Strohmeier57}
{Strohmeier} W.,  {Kippenhahn} R.,   {Geyer} E.,  1957, Ver\"off. Remeis
  Sternw.\ Bamberg, 18

\bibitem[\protect\citeauthoryear{{Sztajno}}{{Sztajno}}{1979}]{Sztajno79}
{Sztajno} M.,  1979, Information Bulletin on Variable Stars, \href
  {https://ui.adsabs.harvard.edu/abs/1979IBVS.1710....1S} {1710, 1}

\bibitem[\protect\citeauthoryear{{Thomas} \& {Wood}}{{Thomas} \&
  {Wood}}{2015}]{Thomas15}
{Thomas} D.~M.,  {Wood} M.~A.,  2015, \mn@doi [\apj]
  {10.1088/0004-637X/803/2/55}, \href
  {https://ui.adsabs.harvard.edu/abs/2015ApJ...803...55T} {803, 55}

\bibitem[\protect\citeauthoryear{{Thorstensen}, {Smak}  \&
  {Hessman}}{{Thorstensen} et~al.}{1985}]{Thorstensen85}
{Thorstensen} J.~R.,  {Smak} J.,   {Hessman} F.~V.,  1985, \mn@doi [\pasp]
  {10.1086/131558}, \href
  {https://ui.adsabs.harvard.edu/abs/1985PASP...97..437T} {97, 437}

\bibitem[\protect\citeauthoryear{{Tremko}, {Andronov}, {Luthardt}, {Pajdosz},
  {Patkos}, {Roessiger}  \& {Zola}}{{Tremko} et~al.}{1992}]{Tremko92}
{Tremko} J.,  {Andronov} I.~L.,  {Luthardt} R.,  {Pajdosz} G.,  {Patkos} L.,
  {Roessiger} S.,   {Zola} S.,  1992, Information Bulletin on Variable Stars,
  \href {https://ui.adsabs.harvard.edu/abs/1992IBVS.3763....1T} {3763, 1}

\bibitem[\protect\citeauthoryear{{Tremko} et~al.,}{{Tremko}
  et~al.}{1996}]{Tremko96}
{Tremko} J.,  et~al., 1996, \aap, \href
  {https://ui.adsabs.harvard.edu/abs/1996A&A...312..121T} {312, 121}

\bibitem[\protect\citeauthoryear{{Udalski}}{{Udalski}}{1988}]{Udalski88}
{Udalski} A.,  1988, \actaa, \href
  {https://ui.adsabs.harvard.edu/abs/1988AcA....38..315U} {38, 315}

\bibitem[\protect\citeauthoryear{{Volpi}, {Natali}  \& {D'Antona}}{{Volpi}
  et~al.}{1988}]{Volpi88}
{Volpi} A.,  {Natali} G.,   {D'Antona} F.,  1988, \aap, \href
  {https://ui.adsabs.harvard.edu/abs/1988A&A...193...87V} {193, 87}

\bibitem[\protect\citeauthoryear{{Walraven} \& {Walraven}}{{Walraven} \&
  {Walraven}}{1960}]{Walraven60}
{Walraven} T.,  {Walraven} J.~H.,  1960, \bain, \href
  {https://ui.adsabs.harvard.edu/abs/1960BAN....15...67W} {15, 67}

\bibitem[\protect\citeauthoryear{{Warner}}{{Warner}}{1995}]{Warner95}
{Warner} B.,  1995, Cambridge Astrophysics Series, \href
  {https://ui.adsabs.harvard.edu/abs/1995CAS....28.....W} {28}

\bibitem[\protect\citeauthoryear{{Warner}}{{Warner}}{2004}]{Warner04}
{Warner} B.,  2004, \pasp, 116, 115

\bibitem[\protect\citeauthoryear{{Weingrill}, {Kleinschuster}, {Kuschnig},
  {Matthews}, {Moffat}, {Rucinski}, {Sasselov}  \& {Weiss}}{{Weingrill}
  et~al.}{2009}]{Weingrill09}
{Weingrill} J.,  {Kleinschuster} G.,  {Kuschnig} R.,  {Matthews} J.~M.,
  {Moffat} A.,  {Rucinski} S.,  {Sasselov} D.,   {Weiss} W.~W.,  2009, \mn@doi
  [Communications in Asteroseismology] {10.1553/cia159s114}, \href
  {https://ui.adsabs.harvard.edu/abs/2009CoAst.159..114W} {159, 114}

\bibitem[\protect\citeauthoryear{{Whitehurst}}{{Whitehurst}}{1988}]{Whitehurst88}
{Whitehurst} R.,  1988, \mn@doi [\mnras] {10.1093/mnras/232.1.35}, \href
  {https://ui.adsabs.harvard.edu/abs/1988MNRAS.232...35W} {232, 35}

\bibitem[\protect\citeauthoryear{{Whitehurst} \& {King}}{{Whitehurst} \&
  {King}}{1991}]{Whitehurst91}
{Whitehurst} R.,  {King} A.,  1991, \mn@doi [\mnras] {10.1093/mnras/249.1.25},
  \href {https://ui.adsabs.harvard.edu/abs/1991MNRAS.249...25W} {249, 25}

\bibitem[\protect\citeauthoryear{{Williams}}{{Williams}}{1966}]{Williams66}
{Williams} J.~O.,  1966, \mn@doi [\pasp] {10.1086/128349}, \href
  {https://ui.adsabs.harvard.edu/abs/1966PASP...78..279W} {78, 279}

\bibitem[\protect\citeauthoryear{{Wood} \& {Burke}}{{Wood} \&
  {Burke}}{2007}]{Wood07}
{Wood} M.~A.,  {Burke} C.~J.,  2007, \mn@doi [\apj] {10.1086/516723}, \href
  {https://ui.adsabs.harvard.edu/abs/2007ApJ...661.1042W} {661, 1042}

\bibitem[\protect\citeauthoryear{{Wu}, {Li}, {Ding}, {Zhang}  \& {Li}}{{Wu}
  et~al.}{2002}]{Wu02}
{Wu} X.,  {Li} Z.,  {Ding} Y.,  {Zhang} Z.,   {Li} Z.,  2002, \mn@doi [\apj]
  {10.1086/339278}, \href
  {https://ui.adsabs.harvard.edu/abs/2002ApJ...569..418W} {569, 418}

\makeatother
\end{thebibliography}


\begin{thebibliography}{99}
\bibitem[\protect\citeauthoryear{Bailer-Jones et al.}{2018}]{Bailer-Jones18} 
        Bailer-Jones C.A.L., Rybizki J., Fouesneau M., Mantelet G., 
        Andrae R., 2018, AJ, 156, 58
\bibitem[\protect\citeauthoryear{Bailey et al.}{1983}]{Bailey83} 
        Bailey J., Axon D.J., Hough J.H. et al., 1983, MNRAS, 205, 1p
\bibitem[\protect\citeauthoryear{Brandi et al.}{2009}]{Brandi09} 
        Brandi E., Quiroga C., Mikolajewska J., Ferrer O.E., 
        Garc\'{\i}a, L.G., 2009, A\&A, 497, 815 
\bibitem[\protect\citeauthoryear{Bruch}{1991a}]{Bruch91} 
        Bruch A., 1991, A\&A, 251, 59
\bibitem[\protect\citeauthoryear{Bruch}{1992a}]{Bruch92} 
        Bruch A., 1992a, A\&A, 266, 237
\bibitem[\protect\citeauthoryear{Bruch}{1992b}]{Bruch92b} 
        Bruch A., 1992b, in Vogt N., ed., ASP Conf.\ Series, Vol.\ 29,
        Vi\~na del Mar Workshop on Cataclysmic Variable Stars, p.\ 47
\bibitem[\protect\citeauthoryear{Bruch}{1992b, 1995}]{Bruch95} 
        Bruch A., 1995, in Greiner J., Duerbeck H.W., Gershberg R.E., eds,
        Proc.\ IAU Coll.\ 151, Flares and Flashes, p.\ 288
\bibitem[\protect\citeauthoryear{Bruch}{1996}]{Bruch96} 
        Bruch A., 1996, A\&A, 312, 97
\bibitem[\protect\citeauthoryear{Bruch}{2015}]{Bruch15} 
        Bruch A., 2015, A\&A, 579, A50
\bibitem[\protect\citeauthoryear{Bruch}{2018}]{Bruch18} 
        Bruch A., 2018, New Astr., 58, 53
\bibitem[\protect\citeauthoryear{Bruch}{2019}]{Bruch19} 
        Bruch A., 2019, MNRAS, 489, 2961
\bibitem[\protect\citeauthoryear{Bruch}{2020}]{Bruch20} 
        Bruch A., 2020, New Astr., 78, 101369
\bibitem[\protect\citeauthoryear{Buckley et al.}{1992}]{Buckley92}
        Buckley D.A.H., O'Donoghue D., Kilkenny D., Stobie S.R., \&
        Remillard R.A., 1992, MNRAS, 258,285
\bibitem[\protect\citeauthoryear{Cropper et al.}{1986}]{Cropper86}
        Cropper M., Menzies J.W., Tapia S., 1986, MNRAS, 218, 201
\bibitem[\protect\citeauthoryear{Costa \& Rodrigues}{2009}]{Costa09}
        Costa J.E.R., Rodrigues C.V., 2009, MNRAS, 398, 240
\bibitem[\protect\citeauthoryear{Dai et al.}{2013}]{Dai13}
        Dai Z.B., Qian S.B., Li L., 2013, ApJ, 774, 153
\bibitem[\protect\citeauthoryear{Deeming}{1975}]{Deeming75} 
        Deeming T.J., 1975, Ap\&SS, 39, 137
\bibitem[\protect\citeauthoryear{de Miguel et al.}{2016}]{deMiguel16}
        de Miguel E., Patterson J., Cejudo D. et al., 2016, MNRAS, 457, 1447
\bibitem[\protect\citeauthoryear{Dobrotka et al.}{2012}]{Dobrotka12} 
        Dobrotka A., Mineshige S., Casares J., 2012, MNRAS, 420, 2467
\bibitem[\protect\citeauthoryear{Dobrotka et al.}{2017}]{Dobrotka17} 
        Dobrotka A., Ness J.-U., Mineshige S., Nucita A.A., 2017, 
        MNRAS, 468, 1183
\bibitem[\protect\citeauthoryear{Dobrotka et al.}{2020}]{Dobrotka20} 
        Dobrotka A., Nogoro H., Konopka P., 2020, A\&A, 640, A55 
\bibitem[\protect\citeauthoryear{Dobrzycka et al.}{1996}]{Dobrzycka96} 
        Dobrzycka D., Kenyon S.J., Milone A.A.E., 1996, AJ, 111,414
\bibitem[\protect\citeauthoryear{Elsworth \& James}{1982, 1986}]{Elsworth82}
        Elsworth Y.P., James J.F., 1982, MNRAS, 198, 889 
\bibitem[\protect\citeauthoryear{Elsworth \& James}{1986}]{Elsworth86}
        Elsworth Y.P., James J.F., 1986, MNRAS, 220, 895
\bibitem[\protect\citeauthoryear{Fritz \& Bruch}{1998}]{Fritz98} 
        Fritz T., Bruch A., 2015, A\&A, 332, 586 
\bibitem[\protect\citeauthoryear{Ferrario \& Wickramasinghe}{1990}]{Ferrario90} 
        Ferrario L., Wickramasinghe D.T., 1990, ApJ, 357, 582
\bibitem[\protect\citeauthoryear{G\"ansicke et al.}{2009}]{Gaensicke09} 
        G\"ansicke B.T., Dhillon M., Southworth J. et al., 2009,
        MNRAS, 397, 2170
\bibitem[\protect\citeauthoryear{Garcia et al.}{1999}]
        {Garcia99} Garcia A., Sodr\'e Jr. L., Jablonski F.J., \&
        Terlevich R.J., 1999, MNRAS, 309, 803
\bibitem[\protect\citeauthoryear{Greiner et al.}{2010}]{Greiner10} 
        Greiner J., Schwarz R., Tappert C. et al., 2010, 
        Astron.\ Nachr., 331, 227
\bibitem[\protect\citeauthoryear{Herbst \& Shevchenko}{1999}]{Herbst99} 
        Herbst W., Shevchenko V.S., 1999, AJ, 118, 1043
\bibitem[\protect\citeauthoryear{Hoard et al.}{1997}]{Hoard97} 
        Hoard D.W., Baptista R., Eracleous M. et al., 1997, MNRAS, 288, 691
\bibitem[\protect\citeauthoryear{Horne \& Baliunas}{1986}]{Horne86} 
        Horne J.H., Baliunas S.L., 1986, ApJ, 302, 757
\bibitem[\protect\citeauthoryear{Horne \& Stiening}{1985}]{Horne85} 
        Horne K., Stiening R.F., 1985, MNRAS, 216, 933
\bibitem[\protect\citeauthoryear{Imamura et al.}{2000}]{Imamura00}
        Imamura J.N., Steiman-Cameron T.Y., Wolff M.T., 2000, 
        PASP, 112, 18
\bibitem[\protect\citeauthoryear{Ka\-lu\.zny}{1989}]{Kaluzny89} 
        Ka\-lu\.zny J., 1989, Acta Astron., 39, 235
\bibitem[\protect\citeauthoryear{Kenyon et al.}{2000}]{Kenyon00} 
        Kenyon S.J., Kolotilov E.A., Ibragimos M.A., Mattei, J.A., 
        2000, ApJ, 531, 1028
\bibitem[\protect\citeauthoryear{Kley \& Papaloizou}{1997}]{Kley97} 
        Kley W., Papaloizou J.C.B., 1997, MNRAS, 285, 239
\bibitem[\protect\citeauthoryear{Lomb}{1976}]{Lomb76} 
        Lomb N.R., 1976, Ap\&SS, 39, 447
\bibitem[\protect\citeauthoryear{Lyubarskii}{1997}]{Lyubarskii97} 
        Lyubarskii Yu.E., 1997, MNRAS 292, 679
\bibitem[\protect\citeauthoryear{Middleditch \& C\'ordova}{1982}]{Middleditch82}
        Middleditch J., C\'ordova F.A., 1982, ApJ, 255, 585
        1991, ApJ, 382, 315
\bibitem[\protect\citeauthoryear{Middleditch et al.}{1991}]{Middleditch91}
        Middleditch J., Imamura J.N., Wolff M.T., Steiman-Cameron T.Y.,
        1991, ApJ, 382, 315
\bibitem[\protect\citeauthoryear{Morales-Rueda et al.}{2002}]{Morales-Rueda02} 
        Morales-Rueda L., Still M.D., Roche R., Wood J.H., Lockley J.J.,
        2002, MNRAS, 329, 597
\bibitem[\protect\citeauthoryear{Papdaki et al.}{2006}]{Papadaki06} 
        Papadaki C., Boffin H.M.J., Sterken C. et al., 2006, A\&A, 456, 599
\bibitem[\protect\citeauthoryear{Paczynksi \& Schwarzenberg-Czerny}{1980}]
        {Paczynski80} 
        Paczynski B. Schwarzenberg-Czerny A., 1980, Acta Astron., 30, 127
\bibitem[\protect\citeauthoryear{Patterson \& Skillman}{1994}]{Patterson94} 
        Patterson J., Skillman D.R., 1994, PASP, 106, 1141
\bibitem[\protect\citeauthoryear{Pavlidou et al.}{2001}]{Pavlidou01} 
        Pavlidou V., Kuijpers J., Vlahos L., Isliker H., 2001, 
        A\&A, 372, 326
\bibitem[\protect\citeauthoryear{Pearson et al.}{2005}]{Pearson05} 
        Pearson K.J., Horne K., Skidmore W., 2005, ApJ, 619, 999
\bibitem[\protect\citeauthoryear{Potter et al.}{2004}]{Potter04} 
        Potter S.B., Romero-Colmenero E., Watson C.A., Buckley D.A.H.,
        Phillips A., 2004, MNRAS, 348, 316 
\bibitem[\protect\citeauthoryear{Ritter \& Kolb}{2003}]{Ritter03} 
        Ritter H., Kolb U., 2003, A\&A, 404, 301
\bibitem[\protect\citeauthoryear{Savitzky \& Golay}{1964}]{Savitzky64} 
        Savitzky A., Golay M.J.E., 1964, Analytical Chemistry, 36, 1627
\bibitem[\protect\citeauthoryear{Scargle}{1982}]{Scargle82} 
        Scargle J.D., 1982, ApJ, 263, 853
\bibitem[\protect\citeauthoryear{Scargle et al.}{1993}]{Scargle93} 
        Scargle J.D., Steiman-Cameron T.Y., Young K., et al., 
        1993, ApJ, 411, L91
\bibitem[\protect\citeauthoryear{Scaringi}{2014}]{Scaringi14a} 
        Scaringi S., 2014, MNRAS, 438, 1233
\bibitem[\protect\citeauthoryear{Scaringi et al.}{2014}]{Scaringi14} 
        Scaringi S., Maccarone T.J., Middleton S., 2014, MNRAS, 445, 1031
\bibitem[\protect\citeauthoryear{Schaefer}{2018}]{Schaefer18}
        Schaefer B.E., 2018, MNRAS, 481, 3033
\bibitem[\protect\citeauthoryear{Schoembs \& Stolz}{1981}]{Schoembs81} 
        Schoembs R., Stolz B., 1981, IBVS, 1986
\bibitem[\protect\citeauthoryear{Schwope et al.}{1993}]{Schwope93}
        Schwope A.D., Thomas H.-C., Beuermann K., Reinsch K., 1993, 
        A\&A, 267, 103
\bibitem[\protect\citeauthoryear{\v{S}imon et al.}{2012}]{Simon12} 
        \v{S}imon V., Pol\'a\v{s}ek C., \v{S}trobl J., Hudec R., 
        Bla\v{z}ek M., 2012, A\&A, 540, A15
\bibitem[\protect\citeauthoryear{Semena et al.}{2014}]{Semena14} 
        Semena A.N., Revnivtsev M.G., Buckley D.A.H. et al., 2014, 
        MNRAS 442, 1123.
\bibitem[\protect\citeauthoryear{Sokoloski et al.}{2001}]{Sokoloski01} 
        Sokoloski J.L., Bildsten L., Ho W.C.G., 2001, MNRAS, 236, 553
\bibitem[\protect\citeauthoryear{Tappert et al.}{2020}]{Tappert20} 
        Tappert C., Vogt N., Ederoclite A., 2020, A\&A, 641, A122
\bibitem[\protect\citeauthoryear{van der Klis}{2004}]{vanderKlis04} 
        van der Klis M., 2004, arXiv e-prints [{\tt arXiv:astro-ph/0410551}]
\bibitem[\protect\citeauthoryear{Walraven \& Walraven}{1960}]{Walraven60} 
        Walraven T., Walraven J.H., 1960, Bull.\ Astron.\ Inst.\ Neth., 15, 67
\bibitem[\protect\citeauthoryear{Warner}{1995}]{Warner95}
        Warner B., 1995, {\it Cataclysmic Variable Stars}, Cambridge 
        University Press, Cambridge
\bibitem[\protect\citeauthoryear{Warner \& Nather}{1971}]{Warner71}
        Warner B., Nather R.E., 1971, MNRAS, 152, 219
\bibitem[\protect\citeauthoryear{Williams \& Maletesta}{2002}]{Williams02} 
        Williams G., Maletesta K., 2000, AJ, 123, 1095
\bibitem[\protect\citeauthoryear{Yonehara et al.}{1997}]{Yonehara97} 
        Yonehara A., Mineshige S., Welsh W.F., 1997, ApJ, 486, 388
\bibitem[\protect\citeauthoryear{Zamanov et al.}{2015}]{Zamanov15} 
        Zamanov R.K., Boeva S., Latev G., Stoyanov K.A., Tsvetkova S.V., 2015,
        AN, 336, 189
\end{thebibliography}

\begin{thebibliography}{99}
\bibitem[\protect\citeauthoryear{Anupama \& Mikolajewska}{1999}]{Anupama99B} 
        Anupama G.C., Mikolajewska J., 1999, A\&A, 344, 177
\bibitem[\protect\citeauthoryear{Bailey}{1981}]{Bailey81B} 
        Bailey J., 1981, MNRAS, 197, 31
\bibitem[\protect\citeauthoryear{Beuermann et al.}{1992}]{Beuermann92B} 
        Beuermann K., Stasiewski U., Schwope A.D., 1992, A\&A, 256, 334
\bibitem[\protect\citeauthoryear{Bianchini \& Sabbadin}{1983}]{Bianchini83B} 
        Bianchini A., Sabbadin F., 1983, A\&A Suppl., 54, 393
\bibitem[\protect\citeauthoryear{Bitner et al.}{2007}]{Bitner07B} 
        Bitner M.A., Robinson E.L., Behr B.B., 2007, ApJ, 662, 564
\bibitem[\protect\citeauthoryear{Bolick et al.}{1987}]{Bolick87B} 
        Bolick K., Beuermann K., Bruch A., Lenzen R., 1987, 
        Ap\&SS, 130, 175\
\bibitem[\protect\citeauthoryear{Bruch}{1987}]{Bruch87B} 
        Bruch A., 1987, A\&A, 172, 187
\bibitem[\protect\citeauthoryear{Bruch}{2003}]{Bruch03B} 
        Bruch A., 2003, A\&A, 409, 647
\bibitem[\protect\citeauthoryear{Bruch}{2016}]{Bruch16B}  
        Bruch A., 2016, New Astr., 46, 60
\bibitem[\protect\citeauthoryear{Bruch \& Diaz}{2017}]{Bruch17B} 
        Bruch A., Diaz M.P., 2017, New Astr., 50, 109
\bibitem[\protect\citeauthoryear{Bruch \& Engel}{1994}]{Bruch94B} 
        Bruch A., Engel A., 1994, A\&AS, 104, 79
\bibitem[\protect\citeauthoryear{Bruch et al.}{1997}]{Bruch97B} 
        Bruch A., Vrielmann S., Hessman F.V., Kochsiek A., Schimpke T.,
        1997, A\&A, 237, 1107
\bibitem[\protect\citeauthoryear{Buckley et al.}{1992}]{Buckley92B} 
        Buckley D.A.H., O'Donoghue D., Kilkenny D., Stobie S.R., \&
        Remillard R.A., 1992, MNRAS, 258,285
\bibitem[\protect\citeauthoryear{Crampton et al.}{1986}]{Crampton86B} 
        Crampton D., Cowley A.P., Fisher W.A., 1986, ApJ, 300, 788
\bibitem[\protect\citeauthoryear{Casares et al.}{2009}]{Casares09B} 
        Casares J., Mart\'{\i}nez-Pais I.G., Rodr\'{\i}guez-Gil P., 2009
        MNRAS, 399, 1534
\bibitem[\protect\citeauthoryear{Cowley et al.}{1980}]{Cowley80B} 
        Cowley A.P., Crampton D., Hutchings J.B., 1980, ApJ, 241, 269
\bibitem[\protect\citeauthoryear{Dhillon et al.}{2000}]{Dhillon00B} 
        Dhillon V.S., Littlefair S.P., Howell S.B. et al., 2000, 
        MNRAS, 314, 826
\bibitem[\protect\citeauthoryear{Friend et al.}{1990}]{Friend90B} 
        Friend M.T., Martin J.S., Smith R.C., Jones D.H.P., 1990, 
        MNRAS, 246, 654
\bibitem[\protect\citeauthoryear{Froning et al.}{1999}]{Froning99B} 
        Froning C.S., Robinson E.L., Welsh W.F., Wood J.H., 1999, 
        ApJ, 523, 399
\bibitem[\protect\citeauthoryear{Gallagher \& Oinas}{1974}]{Gallagher74B} 
        Gallagher J.S., Oinas V., 1974, PASP, 86, 952
\bibitem[\protect\citeauthoryear{Garlick et al.}{1994}]{Garlick94B} 
        Garlick M.A., Mittaz J.P.G., Rosen S.R., Mason K.O., 1994, 
        MNRAS, 269, 517
\bibitem[\protect\citeauthoryear{Gilliland}{1982}]{Gilliland82B} 
        Gilliland R.L., 1982, ApJ, 263, 302 
\bibitem[\protect\citeauthoryear{Harrison et al.}{2004}]{Harrison04B} 
        Harrison T.E., Osborne H.L, Howell S.B., 2004, AJ, 127, 3493
\bibitem[\protect\citeauthoryear{Harrison et al.}{2000}]{Harrison00B} 
        Harrison T.E., McNamara B.J., Szkokdy P., Gilliland R.L., 2000,
        AJ, 120, 2649
\bibitem[\protect\citeauthoryear{Hessman}{1988}]{Hessman88B} 
        Hessman F.V., 1988, A\&AS, 72, 512
\bibitem[\protect\citeauthoryear{Horne et al.}{1986}]{Horne86B} 
        Horne K., Wade, R. A., Szkody P., 1986, MNRAS, 219, 791
\bibitem[\protect\citeauthoryear{Hutchings \& Thomas}{1982}]{Hutchings82B} 
        Hutchings J.B., Thomas B., 1982, PASP, 94, 102 
\bibitem[\protect\citeauthoryear{Kelly et al.}{2014}]{Kelly14B} 
        Kelly P.L., Fox O.D., Filippenko A.V. et al., 2014, ApJ, 790, 3
\bibitem[\protect\citeauthoryear{Knigge}{2006}]{Knigge06B} 
        Knigge C., 2006, MNRAS, 373, 484
\bibitem[\protect\citeauthoryear{Kraft}{1964}]{Kraft64B} 
        Kraft R.P., 1964, ApJ 139, 457
\bibitem[\protect\citeauthoryear{Littlefair et al.}{2001}]{Littlefair01B} 
        Littlefair S.P., Dhillon V.S., Marsh T.R., Harlaftis E.T.,
        2001, MNRAS, 327, 475
\bibitem[\protect\citeauthoryear{Lynden-Bell \& Pringle}{1974}]{Lynden-Bell74B} 
        Lynden-Bell D., Pringle J.E., 1974, MNRAS, 168, 603
\bibitem[\protect\citeauthoryear{Marino \& Walker}{1984}]{Marino84B} 
        Marino B.F., Walker W.S.G., 1984, Southern Stars, 30, 389
\bibitem[\protect\citeauthoryear{Mateo et al.}{1991}]{Mateo91B} 
        Mateo M., Szkody P., Garnavich P., 1991, ApJ, 370, 370
\bibitem[\protect\citeauthoryear{Nauenberg's}{1972}]{Nauenberg72B} 
        Nauenberg M., 1972, ApJ, 175, 417
\bibitem[\protect\citeauthoryear{North et al.}{2002}]{North02B} 
        North R.C., Marsh T.R., Kolb U., Dhillon V.S., Moran C.K.K.,
        2002, MNRAS, 337, 1215
\bibitem[\protect\citeauthoryear{North et al.}{2000}]{North00B} 
        North R.D., Marsh T.R., Moran C.K.J et al., 2000, MNRAS, 313, 383
\bibitem[\protect\citeauthoryear{Patterson et al.}{1992}]{Patterson92B} 
        Patterson J., Schwartz D.A., Pye J.P. et al., 1992, ApJ, 393, 233
\bibitem[\protect\citeauthoryear{Pickles}{1998}]{Pickles98B} 
        Pickles A.J., 1998, PASP, 110, 863
\bibitem[\protect\citeauthoryear{Poole et al.}{2003}]{Poole03B} 
        Poole T., Mason K.O., Ramsay G., Drew J.H. Smith R.C., 2003, 
        MNRAS, 340, 499 
\bibitem[\protect\citeauthoryear{Reinsch}{1994}]{Reinsch94B} 
        Reinsch K., 1994, A\&A, 281, 108
\bibitem[\protect\citeauthoryear{Ribeiro et al.}{2007}]{Ribeiro07B} 
        Ribeiro T., Baptista R., Harlaftis E.T., Dhillon V.S, 
        Rutten R.G.M., 2007, A\&A, 474, 213
\bibitem[\protect\citeauthoryear{Rutten et al.}{1994}]{Rutten94B} 
        Rutten R.G.M., Dhillon V.S., Horne K., Kuulkers E., 1994, 
        A\&A, 283, 441
\bibitem[\protect\citeauthoryear{Schmidt-Kaler}{1982}]{Schmidt-Kaler82B} 
        Schmidt-Kaler Th., 1982, in Schaifers K., Voigt H.H., eds, 
        Landolt B\"ornstein, Numerical Data and Functional Relationships
        in Science and Technology, New Series, Group VI, Vol.\ 2, Subvol.\ b,
        Springer Verlag, Heidelberg, p.\ 1
\bibitem[\protect\citeauthoryear{Sion et al.}{2007}]{Sion07B} 
        Sion E.M., Godon P., Cheng F. Szkody P., 2007, AJ, 134, 886
\bibitem[\protect\citeauthoryear{Sion et al.}{2001}]{Sion01B} 
        Sion E.M., Szkody P., G\"ansicke B.T. et al., 2001, ApJ, 555, 834
\bibitem[\protect\citeauthoryear{Smith et al.}{1998}]{Smith98B} 
        Smith D.A., Dhillon V.S., Marsh T.R., 1998, MNRAS, 296, 465
\bibitem[\protect\citeauthoryear{Stover}{1979}]{Stover79B} 
        Stover R.J., 1979, in van Horn H.M., Weidemann V., eds, 
        Proc.\ IAU Coll.\ 53, White Dwarfs and Variable Degenerate Stars, 
        p.\ 489  
\bibitem[\protect\citeauthoryear{Stover et al.}{1980}]{Stover80B} 
        Stover R.J., Robinson E.L., Nather R.E., Montemayor T.J., 1980,
        ApJ, 20, 597
\bibitem[\protect\citeauthoryear{Szkody \& Mateo}{1986}]{Szkody86B} 
        Szkody P., Mateo M., 1986, AJ, 92, 483
\bibitem[\protect\citeauthoryear{Thoroughgood et al.}{2004}]{Thoroughgood04B} 
        Thoroughgood T.D., Dhillon V.S., Watson C.A. et al., 2004, 
        MNRAS, 253, 1135
\bibitem[\protect\citeauthoryear{Vande Putte et al.}{2003}]{VandePutte03B} 
        Vande Putte D., Smith R.D., Hawkings N.A., Martin J.S., 2003,
        MNRAS, 342, 151
\bibitem[\protect\citeauthoryear{Vogt \& Breysacher}{1980}]{Vogt80B} 
        Vogt N., Breysacher J., 1980, ApJ, 235, 945
\bibitem[\protect\citeauthoryear{Wade}{1979}]{Wade79B} 
        Wade R.A., 1979, AJ, 84, 562
\bibitem[\protect\citeauthoryear{Zamanov \& Bruch}{1998}]{Zamanov98B} 
        Zamanov R.K., Bruch A., 1998, A\&A, 338, 988 
\bibitem[\protect\citeauthoryear{Zhang \& Robinson}{1987}]{Zhang87B} 
        Zhang E.-H., Robinson E.L., 1987, ApJ 321, 813 
\bibitem[\protect\citeauthoryear{Zorotovic et al.}{2011}]{Zorotovic11B} 
        Zorotovic M., Schreiber M.R., G\"ansicke B.T., 2011, 
        A\&A, 536, A42

\end{thebibliography}

\begin{thebibliography}{99}
 
\bibitem[\protect\citeauthoryear{Agrawal et al.}{1983}]{Agrawal83C} 
        Agrawal P.C., Riegler G.R., Rao A.R., 1983, Nature, 301, 318
\bibitem[\protect\citeauthoryear{Ak et al.}{2005}]{Ak05C}
        Ak T., Retter A., Liu A., 2005, New Astr., 11, 147
\bibitem[\protect\citeauthoryear{Alcock}{1967}]{Alcock67C}
        Alcock G.E.D., 1967, IAU Circ. 2022
\bibitem[\protect\citeauthoryear{Andronov et al}{2008}]{Andronov08C}
        Andronov I.L., Chinarova L.L., Han W., Kim Y. Yoon J.-N., 2008,
        A\&A, 486, 855
\bibitem[\protect\citeauthoryear{Aranzana et al.}{2018}]{Aranzana18C}
        Aranzana E., Scaringi S., K\"ording E., Dhillon V.S., \&
        Coppejans D.L., 2018, MNRAS, 481, 2140
\bibitem[\protect\citeauthoryear{Araujo-Betancon et al.}{2005}]
        {Araujo-Betancor05C} 
        Araujo-Betancor S., G\"ansicke B.T., Hagen H.-J. et al., 2005
        A\&A, 430, 629
\bibitem[\protect\citeauthoryear{Arenas et al.}{2000}]{Arenas00C} 
        Arenas J., Catal\'an M.S., Augusteijn T., Retter A., 2000,
        MNRAS, 311, 135
\bibitem[\protect\citeauthoryear{Armstrong et al.}{2013}]{Armstrong13C} 
        Armstrong E., Patterson J., Michelsen E. et al., 2013, 
        MNRAS, 435, 707
\bibitem[\protect\citeauthoryear{Augusteijn}{1994}]{Augusteijn94C} 
        Augusteijn T., 1994, A\&A, 292, 481
\bibitem[\protect\citeauthoryear{Agusteijn et al.}{2010}]{Augusteijn10C} 
        Augusteijn T., Tappert C., Dall T., Maza J., 2010, 
        MNRAS, 405, 621
\bibitem[\protect\citeauthoryear{Barnard}{1921}]{Barnard21C}
        Barnard E.E., 1921, MNRAS, 81, 167
\bibitem[\protect\citeauthoryear{Bailer-Jones et al.}{2018}]{Bailer-Jones18C} 
        Bailer-Jones C.A.L., Rybizki J., Fouesneau M., Mantelet G., 
        Andrae R., 2018, AJ, 156, 58
\bibitem[\protect\citeauthoryear{Bailey}{1979}]{Bailey79C} 
        Bailey J., 1979, MNRAS, 188, 681
\bibitem[\protect\citeauthoryear{Bailey et al.}{1983}]{Bailey83C} 
        Bailey J., Axon D.J., Hough J. et al., 1983, MNRAS, 205, 1p
\bibitem[\protect\citeauthoryear{Balman et al.}{1995}]{Balman95C} 
        Balman \c{S}., Orio M., \"Ogelman H., 1995, ApJ, 449, L47 
\bibitem[\protect\citeauthoryear{Baptista et al.}{2003}]{Baptista03C} 
        Baptista R., Borges B.W., Bond H.E. et al., 2003, MNRAS, 345, 889
\bibitem[\protect\citeauthoryear{Baptista et al.}{2016}]{Baptista16C} 
        Baptista R., Borges B.W., Oliveira A.S., 2016, MNRAS, 463, 3799 
\bibitem[\protect\citeauthoryear{Baptista \& Bortoletto}{2004}]{Baptista04C}
        Baptista R., Bortoletto A., 2004, AJ, 128, 411
\bibitem[\protect\citeauthoryear{Baptista \& Bortoletto}{2008}]{Baptista08C}
        Baptista R., Bortoletto A., 2008, ApJ, 676, 1240
\bibitem[\protect\citeauthoryear{Baptista et al.}{2002}]{Baptista02C} 
        Baptista R., Bortoletto A., Harlaftis E.T., 2002, MNRAS, 335, 665
\bibitem[\protect\citeauthoryear{Baptista et al.}{1998}]{Baptista98C} 
        Baptista R., Catal\'an M.S., Horne K., Zilli D., 1998, 
        MNRAS, 300, 233
\bibitem[\protect\citeauthoryear{Baptista et al.}{1995}]{Baptista95C} 
        Baptista R., Horne K., Hilditch R.W., Mason K.O., Drew J.E.,
        1995, ApJ, 448, 393
\bibitem[\protect\citeauthoryear{Baptista et al.}{1989}]{Baptista89C} 
        Baptista R., Jablonski F.J., Steiner J.E., 1989, MNRAS, 241, 631
\bibitem[\protect\citeauthoryear{Baptista et al.}{1992}]{Baptista92C} 
        Baptista R., Jablonski F.J., Steiner J.E., 1992, AJ, 104, 1557 
\bibitem[\protect\citeauthoryear{Baptista et al.}{1996}]{Baptista96C} 
        Baptista R., Patterson J., O'Donoghue D. et al., 1996, 
        IAU Circ., 6327
\bibitem[\protect\citeauthoryear{Baptista et al.}{2000}]{Baptista00C}
        Baptista R., Silveira C., Steiner J.E., Horne K., 2000, 
        MNRAS, 314, 713
\bibitem[\protect\citeauthoryear{Baptista et al.}{1994}]{Baptista94C} 
        Baptista R., Steiner J.E., Cieslinski D., 1994, ApJ, 433, 332
\bibitem[\protect\citeauthoryear{Barrera \& Vogt}{1989a}]{Barrera89aC}
        Barrera L.H., Vogt N., 1989a, A\&A, 220, 99
\bibitem[\protect\citeauthoryear{Barrera \& Vogt}{1989b}]{Barrera89bC} 
        Barrera L.H., Vogt N., 1989b, Rev.\ Mex.\ A\&A, 19, 99
\bibitem[\protect\citeauthoryear{Barrett et al.}{1988}]{Barrett88C}
        Barrett P., O'Donoghue D., Warner B., 1988, MNRAS, 273, 759
\bibitem[\protect\citeauthoryear{Barwig et al.}{1982}]{Barwig82C} 
        Barwig H., Hunger K., Kudritzki R.P., Vogt N., 1982, 
        A\&A, 114, L11
\bibitem[\protect\citeauthoryear{Barwig et al.}{1992}]{Barwig92C} 
        Barwig H., Mantel K.H., Ritter H., 1992, A\&A, 266, L5
\bibitem[\protect\citeauthoryear{Bateson}{1974}]{Bateson74C} 
        Bateson F.M., 1974, 
        Publ.\ V.S.S. Roy.\ Astron.\ Soc.\ New Zealand, 2, 1
\bibitem[\protect\citeauthoryear{Bateson}{1978}]{Bateson78C}
        Bateson F.M., 1978, 
        Publ.\ V.S.S. Roy.\ Astron.\ Soc.\ New Zealand, 6, 1
\bibitem[\protect\citeauthoryear{Bateson}{1979}]{Bateson79C}
        Bateson F.M., 1979b, 
        Publ.\ V.S.S. Roy.\ Astron.\ Soc.\ New Zealand, 7, 29
\bibitem[\protect\citeauthoryear{Bateson}{1982a, 1982b, 1984}]{Bateson82aC} 
        Bateson F.M., 1982a, Publ.\ V.S.S. Roy.\ Astron.\ Soc.\ New Zealand, 
        10, 12
\bibitem[\protect\citeauthoryear{Bateson}{1982b}]{Bateson82bC} 
        Bateson F.M., 1982b, Publ.\ V.S.S. Roy.\ Astron.\ Soc.\ New Zealand, 
        10, 24
\bibitem[\protect\citeauthoryear{Bateson}{1984}]{Bateson84C} 
        Bateson F.M., 1984, Publ.\ V.S.S. Roy.\ Astron.\ Soc.\ New Zealand, 
        12, 45
\bibitem[\protect\citeauthoryear{Bateson \& McIntosh}{1986}]{Bateson86C} 
        Bateson F.M., McIntosh R., 1986, Publ.\ V.S.S. Roy.\ Aston.\ Soc.\
        New Zealand, 14, 1
\bibitem[\protect\citeauthoryear{Bateson et al.}{1990}]{Bateson90C} 
        Bateson F.M., McIntosh R., Brunt D., 1990, 
        Publ.\ V.S.S. Roy.\ Aston.\ Soc.\ New Zealand, 17, 29
\bibitem[\protect\citeauthoryear{Bearmore \& Osborne}{1977}]{Beardmore97C}
        Beardmore A.P., Osborne J.P., 1997, MNRAS, 290, 145
\bibitem[\protect\citeauthoryear{Beekman et al.}{2000}]{Beekman00C}
        Beekman G., Somers M., Naylor T., Hellier C., 2000, MNRAS, 318, 9
\bibitem[\protect\citeauthoryear{Belczynski \& Mikolajewska}{1998}]
        {Belczynski98C}
        Belczynski K., Mikolajewska J., 1998, MNRAS, 296, 77
\bibitem[\protect\citeauthoryear{Beljawsky}{1926}]{Beljawsky26C}
        Beljawsky P.R., 1926, Beobachtungs-Zirkular der Astron.\ Nachr., 6, 38
\bibitem[\protect\citeauthoryear{Belova et al.}{2013}]{Belova13C}
        Belova A.I., Suleimanov V.F., Bikmaev I.F. et al., 2013, 
        Astron. Lett., 39, 111
\bibitem[\protect\citeauthoryear{Beuermann \& Schope}{1989}]{Beuermann89C} 
        Beuermann K., Schwope A.D., 1989, A\&A, 223, 179
\bibitem[\protect\citeauthoryear{Beuermann et al.}{1992}]{Beuermann92C} 
        Beuermann K., Stasiewski U., Schwope A.D., 1992, A\&A, 256, 433
\bibitem[\protect\citeauthoryear{Beuermann \& Thomas}{1990}]{Beuermann90C} 
        Beuermann K., Thomas H.-C., 1990, A\&A, 230, 326
\bibitem[\protect\citeauthoryear{Bianchini}{1980}]{Bianchini80C} 
        Bianchini A., 1980, MNRAS, 192, 127
\bibitem[\protect\citeauthoryear{Bianchini \& Middleditch}{1976}]{Bianchini76C}
        Bianchini A., Middleditch J., 1976, IBVS, 1151,
\bibitem[\protect\citeauthoryear{Bianchini \& Sabbadin}{1983}]{Bianchini83C} 
        Bianchini A., Sabbadin F., 1983, A\&A Suppl., 54, 393
\bibitem[\protect\citeauthoryear{Bianchini et al.}{2012}]{Bianchini12C} 
        Bianchini A., Saygac T., Orio M., della Valle M., Williams R., 
        2012 A\&A, 539, A94
\bibitem[\protect\citeauthoryear{Biryukov \& Borisov}{1990}]{Biryukov90C}
        Biryukov W., Borisov G.V., 1990, Perem.\ Szvesdy 1544, 1
\bibitem[\protect\citeauthoryear{Bisol et al.}{2012}]{Bisol12C}
        Bisol A.C., Godon P., Sion E.M., 2012, PASP, 124, 158
\bibitem[\protect\citeauthoryear{Bitner et al.}{2007}]{Bitner07C}
        Bitner M.A., Robinson E.L., Behr B.B., 2007, ApJ, 662, 564
\bibitem[\protect\citeauthoryear{Blackman}{2010}]{Blackman10C} 
        Blackman C., 2010, ApJ Suppl., 191, 185
\bibitem[\protect\citeauthoryear{Bloemen et al.}{2010}]{Bloemen10C} 
        Bloemen S., Marsh T.R., Steeghs D., Ostensen R.H., 2010, 
        MNRAS, 407, 1903
\bibitem[\protect\citeauthoryear{Bolick et al}{1987}]{Bolick87C} 
        Bolick K., Beuermann K., Bruch A., Lenzen R., 1987, 
        Ap\&SS, 130, 175
\bibitem[\protect\citeauthoryear{Bond}{1978}]{Bond78C} 
        Bond H.E., 1978, PASP, 90, 216
\bibitem[\protect\citeauthoryear{Bond et al.}{1987}]{Bond87C}
        Bond H.E., Grauer A.D., Burstein D., Marzke R.O., 1987, 
        PASP, 99, 1097
\bibitem[\protect\citeauthoryear{Borges \& Baptista}{2005}]{Borges05C} 
        Borges B.W., Baptista R., 2005, A\&A, 437, 235
\bibitem[\protect\citeauthoryear{Borisov}{1992}]{Borisov92C} 
        Borisov G.V., 1992, A\&A, 261, 154
\bibitem[\protect\citeauthoryear{Brainerd \& Lamb}{1985}]{Brainerd85C}
        Brainerd J.J., Lamb D.Q. 1985, in Lamb D.Q., Patterson J., eds,
        Proc. 7th North American Workshop on Cataclysmic Variables and Low
        Mass X-ray Binaries, Reidel, Dordrecht, p.\ 247
\bibitem[\protect\citeauthoryear{Brandi et al.}{2009}]{Brandi09C} 
        Brandi E., Quiroga C., Mikolajewska J., Ferrer O.E., 
        Garc\'{\i}a L.G., 2009, A\&A, 497, 815 
\bibitem[\protect\citeauthoryear{Bruch}{1982a}]{Bruch82aC}
        Bruch A., 1982a, {\it Kataklysmische Ver\"anderliche in Beobachtung,
        Statistik und Analyse}, Ph.D.\ thesis, M\"unster
\bibitem[\protect\citeauthoryear{Bruch}{1982b}]{Bruch82bC} 
        Bruch A., 1982b, PASP, 94, 916
\bibitem[\protect\citeauthoryear{Bruch}{1986}]{Bruch86C}
        Bruch A., 1986, A\&A, 167, 91
\bibitem[\protect\citeauthoryear{Bruch}{1987}]{Bruch87C}
        Bruch A., 1987, A\&A, 172, 187
\bibitem[\protect\citeauthoryear{Bruch}{1989}]{Bruch89C}
        Bruch A., 1989, A\&AS, 78, 145
\bibitem[\protect\citeauthoryear{Bruch}{1991a}]{Bruch91aC} 
        Bruch A., 1991a, Acta Astron., 41, 101
\bibitem[\protect\citeauthoryear{Bruch}{1991b}]{Bruch91bC} 
        Bruch A., 1991b, IBVS, 3567
\bibitem[\protect\citeauthoryear{Bruch}{1992}]{Bruch92C} 
        Bruch A., 1992, A\&A, 266, 237
\bibitem[\protect\citeauthoryear{Bruch}{1996}]{Bruch96C} 
        Bruch A., 1996, A\&A, 312, 97
\bibitem[\protect\citeauthoryear{Bruch}{2000}]{Bruch00C} 
        Bruch A., 2000, A\&A, 359, 998
\bibitem[\protect\citeauthoryear{Bruch}{2003}]{Bruch03C} 
        Bruch A., 2003, A\&A, 409, 647
\bibitem[\protect\citeauthoryear{Bruch}{2014}]{Bruch14C} 
        Bruch A., 2014, A\&A, 566, A101
\bibitem[\protect\citeauthoryear{Bruch}{2015}]{Bruch15C} 
        Bruch A., 2015, A\&A, 579, A50
\bibitem[\protect\citeauthoryear{Bruch}{2016}]{Bruch16C} 
        Bruch A., 2016, New Astr., 46, 60\
\bibitem[\protect\citeauthoryear{Bruch}{2017a}]{Bruch17aC} 
        Bruch A., 2017a, New Astr., 52, 117
\bibitem[\protect\citeauthoryear{Bruch}{2017b}]{Bruch17bC}
        Bruch A., 2017b, New Astr., 56, 69
\bibitem[\protect\citeauthoryear{Bruch}{2017c}]{Bruch17cC} 
        Bruch A., 2017c, New Astr., 57, 51
\bibitem[\protect\citeauthoryear{Bruch}{2018}]{Bruch18aC} 
        Bruch A., 2018, New Astr., 58, 53
\bibitem[\protect\citeauthoryear{Bruch}{2019a}]{Bruch19aC}
        Bruch A., 2019a, IBVS, 6257
\bibitem[\protect\citeauthoryear{Bruch}{2019b}]{Bruch19bC}
        Bruch A., 2019b, MNRAS, 489, 2961
\bibitem[\protect\citeauthoryear{Bruch}{2020}]{Bruch20C} 
        Bruch A., 2020, New Astr., 78, 101369
\bibitem[\protect\citeauthoryear{Bruch et al.}{1986}]{Bruch96aC} 
        Bruch A., Beele D., Baptista R., 1996, A\&A, 306, 151
\bibitem[\protect\citeauthoryear{Bruch \& Cook}{2018}]{Bruch18bC} 
        Bruch A., Cook L.M., 2018, New Astr., 63,1 
\bibitem[\protect\citeauthoryear{Bruch \& Diaz}{2017}]{Bruch17dC} 
        Bruch A., Diaz M.P., 2017, New Astr., 50, 109
\bibitem[\protect\citeauthoryear{Bruch et al.}{1981}]{Bruch81C} 
        Bruch A., Duerbeck H.W., Seitter W.C., 1981, 
        Mitt.\ Astron.\ Ges., 52, 34
\bibitem[\protect\citeauthoryear{Bruch \& Engel}{1994}]{Bruch94C} 
        Bruch A., Engel A., 1994, A\&AS, 104, 79
\bibitem[\protect\citeauthoryear{Bruch \& Schimpke}{1992}]{Bruch92cC}
        Bruch A., Schimpke T., 1992, A\&AS, 93, 419
\bibitem[\protect\citeauthoryear{Bruch et al.}{2000}]{Bruch00aC} 
        Bruch A., Steiner J.E., Gneiding C.D., 2000, PASP, 112, 327
\bibitem[\protect\citeauthoryear{Bruch et al.}{1997}]{Bruch97C} 
        Bruch A., Vrielmann S., Hessman F.V., Kochsiek A., Schimpke T.,
        1997, A\&A, 237, 1107
\bibitem[\protect\citeauthoryear{Buckley et al.}{2000}]{Buckley00C} 
        Buckley D.A.H., Cropper M., van der Heyden K., Potter S.B., 
        Wickramasinghe D.T., 2000, MNRAS, 318, 187 
\bibitem[\protect\citeauthoryear{Buckley et al.}{1992}]{Buckley92C}
        Buckley D.A.H., O'Donoghue D., Kilkenny D., Stobie S.R., \&
        Remillard R.A., 1992, MNRAS, 258,285
\bibitem[\protect\citeauthoryear{Bullock et al.}{2011}]{Bullock11C} 
        Bullock E., Szkody P., Mukadam A.S. et al., 2011, AJ, 141, 84
\bibitem[\protect\citeauthoryear{Capitano et al.}{2017}]{Capitano17C}
        Capitano L, Lallement R., Vergely J.L., Elyajouri M., 
        Monreal-Ibero A., 2017, A\&A, 606, A65  
\bibitem[\protect\citeauthoryear{Casares et al.}{1996}]{Casares96C} 
        Casares J., Mart\'{\i}nez-Pais I.G., Marsh T.R., Charles P.A.,
        Lazaro C., 1996, MNRAS, 278, 219
\bibitem[\protect\citeauthoryear{Casares et al.}{2009}]{Casares09C} 
        Casares J., Mart\'{\i}nez-Pais I.G., Rodr\'{\i}guez-Gil P., 2009
        MNRAS, 399, 1534
\bibitem[\protect\citeauthoryear{Cassatella et al.}{1985}]{Cassatella85C} 
        Cassatella A., Hassall B.J.M., Harris A., Snijders M.A.J., 1985, 
        in Burke W.R., ed., Recent results on cataclysmic variables, 
        ESA SP-236, p.\ 281
\bibitem[\protect\citeauthoryear{Cassatella et al.}{1982}]{Cassatella82C} 
        Cassatella A., Patriarchi A., Selvelli P. et al., 1982, in 
        Rolfe E., Heck A., Battrick B., eds, Third European IUE 
        Conference, ESA SP-176, p.\ 229
\bibitem[\protect\citeauthoryear{Charles et al.}{1979}]{Charles79C}
        Charles P.A., Thostensen J., Bowyers S., Middledich J., 1979,
        ApJ, 231, L131
\bibitem[\protect\citeauthoryear{Chen et al.}{2001}]{Chen01C} 
        Chen A., O'Donoghue D., Stobie R.S., Kilkenny D., Warner B.,
        2001, MNRAS, 385, 89
\bibitem[\protect\citeauthoryear{Chote \& Sullivan}{2016}]{Chote16C} 
        Chote P., Sullivan D.J., 2016, MNRAS, 458, 1393
\bibitem[\protect\citeauthoryear{Cieslinski \& Steiner}{1997}]{Cieslinski97C} 
        Cieslinski D., Steiner J.E., 1997, MNRAS, 291, 321
\bibitem[\protect\citeauthoryear{Cieslinski et al.}{1998}]{Cieslinski98C} 
        Cieslinski D., Steiner J.E., Jablonski F., 1998, A\&AS, 131, 119
\bibitem[\protect\citeauthoryear{Cook}{1985}]{Cook85C} 
        Cook M.C., 1985, MNRAS, 215, 211
\bibitem[\protect\citeauthoryear{Copperwheat et al.}{2009}]{Copperwheat09C} 
        Copperwheat C.M., Marsh T.R., Dhillon V.S. et al., 2009, 
        MNRAS, 393, 157
\bibitem[\protect\citeauthoryear{Copperwheat et al.}{2010}]{Copperwheat10C}
        Copperwheat C.M., Marsh T.R., Dhillon V.S. et al., 2010,
        MNRAS 402, 1824 
\bibitem[\protect\citeauthoryear{Costa \& Rodrigues}{2009}]{Costa09C} 
        Costa J.E.R., Rodrigues C.V., 2009, MNRAS, 398, 240
\bibitem[\protect\citeauthoryear{Cowley et al.}{1977a}]{Cowley77aC} 
        Cowley A.P., Crampton D., Hesser J.E., 1977a, PASP, 89, 716
\bibitem[\protect\citeauthoryear{Cowley et al.}{1977b}]{Cowley77bC} 
        Cowley A.P., Crampton D., Hesser J.E., 1977b, ApJ, 214, 471
\bibitem[\protect\citeauthoryear{Cowley et al.}{1982}]{Cowley82C} 
        Cowley A.P., Crampton D., Hutchings J.B., 1982, ApJ, 259, 730
\bibitem[\protect\citeauthoryear{Cowley et al.}{1975}]{Cowley75C}
        Cowley A.P., Crampton D., Hutchings J.B., Marlborough J.M., 1975,
        ApJ, 195, 413
\bibitem[\protect\citeauthoryear{Cropper}{1987}]{Cropper87C} 
        Cropper M., 1987, MNRAS, 228, 389
\bibitem[\protect\citeauthoryear{Cropper}{1988}]{Cropper88C} 
        Cropper M., 1988, MNRAS, 231, 508 
\bibitem[\protect\citeauthoryear{Cropper}{1989}]{Cropper89C} 
        Cropper M., 1989, MNRAS, 236, 935 
\bibitem[\protect\citeauthoryear{Cropper et al.}{1986}]{Cropper86C}
        Cropper M., Menzies J.W., Tapia S., 1986, MNRAS, 218, 201
\bibitem[\protect\citeauthoryear{Cropper et al.}{1998}]{Cropper98C}
        Cropper M., Ramsay G., Wu K., 1998, MNRAS, 293, 222
\bibitem[\protect\citeauthoryear{Dahlmark}{1999}]{Dahlmark99C}
        Dahlmark L., 1999, IBVS, 4734
\bibitem[\protect\citeauthoryear{Dai \& Qian}{2009}]{Dai09C} 
        Dai Z.B., Qian S.B., 2009, A\&A, 503, 883 
\bibitem[\protect\citeauthoryear{Dai \& Qian}{2010a}]{Dai10aC} 
        Dai Z.B., Qian S.B., 2010a, New Astron., 15, 380 
\bibitem[\protect\citeauthoryear{Dai \& Qian}{2010b}]{Dai10bC} 
        Dai Z.B., Qian S.B., 2010b, PASJ, 62, 965
\bibitem[\protect\citeauthoryear{Dai et al.}{2009a}]{Dai09aC} 
        Dai Z.B., Qian S.B., Fern\'andez-Laj\'us E., 2009a, ApJ, 703, 109
\bibitem[\protect\citeauthoryear{Dai et al.}{2010}]{Dai10cC} 
        Dai Z.B., Qian S.B., Fern\'andez-Lajus E., Baume G.L., 2010,
        MNRAS, 409, 1195
\bibitem[\protect\citeauthoryear{Dai et al.}{2013}]{Dai13C} 
        Dai Z.B., Qian S.B., Li L., 2013, ApJ, 774, 153
\bibitem[\protect\citeauthoryear{Davey \& Smith}{1996}]{Davey96C} 
        Davey S., Smith R.C., 1996, MNRAS, 280, 481
\bibitem[\protect\citeauthoryear{Della Valle \& Rosino}{1987}]{DellaValle87C}
        Della Valle M., Rosino L., 1987, IBVS, 2995
\bibitem[\protect\citeauthoryear{Della Valle et al.}{1992}]{DellaValle92C} 
        Della Valle M., Duerbeck H.W., Motch C., 1992, 
        IAU Circ., No.\ 5503
\bibitem[\protect\citeauthoryear{de Miguel et al.}{2016}]{deMiguel16C} 
        de Miguel E., Patterson J., Cejudo E. et al., 2016, MNRAS, 457, 1447
\bibitem[\protect\citeauthoryear{d'Esterre}{1912}]{dEsterre12C}
        d'Esterre C.R., 1912, Astron.\ Nachr., 192, 131
\bibitem[\protect\citeauthoryear{Dhillon et al.}{1991}]{Dhillon91C} 
        Dhillon V.S., Marsh T.R., Jones D.H.P., 1991, MNRAS, 252, 342
\bibitem[\protect\citeauthoryear{Diaz}{2001}]{Diaz01C} 
        Diaz M.P., 2001, ApJ, 553, L177
\bibitem[\protect\citeauthoryear{Diaz \& Ribeiro}{2003}]{Diaz03C} 
        Diaz M.P., Ribeiro F.M.A., 2003, AJ, 125, 3359
\bibitem[\protect\citeauthoryear{Diaz \& Steiner}{1991a}]{Diaz91aC}
        Diaz M.P., Steiner J.E., 1991a, AJ, 102, 1417
\bibitem[\protect\citeauthoryear{Diaz \& Steiner}{1991b}]{Diaz91bC} 
        Diaz M.P., Steiner J.E., 1991b, PASP, 103, 964
\bibitem[\protect\citeauthoryear{Diaz \& Steiner}{1995}]{Diaz95C} 
        Diaz M.P., Steiner J.E., 1995, AJ, 110, 1816
\bibitem[\protect\citeauthoryear{Dobrotka et al.}{2010}]{Dobrotka10C}
        Dobrotka A., Hric L., Casares J. et al., 2010, MNRAS, 402, 2567
\bibitem[\protect\citeauthoryear{Dobrotka et al.}{2012}]{Dobrotka12C}
        Dobrotka A., Mineshige S., Casares J., 2012, MNRAS, 420, 2467
\bibitem[\protect\citeauthoryear{Dobrotka et al}{2015}]{Dobrotka15C} 
        Dobrotka A., Mineshige S., Ness J.-U., 2015, MNRAS, 447, 3162
\bibitem[\protect\citeauthoryear{Dobrotka et al.}{2017}]{Dobrotka17C} 
        Dobrotka A., Ness J.-U., Mineshige S., Nucita A.A., 2017, 
        MNRAS, 468, 1183
\bibitem[\protect\citeauthoryear{Dobrzycka et al.}{1996}]{Dobrzycka96C}
        Dobrzycka D., Kenyon S.J., Milone A.A.E., 1996, AJ, 111,414
\bibitem[\protect\citeauthoryear{Downes}{1982}]{Downes82C} 
        Downes R.A., 1982, PASP, 94, 590
\bibitem[\protect\citeauthoryear{Downes}{1990}]{Downes90C} 
        Downes R.A., 1990, AJ, 99, 339
\bibitem[\protect\citeauthoryear{Downes et al.}{1986}]{Downes86C}
        Downes R.A., Szkody P., Jenner D.C., Margon B., 1986, 
        ApJ, 301, 240
\bibitem[\protect\citeauthoryear{Downes et al.}{2001}]{Downes01C}
        Downes R.A., Webbink R.F., Shara M.M. et al., 2001, PASP, 113, 764
\bibitem[\protect\citeauthoryear{Drechsel et al.}{1983}]{Drechsel83C} 
        Drechsel H., Rahe J., Seward F.D., Wargau W., 
        1983, A\&A, 126, 357
\bibitem[\protect\citeauthoryear{Drew et al.}{2003}]{Drew03C} 
        Drew J.E., Hartley L.E., Long K.S., van der Walt J., 2003, 
        MNRAS, 338, 401 
\bibitem[\protect\citeauthoryear{Drew et al.}{1993}]{Drew93C} 
        Drew J.E., Jones D.H.P., Woods J.A., 1993, MNRAS, 260, 803
\bibitem[\protect\citeauthoryear{Duerbeck}{1980}]{Duerbeck80C} 
        Duerbeck H.W., 1980, Habilitation thesis, Univ.\ Bonn
\bibitem[\protect\citeauthoryear{Duerbeck}{1981}]{Duerbeck81C} 
        Duerbeck H.W., 1981, PASP, 93, 165
\bibitem[\protect\citeauthoryear{Duerbeck et al.}{1987}]{Duerbeck87C} 
        Duerbeck H.W., Seitter W.C., Duemmler R., 1987, MNRAS, 229, 653
\bibitem[\protect\citeauthoryear{Dulcin-Hacyan et al.}{1980}]{Dulcin-Hacyan80C} 
        Dulcin-Hacyan D., Andrillat Y., Audouze J. et al., 1980, in
        Battrick B., Mort J., eds, Second European IUE 
        Conference, ESA SP-156, p.\ 87
\bibitem[\protect\citeauthoryear{Eason \& Warden}{1983}]{Eason83C}
        Eason E.L.E. Warden S.P., 1983, PASP, 95, 58 
\bibitem[\protect\citeauthoryear{Echevarr\'{\i}a \& Alvarez}{1993}]
        {Echevarria93C}
        Echevarr\'{\i}a J., Alvarez M., 1993, 275, 187
\bibitem[\protect\citeauthoryear{Echevarr\'{\i}a et al.}{2002}]{Echevarria02C} 
        Echevarr\'{\i}a J., Costero R., Tovmassian G. et al., 2002, 
        Rev.\ Mex.\ A\&A Conf.\ Ser., 12,86
\bibitem[\protect\citeauthoryear{Echevarr\'{\i}a et al.}{1999}]{Echevarria99C} 
        Echevarr\'{\i}a J., Pineda L., Costero R., 1999, 
        Rev.\ Mex.\ A\&A, 35, 135
\bibitem[\protect\citeauthoryear{Echevarr\'{\i}a et al.}{2016}]{Echevarria16C} 
        Echevarr\'{\i}a J., Ram\'{\i}rez-Torres A., Michel R., 
        Hern\'andez Santisteban J.V., 2016, MNRAS, 461, 1576 
\bibitem[\protect\citeauthoryear{Eggen et al.}{1968}]{Eggen68C} 
        Eggen O.J., Freeman K.C., Sandage A., 1968, ApJ, 154, L27
\bibitem[\protect\citeauthoryear{Elsworth \& James}{1986}]{Elsworth86C} 
        Elsworth Y., James J.F., 1986, MNRAS, 220, 895
\bibitem[\protect\citeauthoryear{Evans et al.}{2009}]{Evans09C} 
        Evans P.A., Beardmore A.P., Osborne J.P., Wynn G.A., 2009, 
        MNRAS, 399, 1167
\bibitem[\protect\citeauthoryear{Feinswog et al.}{1988}]{Feinswog88C}
        Feinswog L., Szkody P., Garnavich P., 1988, AP 96, 1702
\bibitem[\protect\citeauthoryear{Feline et al.}{2005}]{Feline05C} 
        Feline W.J., Dhillon V.S., Marsh T.R., Watson C.A., \&
        Littlefair S.P., 2005, MNRAS, 364, 1158
\bibitem[\protect\citeauthoryear{Ferguson et al.}{1984}]{Ferguson84C}
        Ferguson D.H., Green R.F., Liebert J., 1984, ApJ, 287, 320
\bibitem[\protect\citeauthoryear{Ferland et al.}{1982}]{Ferland82C} 
        Ferland G.J., Lambert D.L., McCall M.L., Shields G.A., Slovak M.H.,
        1982, ApJ, 260, 794
\bibitem[\protect\citeauthoryear{Ferrario et al.}{1992}]{Ferrario92C} 
        Ferrario L., Wickramasinghe D.T., Bailey J., Hough J.H., 
        Tuohy I.R., 1992, MNRAS, 256, 252
\bibitem[\protect\citeauthoryear{Friend et al.}{1990}]{Friend90C} 
        Friend M.T., Martin J.S., Smith R.C., Jones D.H.P., 1990, 
        MNRAS, 246, 654
\bibitem[\protect\citeauthoryear{Fritz \& Bruch}{1998}]{Fritz98C} 
        Fritz T., Bruch A., 1998, A\&A, 332, 586
\bibitem[\protect\citeauthoryear{Froning et al.}{2003a}]{Froning03aC} 
        Froning C.S., Long K.S., Baptista R., 2003a, AJ, 126, 964
\bibitem[\protect\citeauthoryear{Froning et al.}{2003b}]{Froning03bC} 
        Froning C.S., Long K.S., Knigge C., 2003b, ApJ, 584, 433
\bibitem[\protect\citeauthoryear{Fu et al.}{2004}]{Fu04C}
        Fu H., Li Y.-Y., Leung K.-C. et al., 2004, Chin.\ J.\ A\&A, 4, 88
\bibitem[\protect\citeauthoryear{Fuentes-Morales et al.}{2018}]
        {Fuentes-Morales18C} 
        Fuentes-Morales I., Vogt N., Tappert C. et al., 2018, 
        MNRAS, 474, 2494
\bibitem[\protect\citeauthoryear{G\"ansicke}{2007}]{Gaensicke07C}
        G\"ansicke B.T., 2007, in Napiwotzki R., Burleigh M.R., eds,
        15th European Workshop on White Dwarfs, ASP Conf.\ Ser., 372, p.\ 597
\bibitem[\protect\citeauthoryear{G\"ansicke et al.}{1998}]{Gaensicke98C}
        G\"ansicke B.T., Hoard D.W., Beuermann K., Sion E.M., 
        Szkody P., 1998, A\&A, 338, 933
\bibitem[\protect\citeauthoryear{G\"ansicke et al.}{2009}]{Gaensicke09C} 
        G\"ansicke B.T., Dhillon M., Southworth J. et al., 2009,
        MNRAS, 397, 2170
\bibitem[\protect\citeauthoryear{Gallagher \& Holm}{1974}]{Gallagher74C} 
        Gallagher J.S., Holm A.V., 1974, ApJ, 189, L123
\bibitem[\protect\citeauthoryear{Garnavich \& Szkody}{1988}]{Garnavich88C}
        Garnavich P., Szkody P., 1988, PASP, 100, 1522
\bibitem[\protect\citeauthoryear{Garrison et al.}{1984}]{Garrison84C} 
        Garrison R.F., Schild R.E., Hiltner W.A., Krzeminski W., 1984, 
        ApJ, 276, L13
\bibitem[\protect\citeauthoryear{Geertsema \& Achterberg}{1992}]{Geertsema92C} 
        Geertsema G.T., Achterberg A., 1992, A\&A, 255, 427
\bibitem[\protect\citeauthoryear{Georgiev et al.}{2019, 2020a, 2020b}]
        {Georgiev19C}
        Georgiev T., Zamanov R.K., Boeva S. et al., 2019, 
        Bulg.\ Astron.\ J., 30, 1
\bibitem[\protect\citeauthoryear{Georgiev et al.}{2020a}]{Georgiev20aC}
        Georgiev T., Zamanov R.K., Boeva S. et al., 2020a, 
        Bulg.\ Astron.\ J., 32, 1
\bibitem[\protect\citeauthoryear{Georgieve et al.}{2020b}]{Georgiev20bC}
        Georgiev T., Zamanov R.K., Boeva S. et al., 2020b, 
        Bulg.\ Astron.\ J., 33, 1
\bibitem[\protect\citeauthoryear{Gerke et al.}{2006}]{Gerke06C} 
        Gerke J.R., Howell S.B., Walter F.M., 2006, PASP, 118, 678
\bibitem[\protect\citeauthoryear{Gill \& O'Brien}{2000}]{Gill00C} 
        Gill C.D., O'Brien T.J., 2000, MNRAS,314,175
\bibitem[\protect\citeauthoryear{Gilliland}{1982a}]{Gilliland82aC} 
        Gilliland R.L., 1982a, ApJ, 254, 653
\bibitem[\protect\citeauthoryear{Gilliland}{1982b}]{Gilliland82bC} 
        Gilliland R.L., 1982b, ApJ, 263, 302 
\bibitem[\protect\citeauthoryear{Gilliland et al.}{1986}]{Gilliland86C} 
        Gilliland R.L., Kemper E., Suntzeff N., 1986, ApJ, 301, 252
\bibitem[\protect\citeauthoryear{Gilliland \& Phillips}{1982}]{Gilliland82C} 
        Gilliland R.L., Phillips M.M., 1982, ApJ, 261, 617
\bibitem[\protect\citeauthoryear{Giovannelli et al.}{1983}]{Giovannelli83C}
        Giovannelli F., Gaudenzi S., Rossi C., Piccioni A., 1983, 
        Acta Astron., 33, 319 
\bibitem[\protect\citeauthoryear{Giovannelli \& Saubau-Graziati}{2012}]
        {Giovannelli12C}
        Giovannelli F. Sabau-Graziati L., 2012, Mem.\ S.A.I., 83, 698
\bibitem[\protect\citeauthoryear{Glasby}{1970}]{Glasby70C}
        Glasby J.S., 1970, The dwarf novae, Constable \& Co., London
\bibitem[\protect\citeauthoryear{Godon et al.}{2007}]{Godon07C} 
        Godon P., Sion E.M., Barrett P., Szkody P., 2007, ApJ, 656, 1092
\bibitem[\protect\citeauthoryear{Godon et al.}{2014}]{Godon14C} 
        Godon P., Sion E.M., Starrfield S. et al., 2014, ApJ, 784, L33
\bibitem[\protect\citeauthoryear{Godon et al.}{2018}]{Godon18C} 
        Godon P., Sion E.M., Williams R.E., Starrfield S., 2018, 
        ApJ, 860, 89
\bibitem[\protect\citeauthoryear{Gonzales \& Maza}{1983}]{Gonzales83C} 
        Gonzales L.E., Maza J., 1983, IAU Circ.\ No.\ 3854
\bibitem[\protect\citeauthoryear{Goranskij et al.}{1985}]{Goranskij85C}
        Goranskij V.P., Shugarov S.Yu., Orlowsky E.I., Rahimov V.Yu.,
        1985, IBVS, 2653
\bibitem[\protect\citeauthoryear{Grauer \& Bond}{1981}]{Grauer81C}
        Grauer A.D., Bond H.E., 1981, PASP, 93, 388
\bibitem[\protect\citeauthoryear{Greeley et al.}{1999}]{Greeley99C}
        Greeley B.W., Blair W.P., Long K.S., Raymond J.C., 1999, 
        ApJ, 513, 491
\bibitem[\protect\citeauthoryear{Green et al.}{1982}]{Green82C} 
        Green R.F., Ferguson D.H., Liebert J., Schmidt M., 1982, 
        PASP, 94, 560
\bibitem[\protect\citeauthoryear{Green et al.}{1986}]{Green86C}
        Green R.F., Schmidt M., Liebert J., 1986, ApJS, 61, 305
\bibitem[\protect\citeauthoryear{Greenhill et al.}{2006}]{Greenhill06C} 
        Greenhill J.G., Hill K.M., Dieters S. et al., 2006, MNRAS, 372, 1129
\bibitem[\protect\citeauthoryear{Greiner}{1998}]{Greiner98C}
        Greiner J., 1998, A\&A, 336, 626 
\bibitem[\protect\citeauthoryear{Greiner et al.}{2010}]{Greiner10C} 
        Greiner J., Schwarz R., Tappert C. et al., 2010, 
        Astron.\ Nachr., 331, 227
\bibitem[\protect\citeauthoryear{Greiner et al.}{1999}]{Greiner99C} 
        Greiner J., Tovmassian G.H., Di Stefano R. et al., 1999, 
        A\&A, 343, 183
\bibitem[\protect\citeauthoryear{Groot et al.}{1998}]{Groot98C} 
        Groot P.J., Augusteijn T., Barziv O., van Paradijs J., 1998,
        A\&A 340, L31
\bibitem[\protect\citeauthoryear{Hachisu \& Kato}{2018}]{Hachisu18C} 
        Hachisu I., Kato M., 2018, ApJS, 237, 4
\bibitem[\protect\citeauthoryear{Hachisu et al.}{2006, 2007}]{Hachisu06C}
        Hachisu I., Kato M., Kiyota S. et al., 2006, ApJ, 651, L141
\bibitem[\protect\citeauthoryear{Hachisu et al.}{2007}]{Hachisu07C}
        Hachisu I., Kato M., Luna G.J.M., 2007, ApJ, 659, L153
\bibitem[\protect\citeauthoryear{Haefner}{1981}]{Haefner81C}
        Haefner R., 1981, IBVS, 2045
\bibitem[\protect\citeauthoryear{Haefner \& Betzenbichler}{1991}]{Haefner91C} 
        Haefner R., Betzenbichler W., 1991, IBVS, 3665
\bibitem[\protect\citeauthoryear{Haefner \& Metz}{1985}]{Haefner85C} 
        Haefner R., Metz K., 1985, A\&A, 145, 311
\bibitem[\protect\citeauthoryear{Haefner \& Schoembs}{1987}]{Haefner87C}
        Haefner R., Schoembs R., 1987, MNRAS, 224, 231
\bibitem[\protect\citeauthoryear{Hagen et al.}{1995}]{Hagen95C}
        Hagen H.J., Groote D., Engels D., Reimers D., 1995, 
        A\&AS, 111, 195
\bibitem[\protect\citeauthoryear{Hameury \& Lasota}{2017}]{Hameury17C}
        Hameury J.-M., Lasota J.-P., 2017, A\&A, 602, A102
\bibitem[\protect\citeauthoryear{Hamuy \& Maza}{1986}]{Hamuy86C}
        Hamuy M., Maza J., 1986, IBVS, 2867
\bibitem[\protect\citeauthoryear{Harlaftis et al.}{1992}]{Harlaftis92C} 
        Harlaftis E.T., Hassall B.J.M., Naylor T., Charles P.A., \&
        Sonneborn G., 1992, MNRAS, 257, 607
\bibitem[\protect\citeauthoryear{Harrison}{2016}]{Harrison16C} 
        Harrison T.E., 2016, ApJ, 816, 4
\bibitem[\protect\citeauthoryear{Harrison et al.}{2009}]{Harrison09C}
        Harrison T.E., Bornak J., Howell S.B. et al., 2009, AJ, 137, 4061
\bibitem[\protect\citeauthoryear{Harrison et al.}{2004}]{Harrison04C} 
        Harrison T.E., Johnson J.J., McArthur B.E. et al., 2004, 
        AJ, 127, 460
\bibitem[\protect\citeauthoryear{Hartley et al.}{2005}]{Hartley05C} 
        Hartley L.E., Murray J.R., Drew J.E., Long K.S., 2005, 
        MNRAS, 363, 285
\bibitem[\protect\citeauthoryear{Hassall}{1985}]{Hassall85C} 
        Hassall B.J.M., 1985, in Lamb D.Q., Patterson J., eds, 
        Cataclysmic Variables and Low Mass X-ray Binaries, Reidel, 
        Dordrecht, p.\ 287
\bibitem[\protect\citeauthoryear{Haswell et al.}{1997}]{Haswell97C} 
        Haswell C.A., Patterson J., Thorstensen J.R., Hellier C., 
        Skillman D.R., 1997, ApJ, 476, 847
\bibitem[\protect\citeauthoryear{Hellier}{1993}]{Hellier93C} 
        Hellier C., 1993, MNRAS, 264, 132
\bibitem[\protect\citeauthoryear{Hellier}{1996}]{Hellier96C}
        Hellier C., 1996, ApJ, 471, 949
\bibitem[\protect\citeauthoryear{Hellier \& Buckley}{1993}]{Hellier93aC}
        Hellier C., Buckley D.A.H., 1993, MNRAS, 265, 766
\bibitem[\protect\citeauthoryear{Hellier et al.}{2000}]{Hellier00C} 
        Hellier C., Kemp J., Naylor T. et al., 2000, MNRAS, 313, 703 
\bibitem[\protect\citeauthoryear{Hellier et al.}{1991}]{Hellier91C} 
        Hellier C., Mason K.O., Mittaz J.P.D., 1991, MNRAS, 248, 5p
\bibitem[\protect\citeauthoryear{Hesser et al.}{1972}]{Hesser72C} 
        Hesser J.E., Lasker B.M., Osmer P.S., 1972, ApJ, 176, L31
\bibitem[\protect\citeauthoryear{Hesser et al.}{1974}]{Hesser74C} 
        Hesser J.E., Lasker B.M., Osmer P.S., 1974, ApJ, 189, 315
\bibitem[\protect\citeauthoryear{Hessman}{1988}]{Hessman88C} 
        Hessman F.V., 1988, A\&AS, 72, 512
\bibitem[\protect\citeauthoryear{Hessman et al.}{1984}]{Hessman84C} 
        Hessman F.V., Robinson E.L., Nather R.E., Zhang E.-H., 1984, 
        ApJ, 286, 747
\bibitem[\protect\citeauthoryear{Hill et al.}{2017}]{Hill17C} 
        Hill C.A., Smith R.C., Hebb L., Skzody P., 2017, MNRAS 472, 2937
\bibitem[\protect\citeauthoryear{Hind}{1948a, 1948b}]{Hind48C}
        Hind J.R., 1848a, Astron.\ Nachr., 27, 191
\bibitem[\protect\citeauthoryear{Hind}{1948b}]{Hind48bC}
        Hind J.R., 1848b, MNRAS, 8, 146
\bibitem[\protect\citeauthoryear{Hind}{1856}]{Hind56C}
        Hind J.R., 1856, MNRAS 16, 56
\bibitem[\protect\citeauthoryear{Hoard et al.}{1997}]{Hoard97C}
        Hoard D.W., Baptista R., Eracleous M. et al., 1997, MNRAS, 288, 691
\bibitem[\protect\citeauthoryear{Hoard et al.}{2014}]{Hoard14C} 
        Hoard D.W., Long K.S., Howell S.B. et al., 2014, ApJ, 786, 68
\bibitem[\protect\citeauthoryear{Hoard et al.}{1998}]{Hoard98C}
        Hoard D.W., Still M.D., Szkody P., Smith R.C., Buckley D.A.H.,
        1998, MNRAS 294, 687
\bibitem[\protect\citeauthoryear{Hoard et al.}{2000}]{Hoard00C} 
        Hoard D.W., Szkody P., Honeycutt R.K. et al., 2000, PASP, 112, 1595
\bibitem[\protect\citeauthoryear{Hoffmeister}{1928}]{Hoffmeister28C} 
        Hoffmeister C., 1928, Astron.\ Nachr., 234, 33
\bibitem[\protect\citeauthoryear{Hoffmeister}{1929}]{Hoffmeister29C}
        Hoffmeister C., 1929, Mitt.\ Sternw.\ Sonneberg, N16
\bibitem[\protect\citeauthoryear{Hoffmeister}{1943}]{Hoffmeister43C} 
        Hoffmeister C., 1943, AN, 274, 36
\bibitem[\protect\citeauthoryear{Hoffmeister}{1949}]{Hoffmeister49C}
        Hoffmeister C., 1949, Astr.\ Abh., 12, No.\ 1 
\bibitem[\protect\citeauthoryear{Hoffmeister}{1962}]{Hoffmeister62C} 
        Hoffmeister C., 1962, VSS, 6, no.1
\bibitem[\protect\citeauthoryear{Hoffmeister}{1963}]{Hoffmeister63C} 
        Hoffmeister C., 1963, Ver\"off.\ Sternw.\ Sonneberg, Vol.\ 6, Nr.\ 1
\bibitem[\protect\citeauthoryear{H{\o}g et al.}{2000}]{Hog00C} 
        H{\o}g E., Fabricius C., Makarov V.V. et al., 2000, A\&A, 355, L27
\bibitem[\protect\citeauthoryear{Hollander et al.}{1993}]{Hollander93C}
        Hollander A., Kraakman H., van Paradijs J., 1993, A\&AS, 101, 87
\bibitem[\protect\citeauthoryear{Homer et al.}{2004}]{Homer04C}
        Homer L., Szkody P., Raymond J.C. et al., 2004, ApJ, 610, 991
\bibitem[\protect\citeauthoryear{Honeycutt}{2001}]{Honeycutt01C} 
        Honeycutt R.K., 2001, PASP, 113, 473
\bibitem[\protect\citeauthoryear{Honeycutt \& Kafka}{2004}]{Honeycutt04C} 
        Honeycutt R.K., Kafka S., 2004, AJ, 128, 1279
\bibitem[\protect\citeauthoryear{Honeycutt \& Robertson}{1998}]{Honeycutt98aC} 
        Honeycutt R.K., Robertson J.W., 1998, AJ, 116, 1961
\bibitem[\protect\citeauthoryear{Honeycutt et al.}{1998}]{Honeycutt98bC}
        Honeycutt R.K., Robertson J.W., Turner G.W., 1998, AJ, 115, 2527
\bibitem[\protect\citeauthoryear{Honeycutt et al.}{1994, 2014}]{Honeycutt94C} 
        Honeycutt R.K., Robertson J.W., Turner G.W., Vesper D.N., 1994,  
        in Shafter A.W., ed., ASP Conf.\ Ser.\, 56, Interacting Binary Stars, 
        p.\ 277 
\bibitem[\protect\citeauthoryear{Honeycutt \& Schlegel}{1985}]{Honeycutt85C} 
        Honeycutt R.K., Schlegel E.M., 1985, PASP, 97, 1189
\bibitem[\protect\citeauthoryear{Honeycutt et al.}{2014}]{Honeycutt14C} 
        Honeycutt R.K., Shears S., Kafka S., Robertson J.W., 
        Henden A.A., 2014, AJ, 147, 105
\bibitem[\protect\citeauthoryear{Horne et al.}{1982}]{Horne82C}
        Horne K., Lanning H.H., Gomer K., 1982, ApJ, 252, 681
\bibitem[\protect\citeauthoryear{Horne et al.}{1986}]{Horne86C} 
        Horne K., Wade R. A., Szkody P., 1986, MNRAS, 219, 791
\bibitem[\protect\citeauthoryear{Horne et al.}{1993}]{Horne93C} 
        Horne K., Welsh W.F., Wade R.A., 1993, ApJ, 410, 307 
\bibitem[\protect\citeauthoryear{Horne et al.}{1991}]{Horne91C} 
        Horne K., Wood J.H., Stiening R.F., 1991, ApJ 378, 271
\bibitem[\protect\citeauthoryear{Howell et al.}{2006}]{Howell06C} 
        Howell S.B., Harrison T.E., Campbell R.K., C\'ordova F.A., 
        Szkody P., 2006, AJ, 131, 2216
\bibitem[\protect\citeauthoryear{Howell et al.}{1991}]{Howell91C} 
        Howell S.B., Szkody P., Kreidl T.J., Dobrzycka D., 1991,
        PASP, 103, 300,
\bibitem[\protect\citeauthoryear{Howell et al.}{1992}]{Howell92C} 
        Howell S.B., Wagner R.M., Bertram R., Szkody P., 1992, 
        IAU Circ., No.\ 5505
\bibitem[\protect\citeauthoryear{Hric et al.}{1998}]{Hric98C}
        Hric L., Petrik K., Urban Z., Niarchos P., Anupama G.C., 
        1998, A\&A, 329, 449
\bibitem[\protect\citeauthoryear{Hudec et al.}{2005}]{Hudec05C}
        Hudec R., \v{S}imon V., Skalick\'y J., 2005, ASPC, 330, 405
\bibitem[\protect\citeauthoryear{Hunger et al.}{1985}]{Hunger85C}
        Hunger K., Heber U., Koester D., 1985, A\&A, 149, L6
\bibitem[\protect\citeauthoryear{Hutchings et al.}{1981}]{Hutchings81C}
        Hutchings J.B., Crampton D., Cowley A.P., Thorstensen J., \&
        Charles P.A., 1981, ApJ, 249, 680
\bibitem[\protect\citeauthoryear{Hutchings et al.}{1983}]{Hutchings83C} 
        Hutchings J.B., Link R., Crampton D., 1983, PASP, 95, 265
\bibitem[\protect\citeauthoryear{Hutchings \& Cowley}{1984}]{Hutchings84C} 
        Hutchings J.B., Cowley A.P., 1984, PASP, 96, 559
\bibitem[\protect\citeauthoryear{Hutchings et al.}{1985}]{Hutchings85C} 
        Hutchings J.B., Cowley A.P., Crampton D., 1985, PASP, 97, 423
\bibitem[\protect\citeauthoryear{Ianna}{1964}]{Ianna64C}
        Ianna P.A., 1964, ApJ, 139, 780
\bibitem[\protect\citeauthoryear{Itagai}{2009}]{Itagai09C}
        Itagaki K., 2009, CBET, 2050, 1
\bibitem[\protect\citeauthoryear{I\-lkiewics et al.}{2016}]{Ilkiewics16C}
        I\-lkiewics K., Mikolajewska J., Stoyanov K., Manousakis A., \&
        Miszalski B., 2016, MNRAS, 462, 2695
\bibitem[\protect\citeauthoryear{Imada et al.}{2013}]{Imada13C} 
        Imada A., Izumiura H., Kuroda D. et al., 2013, PASJ, 65, 87
\bibitem[\protect\citeauthoryear{Imamura et al.}{2000}]{Imamura00C} 
        Imamura J.N., Steiman-Cameron T.Y., Wolff M.T., 2000, 
        PASP, 112, 18
\bibitem[\protect\citeauthoryear{Ingram \& Szkody}{1992}]{Ingram92C} 
        Ingram D., Szkody P., 1992, IBVS 3810
\bibitem[\protect\citeauthoryear{Ishioka et al.}{2003}]{Ishioka03C} 
        Ishioka R., Kato T., Uemura M. et al., 2003, PASJ, 55, 683
\bibitem[\protect\citeauthoryear{Jablonski \& Steiner}{1987}]{Jablonski87C} 
        Jablonski F.J., Steiner J.E., 1987, ApJ, 313, 376
\bibitem[\protect\citeauthoryear{Jensen et al.}{1982}]{Jensen82C} 
        Jensen K.A., Nousek J.A., Nugent J.J., 1982, ApJ, 261, 625
\bibitem[\protect\citeauthoryear{Johnson et al.}{2014}]{Johnson14C} 
        Johnson C.B., Schaefer B.E., Kroll P., Henden A.A., 2014, 
        ApJ, 780, L25
\bibitem[\protect\citeauthoryear{Jurdana-\v{S}epi\'c et al.}{2012}]
        {Jurdana-Sepic12C}
        Jurdana-\v{S}epi\'c R., Ribeiro V.R.A.M., Darnley M.J., Munari U.,
        Bode M.F., 2012, A\&A, 537, A34
\bibitem[\protect\citeauthoryear{Kafka et al.}{2008}]{Kafka08C} 
        Kafka S., Anderson R., Honeycutt R.K., 2008, AJ, 135, 1649 
\bibitem[\protect\citeauthoryear{Kafka \& Honeycutt}{2004}]{Kafka04C}
        Kafka S., Honeycutt R.K., 2004, 
        Rev.\ Mex.\ A\&A (Conf. Ser.), 20, 238
\bibitem[\protect\citeauthoryear{Kafka \& Honeycutt}{2005}]{Kafka05C}
        Kafka S., Honeycutt R.K., 2005, IBVS, 5597
\bibitem[\protect\citeauthoryear{Kaitchuck}{1989}]{Kaitchuck89C} 
        Kaitchuck R.H., 1989, PASP, 101, 1129
\bibitem[\protect\citeauthoryear{Kaitchuck et al.}{1998}]{Kaitchuck98C}
        Kaitchuck R.H., Schlegel E.M., White J.C., Mansperger C.S., 1998
        ApJ, 499, 444
\bibitem[\protect\citeauthoryear{Kaitchuck et al.}{1983}]{Kaitchuck83C} 
        Kaitchuck R.H., Honeycutt R.K., Schlegel E.M., 1983, ApJ, 267, 239
\bibitem[\protect\citeauthoryear{Ka\-lu\.zny}{1989}]{Kaluzny89C} 
        Ka\-lu\.zny J., 1989, Acta Astron., 39, 235
\bibitem[\protect\citeauthoryear{Kato et al.}{2014}]{Kato14C} 
        Kato T., Dubovsky P.A., Kudzej I. et al., 2014, PASJ, 60, 30
\bibitem[\protect\citeauthoryear{Kato et al.}{2016}]{Kato16C}
        Kato T., Hambsch F.-J., Monard B. et al., 2016, PASJ, 68, 65
\bibitem[\protect\citeauthoryear{Kato et al.}{1998}]{Kato98C} 
        Kato T., Haseda K., Takamizawa K., Kazarovets E.V., Samus N.N.,
        1998, IBVS, 4584
\bibitem[\protect\citeauthoryear{Kato et al.}{1992}]{Kato92C} 
        Kato T., Hirata R., Mineshige S., 1992, PASJ, 44, L215
\bibitem[\protect\citeauthoryear{Kato et al.}{2009, 2014}]{Kato09C} 
        Kato T., Imada A., Uemura M. et al., 2009, PASJ, 61, S395
\bibitem[\protect\citeauthoryear{Kato et al.}{2002}]{Kato02C}
        Kato T., Ishioka R., Uemura M., 2002, PASJ, 54, 1003
\bibitem[\protect\citeauthoryear{Kato et al.}{2017}]{Kato17C} 
        Kato T., Isogai K., Hambsch F.-J. et al., 2017, PASJ, 69, 75
\bibitem[\protect\citeauthoryear{Kato et al.}{2003}]{Kato03aC} 
        Kato T., Santallo R., Bolt G. et al., 2003, MNRAS, 339, 861 
\bibitem[\protect\citeauthoryear{Kato et al.}{2001}]{Kato01C} 
        Kato T., Sekine Y., Hirata R., 2001, PASJ, 53, 1191
\bibitem[\protect\citeauthoryear{Kato \& Stubbings}{2003}]{Kato03C} 
        Kato T., Stubbings R., 2003, IBVS, 5426
\bibitem[\protect\citeauthoryear{Kato \& Uemura}{1999}]{Kato99C}
        Kato T., Uemura M., 1999, IBVS, 4786
\bibitem[\protect\citeauthoryear{Kelly et al.}{1981}]{Kelly81C}
        Kelly B.D., Kilkenny D., Cooke J.A., 1981, MNRAS, 196, 91p
\bibitem[\protect\citeauthoryear{Kenyon \& Garcia}{1986}]{Kenyon86C}
        Kenyon S.J., Garcia M.R., 1986, AJ, 91, 125
\bibitem[\protect\citeauthoryear{Khangale et al.}{2020}]{Khangale20C} 
        Khangale Z.N., Woudt P.A., Potter S.B., et al., 2020, MNRAS, 495, 637
\bibitem[\protect\citeauthoryear{Kholopov \& Samus}{1988}]{Kholopov88C}
        Kholopov P.N., Samus N.N., 1988, IBVS, 3154
\bibitem[\protect\citeauthoryear{Kholopov et al.}{1985}]{Kholopov85C} 
        Kholopov P.N., Samus N.N., Frolov M.S. et al., 1985, {\it General
        Catalogue of Variable Stars}, $4^{\rm th}$ edition, Moscow, 
        Nauka Publishing House
\bibitem[\protect\citeauthoryear{Kilkenny \& Lloyd Evans}{1989}]{Kilkenny89C} 
        Kilkenny D., Lloyd Evans T., 1989, Obs., 109, 89
\bibitem[\protect\citeauthoryear{Klare et al.}{1982}]{Klare82C} 
        Klare G., Krautter J., Wolf B. et al., 1982, A\&A, 113, 76
\bibitem[\protect\citeauthoryear{Knigge}{2006}]{Knigge06C} 
        Knigge C., 2006, MNRAS, 373, 484
\bibitem[\protect\citeauthoryear{Knigge et al.}{2000}]{Knigge00C} 
        Knigge C., King A.R., Patterson J., 2000, A\&A, 364, L75 
\bibitem[\protect\citeauthoryear{Kozhevnikov}{2007}]{Kozhevnikov07C}
        Kozhevnikov V.P., 2007, MNRAS, 378, 957
\bibitem[\protect\citeauthoryear{Kozhevnikov}{2015}]{Kozhevnikov15C}
        Kozhevnikov V.P., 2015, New Astron., 41, 59
\bibitem[\protect\citeauthoryear{Kraft}{1958}]{Kraft58C}
        Kraft R.P., 1958, ApJ 127, 625
\bibitem[\protect\citeauthoryear{Kraft}{1964a}]{Kraft64aC} 
        Kraft R.P., 1964a, ApJ 139, 457
\bibitem[\protect\citeauthoryear{Kraft}{1964b}]{Kraft64bC}
        Kraft R.P., 1964b, in Luyten W.J., ed., 1$^{st}$ Conference on 
        Faint Blue Stars, Univ.\ of Minnesota Press, p.\ 77
\bibitem[\protect\citeauthoryear{Kraicheva et al.}{1999a}]{Kraicheva99aC} 
        Kraicheva Z., Stanishev V., Genkov V., 1999a, A\&AS, 134, 263
\bibitem[\protect\citeauthoryear{Kraicheva et al.}{1999b}]{Kraicheva99bC}
        Kraicheva Z., Stanishev V., Genkov V., Iliev L., 1999b, 
        A\&A 351, 607
\bibitem[\protect\citeauthoryear{Krautter et al.}{1981a}]{Krautter81aC} 
        Krautter J., Klare G., Wolf B. et al., 1981a, A\&A 98, 27
\bibitem[\protect\citeauthoryear{Krautter et al.}{1981b}]{Krautter81bC} 
        Krautter J., Klare G., Wolf B. et al., 1981b, A\&A 102, 337
\bibitem[\protect\citeauthoryear{Kremp\'ec}{1970}]{Krempec70C}
        Kremp\'ec J., 1970, Acta Astr., 20, 267
\bibitem[\protect\citeauthoryear{Kruszewski et al.}{1981}]{Kruszewski81C} 
        Kruszewski A., Mewe R., Heise J. et al., 1981, Space Science Rev., 
        30, 221
\bibitem[\protect\citeauthoryear{Krzeminski}{1962}]{Krzeminski62C} 
        Krzeminski W., 1962, PASP, 74, 66
\bibitem[\protect\citeauthoryear{Kubiak}{1984}]{Kubiak84C} 
        Kubiak K., 1984, Acta Astron., 34, 331,
\bibitem[\protect\citeauthoryear{K\"urster \& Barwig}{1988}]{Kuerster88C} 
        K\"urster M., Barwig H., 1988, A\&A, 199, 201
\bibitem[\protect\citeauthoryear{Kurochkin}{1960}]{Kurochkin60C} 
        Kurochkin N.E., 1960, Astron.\ Tsirk., 212, 9
\bibitem[\protect\citeauthoryear{Kurochkin \& Shugarov}{1980}]{Kurochkin80C} 
        Kurochkin N.E., Shugarov S.Yu., 1980, Astron.\ Tsirk., 1114, 1
\bibitem[\protect\citeauthoryear{Krzeminski}{1966}]{Krzeminski66C} 
        Krzeminski W., 1966, IBVS 160
\bibitem[\protect\citeauthoryear{Kuulkers et al.}{1991}]{Kuulkers91C} 
        Kuulkers E., Hollander A., Oosterbroek T. van Paradijs J., 1991,
        A\&A, 242, 401
\bibitem[\protect\citeauthoryear{la Dous}{1991}]{laDous91C} 
        la Dous C., 1991, A\&A, 252, 100
\bibitem[\protect\citeauthoryear{la Dous et al.}{1985}]{laDous85C} 
        la Dous C., Verbunt F., Schoembs R. et al., 1985, MNRAS, 212, 231
\bibitem[\protect\citeauthoryear{Lanning}{1973}]{Lanning73C}
        Lanning H.H., 1973, PASP, 85, 70
\bibitem[\protect\citeauthoryear{Lanning et al.}{1981}]{Lanning81C}
        Lanning H.H., Horne K., Gomer R., 1981, IAU Circ., 3567
\bibitem[\protect\citeauthoryear{Lasota}{2001}]{Lasota01C} 
        Lasota J.P., 2001, New Astr.\ Rev., 45, 449
\bibitem[\protect\citeauthoryear{Lawrence et al.}{1967}]{Lawrence67C}
        Lawrence G.M., Ostriker J.P., Hesser J.E., 1967, ApJ, 148, 161
\bibitem[\protect\citeauthoryear{Larsson}{1985, 1992}]{Larsson85C} 
        Larsson S., 1985, A\&A, 145, L1
\bibitem[\protect\citeauthoryear{Larsson}{1992}]{Larsson92C} 
        Larsson S., 1992, A\&A, 265, 133
\bibitem[\protect\citeauthoryear{Leach et al.}{1999}]{Leach99C} 
        Leach R., Hessman F.V., King A.R., Stehle R., Mattes J., 1999
        MNRAS, 305, 225
\bibitem[\protect\citeauthoryear{Leavitt \& Mackie}{1919}]{Leavitt19C} 
        Leavitt H.S., Mackie J.C., 1919, Harvard Circ., 219
\bibitem[\protect\citeauthoryear{Leibowitz et al.}{1994}]{Leibowitz94C} 
        Leibowitz E.M., Mendelson H., Bruch A. et al., 1994, 421, 771
\bibitem[\protect\citeauthoryear{Li et al.}{2017}]{Li17C} 
        Li K., Hu S.-M., Zhou J.-L. et al., 2017, PASJ, 69, 28
\bibitem[\protect\citeauthoryear{Liller}{1980}]{Liller80C} 
        Liller M.H., 1980, IBVS, 1743
\bibitem[\protect\citeauthoryear{Lines et al.}{1987}]{Lines87C} 
        Lines H.C., Lines R.D., McFaul T.G., 1987, AJ, 95, 1505
\bibitem[\protect\citeauthoryear{Linnell et al.}{2007}]{Linnell07C} 
        Linnell A.P., Godon P., Hubeny I., Sion E.M., Szkody P., 2007, 
        ApJ, 662, 1204 
\bibitem[\protect\citeauthoryear{Linnell et al.}{2008a}]{Linnell08aC} 
        Linnell A.P., Godon P., Hubeny I., 2008a, ApJ, 676, 1226
\bibitem[\protect\citeauthoryear{Linnell et al.}{2008b}]{Linnell08bC} 
        Linnell A.P., Godon P., Hubeny I., Sion E.M., Szkody P., 2008b,
        ApJ, 688, 568.
\bibitem[\protect\citeauthoryear{Linnell et al.}{2009}]{Linnell09C} 
        Linnell A.P., Godon P., Hubeny I. et al., 2009, ApJ, 703, 1839
\bibitem[\protect\citeauthoryear{Lipkin et al.}{2000}]{Lipkin00C}
        Lipkin Y., Leibowitz E., Retter A., 2000, IAU Circ., 7433
\bibitem[\protect\citeauthoryear{Littlefair et al.}{2000}]{Littlefair00C}
        Littlefair S.P., Dhillon V.S., Howell S.B. Ciardi D.R., 2000
        MNRAS, 313, 117
\bibitem[\protect\citeauthoryear{Liu et al.}{1999}]{Liu99C}
        Liu W., Hu J.Y., Zhu X.H., Li Z.Y., 1999, ApJS, 122, 243
\bibitem[\protect\citeauthoryear{L\"ochel}{1965}]{Loechel65C} 
        L\"ochel K., 1965, MVS, 3, 107
\bibitem[\protect\citeauthoryear{Longa-Pe\~na et al.}{2015}]{Longa-Pena15C}
        Longa-Pe\~na P., Steeghs D., Marsh T., 2015, MNRAS, 447, 149 
\bibitem[\protect\citeauthoryear{Lopes de Oliveira et al.}{2020}]{Lopes20C} 
        Lopes de Oliveira R., Bruch A., Rodrigues C.V., Oliveira A.S.,
        Mukai K., 2020, ApJL, 898, L40
\bibitem[\protect\citeauthoryear{Luyten}{1932}]{Luyten32C} 
        Luyten W.J., 1932, AN, 245, 211
\bibitem[\protect\citeauthoryear{Luyten}{1934}]{Luyten34C}
        Luyten W.J., 1934, AN, 253, 135
\bibitem[\protect\citeauthoryear{Luyten \& Haro}{1959}]{Luyten59C} 
        Luyten W.J., Haro G., 1959, PASP, 71, 469
\bibitem[\protect\citeauthoryear{Mantel et al.}{1987}]{Mantel87C} 
        Mantel K.H., Barwig H., H., Haefner R., Schoembs R., 1987 
        in Drechsel H., Kondo Y., Rahe J., eds, Cataclysmic
        Variables -- Recent Multifrequency Observations and Theoretical
        Developments, ApSS, 130, p.\ 501
\bibitem[\protect\citeauthoryear{Mantel et al.}{1988}]{Mantel88C}
        Mantel K.H., Marschh\"ausser H., Schoembs R., Haefner R., \&
        la Dous C., 1988, A\&A 193, 101
\bibitem[\protect\citeauthoryear{Margon \& Downes}{1981}]{Margon81C}
        Margon B., Downes R., 1981, AJ, 86, 747
\bibitem[\protect\citeauthoryear{Marin et al.}{2007}]{Marin07C} 
        Marin E., Shafter A.W., Misselt K.A., 2007, AAS, 211, 5108 
\bibitem[\protect\citeauthoryear{Marino \& Walker}{1984}]{Marino84C} 
        Marino B.F., Walker W.S.G., 1984, Southern Stars, 30, 389
\bibitem[\protect\citeauthoryear{Marsh}{1988}]{Marsh88C}
        Marsh T.R., 1988, MNRAS 231, 1117
\bibitem[\protect\citeauthoryear{Marsh et al.}{1990}]{Marsh90C} 
        Marsh T.R., Horne K., Schlegel E.M., Honeycutt R.K., 
        Kaitchuck R.H., 1990, ApJ, 364, 637
\bibitem[\protect\citeauthoryear{Martin et al.}{1989}]{Martin89C}
        Martin J.S., Friend M.T., Smith R.C., Jones D.H.P., 1989,
        MNRAS, 240, 519
\bibitem[\protect\citeauthoryear{Martin et al.}{1987, 1989}]{Martin87C}
        Martin J.S., Jones D.H.P., Smith R.C., 1987, MNRAS, 224, 1031
\bibitem[\protect\citeauthoryear{Mart\\{\i}nez-Pais et al.}{2000}]
        {Martinez-Pais00C} 
        Mart\'{\i}nez-Pais I.G., Mart\'{\i}m-Hern\'andez N.L., Casares J., 
        Rodr\'{\i}guez-Gil P., 2000, ApJ 538, 315
\bibitem[\protect\citeauthoryear{Mason et al.}{1997}]{Mason97C} 
        Mason K.O., Drew J.E., Knigge C., 1997, MNRAS, 290, L23 
\bibitem[\protect\citeauthoryear{Mason et al.}{1983}]{Mason83C} 
        Mason K.O., Middleditch J., C\'ordova F.A. et al., 1983, 
        ApJ, 264, 575
\bibitem[\protect\citeauthoryear{Mason \& Howell}{2003}]{Mason03C}
        Mason E., Howell S.B., 2003, A\&A, 403, 707
\bibitem[\protect\citeauthoryear{Mason et al.}{2013}]{Mason13C} 
        Mason E., Orio M., Mukai K. et al., 2013, MNRAS 436, 212
\bibitem[\protect\citeauthoryear{Mason et al.}{2001}]{Mason01C} 
        Mason E., Skidmore W., Howell S.B., Mennickent R.E., 2001, 
        ApJ, 563, 351
\bibitem[\protect\citeauthoryear{Mateo et al.}{1991}]{Mateo91C}
        Mateo M., Szkody P., Garnavich P., 1991, ApJ, 370, 370
\bibitem[\protect\citeauthoryear{Mateo et al.}{1985}]{Mateo85C} 
        Mateo M., Szkody P., Hutchings J., 1985, ApJ, 288, 292
\bibitem[\protect\citeauthoryear{Matsui et al.}{2009}]{Matsui09C}
        Matsui R., Uemura M., Arai A. et al., 2009, PASJ, 61, 1081
\bibitem[\protect\citeauthoryear{Matsumoto et al.}{2000}]{Matsumoto00C} 
        Matsumoto K., Mennickent R.E., Kato T., 2000, A\&A, 363, 1029
\bibitem[\protect\citeauthoryear{Mayall}{1946}]{Mayall46C} 
        Mayall M.W., 1946, Harvard. Bull., 918, 3
\bibitem[\protect\citeauthoryear{Maza et al.}{1989}]{Maza89C} 
        Maza J., Ruiz M.T., Gonz\'alez E., Wischnjewsky M., 1989,
        ApJ Suppl., 69, 349
\bibitem[\protect\citeauthoryear{Mazeh et al.}{1985}]{Mazeh85C} 
        Mazeh T., Tal Y., Shaviv G., Bruch A., Budell R., 1985, 
        A\&A 149, 470
\bibitem[\protect\citeauthoryear{McAllister et al.}{2019}]{McAllister19C} 
        McAllister M., Littlefair S.P., Parsons S.G. et al., 2019, 
        MNRAS, 486, 5535
\bibitem[\protect\citeauthoryear{McIntosh}{1989}]{McIntosh89C} 
        McIntosh R., 1989, Publ.\ Var.\ Star Sect.\, RASNZ, 16, 83
\bibitem[\protect\citeauthoryear{McLaughlin}{1960}]{McLaughlin60C} 
        McLaughlin D.B., 1960, in Greenstein J.E., ed., Stars and Stellar 
        Systems, Vol.\ VI, Stellar Atmospheres, University of
        Chicago Press, Chicago, p.\ 585
\bibitem[\protect\citeauthoryear{McQuillin et al.}{2012}]{McQuillin12C} 
        McQuillin R., Evans A., Wilson D. et al., 2012, MNRAS, 419, 330
\bibitem[\protect\citeauthoryear{Mennickent}{1994}]{Mennickent94C} 
        Mennickent R.E., 1994, A\&A, 250, 363 
\bibitem[\protect\citeauthoryear{Mennickent \& Diaz}{2002}]{Mennickent02C}
        Mennickent R.E., Diaz M.P., 2002, MNRAS, 326, 765
\bibitem[\protect\citeauthoryear{Mennickent et al.}{1999}]{Mennickent99C} 
        Mennickent R.E., Diaz M.P., Arenas J., 1999, A\&A, 352, 167
\bibitem[\protect\citeauthoryear{Mennickent \& Sterken}{1998}]{Mennickent98C} 
        Mennickent R.E., Sterken C., 1998, PASP, 110, 1039
\bibitem[\protect\citeauthoryear{Mennickent et al.}{2006}]{Mennickent06C} 
        Mennickent R.E., Unda-Sanzana E., Tappert C., 2006, A\&A, 451, 619
\bibitem[\protect\citeauthoryear{Metik}{1981}]{Metik61C}
        Metik L.P., 1981, Perem.\ Zvesdy 13, 364
\bibitem[\protect\citeauthoryear{Middleditch \& C\'ordova}{1982}]
        {Middleditch82C}
        Middleditch J., C\'ordova F.A., 1982, ApJ, 255, 585
\bibitem[\protect\citeauthoryear{Middleditch et al.}{1991}]{Middleditch91C} 
        Middleditch J., Imamura J.N., Wolff M.T., Steiman-Cameron T.Y.,
        1991, ApJ, 382, 315
\bibitem[\protect\citeauthoryear{Mikolajewski et al.}{1997}]{Mikolajewski97C}
        Mikolajewski M., Tomov T., Kolev D., 1997, IBVS, 4428
\bibitem[\protect\citeauthoryear{Miroshnichenko}{1988}]{Miroshnichenko88C} 
        Miroshnichenko A.S., 1988, Astron.\ Zh., 65, 582
\bibitem[\protect\citeauthoryear{Mizusawa e al.}{2010}]{Mizusawa10C} 
        Mizusawa T., Merritt J., Ballous R.-L. et al., 2010, PASP, 122, 299
\bibitem[\protect\citeauthoryear{Moffat \& Shara}{1984}]{Moffat84C}
        Moffat A.F.J., Shara M.M., 1984, PASP, 96, 552
\bibitem[\protect\citeauthoryear{Moffett \& Barnes}{1974}]{Moffett74C} 
        Moffett T., Barnes T.G., 1974, ApJ, 194, 141
\bibitem[\protect\citeauthoryear{Moreales-Rueda et al.}{2002}]
        {Morales-Rueda02C} 
        Morales-Rueda L., Still M.D., Roche R., Wood J.H., 
        Lockley J.J., 2002, MNRAS, 329, 597
\bibitem[\protect\citeauthoryear{Morgenroth}{1934}]{Morgenroth34C}
        Morgenroth O., 1934, Astron.\ Nachr., 253, 443 
\bibitem[\protect\citeauthoryear{Motch et al.}{1996}]{Motch96C} 
        Motch C., Haberl F., Guillot P. et al., 1996, A\&A, 307, 459
\bibitem[\protect\citeauthoryear{Mouchet}{1983}]{Mouchet83C}
        Mouchet M., 1983, in Livio M., Shaviv G., eds, Proc.\ IAU Coll. 72,
        Cataclysmic Variables and Related Objects,  p.\ 173  
\bibitem[\protect\citeauthoryear{Mouchet et al.}{1981}]{Mouchet81C} 
        Mouchet M., Bonnet-Bidaud J.-M., Ilovaisky S.A., Chevalier C.,
        1981, A\&A, 102, 31
\bibitem[\protect\citeauthoryear{Mouchet et al.}{2017}]{Mouchet17C} 
        Mouchet M., Bonnet-Bidaud J.-M., Van Box Som L. et al., 2017, 
        A\&A, 600, A53
\bibitem[\protect\citeauthoryear{Moyer et al.}{2003}]{Moyer03C} 
        Moyer E., Sion E.M., Szkody P. et al., 2003, AJ, 125, 288
\bibitem[\protect\citeauthoryear{Mukadam et al.}{2016}]{Mukadam16C}
        Mukadam A.S., Pyrzas S., Townsley D.M. et al., 2016. ApJ, 821, 14
\bibitem[\protect\citeauthoryear{Mukai \& Charles}{1987}]{Mukai87C}
        Mukai K., Charles P.A., 1987, MNRAS 226, 209
\bibitem[\protect\citeauthoryear{Mukai et al.}{2009}]{Mukai09C} 
        Mukai K., Zietsman E., Still M., 2009, ApJ, 707, 652
\bibitem[\protect\citeauthoryear{Mumford}{1964}]{Mumford64C} 
        Mumford G.S., 1964, PASP, 76, 57
\bibitem[\protect\citeauthoryear{Mumford}{1966}]{Mumford66C}
        Mumford G.S., 1966, ApJ, 146, 411
\bibitem[\protect\citeauthoryear{Mumford \& Krzeminski}{1969}]{Mumford69C}
        Mumford G.S., Krzeminski W., 1969, ApJS, 18, 429
\bibitem[\protect\citeauthoryear{Munari \& Dallaporte}{2014}]{Munari14C}
        Munari U., Dallaporte S., 2014, New Astr., 27, 25
\bibitem[\protect\citeauthoryear{Nadalin \& Sion}{2001}]{Nadalin01C} 
        Nadalin I., Sion E.M., 2001, PASP, 113, 829
\bibitem[\protect\citeauthoryear{Nassau \& Stephenson}{1963}]{Nassau63C} 
        Nassau J.J., Stephenson C.B., 1963, Hamburger Sternw., 
        Warner \& Swasey Obs.
\bibitem[\protect\citeauthoryear{Naylor}{2005}]{Naylor05C}
        Naylor T., Allen A., Long K., 2005, MNRAS, 361,1091
\bibitem[\protect\citeauthoryear{Neustroev et al.}{2011}]{Neustroev11C} 
        Neustroev V.V., Suleimanov V.F., Borisov N.V., Belyakov K.V., 
        Shearer A., 2011, MNRAS, 410, 963
\bibitem[\protect\citeauthoryear{Nogami et al.}{2009}]{Nogami09C}
        Nogami D., Hiroi K., Suzuki K. et al. 2009, in Soonthornthum B.,
        Komonjinda S., Cheng K.S. et al.\, eds, ASP Con.\ Ser. 404, 
        The 8th Pacific Rim Conference on Stellar Astrophysics, p.\ 52
\bibitem[\protect\citeauthoryear{Nogami \& Kato}{1995}]{Nogami95aC} 
        Nogami D., Kato T., 1995, IBVS, 4227
\bibitem[\protect\citeauthoryear{Nogami et al.}{1995}]{Nogami95bC} 
        Nogami D., Masuda S., Kato T., 1995, IBVS, 4218
\bibitem[\protect\citeauthoryear{Nogami et al.}{1998}]{Nogami98C} 
        Nogami D., Baba H., Kato T., Nov\'ak R., 1998, PASJ, 50, 297
\bibitem[\protect\citeauthoryear{North et al.}{2000}]{North00C} 
        North R.D., Marsh T.R., Moran C.K.J. et al., 2000, MNRAS, 313, 383
\bibitem[\protect\citeauthoryear{Norton \& Watson}{1989}]{Norton89C}
        Norton A.J., Watson M.G., 1989, MNRAS, 327, 853
\bibitem[\protect\citeauthoryear{O'Donoghue}{1987}]{ODonogue87C} 
        O'Donoghue D., 1987, ApSS, 136, 297
\bibitem[\protect\citeauthoryear{O'Donoghue et al.}{1987}]{ODonoghue87aC} 
        O'Donoghue D., Fairall A.P., Warner B., 1987, MNRAS, 225, 43
\bibitem[\protect\citeauthoryear{O'Donoghue et al.}{1991}]{ODonoghue91C} 
        O'Donoghue D., Chen A., Marang F. et al., 1991, MNRAS, 250, 363
\bibitem[\protect\citeauthoryear{O'Donoghue et al.}{1987}]{ODonoghue87C} 
        O'Donoghue D., Fairall A.P., Warner B., 1987, MNRAS, 225, 43
\bibitem[\protect\citeauthoryear{O'Donoghue \& Soltynski}{1992}]{ODonoghue92C} 
        O'Donoghue D., Soltynski M.G., 1992, MNRAS 254, 9
\bibitem[\protect\citeauthoryear{O'Donoghue et al.}{1989}]{ODonoghue89C} 
        O'Donoghue D., Warner B., Wargau W., Grauer A.D., 1989, MNRAS, 
        240, 41
\bibitem[\protect\citeauthoryear{Oliveira et al.}{2014}]{Oliveira14C} 
        Oliveira A.S., Lima H.J.F., Steiner J.E., Borges B.W., 
        Cieslinski D., 2014, MNRAS, 444, 2682
\bibitem[\protect\citeauthoryear{Oliveira \& Steiner}{2004}]{Oliveira04C} 
        Oliveira A.S., Steiner J.E., 2004, MNRAS, 351, 685
\bibitem[\protect\citeauthoryear{Oskanian}{1983}]{Oskanian83C}
        Oskanian A.V., 1983, IBVS 2349
\bibitem[\protect\citeauthoryear{Okazaki et al.}{1982}]{Okazaki82C} 
        Okazaki A., Kitamura M., Yamasaki A., 1982, PASP, 94, 162
\bibitem[\protect\citeauthoryear{Paczynski}{1965}]{Paczynski65C}
        Paczynski B., 1965, Acta Astr., 15, 197
\bibitem[\protect\citeauthoryear{Panek}{1980}]{Panek80C}
        Panek R.J., 1980, ApJ, 241, 1077
\bibitem[\protect\citeauthoryear{Panek \& Holm}{1984}]{Panek84C} 
        Panek R.J., Holm A.V., 1984, ApJ, 227, 700
\bibitem[\protect\citeauthoryear{Papadaki et al.}{2009}]{Papadaki09C} 
        Papadaki C., Boffin H.M.J., Stanishev V. et al., 2009, 
        J.\ Astron.\ Data 15, 1
\bibitem[\protect\citeauthoryear{Papadaki et al.}{2006}]{Papadaki06C} 
        Papadaki C., Boffin H.M.J., Sterken C. et al., 2006, A\&A, 456, 599
\bibitem[\protect\citeauthoryear{Patterson}{1979a}]{Patterson79aC} 
        Patterson J., 1979a, AJ, 84, 804
\bibitem[\protect\citeauthoryear{Patterson}{1979b}]{Patterson79bC} 
        Patterson J., 1979b, ApJ, 233, L13 
\bibitem[\protect\citeauthoryear{Patterson}{1980}]{Patterson80C} 
        Patterson J., 1980, ApJ, 241, 235
\bibitem[\protect\citeauthoryear{Patterson}{1981}]{Patterson81C}
        Patterson J., 1981, ApJ Suppl., 45, 517
\bibitem[\protect\citeauthoryear{Patterson \& Eisenmann}{1987}]{Patterson87C}
        Patterson J. Eisenman N., 1987, IBVS, 3079
\bibitem[\protect\citeauthoryear{Patterson et al.}{2002}]{Patterson02C}
        Patterson J., Fenton W.H., Thorstensen J.R. et al., 2002, 
        PASP, 114, 1364
\bibitem[\protect\citeauthoryear{Patterson et al.}{2005}]{Patterson05C}
        Patterson J., Kemp J., Harvey D.A. et al., 2005, PASP, 117, 1204 
\bibitem[\protect\citeauthoryear{Patterson et al.}{1981}]{Patterson81aC} 
        Patterson J., McGraw J.T., Coleman L., Africano J.L., 1981,
        ApJ 248, 1067
\bibitem[\protect\citeauthoryear{Patterson et al.}{1998}]{Patterson97C} 
        Patterson J., Kemp J., Shambrook A. et al., 1997, PASP, 109, 1100 
\bibitem[\protect\citeauthoryear{Patterson et al.}{1998}]{Patterson98C} 
        Patterson J., Kemp J., Shambrook A. et al., 1998, PASP, 110, 380
\bibitem[\protect\citeauthoryear{Patterson et al.}{2017}]{Patterson17C} 
        Patterson J., Oksanen A., Kemp J. et al., 2017, MNRAS, 466, 581
\bibitem[\protect\citeauthoryear{Patterson et al.}{1992}]{Patterson92C}
        Patterson J., Schwartz D.A., Pye J.P. et al., 1992, ApJ, 393, 233
\bibitem[\protect\citeauthoryear{Patterson \& Skillman}{1994}]{Patterson94C} 
        Patterson J., Skillman D.R., 1994, PASP, 106, 1141
\bibitem[\protect\citeauthoryear{Patterson et al.}{2018}]{Patterson18C} 
        Patterson J., Stone G., Kemp J. et al., 2018, PASP, 130, 4202
\bibitem[\protect\citeauthoryear{Patterson \& Szkody}{1993}]{Patterson93aC}
        Patterson J., Szkody P., 1993, PASP, 105, 1116 
\bibitem[\protect\citeauthoryear{Patterson et al.}{1993}]{Patterson93C}
        Patterson J., Thomas G., Skillman D.R., Diaz M., 1993, 
        ApJ Suppl., 83, 235
\bibitem[\protect\citeauthoryear{Patterson et al}{2001}]{Patterson01C}
        Patterson J., Thorstensen J.R., Fried R. et al., 2001, 
        PASP, 113, 72
\bibitem[\protect\citeauthoryear{Patterson et al.}{2003}]{Patterson03C} 
        Patterson J., Thorstensen J.R., Kemp J. et al., 2003, PASP, 115, 1308
\bibitem[\protect\citeauthoryear{Pavlenko \& Shugarov}{1998}]{Pavlenko98C} 
        Pavlenko E.P., Shugarov S.Yu., 1998, AA Trans., 15, 89
\bibitem[\protect\citeauthoryear{Pavlenko et al.}{2000}]{Pavlenko00C} 
        Pavlenko E.P., Shugarov S.Yu., Katysheva N.A., 2000, 
        Astrophys., 43, 419
\bibitem[\protect\citeauthoryear{Peters \& Thorstensen}{2006}]{Peters06C} 
        Peters C.S., Thorstensen J., 2006, PASP, 118, 687
\bibitem[\protect\citeauthoryear{Pezzuto et al.}{1996}]{Pezzuto96C} 
        Pezzuto S., Bianchini A., Stagni R., 1996, A\&A, 312, 865
\bibitem[\protect\citeauthoryear{Philip \& Stock}{1972}]{Philip72C} 
        Philip A.G.D., Stock J., 1972, Bull.\ Obs.\ Ton.\ Tac., 38, 201
\bibitem[\protect\citeauthoryear{Pickering}{1906}]{Pickering06C} 
        Pickering E.C., 1906, Harv.\ Circ., 120
\bibitem[\protect\citeauthoryear{Pickles \& Visvanathan}{1983}]{Pickles83C} 
        Pickles A.J., Visvanathan N., 1983, MNRAS, 204, 463
\bibitem[\protect\citeauthoryear{Pilyugin}{1985}]{Pilyugin85C}
        Pilyugin L.S., 1985, Astrofiz., 23, 277
\bibitem[\protect\citeauthoryear{Poole et al.}{2003}]{Poole03C} 
        Poole T., Mason K.O., Ramsay G., Drew J.H. Smith R.C., 2003, 
        MNRAS, 340, 499 
\bibitem[\protect\citeauthoryear{Popowa}{1961}]{Popowa61C}
        Popowa M., 1961, Astron.\ Nachr., 286, 81
\bibitem[\protect\citeauthoryear{Potter et al.}{2004}]{Potter04C} 
        Potter S.B., Romero-Colmerero E., Watson C.A., Buckley D.A.H., 
        Phillips A., 2004, MNRAS, 348, 316
\bibitem[\protect\citeauthoryear{Pretorius et al.}{2006}]{Pretorius06C} 
        Pretorius M.L., Warner B., Woudt P.A., 2006, MNRAS, 368, 361
\bibitem[\protect\citeauthoryear{Qian et al.}{2007}]{Qian07C} 
        Qian S.-B., Dai Z.-B., He J.-J. et al., 2007, A\&A, 466, 589 
\bibitem[\protect\citeauthoryear{Qian et al.}{2015}]{Qian15C}
        Qian S.-B., Han Z.T., Fern\'andez Laj\'us R. et al., 2015, 
        ApJ Suppl., 221, 17
\bibitem[\protect\citeauthoryear{Qian et al.}{2013}]{Qian13C} 
        Qian S.-B., Shi G., Fern\'andez Laj\'us R. et al., 2013, 
        ApJ, 771, L18 
\bibitem[\protect\citeauthoryear{Raikova \& Antov}{1986}]{Raikova86C}
        Raikova D., Antov A., 1986, IBVS, 2260
\bibitem[\protect\citeauthoryear{Ramsay}{2000}]{Ramsay00C} 
        Ramsay G., 2000, MNRAS, 314, 403
\bibitem[\protect\citeauthoryear{Rana et al.}{2004}]{Rana04C}
        Rana V.R., Singh K.P., Schlegel E.M., 2004, AJ, 127, 489
\bibitem[\protect\citeauthoryear{Ratering et al.}{1993}]{Ratering93C} 
        Ratering C., Bruch A., Diaz M.P., 1993, A\&A, 268, 694
\bibitem[\protect\citeauthoryear{Rayne \& Whelan}{1981}]{Rayne81C} 
        Rayne M.W., Whelan J.A.J., 1981, MNRAS, 196, 73
\bibitem[\protect\citeauthoryear{Reinmuth}{1925}]{Reinmuth25C} 
        Reinmuth K., 1925, Astron. Nachr., 225, 385
\bibitem[\protect\citeauthoryear{Reinsch \& Beuermann}{1990}]{Reinsch90C} 
        Reinsch K., Beuermann K., 1990, A\&A, 240, 360
\bibitem[\protect\citeauthoryear{Ribeiro \& Diaz}{2006}]{Ribeiro06C} 
        Ribeiro F.M.A., Diaz M.P., 2006, PASP, 118, 84
\bibitem[\protect\citeauthoryear{Ribeiro \& Diaz}{2007}]{Ribeiro07C} 
        Ribeiro F.M.A., Diaz M.P., 2007, AJ, 133, 2659
\bibitem[\protect\citeauthoryear{Ribeiro \& Diaz}{2007, 2009}]{Ribeiro09C} 
        Ribeiro F.M.A., Diaz M.P., 2009, PASJ, 61, 137
\bibitem[\protect\citeauthoryear{Ribeiro et al.}{2007}]{Ribeiro07aC}
        Ribeiro T., Baptista R., Harlaftis E.T., Dhillon V.S., \&
        Rutten R.G.M., 2007, A\&A, 474, 213
\bibitem[\protect\citeauthoryear{Ringwald}{1995}]{Ringwald95C}
        Ringwald F.A., 1995, MNRAS, 274, 127
\bibitem[\protect\citeauthoryear{Ringwald et al.}{1996}]{Ringwald96aC} 
        Ringwald F.A., Naylor T., Mukai K., 1996a, MNRAS, 281, 192
\bibitem[\protect\citeauthoryear{Ringwald et al.}{1994}]{Ringwald94C}
        Ringwald F.A., Thorstensen J.R., Hamvey R.M., 1994, 
        MNRAS, 271, 323
\bibitem[\protect\citeauthoryear{Ringwald et al.}{1996b}]{Ringwald96bC} 
        Ringwald F.A., Thorstensen J.R., Honeycutt R.K., Smith R.C.,
        1996b, AJ, 111, 2077
\bibitem[\protect\citeauthoryear{Ritter \& Kolb}{2003}]{Ritter03C} 
        Ritter H., Kolb U., 2003, A\&A, 404, 301
\bibitem[\protect\citeauthoryear{Robinson}{1975}]{Robinson75C}
        Robinson E.L., 1975, AJ, 80, 515
\bibitem[\protect\citeauthoryear{Robinson \& Nather}{1979}]{Robinson79C} 
        Robinson E.L., Nather R.E., 1979, ApJS, 39, 401
\bibitem[\protect\citeauthoryear{Robinson \& Nather}{1983}]{Robinson83C} 
        Robinson E.L., Nather R.E., 1983, ApJ, 273, 255
\bibitem[\protect\citeauthoryear{Robinson et al.}{1982}]{Robinson82C} 
        Robinson E.L., Nather R.E., Kepler S.O., 1982, ApJ, 254, 646
\bibitem[\protect\citeauthoryear{Robinson et al.}{1991}]{Robinson91C} 
        Robinson E.L., Shretone M.D., Africano J.L., 1991, AJ, 102, 1176
\bibitem[\protect\citeauthoryear{Robinson \& Warner}{1984}]{Robinson84C} 
        Robinson E.W., Warner B., 1984, ApJ, 277, 250
\bibitem[\protect\citeauthoryear{Robinson et al.}{1987}]{Robinson87C} 
        Robinson E.L., Shafter A.W., Hill J.A., Wood M.A., Mattei J.A.,
        1987, ApJ, 313, 772
\bibitem[\protect\citeauthoryear{Rodrigues et al.}{1998}]{Rodrigues98C} 
        Rodrigues C.V., Cieslinski D., Steiner J.E., 1998, A\&A, 335, 939
\bibitem[\protect\citeauthoryear{Rodr\'{\i}guez-Gil \& Mart\'{\i}nez-Pais}
        {2002}]{Rodriguez-Gil02C} 
        Rodr\'{\i}guez-Gil P., Mart\'{\i}nez-Pais I.G., 2002, 
        MNRAS, 337, 209
\bibitem[\protect\citeauthoryear{Rodr\'{\i}guez-Gil et al.}{2001}]
        {Rodriguez-Gil01C} 
        Rodr\'{\i}guez-Gil P., Mart\'{\i}nez-Pais I.G., Casares J., 
        Villada M., van Zyl L., 2001, MNRAS, 238, 903
\bibitem[\protect\citeauthoryear{Rodr\'{\i}guez-Gil et al.}{2007}]
        {Rodriguez-Gil07C} 
        Rodr\'{\i}guez-Gil P., Schmidtobreick L., G\"ansicke B.T., 2007,
        MNRAS, 374, 1359
\bibitem[\protect\citeauthoryear{Rodr\'{\i}guez-Gil et al.}{2020}]
        {Rodriguez-Gil20C}
        Rodr\'{\i}guez-Gil P., Shahbaz T., Torres M.P.A. et al., 2020, 
        MNRAS, 494, 425
\bibitem[\protect\citeauthoryear{Rodr\'{\i}guez-Gil \& Torres}{2005}]
        {Rodriguez-Gil05C}
        Rodr\'{\i}guez-Gil P., Torres M.P.A., 2005, A\&A, 431, 289
\bibitem[\protect\citeauthoryear{Rogoziecki \& Schwarzenberg-Czerny}{2001}]
        {Rogoziecki01C} 
        Rogoziecki P. Schwarzenberg-Cerny A., 2001, MNRAS, 323, 850
\bibitem[\protect\citeauthoryear{Rogoziecki \& Schwarzenberg-Czerny}{2003}]
        {Rogoziecki03C} 
        Rogoziecki P. Schwarzenberg-Cerny A., 2003, A\&A, 397, 961 
\bibitem[\protect\citeauthoryear{Rolfe et al.}{2000}]{Rolfe00C} 
        Rolfe D.J., Haswell C.A., Patterson J., 2000, MNRAS, 317, 759
\bibitem[\protect\citeauthoryear{Romano}{1958}]{Romano58C}
        Romano G., 1958, Mem. S.A.I., 29, 2 
\bibitem[\protect\citeauthoryear{Rosen et al.}{1994}]{Rosen94C} 
        Rosen S.R., Clayton K.L., Osborne J.P., McGale P.A., 1994,
        MNRAS, 269, 913
\bibitem[\protect\citeauthoryear{Rosen et al.}{1987}]{Rosen87C} 
        Rosen S.R., Mason K.O., C\'ordova F.A., 1987, MNRAS, 224, 987
\bibitem[\protect\citeauthoryear{Rosen et al.}{1995}]{Rosen95C} 
        Rosen S.R., Watson K.T., Robinson E.L. et al., 1995, A\&A, 300, 392
\bibitem[\protect\citeauthoryear{Rosenzweig et al.}{2000}]{Rosenzweig00C} 
        Rosenzweig P., Mattei J.A., Kafka S., Turner G.W., 
        Honeycutt R.K., 2000, PASP, 112, 632
\bibitem[\protect\citeauthoryear{Rosino et al.}{1993}]{Rosino93C} 
        Rosino L., Romano G., Marziani P., 1993, PASP, 105, 51,
\bibitem[\protect\citeauthoryear{Rude \& Ringwald}{2012}]{Rude12C}
        Rude G.D., Ringwald F.A., 2012, New Astr., 17, 453
\bibitem[\protect\citeauthoryear{Rutten et al.}{1992}]{Rutten92C} 
        Rutten R.G.M., van Paradijs J., Tinbergen J., 1992, A\&A, 260, 213
\bibitem[\protect\citeauthoryear{Sahman et al.}{2018}]{Sahman18C}
        Sahman D.I., Dhillon V.S., Littlefair S.P., Hanninan G., 2018,
        MNRAS, 477, 4483
\bibitem[\protect\citeauthoryear{Saito \& Baptista}{2016}]{Saito16C} 
        Saito R., Baptista R., 2016, MNRAS, 457, 198
\bibitem[\protect\citeauthoryear{Sambruna et al.}{1991}]{Sambruna91C} 
        Sambruna R.M., Chiapetti L., Treves A. et al., 1991, ApJ, 374, 744
\bibitem[\protect\citeauthoryear{Sanduleak}{1972}]{Sanduleak72C}
        Sanduleak N., 1972, IBVS, 663
\bibitem[\protect\citeauthoryear{Sanford}{1949}]{Sanford49C}
        Sanford R.F., 1949, ApJ, 109, 81
\bibitem[\protect\citeauthoryear{Satyvoldiev}{1972}]{Satyvoldiev72C} 
        Satyvoldiev V., 1972, Astron. Tsirk., 711, 1
\bibitem[\protect\citeauthoryear{Scaringi}{2014}]{Scaringi14C} 
        Scaringi S., 2014, MNRAS, 438, 1233
\bibitem[\protect\citeauthoryear{Scaringi et al.}{2012a}]{Scaringi12aC} 
        Scaringi S., K\"ording E., Uttley P. et al., 2012a, MNRAS, 427, 2854
\bibitem[\protect\citeauthoryear{Scaringi et al.}{2012b}]{Scaringi12bC} 
        Scaringi S., K\"ording E., Uttley P. et al., 2012b, MNRAS, 427, 3396
\bibitem[\protect\citeauthoryear{Schaefer}{2010}]{Schaefer10C}
        Schaefer B.E., 2010, ApJS, 187, 275
\bibitem[\protect\citeauthoryear{Schaefer}{2018}]{Schaefer18C}
        Schaefer B.E., 2018, MNRAS, 481, 3033
\bibitem[\protect\citeauthoryear{Schaefer et al.}{1992}]{Schaefer92C} 
        Schaefer B.E., Landolt A.U., Vogt N. et al., 1992, ApJS, 81, 321
\bibitem[\protect\citeauthoryear{Schaefer et al.}{2010}]{Schaefer10aC} 
        Schaefer B.E., Pagnotta A., Shara M.M., 2010, ApJ, 708, 381
\bibitem[\protect\citeauthoryear{Schild}{1969}]{Schild69C} 
        Schild R.E., 1969, ApJ, 157, 709
\bibitem[\protect\citeauthoryear{Schlegel et al.}{1986}]{Schlegel86C}
        Schlegel E.M., Honeycutt R.K., Kaitchuck R.H., 1986, ApJ, 307, 760
\bibitem[\protect\citeauthoryear{Schlindwein}{2017}]{Schlindwein17C} 
        Schlindwein W. 2017, Mapeando o disco de acr\'ecimo em vari\'aveis
        catacl\'{\i}smicas: SDSS J0926+3624 and OY~Carinae, Master thesis,
        Fed.\ Univ.\ Santa Catarina, Florian\'opolis
\bibitem[\protect\citeauthoryear{Schmeer}{1992}]{Schmeer92C} 
        Schmeer P., 1992, IAU Circ., No.\ 5502
\bibitem[\protect\citeauthoryear{Schmidke et al.}{2002}]{Schmidke02C}
        Schmidke P.C., Ciudin G.A., Indlekofer U.R. et al., 2002, in 
        G\"ansicke B., Beuermann K., eds, ASP Conf.\ Ser., 261, Physics of 
        Cataclysmic Variables and Related Objects,  p.\ 539
\bibitem[\protect\citeauthoryear{Schmidtobreick et al.}{2018}]
        {Schmidtobreick18C} 
        Schmidtobreick L., Mason E., Howell S.B. et al., 2018, 
        A\&A, 617, A16
\bibitem[\protect\citeauthoryear{Schmidtobreick et al.}{2008}]
        {Schmidtobreick08C} 
        Schmidtobreick L., Papadaki C., Tappert C., Ederoclite A., 2008,
        MNRAS, 389, 1345
\bibitem[\protect\citeauthoryear{Schmidtobreick et al.}{2005}]
        {Schmidtobreick05C} 
        Schmidtobreick L., Tappert C., Galli L., Whiting A., 2005, 
        IBVS, 5627
\bibitem[\protect\citeauthoryear{Schmidtobreick et al.}{2003}]
        {Schmidtobreick03C} 
        Schmidtobreick L., Tappert C., Saviane I., 2003, MNRAS, 342, 142
\bibitem[\protect\citeauthoryear{Schneider \& Young}{1980}]{Schneider80C} 
        Schneider D.P., Young P., 1980, ApJ, 240, 821
\bibitem[\protect\citeauthoryear{Schneider et al.}{1981}]{Schneider81C} 
        Schneider D.P., Young P., Shectman S.A., 1981, ApJ, 245, 644
\bibitem[\protect\citeauthoryear{Schneller}{1931}]{Schneller31C} 
        Schneller H., 1931, Astron.\ Nachr., 243, 335
\bibitem[\protect\citeauthoryear{Schoembs}{1982}]{Schoembs82C}
        Schoembs R., 1982, A\&A, 115, 190
\bibitem[\protect\citeauthoryear{Schoembs et al.}{1987}]{Schoembs87C} 
        Schoembs R., Dreier H., Barwig H., 1987, A\&A, 181, 50
\bibitem[\protect\citeauthoryear{Schoembs \& Stolz}{1981}]{Schoembs81aC} 
        Schoembs R., Stolz B., 1981, IBVS, 1986
\bibitem[\protect\citeauthoryear{Schoembs \& Vogt}{1981}]{Schoembs81bC} 
        Schoembs R., Vogt N., 1981, A\&A, 97, 185 
\bibitem[\protect\citeauthoryear{Schrijver et al.}{1987}]{Schrijver87C}
        Schrijver J., Brinkman A.C., van der Woerd H., 1987, 
        ApSS, 130, 261 
\bibitem[\protect\citeauthoryear{Schrijver et al.}{1985}]{Schrijver85C}
        Schrijver J., Brinkman A.C., van der Woerd H. et al., 1985, 
        Sp.\ Sc.\ Rev., 40, 121
\bibitem[\protect\citeauthoryear{Schwarzenberg-Czerny et al.}{1992}]
        {Schwarzenberg-Czerny92C} 
        Schwarzenberg-Czerny A., Udalski A., Monier R., 1992, 
        ApJ, 401, L19
\bibitem[\protect\citeauthoryear{Schwope et al.}{1993}]{Schwope93C} 
        Schwope A.D., Thomas H.-C., Beuermann K., Reinsch K., 1993, 
        A\&A, 267, 103
\bibitem[\protect\citeauthoryear{Selvelli}{2004}]{Selvelli04C} 
        Selvelli P., 2004, Baltic Astr., 13, 93
\bibitem[\protect\citeauthoryear{Selvelli \& Gilmozzi}{2013}]{Selvelli13C} 
        Selvelli P., Gilmozzi R., 2013, A\&A, 560, A49
\bibitem[\protect\citeauthoryear{Selvelli et al.}{1995}]{Selvelli95C} 
        Selvelli P.L., Gilmozzi R., Cassatella A., 1995, in 
        Bianchini A., Della Valle M., Orio M., eds, Cataclysmic 
        Variables, Kluwer Academic Publishers, Dordrecht, p.\ 182
\bibitem[\protect\citeauthoryear{Semena et al.}{2014}]{Semena14C} 
        Semena A.N., Revnivtsev M.G., Buckley D.A.H. et al., 2014, 
        MNRAS 442, 1123.
\bibitem[\protect\citeauthoryear{Semeniuk}{1980}]{Semeniuk80C} 
        Semeniuk I., 1980, A\&AS, 39, 29
\bibitem[\protect\citeauthoryear{Semeniuk et al.}{1997}]{Semeniuk97C} 
        Semeniuk I., Olech A., Kwast T., Nale\.zyty M., 1997,
        Acta Astron., 47, 201
\bibitem[\protect\citeauthoryear{Shafter}{1983a}]{Shafter83aC} 
        Shafter A.W., 1983a, PhD thesis, UCLA
\bibitem[\protect\citeauthoryear{Shafter}{1983b}]{Shafter83bC} 
        Shafter A.W., 1983b, ApJ, 267, 222
\bibitem[\protect\citeauthoryear{Shafter}{1983d, 1985}]{Shafter83dC}
        Shafter A.W., 1983d, IBVS, 2377
\bibitem[\protect\citeauthoryear{Shafter}{1985}]{Shafter85C}
        Shafter A.W., 1985, AJ, 90, 643
\bibitem[\protect\citeauthoryear{Shafter \& Harkness}{1986}]{Shafter86aC} 
        Shafter A.W., Harkness R.P., 1986, AJ, 93, 659
\bibitem[\protect\citeauthoryear{Shafter \& Hessman}{}]{Shafter88C} 
        Shafter A.W., Hessmann F.V., 1988, AJ, 95, 178
\bibitem[\protect\citeauthoryear{Shafter et al.}{1990}]{Shafter90C} 
        Shafter A.W., Robinson E.L., Crampton D., Warner B.,
        Prestage R.M., 1990, ApJ, 354, 719
\bibitem[\protect\citeauthoryear{Shafter et al.}{1986}]{Shafter86bC} 
        Shafter A.W., Szkody P., Thorstensen J.R., 1986, ApJ, 308 765
\bibitem[\protect\citeauthoryear{Shafter et al.}{1995}]{Shafter95C}
        Shafter A.W., Veal J.M., Robinson E.L., 1995, ApJ, 440, 853
\bibitem[\protect\citeauthoryear{Shara et al.}{2018}]{Shara18C} 
        Shara M.M., Prialnik D., Hillman Y., Kovetz A., 2018, 
        ApJ, 860, 110
\bibitem[\protect\citeauthoryear{Shears \& Poyner}{2009}]{Shears09C} 
        Shears J., Poyner G., 2009, JBAA, 120, 169
\bibitem[\protect\citeauthoryear{Sheets et al.}{2007}]{Sheets07C} 
        Sheets H.A., Thorstensen J.R., Peters C.J., Kapusta A.B., 2007
        PASP, 119, 494
\bibitem[\protect\citeauthoryear{Sherrington et al.}{1984}]{Sherrington84C} 
        Sherrington M.R., Bailey J., Jameson R.F., 1984, MNRAS, 206, 859
\bibitem[\protect\citeauthoryear{Shore et al.}{2013}]{Shore13C} 
        Shore S.N. Schwarz G.J., de Gennaro Aquino I. et al., 2013,
        A\&A, 549, 140
\bibitem[\protect\citeauthoryear{Shugarov}{1983}]{Shugarov83C}
        Shugarov S.Yu., 1983, Pis'ma Astron.\ Zh., 9, 31
\bibitem[\protect\citeauthoryear{Shugarov et al.}{2005}]{Shugarov05C}
        Shugarov S.Yu., Katysheva N.A., Seregina T.M., VolkovI.M.,
        2005, in Hameury J.-M., Lasota J.-P., eds, ASP Conf.\ Series, 330,
        The Astrophysics of Cataclysmic Variables and Related Objects, p.\ 495
\bibitem[\protect\citeauthoryear{\v{S}imon}{2000}]{Simon00C}
        \v{S}imon V., 2000, A\&A, 364, 694
\bibitem[\protect\citeauthoryear{\v{S}imon et al.}{2004}]{Simon04C}
        \v{S}imon V., Hudec R., Hroch F., 2004, IBVS, 5562
\bibitem[\protect\citeauthoryear{\v{S}imon}{2012}]{Simon12C} 
        \v{S}imon V., Pol\'a\v{s}ek C., \v{S}trobl J., Hudec R., 
        Bla\v{z}ek M., 2012, A\&A, 540, A15
\bibitem[\protect\citeauthoryear{Simonsen}{2011}]{Simonsen11aC} 
        Simonsen M., 2011, JAAVSO, 39, 1
\bibitem[\protect\citeauthoryear{Simonsen et al.}{2014a}]{Simonsen14aC}
        Simonsen M., Bohlsen T., Hambsch F.-J., Stubbings R., 2014a
        JAAVSO, 42
\bibitem[\protect\citeauthoryear{Simonsen et al.}{2014b}]{Simonsen14bC}
        Simonsen M., Boyd D., Goff W. et al., 2014b, JAAVSO, 42, 177
\bibitem[\protect\citeauthoryear{Simonsen \& Stubbings}{2011}]{Simonsen11C} 
        Simonsen M., Stubbings R., 2011, JAAVSO, 39, 73\
\bibitem[\protect\citeauthoryear{Singh et al.}{1993}]{Singh93C}
        Singh J., Rao P.V., Agrawal P.C. et al., 1993, ApJ, 419, 337
\bibitem[\protect\citeauthoryear{Sion \& Guinan}{1982}]{Sion82C} 
        Sion E.M., Guinan E.F., 1982, in Kondo Y., Mead J.M., 
        Chapman R.D., eds, Advances in ultraviolet astronomy: Four years of 
        IUE Research, NASA CP-2238., p.\ 460,
\bibitem[\protect\citeauthoryear{Sirk \& Howell}{1998}]{Sirk98C} 
        Sirk M., Howell S.B., 1998, ApJ, 506, 824
\bibitem[\protect\citeauthoryear{Skidmore et al.}{2000, 2002}]{Skidmore00C} 
        Skidmore W., Mason E., Howell S.B. et al., 2000, MNRAS, 318, 429
\bibitem[\protect\citeauthoryear{Skidmore et al}{2002}]{Skidmore02C} 
        Skidmore W., Wynn G.A., Leach R., Jameson R.F., 2002,
        MNRAS, 336, 1223
\bibitem[\protect\citeauthoryear{Skillman et al.}{1995}]{Skillman95C} 
        Skillman D.R., Patterson J., Thorstensen J.R., 1995, PASP, 107, 545
\bibitem[\protect\citeauthoryear{Smak}{1993}]{Smak93C} 
        Smak J., 1993, Acta Astron., 43, 212
\bibitem[\protect\citeauthoryear{Smak}{1994}]{Smak94C} 
        Smak J., 1994, Acta Astron., 44, 59 
\bibitem[\protect\citeauthoryear{Smak}{1995, 2019}]{Smak95C} 
        Smak J., 1995, Acta Astr. 45, 259
\bibitem[\protect\citeauthoryear{Smak}{2002}]{Smak02C}
        Smak J., 2002, Acta Astr., 52, 189
\bibitem[\protect\citeauthoryear{Smak}{2019}]{Smak19C}
        Smak J., 2019, Acta Astr., 69, 79
\bibitem[\protect\citeauthoryear{Smith et al.}{2006}]{Smith06C} 
        Smith A.J., Haswell C.A., Hynes R.I., 2006, MNRAS, 369, 1546
\bibitem[\protect\citeauthoryear{Smith et al.}{1998}]{Smith98C} 
        Smith D.A., Dhillon V.S., Marsh T.R., 1998, MNRAS, 296, 465
\bibitem[\protect\citeauthoryear{Smith et al.}{1993}]{Smith93C}
        Smith R.C., Fiddik R.J., Nawkins N.A., Catal\'an M.S., 1993,
        MNRAS, 264, 619
\bibitem[\protect\citeauthoryear{Smith et al.}{2005}]{Smith05C}
        Smith R.C., Mehes O., Vande Putte D., Hawkins N.A., 2005,
        MNRAS, 360, 374
\bibitem[\protect\citeauthoryear{Snijders}{1987}]{Snijders87C} 
        Snijders M.A.J., 1987, in Bode M.F., ed., RS~Ophiuchi (1985) 
        and the recurrent nova phenomenon, VNU Science Press, p.\ 51
\bibitem[\protect\citeauthoryear{Soejima et al.}{2009}]{Soejima09C} 
        Soejima Y., Nogami D., Kato T. et al., 2009, PASJ, 61, 659
\bibitem[\protect\citeauthoryear{Sokolov et al.}{1996}]{Sokolov96C}
        Sokolov D.A., Shugarov S.Yu., Pavlenko E.P., 1996, in: 
        Evans A., Wood J.H., eds, Cataclysmic Variables and Related 
        Objects, Kluwer Academic Publishers, p.\ 219
\bibitem[\protect\citeauthoryear{Solf}{1983}]{Solf83C}
        Solf J., 1983, ApJ, 273, 647
\bibitem[\protect\citeauthoryear{Somero et al.}{2013}]{Somero13C}
        Somero A., Hakala P., Wynn G.A., 2013, MNRAS, 2784
\bibitem[\protect\citeauthoryear{Southwell et al.}{1995}]{Southwell95C} 
        Southwell K.A., Still M.D., Smith R.C., Martin J.S., 1995,
        A\&A, 302, 90
\bibitem[\protect\citeauthoryear{Spogli \& Claudi}{1994}]{Spogli94C} 
        Spolgi C., Claudi R., 1994, A\&A, 281, 808 
\bibitem[\protect\citeauthoryear{Spruit \& Rutten}{1998}]{Spruit98C} 
        Spruit H., Rutten R.G.M., 1998, MNRAS, 299, 768
\bibitem[\protect\citeauthoryear{Stanishev et al.}{2004}]{Stanishev04C}   
        Stanishev V., Zamanov R.K., Tomov N., Marziani T.G., 2004, 
        A\&A, 415, 609
\bibitem[\protect\citeauthoryear{Stark et al.}{2008}]{Stark08C} 
        Stark M.A., Wade R.A., Thorstensen J.R. et al., 2008, AJ, 135, 991
\bibitem[\protect\citeauthoryear{Starrfield et al.}{1985}]{Starrfield85C} 
        Starrfield S., Sparks W.M., Truran J.W., 1985, 291, 136
\bibitem[\protect\citeauthoryear{Steavenson}{1923}]{Steavenson23C}
        Steavenson W.H., 1923, MNRAS, 83, 160
\bibitem[\protect\citeauthoryear{Steeghs et al.}{2007}]{Steeghs07C} 
        Steeghs D., Howell S.B., Knigge C. et al., 2007, ApJ, 667, 442
\bibitem[\protect\citeauthoryear{Steiner \& Diaz}{1998}]{Steiner98C} 
        Steiner J.E., Diaz M.P., 1998, PASP, 110, 276
\bibitem[\protect\citeauthoryear{Stepanyan}{1979}]{Stepanyan79C} 
        Stepanyan J.A., 1979, IVBS, 1630
\bibitem[\protect\citeauthoryear{Sterken et al.}{2007}]{Sterken07C} 
        Sterken C., Vogt N., Schreiber M.R., Uemura M., Tuvikene T.,
        2007, A\&A, 463, 1053
\bibitem[\protect\citeauthoryear{Stickland et al.}{1984}]{Stickland84C}
        Stickland D.J., Kelly B.D., Cooke J.A., 1984, MNRAS 206, 819
\bibitem[\protect\citeauthoryear{Still et al.}{1995}]{Still95C} 
        Still M.D., Dhillon V.S., Jones D.H.P., 1995, MNRAS, 273, 849
\bibitem[\protect\citeauthoryear{Stobie et al.}{1995}]{Stobie95C} 
        Stobie R.S., Kilkenny D., O'Donoghue D., 1995, ApSS, 230, 101
\bibitem[\protect\citeauthoryear{Stobie et al.}{1997}]{Stobie97C} 
        Stobie R.S., Kilkenny D., O'Donoghue D. et al., 1997, 
        MNRAS, 287, 848
\bibitem[\protect\citeauthoryear{Stockman Sargent}{1979}]{Stockman79C}
        Stockman H.S., Sargent T.A., 1979, ApJ, 227, 197
\bibitem[\protect\citeauthoryear{Stover}{1981}]{Stover81C}
        Stover R.E., 1981, ApJ 248, 684 
\bibitem[\protect\citeauthoryear{Strope et al.}{2010}]{Strope10C} 
        Strope R.J., Schaefer B.E., Henden A.A., 2010, AJ, 140, 134
\bibitem[\protect\citeauthoryear{Svolopoulos}{1966}]{Svolopoulos66C} 
        Svolopoulos S.N., 1966, PASP, 78, 157
\bibitem[\protect\citeauthoryear{Szkody}{1981}]{Szkody81aC} 
        Szkody P., 1981, ApJ, 247, 577
\bibitem[\protect\citeauthoryear{Szkody}{1985}]{Szkody85C} 
        Szkody P., 1985, AJ, 90, 1837
\bibitem[\protect\citeauthoryear{Szkody et al.}{2002}]{Szkody02C} 
        Szkody P., G\"ansicke B.T., Sion E.M., Howell S.B., 2002, 
        ApJ, 574, 950
\bibitem[\protect\citeauthoryear{Szkody et al.}{1980}]{Szkody80aC}
        Szkody P., C\'ordova F.A., Tuohy I.R. et al., 1980, ApJ, 241,1070
\bibitem[\protect\citeauthoryear{Szkody Crosa}{1981}]{Szkody81bC} 
        Szkody P., Crosa L., 1981, ApJ, 251, 620
\bibitem[\protect\citeauthoryear{Szkody et al.}{2000}]{Szkody00C} 
        Szkody P., Desai V., Hoard D.W., 2000, AJ, 119, 365
\bibitem[\protect\citeauthoryear{Szkody et al.}{1984, 1989}]{Szkody89C}
        Szkody P., Howell S.B., Mateo M., Kreidl T.J., 1989, 
        PASP, 101, 899
\bibitem[\protect\citeauthoryear{Szkody \& Margon}{1980}]{Szkody80C}
        Szkody P., Margon B., 1980, ApJ, 236, 862
\bibitem[\protect\citeauthoryear{Szkody \& Mateo}{1984}]{Szkody84aC}
        Szkody P., Mateo M., 1984, ApJ, 280, 729
\bibitem[\protect\citeauthoryear{Szkody \& Mateo}{1988}]{Szkody88C} 
        Szkody P., Mateo M., 1988, PASP, 100, 1111
\bibitem[\protect\citeauthoryear{Szkody \& Mattei}{1984}]{Szkody84bC}
        Szkody P., Mattei J.A., 1984, PASP, 96, 988 
\bibitem[\protect\citeauthoryear{Szkody et al.}{2016}]{Szkody16C} 
        Szkody P., Mukadam A.S., G\"ansicke B.T. et al., 2016, AJ, 152, 48
\bibitem[\protect\citeauthoryear{Szkody \& Pich\'e}{1990}]{Szkody90C}
        Szkody P., Pich\'e F., 1990, ApJ, 361, 235
\bibitem[\protect\citeauthoryear{Szkody \& Shafter}{1983}]{Szkody83C} 
        Szkody P., Shafter A.W., 1983, PASP, 95, 509
\bibitem[\protect\citeauthoryear{Szkody et al.}{1984}]{Szkody84C} 
        Szkody P., Shafter A.W., Cowley A.P., 1984, ApJ, 282, 236 
\bibitem[\protect\citeauthoryear{Szkody \& Wade}{1980}]{Szkody80bC}
        Szkody P., Wade R.A., 1980, PASP, 92,806
\bibitem[\protect\citeauthoryear{Tapia}{1977a}]{Tapia77aC}
        Tapia S., 1977a, ApJ, 212, L125
\bibitem[\protect\citeauthoryear{Tapia}{1977b}]{Tapia77bC} 
        Tapia S., 1977b, IAU Circ., 3954
\bibitem[\protect\citeauthoryear{Tapia}{1982}]{Tapia82C} 
        Tapia S., 1982, IAU Circ., 3685
\bibitem[\protect\citeauthoryear{Tappert et al.}{1998}]{Tappert98C}
        Tappert C., Hanuschik R.W., Wargau W.F., 1998, 
        AG Astr.\ Series, 14, 126
\bibitem[\protect\citeauthoryear{Tappert et al.}{1997}]{Tappert97C} 
        Tappert C., Wargau W.F., Hanuschik R.W., Vogt N., 1997, 
        A\&A, 327, 231
\bibitem[\protect\citeauthoryear{Taylor \& Thorstensen}{1996}]{Taylor96C} 
        Taylor C.J., Thorstensen J.R., 1996, PASP, 108, 894
\bibitem[\protect\citeauthoryear{Thoroughgood et al.}{2004}]{Thoroughgood04C} 
        Thoroughgood T.D., Dhillon V.S., Watson C.A. et al., 2004, 
        MNRAS, 253, 1135
\bibitem[\protect\citeauthoryear{Thorstensen}{1996}]{Thorstensen96aC} 
        Thorstensen J.R., 1996, AJ, 91, 940
\bibitem[\protect\citeauthoryear{Thorstensen \& Freed}{1985}]{Thorstensen85C} 
        Thorstensen J.R., Freed I.W., 1985, AJ, 90, 2082
\bibitem[\protect\citeauthoryear{Thorstensen \& Taylor}{1997}]{Thorstensen97C} 
        Thorstensen J.R., Taylor C.J., 1997, PASP, 109, 1359
\bibitem[\protect\citeauthoryear{Thorstensen \& Taylor}{2000}]{Thorstensen00C} 
        Thorstensen J.R., Taylor C.J., 2000, MNRAS, 312, 629
\bibitem[\protect\citeauthoryear{Thorstensen et al.}{2002}]{Thorstensen02C} 
        Thorstensen J.R., Patterson J., Kemp J., Vennes S., 2002, 
        PASP, 114, 1108
\bibitem[\protect\citeauthoryear{Thorstensen et al.}{1996}]{Thorstensen96C} 
        Thorstensen J.R., Patterson J.O., Shambrook A., Thomas G., 1996,
        PASP, 108, 73. 
\bibitem[\protect\citeauthoryear{Thorstensen \& Ringwald}{1997}]
        {Thorstensen97aC}
        Thorstensen J.R., Ringwald F.A., 1997, PASP 109, 483
\bibitem[\protect\citeauthoryear{Thorstensen et al.}{1986}]{Thorstensen86C} 
        Thorstensen J.R., Wade R.A., Oke J.B., 1986, ApJ, 309, 721
\bibitem[\protect\citeauthoryear{Toloza et al.}{2016}]{Toloza16C} 
        Toloza O., G\"ansicke B.T., Hermes J.J. et al., 2016, 
        MNRAS, 459, 1393
\bibitem[\protect\citeauthoryear{Townsley et al.}{2004}]{Townsley04C} 
        Townsley D.M., Arras P., Bildsten H., 2004, ApJ, 608, L105
\bibitem[\protect\citeauthoryear{Tsessevich}{1969}]{Tsessevich69C} 
        Tsessevich V.P., 1969, Astron. Tsirk., 529, 7
\bibitem[\protect\citeauthoryear{Tuohy et al.}{1990}]{Tuohy90C} 
        Tuohy I.R., Remillard R.A., Brissenden R.J.V., Bradt H.V., 1990, 
        ApJ, 359, 204
\bibitem[\protect\citeauthoryear{Tuohy et al.}{1985}]{Tuohy85C} 
        Tuohy I.R., Visvanathan N., Wickramasinghe D.T., 1985, 
        ApJ, 289, 721
\bibitem[\protect\citeauthoryear{Turner}{1921}]{Turner21C}
        Turner H.H., 1921, MNRAS, 81, 426
\bibitem[\protect\citeauthoryear{Udalski}{1990}]{Udalski90C} 
        Udalski A., 1990, AJ, 100, 226
\bibitem[\protect\citeauthoryear{Udalski \& Schwarzenberg-Czerny}{1989}]
        {Udalski89C} 
        Udalski A., Schwarzenberg-Czerny A., 1989, Acta Astron.,39, 125
\bibitem[\protect\citeauthoryear{Udalksi \& Szymanski}{1988}]{Udalski88C} 
        Udalski A., Szymanski M., 1988, Acta Astron., 38, 215
\bibitem[\protect\citeauthoryear{Uemura et al.}{2001}]{Uemura01C} 
        Uemura M., Kato T., Pavlenko E., Baklanov A., Pietz J., 2001
        PASJ, 53, 539
\bibitem[\protect\citeauthoryear{Unda-Sanzana et al.}{2006}]{Unda-Sanzana06C}
        Unda-Sanzana E., Marsh T.R., Morales-Rueda L., 2006, 
        MNRAS, 269, 805
\bibitem[\protect\citeauthoryear{Urban \& Sion}{2006}]{Urban06C} 
        Urban J.A., Sion E.M., 2006, ApJ, 642, 1029
\bibitem[\protect\citeauthoryear{Uthas et al.}{2010}]{Uthas10C} 
        Uthas H., Knigge C., Steeghs D., 2010, MNRAS, 409, 237
\bibitem[\protect\citeauthoryear{van Paradijs et al.}{1994}]{vanParadijs94C} 
        van Paradijs J., Charles P.A., Harlaftis E.T. et al., 1994, 
        MNRAS, 267, 465
\bibitem[\protect\citeauthoryear{van Spaandonk et al.}{2010}]{vanSpaandonk10C} 
        van Spaandonk L., Steeghs D., Marsh T.R., Parsons S.G., 2010, 
        ApJ 715, L109
\bibitem[\protect\citeauthoryear{van Zyl et al.}{2000}]{vanZyl00C} 
        van Zyl L., Warner B. O'Donoghue D. et al., 2000, 
        Baltic Astr., 9, 231
\bibitem[\protect\citeauthoryear{van Zyl et al.}{2004}]{vanZyl04C} 
        van Zyl L., Warner B. O'Donoghue D. et al., 2004, MNRAS, 350, 307
\bibitem[\protect\citeauthoryear{Vande Putte et al.}{2003}]{VandePutte03C} 
        Vande Putte D., Smith R.D., Hawkings N.A., Martin J.S., 2003,
        MNRAS, 342, 151
\bibitem[\protect\citeauthoryear{Vennes et al.}{1995}]{Vennes95C} 
        Vennes S., Szkody P., Sion E.M., Long K.S., 1995, ApJ, 445, 921
\bibitem[\protect\citeauthoryear{Verbunt}{1987}]{Verbunt87C} 
        Verbunt F., 1987, A\&AS, 71, 339
\bibitem[\protect\citeauthoryear{Verbunt et al.}{1984}]{Verbunt84C} 
        Verbunt F., Pringle J.E., Wade R.A. et al., 1984, MNRAS, 210, 197
\bibitem[\protect\citeauthoryear{Vican et al.}{2011}]{Vican11C} 
        Vican L., Patterson J., Allen W. et al., 2011, PASP, 123, 1156
\bibitem[\protect\citeauthoryear{Visvanathan \& Tuohy}{1983}]{Visvanathan83C} 
        Visvanathan N., Tuohy I., 1983, ApJ, 275, 709
\bibitem[\protect\citeauthoryear{Vogt}{1974}]{Vogt74C} 
        Vogt N., 1974, A\&A, 36, 369
\bibitem[\protect\citeauthoryear{Vogt}{1975}]{Vogt75C} 
        Vogt N., 1975, A\&A, 41, 15
\bibitem[\protect\citeauthoryear{Vogt}{1976}]{Vogt76C}
        Vogt N., 1976, in Eggleton P., Mitton S., Whelan J., eds, 
        Proc.\ IAU Symp.\ 73, Structure and evolution of close binary systems,
        Reidel, Dordrecht, p.\ 147
\bibitem[\protect\citeauthoryear{Vogt \& Breysacher}{1980}]{Vogt80C} 
        Vogt N., Breysacher J., 1980, ApJ, 235, 945
\bibitem[\protect\citeauthoryear{Vogt et al.}{1980}]{Vogt80aC} 
        Vogt N., Krzeminski W., Sterken C., 1980, A\&A 85, 106
\bibitem[\protect\citeauthoryear{Vogt et al.}{2017}]{Vogt17C} 
        Vogt N., Schreiber M.R., Hambsch F.-J. et al., 2017, PASP, 129, 4201
\bibitem[\protect\citeauthoryear{Voikhanskaya}{1988}]{Voikhanskaya88C} 
        Voikhanskaya N.F., 1988, A\&A, 192, 128
\bibitem[\protect\citeauthoryear{Voikhanskaya}{1996}]{Voikhanskaya96C} 
        Voikhanskaya N.F., 1996, Astron.\ Rep., 40, 674 
\bibitem[\protect\citeauthoryear{Voikhanskaya \& Nazarenko}{1983}]
        {Voikhanskaya83C} 
        Voikhanskaya N.F., Nazarenko I.I., 1983, Astron.\ Zh. 62, 81
\bibitem[\protect\citeauthoryear{Volkov et al.}{1986}]{Volkov86C}
        Volkov I.M., Shugarov S.Yu., Seregina T.M., 1986, 
        Astr.\ Tsirk., 1418, 3
\bibitem[\protect\citeauthoryear{Voloshina \& Khruzina}{2000}]{Voloshina00C}
        Voloshina I.B., Khruzina T.S., 2000, Astron.\ Rep., 44, 89
\bibitem[\protect\citeauthoryear{Wada et al.}{2018}]{Wada18C} 
        Wada Y., Yuasa T., Nakazawam K. et al., 2018, MNRAS, 474, 1564
\bibitem[\protect\citeauthoryear{Wade \& Horne}{1988}]{Wade88C} 
        Wade R.A., Horne K., 1988, ApJ 324, 411
\bibitem[\protect\citeauthoryear{Walker}{1977}]{Walker77C}
        Walker A.R., 1977, MNRAS, 179, 587
\bibitem[\protect\citeauthoryear{Walker}{1954}]{Walker54C}
        Walker M.F., 1954, PASP, 66, 71
\bibitem[\protect\citeauthoryear{Walker}{1956}]{Walker56C} 
        Walker M.F., 1956, ApJ, 123, 68
\bibitem[\protect\citeauthoryear{Walker}{1957}]{Walker57C}
        Walker M.F., 1957, in Herbig G.H., ed., Proc.\ IAU Symp.\ 3, 
        Non Stable Stars, p.\ 46
\bibitem[\protect\citeauthoryear{Walker}{1963}]{Walker63C}
        Walker M.F., 1963, ApJ, 138, 313
\bibitem[\protect\citeauthoryear{Walker}{1965}]{Walker65C} 
        Walker M.F., 1965, Comm.\ Konkoly Obs., No. 57, p.\ 1
\bibitem[\protect\citeauthoryear{Walker et al.}{1976}]{Walker76C} 
        Walker W.S.G., Marino B.F., Freeth G., 1976, IBVS, 1185
\bibitem[\protect\citeauthoryear{Warner}{1973}]{Warner73C} 
        Warner B., 1973, MNRAS, 163, 25p
\bibitem[\protect\citeauthoryear{Warner}{1974}]{Warner74C} 
        Warner B., 1974, MNRAS, 168, 235
\bibitem[\protect\citeauthoryear{Warner}{1975a}]{Warner75aC} 
        Warner B., 1975a, MNRAS, 170, 219
\bibitem[\protect\citeauthoryear{Warner}{1975b}]{Warner75bC} 
        Warner B., 1975b, MNRAS, 173, 37p
\bibitem[\protect\citeauthoryear{Warner}{1976a}]{Warner76aC} 
        Warner B., 1976a, in Eggleton P., Mitton S., Whelan J., eds,
        Structure and Evolution of Close Binary Systems, Reidel,
        Dordrecht, p.\ 85
\bibitem[\protect\citeauthoryear{Warner}{1976b}]{Warner76bC}
        Warner B., 1976b, in Fitch W.S., ed., Proc.\ IAU Coll.\ No.\ 29,
        Multiple periodic variable stars,  Budapest, p.\ 247
\bibitem[\protect\citeauthoryear{Warner}{1981}]{Warner81C} 
        Warner B., 1981, MNRAS, 195, 101
\bibitem[\protect\citeauthoryear{Warner}{1982}]{Warner82C} 
        Warner B., 1982, in Livio M., Shaviv G., eds, Proc.\ IAU Coll.\ 72,
        Cataclysmic Variables and Related Objects,  p.\ 155 
\bibitem[\protect\citeauthoryear{Warner}{1985}]{Warner85C} 
        Warner B., 1985, MNRAS, 217, 1p
\bibitem[\protect\citeauthoryear{Warner}{1986}]{Warner86C} 
        Warner B., 1986, MNRAS, 219, 751,
\bibitem[\protect\citeauthoryear{Warner}{1987}]{Warner87C} 
        Warner B., 1987, MNRAS, 227, 23
\bibitem[\protect\citeauthoryear{Warner}{1995}]{Warner95C} 
        Warner B., 1995, Cataclysmic Variable Stars, Cambridge University 
        Press, Cambridge
\bibitem[\protect\citeauthoryear{Warner \& Brickhill}{1978}]{Warner78C} 
        Warner B., Brickhill A.J., 1978, MNRAS 182, 777 
\bibitem[\protect\citeauthoryear{Warner \& Cropper}{1983}]{Warner83C}
        Warner B., Cropper M., 1983, MNRAS, 203, 909
\bibitem[\protect\citeauthoryear{Warner et al.}{1995}]{Warner95aC}
        Warner B., Martinez P., O'Donoghue D., 1995, ApSS, 226, 27
\bibitem[\protect\citeauthoryear{Warner \& Nather}{1972}]{Warner72C} 
        Warner B., Nather R.E., 1972, MNRAS, 159, 429
\bibitem[\protect\citeauthoryear{Warner \& Nather}{1988}]{Warner88aC}
        Warner B., Nather R.E., 1988, IBVS, 3140 
\bibitem[\protect\citeauthoryear{Warner \& O'Donoghue}{1987}]{Warner87aC}
        Warner B., O'Donoghue D., 1987, MNRAS, 224, 733
\bibitem[\protect\citeauthoryear{Warner \& O'Donoghue}{1988}]{Warner88C} 
        Warner B., O'Donoghue D., 1988, MNRAS, 233, 705
\bibitem[\protect\citeauthoryear{Warner et al.}{1985}]{Warner85aC} 
        Warner B., O'Donoghue D., Allen S., 1985, MNRAS, 212, 9p
\bibitem[\protect\citeauthoryear{Warner et al.}{1989}]{Warner89C} 
        Warner B., O'Donoghue D., Wargau W., 1989, MNRAS, 238, 73 
\bibitem[\protect\citeauthoryear{Warner \& Woudt}{2006}]{Warner06C} 
        Warner B., Woudt P.A., 2006, MNRAS, 367, 1562
\bibitem[\protect\citeauthoryear{Warner et al.}{2003}]{Warner03C} 
        Warner B., Woudt P.A., Pretorius M.L., 2003, MNRAS, 344, 1193
\bibitem[\protect\citeauthoryear{Watanabe}{1999}]{Watanabe99C}
        Watanabe T., 1999, SVOLJ Var.\ Star Bull., 34, 3
\bibitem[\protect\citeauthoryear{Watson et al.}{2003}]{Watson03C} 
        Watson C.A., Dhillon V.S.,Rutten R.G.M., Schwope A.D., 2003, 
        MNRAS, 341, 129
\bibitem[\protect\citeauthoryear{Watson et al.}{2007}]{Watson07C} 
        Watson C.A., Steeghs D., Shahbaz T., Dhillon V., 2007, 
        MNRAS, 282, 1105
\bibitem[\protect\citeauthoryear{Watson et al.}{1985}]{Watson85C} 
        Watson M.G., King A.R., Osborne J., 1985, MNRAS, 212, 917
\bibitem[\protect\citeauthoryear{Webbink}{1978}]{Webbink78C} 
        Webbink R.F., 1978, PASP, 90, 57
\bibitem[\protect\citeauthoryear{Weight et al.}{1994}]{Weight94C} 
        Weight A., Evans A., Naylor T., Wood J.H., Bode M.F., 1994,
        MNRAS, 266, 761
\bibitem[\protect\citeauthoryear{Weil et al.}{2018}]{Weil18C} 
        Weil K.E., Thorstensen J.R., Haberl F., 2018, AJ, 156, 231
\bibitem[\protect\citeauthoryear{Welsh et al.}{2007}]{Welsh07C} 
        Welsh F.W., Froning C.S., Marsh T.R. et al., 2007, in Kang Y.W.
        et al., eds, ASP Conf.\ Series, 362, The Seventh Pacific Rim 
        Conference on Stellar Astrophysics, p.\ 241
\bibitem[\protect\citeauthoryear{Wenzel}{1980}]{Wenzel80C} 
        Wenzel W., 1980, IBVS, 1810
\bibitem[\protect\citeauthoryear{White \& Honeycutt}{1993}]{White93C} 
        White J.C., Honeycutt R.K., 1993, ApJ, 412, 278
\bibitem[\protect\citeauthoryear{Wickramasinghe et al.}{1991}]
        {Wickramasinghe91C}
        Wickramasinghe D.T., Bailey J., Meggitt S.M.A. et al., 1991,
        MNRAS 251, 28
\bibitem[\protect\citeauthoryear{Williams \& Hiltner}{1984}]{Williams84C} 
        Williams G., Hiltner W.A., 1984, MNRAS, 211, 629
\bibitem[\protect\citeauthoryear{Williger et al.}{1988}]{Williger88C} 
        Williger G., Berriman G., Wade R.A., Hassall B.J.M., 1988, 
        ApJ, 333, 277
\bibitem[\protect\citeauthoryear{Wolfe et al.}{2013}]{Wolfe13C} 
        Wolfe A., Sion E.M., Bond H.E., 2013, AJ, 145, 168
\bibitem[\protect\citeauthoryear{Wolff et al.}{1999}]{Wolff99C} 
        Wolff M.T., Wood K.S., Imamura J.N., Middleditch J., \&
        Steiman-Cameron T.Y., 1999, ApJ, 526, 435
\bibitem[\protect\citeauthoryear{Stephenson et al.}{1968}]{Stephenson68C} 
        Stephenson C.B., Sanduleak N., Schild R.E., 1968, ApL, 1, 247
\bibitem[\protect\citeauthoryear{Wood et al.}{1986}]{Wood86C} 
        Wood J.H., Horne L., Berriman G. et al., 1986, MNRAS, 219, 629
\bibitem[\protect\citeauthoryear{Wood et al.}{1989}]{Wood89C} 
        Wood J.H., Horne K., Berriman G., Wade R.A., 1989, ApJ, 341, 974
\bibitem[\protect\citeauthoryear{Wood et al.}{2005}]{Wood05C} 
        Wood M.A., Robertson J.R., Simpson J.C. et al., 2005, ApJ, 634, 570
\bibitem[\protect\citeauthoryear{Would \& Warner}{2002}]{Woudt02C} 
        Woudt P., Warner B., 2002, ApSS, 282, 433
\bibitem[\protect\citeauthoryear{Woudt}{2010}]{Woudt10C} 
        Woudt P.A., Warner B., O'Donoghue D. et al., 2010, MNRAS, 401, 500
\bibitem[\protect\citeauthoryear{Wu et al.}{1989}]{Wu89C} 
        Wu C.C., Panek R.H., Holm A.V., 1989, ApJ, 339, 443 
\bibitem[\protect\citeauthoryear{Wu et al.}{1995}]{Wu95C}
        Wu K., Chanmugam G., Shaviv G., 1995, ApJ, 455, 260 
\bibitem[\protect\citeauthoryear{Wu et al.}{2002}]{Wu02C} 
        Wu X., Li Z., Ding Y., Zhang Z., Li Z., 2002, ApJ, 569, 418
\bibitem[\protect\citeauthoryear{Young et al.}{1981a}]{Young81aC} 
        Young P., Schneider D.P., Shectman S.A., 1981a, ApJ, 244, 259
\bibitem[\protect\citeauthoryear{Young et al.}{1981}]{Young81bC} 
        Young P., Schneider D.P., Shectman S.A., 1981b, ApJ, 244, 259
\bibitem[\protect\citeauthoryear{Zamanov et al.}{2004}]{Zamanov04C}
        Zamanov R.K., Bode M.F., Stanishev V., Mart\i{\i}, J., 2004, 
        MNRAS, 350, 1477
\bibitem[\protect\citeauthoryear{Zamanov et al.}{2010}]{Zamanov10C}
        Zamanov R.K., Boeva S., Bachev R., 2010, MNRAS, 404, 381
\bibitem[\protect\citeauthoryear{Zamanov et al.}{2016}]{Zamanov16C} 
        Zamanov R.K., Boeva S., Latev G. et al., 2016, MNRAS, 457, L10
\bibitem[\protect\citeauthoryear{Zamanov \& Bruch}{1998}]{Zamanov98C}
        Zamanov R.K. Bruch A., 1998, A\&A, 338, 988 
\bibitem[\protect\citeauthoryear{Zamanov et al.}{2015}]{Zamanov15C}
        Zamanov R.K., Latev G., Boeva S. et al., 2015, MNRAS, 450, 3958
\bibitem[\protect\citeauthoryear{Zhang \& Robinson}{1987}]{Zhang87C} 
        Zhang E.-H., Robinson E.L., 1987, ApJ 321, 813 
\bibitem[\protect\citeauthoryear{Zhang et al.}{1991}]{Zhang91C} 
        Zhang E.-H., Robinson E.L., Ramseyer T.F., Shetrone M.D. 
        Stiening R.F., 1991, ApJ, 381, 534
\bibitem[\protect\citeauthoryear{Zuckemann}{1961}]{Zuckermann61C}
        Zuckermann M.C., 1961, Ann.\ Astroph., 24, 431
\bibitem[\protect\citeauthoryear{Zwitter \& Munari}{1994}]{Zwitter94C} 
        Zwitter T., Munari U., 1994, A\&AS, 107, 503
\bibitem[\protect\citeauthoryear{Zwitter \& Munari}{1995}]{Zwitter95C} 
        Zwitter T., Munari U., 1995, A\&AS, 114, 575

\end{thebibliography}

\begin{thebibliography}{99}
\bibitem[\protect\citeauthoryear{Frank et al.}{2002}]{Frank02D} 
        Frank J., King A.R., Raine D., 2002, {\it Accretion 
        Power in Astrophysics, Third Edition}, Cambridge University Press
\bibitem[\protect\citeauthoryear{Nauenberg}{1972}]{Nauenberg72D} 
        Nauenberg M., 1972, ApJ, 175, 417
\bibitem[\protect\citeauthoryear{Shakura \& Sunyaev}{1973}]{Shakura73D} 
        Shakura N.I., Sunyaev R.A., 1973, A\&A, 24, 337
\end{thebibliography}

\begin{thebibliography}{99}
\bibitem[\protect\citeauthoryear{Bruch}{2014}]{Bruch14A} 
        Bruch A., 2014, A\&A, 566, A101
\bibitem[\protect\citeauthoryear{Bruch}{2016}]{Bruch16A} 
        Bruch A., 2016, New Astr., 46, 60
\bibitem[\protect\citeauthoryear{Bruch}{2018}]{Bruch17A} 
        Bruch A., 2017, New Astr., 52, 117
\bibitem[\protect\citeauthoryear{Bruch}{2018}]{Bruch18A} 
        Bruch A., 2018, New Astr., 58, 53
\bibitem[\protect\citeauthoryear{Chiapetti et al.}{1989}]{Chiapetti89A} 
        Chiapetti L., Belloni T., Bonnet-Bidaud J.-M., del Gratta C., 
        de Martino D. et al. 1989, ApJ 342, 493 
\bibitem[\protect\citeauthoryear{Ka\-lu\.zny}{1989}]{Kaluzny89A} 
        Ka\-lu\.zny J., 1989, Acta Astron., 39, 235
\bibitem[\protect\citeauthoryear{Leibowitz et al.}{1994}]{Leibowitz94A} 
        Leibowitz E.M., Mendelson H., Bruch A. et al. 1994, 421, 771
\bibitem[\protect\citeauthoryear{O'Donoghue}{1987}]{ODonoghue87A} 
        O'Donoghue, D., 1987, ApSS, 136, 297
\bibitem[\protect\citeauthoryear{Patterson}{1979}]{Patterson79A} 
        Patterson J., 1979, AJ, 84, 804
\bibitem[\protect\citeauthoryear{Patterson}{1980}]{Patterson80A} 
        Patterson J., 1980, ApJ, 241, 235
\bibitem[\protect\citeauthoryear{Patterson}{1981}]{Patterson81A} 
        Patterson J., 1981, ApJ Suppl., 45, 517
\bibitem[\protect\citeauthoryear{Patterson et al.}{1981}]{Patterson81Ab} 
        Patterson J., McGraw J.T., Coleman L., Africano J.L., 1981,
        ApJ 248, 1067
\bibitem[\protect\citeauthoryear{Warner}{1974}]{Warner74A} 
        Warner B., 1974, MNRAS, 168, 235
\bibitem[\protect\citeauthoryear{Warner}{1975}]{Warner75A} 
        Warner B., 1975, MNRAS, 170, 219
\bibitem[\protect\citeauthoryear{Warner}{1985}]{Warner85A} 
        Warner B., 1985, MNRAS, 217, 1p
\bibitem[\protect\citeauthoryear{Warner}{1986}]{Warner86A} 
        Warner B., 1986, MNRAS 219, 751,
\bibitem[\protect\citeauthoryear{Warner \& O'Donoghue}{1987}]{Warner87A} 
        Warner B., O'Donoghue D., 1987, MNRAS, 224, 733
\bibitem[\protect\citeauthoryear{Warner \& O'Donoghue}{1988}]{Warner88A} 
        Warner B., O'Donoghue D., 1988, MNRAS, 233, 705
\bibitem[\protect\citeauthoryear{Warner et al.}{1981}]{Warner81A} 
        Warner B., O'Donoghue D., Fairall A.P., 1981, MNRAS, 196, 705 
\bibitem[\protect\citeauthoryear{Warner et al.}{1989}]{Warner89A} 
        Warner B., O'Donoghue D., Wargau W., 1989, MNRAS, 238, 73 
\bibitem[\protect\citeauthoryear{Zamanov \& Bruch}{1998}]{Zamanov98A} 
        Zamanov R.K., Bruch A., 1998, A\&A, 338, 988 
\end{thebibliography}
\end{document}